\documentclass{article}
\usepackage[utf8]{inputenc}
\usepackage{comment}

\pdfoutput=1
\usepackage{amssymb}
\usepackage{amsmath}
\usepackage[dvips]{graphicx}
\usepackage{setspace}
\usepackage{slashed}
\usepackage{mathtools}
\usepackage{amsfonts}
\usepackage{fancyhdr}
\usepackage{xcolor}
\usepackage{physics}
\usepackage{graphicx}
\usepackage{rotating}
\usepackage{comment}
\usepackage{color}
\usepackage{subcaption}
\usepackage[percent]{overpic}
\usepackage{cite}
\usepackage{physics}
\usepackage{moresize}
\usepackage{dsfont}
\usepackage{dsdshorthand}
\usepackage{bbm}
\usepackage{booktabs}

\definecolor{darkgreen}{rgb}{0,0.5,0}
\definecolor{darkblue}{rgb}{0,0,0.6}
\definecolor{purple}{rgb}{0.4,.2,0.7}

\usepackage[colorlinks=true,citecolor=darkgreen,linkcolor=black,urlcolor=purple]{hyperref}

\usepackage{pdfsync}

\makeatletter
\newcommand*{\defeq}{\mathrel{\rlap{%
                     \raisebox{0.3ex}{$\m@th\cdot$}}%
                     \raisebox{-0.3ex}{$\m@th\cdot$}}%
                     =} 
\makeatother

\def\be{\begin{eqnarray}}
\def\ee{\end{eqnarray}}

\newcommand{\bea}{\begin{eqnarray}}
\newcommand{\eea}{\end{eqnarray}}
\def\ben{\begin{equation}}
\def\een{\end{equation}}

 \let\m=\mu   \let\p=\phi \let\r=v

\def\nn{\nonumber}

\def\be{\begin{equation}}
\def\ee{\end{equation}}
\def\ba{\begin{eqnarray}}
\def\ea{\end{eqnarray}}

\def\bal#1\eal{\begin{align}#1\end{align}}
\def\bs#1\es{\begin{split}#1\end{split}}

\newcommand\jeevan[1]{\textcolor{blue}{{\it #1 -Jeevan}}}
\newcommand{\aia}[1]{\textcolor{teal}{[#1 -AA ]}}
\renewcommand{\p}{\partial}

\allowdisplaybreaks  

\numberwithin{equation}{section}

\def\bx{\bar{X}}

\def\m{{\mu}}

\def\ep{{\epsilon}}

\def\p{{\phi}}

\def\be{\begin{equation}}
\def\ee{\end{equation}}
\def\ba{\begin{eqnarray}}
\def\ea{\end{eqnarray}}
\def\bal#1\eal{\begin{align}#1\end{align}}

\def\r{\rightarrow}

\def\r{\right}
\def\ep{\epsilon}

\def\lr{\leftrightarrow}

\def\br{\rbrace}

\usepackage{tikz}
\usetikzlibrary{positioning,arrows}
\usetikzlibrary{decorations.pathmorphing}
\usetikzlibrary{decorations.markings}

\usetikzlibrary{%
  arrows,
  knots,
  calc,
}

\tikzset{
  every path/.style={
    red,
    line width=2pt
  },
  every node/.style={
    transform shape,
    knot crossing,
    inner sep=1.5pt
  }
}

\def\YW#1{{\color{red!80} [YW: #1]}}
 
\def\ie{\begin{equation}\begin{aligned}}
\def\fe{\end{aligned}\end{equation}}

\usepackage{multirow}

\def \be {\begin{equation}}
\def \ee {\end{equation}}

\renewcommand{\p}{\partial}
\renewcommand{\min}{{\rm min}}

\newcommand{\B}{\mathcal{B}}

\newcommand{\id}{\mathbf{1}}

\newcommand{\kett}[1]{
    \ket{#1}\rangle
}
\newcommand{\inlinefig}[2][10]{
    \raisebox{-0.5\totalheight}{\includegraphics[height=#1\fontcharht\font`\B]{#2}}
}
\usepackage{framed}

\begin{document}

\onehalfspacing


\begin{center}

~
\vskip5mm


{ {\LARGE
Line Defects in Liouville Conformal Field Theory: }
\\
\large Localized Cosmological Constants and Decohered Hyperbolic Geometries
}



\title{Line Defects in Liouville CFT:\ Localized Cosmological Constants and Decohered Hyperbolic~Geometries}

\vskip7mm
Ahmed I. Abdalla$^1$, Jeevan Chandra$^1$, and Yifan Wang$^2$

\vskip5mm

{[1] \textit{Leinweber Institute for Theoretical Physics and Department of Physics, \\ University of California, Berkeley, USA
}
\\
{[2] \textit{Center for Cosmology and Particle Physics, New York University, New York, NY 10003, USA}
}}

\vskip5mm

{\tt aiabdalla@berkeley.edu, jcn1998@berkeley.edu, yifan.wang@nyu.edu}

\end{center}

\vspace{2mm}

\begin{abstract}

The study of quantum impurities has long been a central and inspiring theme in quantum many-body physics. Localized impurities are modeled by line defects in quantum field theory. We describe a line defect in Liouville CFT realized as a ``localized cosmological constant'': a non-topological line insertion into the Liouville path integral that is tractable at both weak and strong defect coupling. At weak coupling, we analyze the defect perturbatively and characterize it through its correlations with local operators, energy and information transport, the Casimir energies associated with fusion, and corrections to the open string channel spectrum. We also study the effect of a cuspidal deformation of the defect locus on these observables and describe novel monotonicity properties as the cusp angle is varied. These results derived using perturbation theory are more generally applicable to pinning defects constructed from scalar primary operators in compact $2d$ CFTs. At strong coupling, in a semiclassical limit, the defect admits a geometric interpretation in terms of a discontinuity in the extrinsic curvature of the $1d$ defect locus embedded in $2d$ hyperbolic geometries. The observables characterizing the defect in this regime are computed by gluing hyperbolic surfaces across the defect, and are compared with the corresponding weak coupling results. The correlations across the defect, both at weak and strong coupling, can also be realized by an effective ``decohered FZZT interface'' constructed by diagonal gluing of two copies of the fixed-length FZZT boundary state. These line defects also have
interesting interpretations in other models, in terms of end-of-the-world branes in Jackiw-Teitelboim gravity,  dust shells in AdS$_3$ gravity, and interfaces with a proliferation of non-abelian Wilson loops in $4d$ $\mathcal{N}=2$ gauge theories.

\end{abstract}

\pagebreak

\tableofcontents

\section{Introduction} \label{secIntro}

The study of quantum impurities has long been a central and inspiring theme in quantum many-body physics (see, e.g., \cite{Affleck:1995ge,Saleur:1998hq,Andrei:2018die} for a review). Within the framework of quantum field theory (QFT), such objects are naturally formalized as defects, also referred to as extended operators, which are supported on submanifolds of spacetime.
It has become increasingly clear that defects play an essential role in examining both the symmetry content and the phase structure of a bulk QFT \cite{Gaiotto:2014kfa}. This is made possible by their nontrivial dynamics, which effectively place the bulk degrees of freedom in backgrounds with singularities and/or nontrivial topology, thereby inducing responses that are invisible to purely local operator insertions. As a result, a substantial body of recent work has focused on elucidating these dynamical signatures of defects.

One prominent example concerns the renormalization group (RG) flows of defect couplings, which obey monotonicity properties closely analogous to those governing bulk RG flows \cite{Affleck:1991tk,Friedan:2003yc,Jensen:2015swa,Casini:2016fgb,Casini:2018nym,Kobayashi:2018lil,Wang:2020xkc,Wang:2021mdq,Cuomo:2021rkm,Shachar:2022fqk,Casini:2022bsu,Casini:2023kyj}. Such defect monotonicity theorems provide universal and powerful constraints on the long-distance behavior of defects and, through them, yield valuable insight into the physical properties of the bulk theory.
Moreover, the study of defect fusion and, relatedly, cusped defects reveals new universal structures, such as defect-induced Casimir energies (and cusp anomalous dimensions), that are highly sensitive to bulk dynamics \cite{Bachas:2007td,Bachas:2013ora,Konechny:2015wla,Soderberg:2021kne,Bissi:2022bgu,SoderbergRousu:2023zyj,Diatlyk:2024defect,Kravchuk:2024qoh,Cuomo:2024cusp,Giombi:2025cuspFermions}. Finally, interfaces between distinct QFTs offer a complementary perspective on bulk phase transitions: universally defined interface observables, including the energy transmission coefficient \cite{Bachas:2001vj,Quella:2006de} and interface entanglement entropy \cite{Calabrese:2009qy}, are  beginning to be used as quantitative probes of the bulk phase diagram (see e.g. \cite{
	Gaiotto:2012np,Brehm:2015lja,Gliozzi:2015qsa,Meineri:2019ycm,Konechny:2020jym,Karch:2021qhd,Karch:2023evr,Karch:2024udk,Giombi:2024qbm,
	Komatsu:2025cai}).

Conformal Field Theory (CFT) is a particularly interesting setting for studying defect dynamics. The axiomatic bootstrap approach to CFT has a natural extension in the presence of defects that preserve a nontrivial conformal subgroup and obey locality constraints (i.e. consistency of cutting and gluing observables with defect insertions) \cite{Cardy:1989ir,Lewellen:1991tb,Liendo:2012hy,Gaiotto:2013nva,Billo:2016cpy}. This bootstrap approach provides a non-perturbative method to constrain and, in favorable cases, solve defect dynamics. Such conformal defects describe the fixed points of RG flows of defect couplings, and their correlation functions with bulk local operators capture dynamical signatures of the defect.
In $d=2$, the relevant bootstrap equations include the familiar Cardy condition (also known as open-closed duality) for the annulus partition function with conformal boundary conditions \cite{Cardy:1989ir}, as well as the Cardy-Lewellen equations \cite{Lewellen:1991tb} that impose associativity of the bulk operator algebra in the presence of a conformal defect. These equations have been explored to yield the complete classification of boundary conditions in minimal models \cite{Cardy:1989ir} and stringent constraints on boundaries in more general settings \cite{Friedan:2012jk,Collier:2021ngi,Meineri:2025qwz}.
However, exact results on defects that are not boundaries are rare even in $d=2$. For example, even though topological line defects in minimal models are classified \cite{Petkova:2000ip}, general conformal line defects are only classified in the Ising CFT \cite{Oshikawa:1996ur, Oshikawa:1997np}. By the folding trick, conformal lines are equivalent to conformal boundaries of the folded CFT with doubled central charge $c_{\rm fold}=2c$. The obvious obstruction then has to do with the Virasoro representation theory becoming much more complicated at $c>\frac{1}{2}$.

There is an elementary yet powerful way to produce nontrivial defects in CFT, for which many exact properties have been deduced in various cases. This is the pinning field construction, first introduced in the condensed matter literature \cite{Allais:2014wca, ParisenToldin:2016nsz}, and recently studied in field theories in various dimensions \cite{Cuomo:2021kfm, Rodriguez-Gomez:2022gif,Giombi:2022vnz,GimenezGrau:2022hqc,Popov:2022nfq,Pannell:2023pwz}. In a slightly generalized setting (compared to the original definition in \cite{ParisenToldin:2016nsz}), pinning defects are constructed by perturbing the theory with bulk local operators supported on a subspacetime \cite{Popov:2025cha}. The most basic setup is a line defect defined using a bulk operator $\cO$ integrated along the (Euclidean) time $t$ at $\vec x=0$,
\ie
S_{\rm defect}=S_{\rm CFT} + h\int dt \, \cO(t,\vec x=0)\,,
\fe
where $h$ is the pinning field (also known as a localized magnetic field). For $\cO$ of scaling dimension $\Delta<1$, such a deformation is guaranteed to produce a conformal line defect in the IR that is not screened, as a consequence of the $g$-theorem, where $
\log g$ is also known as the defect entropy \cite{Affleck:1991tk,Friedan:2003yc,Casini:2016fgb,Casini:2018nym,Cuomo:2021rkm,Casini:2022bsu}.\footnote{In non-generic situations (such as in the free scalar theory), the flow may be runaway and does not terminate at a conformal defect \cite{Raviv-Moshe:2023yvq, Cuomo:2022xgw}.} Furthermore, for codimension-one pinning defects (interfaces), an IR factorization property was established in \cite{Popov:2025cha}, which states that the interface at long distances is described by factorized boundary conditions for the CFT, possibly dressed by emergent topological and gapless interface modes. In $d=2$, this factorization property, together with the $g$-theorem and symmetry constraints, immediately leads to exact solutions for numerous pinning defects in minimal models \cite{Popov:2025cha}.

In this work, we build on these previous developments to study line defects  that are not necessarily conformal  in the $d=2$ Liouville CFT of central charge $c=1+6Q^2$, defined by the following action for the Liouville field $\phi$ on flat spacetime,
\ie
S_{\rm Liouville} = \frac{1}{\pi}\int d^2z \, \left( \partial\phi\bar\partial \phi + \pi\mu_{\rm bulk} e^{2b\phi}\right)\,,
\label{liouvilleaction}
\fe
with $Q\equiv b+\frac{1}{b}$ and cosmological constant $\mu_{\rm bulk}$. Our investigation is heavily motivated by the special status of Liouville theory among $2d$ CFTs, as the best-known \textit{irrational} CFT, which is unique and universal in many respects (see \cite{Seiberg:1990eb,Teschner:2001rv,Nakayama:2004vk,rhodes2025decadesprobabilisticapproachliouville} for reviews). One ambitious goal is that by studying defects in this theory, one can uncover new universal patterns among defect observables in general $2d$ CFTs and we will comment in the discussion on how the results of this paper may be used toward this goal.

The Liouville CFT, on the one hand, does not possess a normalizable conformally invariant ground state or a discrete operator spectrum, unlike ordinary \textit{compact} CFTs. On the other hand, basic structure constants, such as the DOZZ formula for three-point functions \cite{Dorn:1994xn, Zamolodchikov:1995aa}, are known to govern universal averaged properties of heavy operators in generic compact CFTs \cite{Chang:2016ftb,Collier:2017shs,Collier:2018exn}.
More generally, Liouville CFT is expected to provide a universal ``effective'' description of the random statistics of heavy operators in $2d$ CFTs \cite{Collier:2019weq} and, via holography, of semi-classical geometries (e.g.\ wormholes) in AdS$_3$ \cite{Chandra:2022bqq, Chandra:2024vhm, Abajian:2023bqv}. This connection has been sharpened recently through formulations of pure gravity in AdS$_3$ in terms of Virasoro TQFT \cite{Collier:2023fwi} and the related conformal Turaev-Viro theory \cite{Hartman:2025cyj}, borrowing insights from the better understood setting of finite TQFTs and rational CFTs \cite{Moore:1988qv}.
Furthermore, Liouville CFT also plays a crucial role in the formulation of non-critical string theory on the worldsheet and, relatedly, Liouville quantum gravity as a toy model for $2d$ gravity \cite{DHoker:1990prw,Klebanov:1991qa,Ginsparg:1993is,DiFrancesco:1993cyw,Seiberg:2004at}  that implements a ``quantum deformation" of the Jackiw--Teitelboim (JT) gravity (see e.g. \cite{ Mertens:2017mtv, Mertens:2018fds,Mertens:2020hbs} for recent works). This is related to the $3d$ gravity context mentioned above via dimensional reduction \cite{Brown:1986, Coussaert:1995} (see \cite{Donnay:2016iyk} for a modern review). Here the Liouville field captures the quantum hyperbolic geometry of Riemann surfaces by,
\ie 
ds^2=e^{2b\phi} dz d\bar z\,.
\label{liouvillegeometry}
\fe
More precisely, the conformal blocks of Liouville CFT form a Hilbert space that describes the quantization of the Teichm\"uller space of Riemann surfaces \cite{Verlinde:1989ua,Teschner:2002vx}. Furthermore, the connection between Liouville CFT and random $2d$ geometry has led to a mathematically rigorous construction of the Liouville path integral in terms of random measures arising from Gaussian Multiplicative Chaos (GMC) in probability theory \cite{David:2014aha,Huang:2018ncv}. Concretely, GMC provides a renormalized definition of the Liouville interaction term $e^{2b\phi} d^2 z$ as a random area measure associated with the Gaussian free field. 
This probabilistic framework has subsequently enabled rigorous derivations of a variety of Liouville observables and has clarified several previously mysterious aspects of the theory (see \cite{Chatterjee:2024phq} for a recent review).

Conformal boundary conditions of the Liouville CFT have been classified long ago into Dirichlet-type and (mixed) Neumann-type, also known as ZZ \cite{Zamolodchikov:2001ah} and FZZT boundaries \cite{Fateev:2000ik,Teschner:2000md}, respectively. In particular, the FZZT boundary comes with an exactly marginal parameter $\m_B$, known as the boundary cosmological constant (to be compared with $\m_{\rm bulk}$ in the bulk Liouville action). However, very little is known about defects in this theory beyond the topological defects, which correspond to Verlinde loops \cite{Verlinde:1988sn}. These are again classified into ZZ and FZZT types \cite{Drukker:2010jp} and play a crucial role in the quantization of the Teichm\"uller space \cite{Teschner:2002vx}. Moreover, all of the FZZT and ZZ boundary conditions can be obtained by fusing these topological defects with the identity ZZ brane $|{\rm ZZ}_{(1,1)}\rangle$ \cite{Drukker:2010jp}.

Motivated by the success in the study of pinning defects and the connection between Liouville CFT and random geometry, 
the first Liouville defect we study is constructed by a pinning field for the operator $e^{b\phi}$ on a line, 
\ie
S_{\rm defect}=S_{\rm Liouville} + \m_D\int_{x=0} dt\,  e^{b\phi} 
\,.
\label{liouvilledefect}
\fe
Noting the similarity to the boundary action describing the FZZT brane, we will refer to $\mu_D$ as the \textit{defect cosmological constant} (localized on the line).
Correspondingly, we refer to the defect defined by \eqref{liouvilledefect} as the \emph{localized cosmological constant} defect.
Furthermore, as is already visible from the $2d$ geometry described by the Liouville field in \eqref{liouvillegeometry}, the deformation term in \eqref{liouvilledefect} is the (proper) length operator $\ell\equiv \int_{x=0} dt\,  e^{b\phi} $. In the FZZT context, this is the boundary length operator introduced in \cite{Fateev:2000ik}. 
This connection leads to an elegant geometric interpretation of our defect in Liouville CFT: the defect induces a random length measure on the underlying random surface encoded by the Liouville field. In probabilistic terms, it modifies the Liouville path integral by a Gaussian multiplicative chaos length measure associated with $e^{b\phi}dt$. The probabilistic approach also gives a natural definition of the renormalized localized cosmological constant defect nonperturbatively. 

Moreover, when this defect is fused transversely with an FZZT boundary condition of boundary cosmological constant $\mu_B$, we expect, after appropriate renormalization in the fusion limit \cite{Bachas:2007td,Bachas:2013ora,Konechny:2015wla,Soderberg:2021kne,Bissi:2022bgu,SoderbergRousu:2023zyj,Diatlyk:2024defect,Kravchuk:2024qoh,Cuomo:2024cusp}, that the boundary coupling is shifted as $\m_B\to \m_B+\m_D$.\footnote{The bulk operator $e^{b\phi}$, upon renormalization and taking the limit in which it approaches the FZZT boundary, reduces to the exactly marginal boundary operator. See \cite{Graham:2006gca} and \cite{Prochazka:2019fah} for detailed discussions.}
Naively one might anticipate that the topological FZZT Verlinde loop would implement such a shift via fusion with the FZZT boundary. However, this topological construction suffers from divergences stemming from the definition of the FZZT defect itself \cite{Sarkissian:2009aa}. 
By contrast, the non-topological defect~\eqref{liouvilledefect} provides a finite realization of this operation (after renormalization) by transporting the FZZT brane along its boundary conformal modulus $\mu_B$.

While these are the main reasons we focus on this particular pinning defect in the Liouville CFT, there are obvious generalizations given by considering  Liouville operators other than $e^{b\phi}$. One possibility is the marginal pinning deformation by $e^{2\alpha_c\phi}$ with
\ie 
\alpha_c={Q-\sqrt{Q^2-2}\over 2}\,,
\label{marginalalpha}
\fe
which turns out to be exactly marginal perturbatively. Such a defect is indistinguishable from the localized cosmological constant defect \eqref{liouvilledefect} in the leading semiclassical limit $b\to 0$ and so our semiclassical analysis for \eqref{liouvilledefect} would also apply to \eqref{marginalalpha}. Furthermore, we will also take initial steps to studying this defect via conformal perturbation theory which needs to be handled with care due to peculiarities of the Liouville CFT compared to usual compact CFTs.
Another set of possibilities are relevant pinning deformations by $e^{kb\phi}$ with $0<k<1$ which trigger nontrivial defect RG flows even in the $b\to 0$ limit. They give rise to a large zoo of unexplored defects in the Liouville CFT. We will comment on these generalizations in the discussion section.  The space of defects in the Liouville CFT is immensely rich and deserves to be further investigated.

One important difference of our defect \eqref{liouvilledefect}
in contrast to the FZZT boundary is that the cosmological constant
operator $e^{b\phi}$ is not marginal and has dimension $\Delta=1+{b^2\over 2}$, thus slightly irrelevant in the semiclassical limit of the Liouville CFT (i.e. $b\to 0$). Therefore, one naively expects a perturbative UV fixed point in $\m_D$ in the semiclassical regime. However as we will comment on later, this naive conformal perturbation theory analysis fails. Nonetheless, we will solve the defect exactly for arbitrarily large $\mu_D$ in a double-scaling limit. We will see that the ``heavy'' defect backreacts on the Liouville field to create kinks on the otherwise smooth hyperbolic geometry realized by $\phi$ and the location of the kink is determined by the defect cosmological constant $\mu_D$ (suitably rescaled in the semiclassical limit). See Figure~\ref{fig:scsl_limit} for an example. This is reminiscent of the study of pinning defect in the ${\rm O}(N)$ model at large $N$ in \cite{Cuomo:2021kfm} and of the large spin impurities in \cite{Cuomo:2022xgw} where the defect modifies the saddle point of the path integral and semi-classics is used to deduce defect observables. This geometric perspective on the Liouville CFT in the semiclassical limit is very useful to study this conformal defect and its interesting physical properties. For example, one finds that this defect is purely reflective in the semiclassical limit.

The semiclassical results suggest that a full quantum  analog of the defect \eqref{liouvilledefect} maybe described by gluing $2d$ hyperbolic geometries along boundaries of fixed-length $\ell$ weighted by $e^{\tilde \m_D \ell}$ where $\tilde \m_D$ is the renormalized defect cosmological constant. These hyperbolic geometries with fixed-length boundaries at the quantum level are naturally described by the fixed-length boundary conditions $|\ell\rangle$ obtained by a Laplace transform from the usual FZZT boundaries \cite{Fateev:2000ik}.
We therefore arrive at the following conformal factorized defect, which is defined for finite $b$,
\ie 
L(\Tilde{\mu}_D)=2\sqrt{2}\pi b\int_0^{\infty}\frac{d\ell}{\ell}\ e^{\Tilde{\mu}_D \ell}\ketbra{\ell}\,.
\label{decoheredefect}\fe
In general, conformal interfaces can be thought of as density operators in the CFT Hilbert space (after suitable regularization). 
Here \eqref{decoheredefect} represents a decohered mixed state in the length basis, and thus we refer to the (length) \textit{decohered FZZT interface}. This defect is purely reflective since stress tensor two-point functions vanish as a consequence of the factorization into conformal boundaries $|\ell\rangle$.

A natural question is how the two defects introduced in \eqref{liouvilledefect} and \eqref{decoheredefect} are related in the space of defects in the Liouville CFT. We will not have a complete answer to this question, partly due to the lack of non-perturbative understanding of the space of defect RG flows in the Liouville CFT (see \cite{Teschner:2003qk,Graham:2003nw,Konechny:2008tm} for progress made in the context of Liouville boundary RG flows perturbatively and also in \cite{Caetano:2020dyp} using integrability).\footnote{We mention in passing that usual RG analysis in Liouville deformed by bulk operators is already quite subtle, given the continuous spectrum of operators and various IR divergences. For example, it is not understood what is the fate of an RG flow triggered by a general Liouville operator $V_\alpha$. The special case of  deformation by $e^{-2b\phi}$ is known to flow to the massive sinh-Gordon theory, which intuitively follows from the two-sided potential. Although not proven, it is expected that non-compact CFTs obey a version of the $c$-theorem where the monotonic function evaluates to $c_{\rm effective}\equiv c-24h_{\rm min}$ at the fixed points \cite{Hori:2001ax,Adams:2001sv,Harvey:2001wm}. For Liouville CFT, $c_{\rm effective}=1$ always. Therefore, assuming the $c_{\rm effective}$ theorem, RG flows from the Liouville CFT have to either lift its non-compactness (as in the sinh-Gordon case), or 
	exhibit run-away behavior (i.e. never-ending flow).
} Nonetheless, we can perform explicit computations of observables in the Liouville CFT involving these defects and compare their features. We will find that, in the semiclassical limit $b\to 0$, the two defects \eqref{liouvilledefect} and \eqref{decoheredefect} are identical at strong coupling with the simple identification $\m_D=\tilde \m_D$, whereas at weak coupling, they only agree for matrix elements involving operators near the so-called black hole threshold (known as the Schwarzian limit). These findings indicate that the length decohered defect \eqref{decoheredefect} may not exactly describe the putative UV fixed point of the localized cosmological constant defect, but at least provide an ``effective'' description for it at strong defect coupling (and in the Schwarzian sector at weak coupling).\footnote{There is also the logical possibility that the defect defined in \eqref{liouvilledefect} is a ``run-away'' defect and never reaches a fixed point in the UV. Similar phenomena for IR flows were discussed in \cite{Cuomo:2021kfm,Cuomo:2022xgw}.} We will also discuss modifications of \eqref{decoheredefect} that would achieve a matching in the semiclassical limit at weak defect coupling. This will necessarily involve off-diagonal contributions in the length basis and thus the decohered factorization in \eqref{decoheredefect} should be thought of as a strong defect coupling effect. This  is reminiscent of the IR factorization phenomena discussed in \cite{Popov:2025cha} but applied for an irrelevant pinning deformation. 

In addition to its special role among $2d$ CFTs, Liouville CFT also appears in interesting ways in quantum theories in higher dimensions. This includes a Schwarzian quantum mechanics that computes amplitudes in JT gravity, the aforementioned pure AdS$_3$ gravity, and also the miraculous AGT relation of Liouville CFT to $4d$ $\cN=2$ supersymmetric gauge theories with SU$(2)$ gauge group \cite{Alday:2009aq}.\footnote{In all these cases, there is an ${\rm SL}(2,\mathbb{C})$ Chern-Simons theory (and its various cousins) at play and the Hamiltonian reduction or Drinfeld-Sokolov reduction that relates Liouville CFT and the corresponding WZNW model is key to these correspondences  \cite{Verlinde:1989ua,Witten:1989ip,Carlip:1991zm,Harlow:2011ny,Gaiotto:2011nm}.} Given these $2d{-}2d$, $3d{-}
2d$, and $4d{-}2d$ correspondences, it is natural to consider the dual of the defects \eqref{liouvilledefect} and \eqref{decoheredefect} in these various settings. In JT gravity, thermal one-point functions of our defect become disk amplitudes in the presence of end-of-the-world branes. In AdS$_3$ gravity, we find that the Liouville solutions with the defect \eqref{liouvilledefect} describe the geometry of wormholes sourced by dust shells similar to those constructed in \cite{Chandra:2024vhm,Chandra:2022fwi,Sasieta:2022ksu}. In the context of the AGT correspondence, our defect, described in terms of decohered geometries in \eqref{decoheredefect}, corresponds to a symmetry breaking interface with a proliferation of certain non-abelian Wilson loops. 

\subsection{Summary of results}

\begin{figure}
    \centering
    \includegraphics[width=0.7\linewidth]{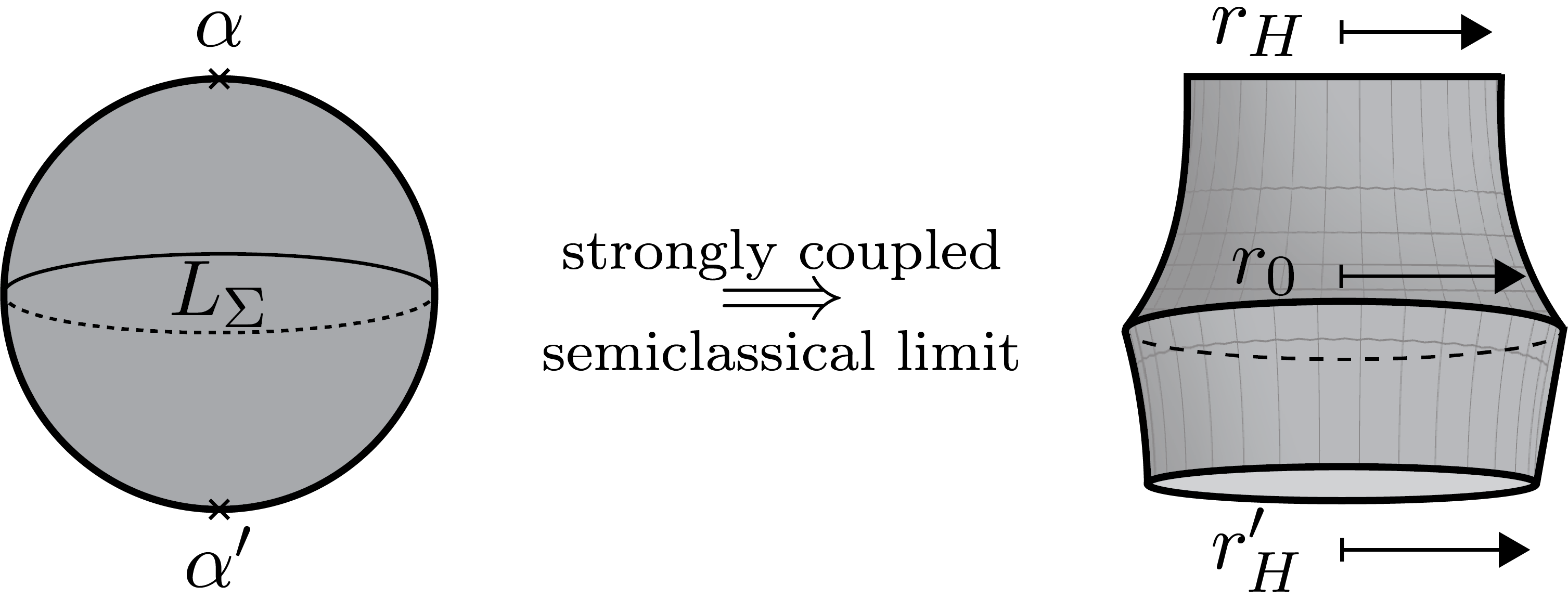}
    \caption{The figure on the left describes the kinematics used to compute the matrix elements of a circular defect between two primary states. The figure on the right shows that in the strongly coupled semiclassical limit, the matrix element can be computed by allowing the local operators and the defect to backreact on the sphere giving a hyperbolic surface that with a kink along the defect. The circular boundaries on the top and bottom are geodesics with proper lengths $2\pi r_H$ and $2\pi r_H'$ determined by the conformal weights of the corresponding operators. Located at the waist in the right figure, $2\pi r_0$ is the proper length (non-geodesic) of the defect determined dynamically in terms of the defect cosmological constant and the two geodesic lengths.}
    \label{fig:scsl_limit}
\end{figure}

Here we provide a more technical summary of the main results in the paper. As mentioned earlier in the general introduction, this paper studies a non-topological line insertion in Liouville CFT that can be viewed as a
\emph{localized cosmological constant} supported on a closed curve $\Sigma$.
Concretely, we define a defect operator by exponentiating an integrated Liouville vertex operator with scaling dimension $\Delta_{b/2}$,
\begin{equation}
	\label{eq:defect_def}
	{\bm L}_\Sigma\;=\;\exp\!\left( \mu_D \int_\Sigma ds\,V_{b/2}(s)\right)\,
	\quad{\rm with}\quad
	\Delta_{b/2}=1+\frac{b^2}{2}\,.
\end{equation}
Here, $b$ is the Liouville coupling.
This defect sources an exactly marginal deformation along the line in the $b\to 0$ limit.
We describe another closely related defect which we refer to as being manifestly conformal since it sources an exactly marginal deformation for any $b$ and fine tuned $\alpha_c$ as in \eqref{alphac},
\begin{equation} \label{eq:exactmarginal}
	{\bm L}_\Sigma\;=\;\exp\!\left( \mu_D \int_\Sigma ds\,V_{\alpha_c}(s)\right)\,
	\quad{\rm with}\quad
	\Delta_{\alpha_c}=1\,,
\end{equation}
which reduces to (\ref{eq:defect_def}) in the $b\to 0$ limit.
A central theme of the paper is that the defects (\ref{eq:defect_def}) and (\ref{eq:exactmarginal}) admit a complementary,
geometric description in a strongly-coupled semiclassical regime, where they can both be analyzed by the same set of hyperbolic Liouville saddles. This provides a concrete bridge between
(i) local CFT diagnostics of transmission and operator mixing across the line at weak coupling,
and (ii) emergent geometry, defect dynamics, and fusion data in the semiclassical limit. As a concrete illustration, the matrix elements across a circular defect between two distinct primary states at weak coupling in the $b\to 0$ limit where the defect becomes conformal are
\begin{equation} \label{matweaksumm}
   \left| \bra{ V_{\frac{Q}{2}+ibk}} {\bm L}_\Sigma \ket{V_{\frac{Q}{2}+ibk'} }\right | \overset{\mu_D \to 0}{=}\mu_D\left(\frac{16\pi^3 kk' \sinh(2\pi k)\sinh(2\pi k')}{\cosh^2(\pi(k+k'))\cosh^2(\pi(k-k'))}\right)^{\frac{1}{2}}\,.
\end{equation}
In the above expression, the bulk cosmological constant is set to $1$.
The conformal weights of the primary states are tuned so that they are close to the edge of the normalizable spectrum. This ensures that both the defect and the external states are ``light". In the complementary strong coupling regime, the result for the matrix elements in the $b\to 0$ limit is given by
\begin{equation} \label{matstrongsumm}
   \left | \bra{V_{\frac{Q}{2}+i\frac{r_H}{2b}}} {\bm L}_\Sigma \ket{V_{\frac{Q}{2}+i\frac{r_H'}{2b}}}  \right |\overset{\mu_D \to \infty}{=}\exp\left[\frac{c}{6}\left(r_H\cos^{-1}\left(\frac{r_H}{r_0}\right)+r'_H\cos^{-1}\left(\frac{r'_H}{r_0}\right)\right)\right]\,.
\end{equation}
Here, $c\sim \frac{6}{b^2}$ is the central charge in the $b\to 0$ limit. The defect coupling and the conformal weights of the external states are tuned to scale with $c$ as $\mu_D=\frac{c\mu}{12\pi }$ and $\Delta_\alpha=\frac{c}{12}(1+r_H^2)$, $\Delta_{\alpha'}=\frac{c}{12}(1+r_H'^2)$. This ensures that both the defect and the local operators are ``heavy" and can backreact on the geometry. The parameter $r_0$ is determined by solving the equation $\sqrt{r_0^2-r_H^2}+\sqrt{r_0^2-r_H'^2}=\mu r_0$. See Figure~\ref{fig:scsl_limit} for a geometrical interpretation of the 3 parameters $r_H, r_H', r_0$ in terms of proper lengths on a hyperbolic surface. 

In addition, we show that the correlation induced by the defect at both weak and strong coupling as captured by (\ref{matweaksumm}) and (\ref{matstrongsumm}) can be seen as a result of ``decohering'' the FZZT boundaries (in a way described by the interface $L(\Tilde{\mu}_D)$ in (\ref{decoheredefect})) on a pair of disks with the local operators inserted at the center of either disk,
\begin{equation}
    2\sqrt{2}\pi b \left|\int_0^{\infty}\frac{d\ell}{\ell}e^{\mu_D \ell}\bra{V_{\alpha}}\ket{\ell}\bra{\ell}\ket{V_{\alpha'}}\right|=
    \begin{cases}
          \mu_D\left(\frac{16\pi^3 kk' \sinh(2\pi k)\sinh(2\pi k')}{\cosh^2(\pi(k+k'))\cosh^2(\pi(k-k'))}\right)^{\frac{1}{2}} &{\rm as}\ \mu_D\to 0\,, \\
          \exp\left[\frac{c}{6}\left(r_H\cos^{-1}\left(\frac{r_H}{r_0}\right)+r_H'\cos^{-1}\left(\frac{r_H'}{r_0}\right)\right)\right] &{\rm as}\ \mu_D\to \infty\,.
    \end{cases}
\end{equation}
Since the overlaps of the length states $\ket{\ell}$ with the primary states are known in closed form in terms of the Bessel-K function \cite{Fateev:2000ik} reviewed in Section \ref{SeclenFT}, we compute the matrix elements of the decohered FZZT interface exactly in terms of the interface cosmological constant $\Tilde{\mu}_D$ the Liouville coupling $b$ in Section \ref{Secmat}. 

\begin{figure}
    \centering
    \includegraphics[width=0.25\linewidth]{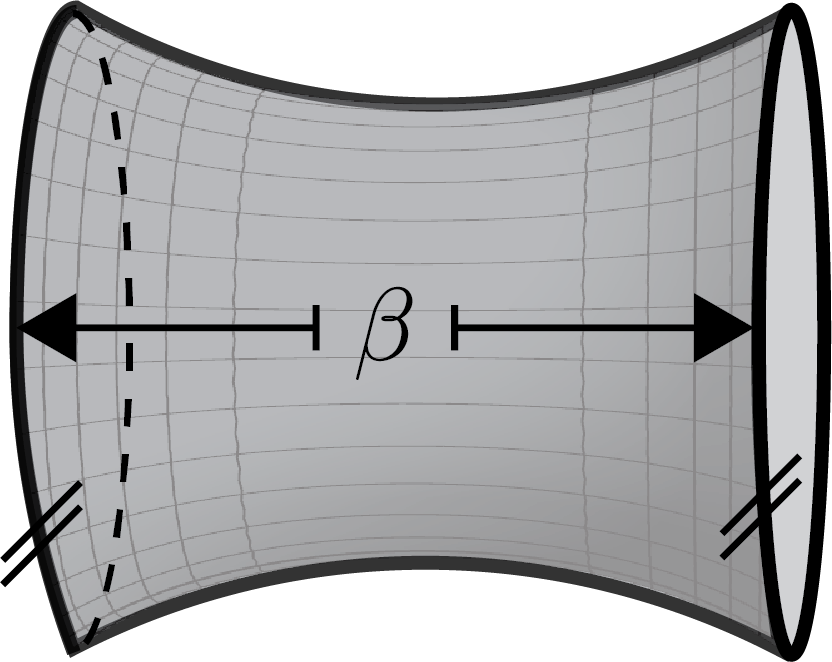}
    \caption{The figure shows the Liouville saddle used to compute the vacuum energy of the defect Hilbert space. It is a torus with the defect, constructed by gluing the ends of a hyperbolic cylinder of height $\beta$.}
    \label{fig:thermal_onept_fcn}
\end{figure}

The decohered FZZT interface has interesting interpretations in other models due to their close relation to Liouville. For example, in Section \ref{secJT}, we interpret the decohered FZZT interface in terms of an end-of-the-world (EOW) brane in JT gravity. Specifically, we show that when the interface cosmological constant $\Tilde{\mu}_D=0$, the thermal 1-point function of the decohered FZZT interface reduces in the Schwarzian limit to the partition function with a tensionless EOW brane in JT gravity (See Figure~\ref{fig:thermal_onept_fcn}),
\begin{equation}
    \text{Tr}\left[e^{-\pi t H}L(\Tilde{\mu}_D=0)\right]\underset{\rm schw}{=} Z_{\rm EOW}(\beta)\equiv\int_0^{\infty} \frac{d\lambda}{2\sinh\left(\frac{\lambda}{2}\right)}Z_{\rm trumpet}(\beta, \lambda)\,.
\end{equation}
Here $\beta$ is the renormalized boundary length in JT and is related to the temperature parameter $t$ in Liouville by $t=\frac{2\beta}{\pi b^2}$. The $\lambda$ integral is over the length of the brane which is along a geodesic and $Z_{\rm trumpet}$ is given in (\ref{eq:jt_eow_divergent}). In Section (\ref{sec:AGT}), we interpret the decohered FZZT interface in $4d$ $\mathcal{N}=2$ gauge theory using the AGT conjecture in terms of an interface in the $4d$ theory on the squashed 4-sphere $S_b^4$. To this end, we show that the fixed-length wavefunction of the FZZT state denoted $\psi_\ell(P)$ (\ref{flwavefunction}) can be expressed using results from supersymmetric localization as
\ie 
\psi_\ell(P)
=
{b\kappa\ell\over 4\sqrt{2} } \int_{\mathbb{R}} ds \,e^{-\kappa\ell \langle W_{\rm fund}\rangle (s)}
 \rho_0(s) 
 Z_{{\rm S}^3_b}\big[T[{\rm SU}(2)]_m\big](s,P)\,,
\fe
where $\langle W_{\rm fund}\rangle (s)=\cosh(2\pi b s)$ is the expectation value of a fundamental ${\rm SU}(2)$-Wilson loop wrapping the BPS circle of length $2\pi b$ on $S^3_b$ and coupled to a scalar whose value after localization is $s$; $e^{-\kappa\ell \langle W_{\rm fund}\rangle (s)}$ is its proliferation. $ Z_{{\rm S}^3_b}\big[T[{\rm SU}(2)]_m\big](s,P)$ is the partition function of the $3d$ $\mathcal{N}=2$ theory denoted $T[{\rm SU}(2)_m]$ on $S^3_b$ which is known to evaluate to the torus 1-point modular-S kernel. The parameter $m$ corresponding to the hypermultiplet mass is chosen to be $m=\frac{i}{2}\left(b-\frac{1}{b}\right)$. The decohered FZZT interface is then described by an interface on the equatorial $S^3_b$ of $S_b^4$ with boundary conditions corresponding to the fixed-length FZZT state on the two hemispheres of $S_b^4$ described in Figure~\ref{fig:AGT} and a correlated proliferation of the above Wilson loops with a distribution specified by (\ref{decoheredefect}).

In the following, we summarize the other results for defects (\ref{eq:defect_def}) and (\ref{eq:exactmarginal}) that we derived at weak and strong coupling. 

\vspace{0.5em}
\noindent\textbf{Weak defect coupling: Perturbative CFT data.}

At weak coupling, we derive results to leading non-trivial order in perturbation theory. At this order, we derive the defect data from the integrated two-point or three-point functions of local operators. Since the position dependence of the two-point or three-point functions is fixed by conformal symmetry alone, the results we derive apply universally to pinning defects obtained by integrating scalar primary operators in \textit{general} $2d$ CFT. In addition to computing the correlations of local primary operators across the defect illustrated above, we also compute correlations of the stress tensor across the defect. This is a measure of energy transmission across the defect. For an infinite line defect obtained by integrating a scalar primary $\mathcal{O}$ of dimension $\Delta$ along the real axis, we show that the normalized stress tensor two-point function is given by
\begin{equation}
    \frac{\langle T(i)\exp\left(\lambda \int_\Sigma \mathcal{O}\right) T(-i) \rangle}{\langle \exp\left(\lambda \int_\Sigma \mathcal{O}\right)\rangle} =\frac{c}{32}+\pi^2\lambda^2\frac{\Delta(\Delta^2(2\Delta-3)+2)}{4^{\Delta+1}\cos(\pi \Delta)}+O(\lambda^3)\,, \quad{\rm if}\ \frac{1}{2}<\Delta<\frac{3}{2}\,.
\end{equation}
The expression is only valid for $\Delta$ in the range indicated above and in this range, the second term is negative indicating that the defect reflects a fraction of the incident energy, and for the marginal ($\Delta=1$) case, the reflection coefficient measuring the fraction of incident energy reflected by the defect,
\begin{equation} \label{refsumm}
    \mathcal{R}=\frac{2\pi^2 \lambda^2}{c}+O(\lambda^3)\,,
\end{equation}
which agrees with the result computed previously in the literature \cite{Makabe:2017uch,Brehm:2020agd}. We will comment on subtleties in applying this to the Liouville CFT due to IR divergences.

Finally, in Section~\ref{Seccusp} we study a sharp geometric deformation of curve on which the defect is placed by introducing a \emph{cusp} \cite{Calabrese:2009qy}. In particular, we study the effect of introducing a cusp on the line defect in the presence of a conformal boundary or a pair of local operators. This is a generalization of the well-studied setup in the literature used to compute the cusp anomalous dimension of a conformal defect in the vacuum state (see recent works \cite{Kravchuk:2024qoh,Cuomo:2024cusp} and references therein). For the configuration with a conformal boundary described in Section~\ref{Seccuspbound} (see Figure~\ref{fig:theta} for the setup), we find a logarithmic divergence at linear order in the defect coupling,
\begin{equation} 
    \log\left[\frac{\left\langle \exp\left(\lambda \int_{\text{cusp}(\theta_1,\theta_2)}\mathcal{O}\right)\right\rangle_B}{\langle \id \rangle_B}\right]=\frac{\lambda}{2} \frac{\langle \mathcal{O}\rangle_B }{\langle \id \rangle_B}\left(\csc(\theta_1)+\csc(\theta_2)\right)\log\left(\frac{L}{y}\right) + O(\lambda^2)\,.
\end{equation}
Interestingly, the angular dependence is given by a sum of two \textit{completely monotonic} functions in the non-compact angle variables $x_{1,2}\equiv \tan^2(\theta_{1,2})$. While the above result applies for the marginal ($\Delta=1$) case, we observe that the irrelevant ($\Delta>1$) case displays no IR divergence while for the relevant ($\Delta<1$) case, there is a power-law divergence, thereby the cusp free energy is a sensitive probe of the nature of the deformation. The analogous result (for the marginal case) in the presence of a pair of local operators computed in Section~\ref{Secloccusp} (See Figure~\ref{fig:thetaloc} for the setup) is given by
\begin{equation} 
    \log\left[\frac{ \langle \mathcal{O}_i(iy) \exp\left(\lambda \int_{\text{cusp}(\theta_1,\theta_2)}\mathcal{O}\right)\mathcal{O}_i(-iy) \rangle}{\langle \mathcal{O}_i(iy)\mathcal{O}_i(-iy)\rangle}\right]=\frac{1}{2}\lambda C_{ii\mathcal{O}}\left(K\left(\sin^2(\theta_1)\right)+K\left(\sin^2(\theta_2)\right)\right)+O(\lambda^2)\,,
\end{equation}
where $C_{ii\mathcal{O}}$ is the OPE coefficient between the external operators $\mathcal{O}_i$ and the operator $\mathcal{O}$. When the external operators are stress tensors considered in Section~\ref{energytrans}, then the setup described in Figure~\ref{fig:cuspreflection} computes the energy transmission across the defect in the presence of the cusp. The reflection coefficient of the defect when a cusp of opening angle $\phi$ is introduced is 
\begin{equation} 
    \mathcal{R}(\phi)=\frac{\lambda^2}{c}\left(\frac{1}{2}(\pi^2+4)+8 I(\phi)\right)\,,
\end{equation}
where $I(\phi)$ is the following integral,
\begin{equation}
    I(\phi)\equiv \int_0^{\infty}dt \left(\frac{2e^{i\phi} t \log(t)+2ite^{i\phi}(\pi-\phi)+t^2-e^{2i\phi}}{(t+e^{i\phi})^3(t-e^{-i\phi})}\right)\,.
\end{equation}
The defect becomes more transmitting as the opening angle at the cusp is increased and the reflection coefficient has the following bounds as we tune the cusp angle,
\begin{equation}
    \frac{2\pi^2 \lambda^2}{c} \leq \mathcal{R}(\phi) <\frac{4\lambda^2}{c}(\pi^2+2)\,.
\end{equation}
In Section \ref{secperinf}, we describe a similar monotonicity property for the information transmitted across the defect i.e., the entanglement entropy across the defect (See Figure~\ref{fig:cusp_replicas}) increases monotonically as the opening angle at the cusp is increased. 
It would be interesting to study these angle-dependent energy and information reflection coefficients and their non-perturbative properties further, and to explore the refined inequalities involving them, generalizing those in \cite{Karch:2024udk}.

\vspace{0.5em}
\noindent\textbf{Strong defect coupling: Semiclassical hyperbolic geometry.}
A key result of the paper is that the Liouville defect becomes tractable in a complementary strongly coupled limit where perturbation theory fails.
This strongly coupled limit is described by a double-scaling between the defect cosmological constant and Liouville coupling,
\begin{equation}
\label{eq:double_scaling}
b\to 0\,, \quad \mu_D \to \infty\,,
\qquad
{\rm with}\ \mu\equiv 2\pi b^2 \mu_D \ \rm{fixed}\,.
\end{equation}
In this regime the defect is conformal and admits an interpretation directly in terms of a discontinuity in the extrinsic curvature of hyperbolic surfaces across the defect locus,
\begin{equation}
\label{eq:jump_condition}
K_+-K_-=-\mu\,,
\end{equation}
where $K_\pm$ are the extrinsic curvatures calculated in the hyperbolic metric on either side of the defect.

With these saddles in hand, we compute several intrinsic pieces of \emph{defect data} geometrically.
First, we evaluate the vacuum expectation value of the defect (the `$g$-function')
from the on-shell Liouville action of the sphere saddle
(two hyperbolic disks glued across the defect) in Section~\ref{Secvev},
\begin{equation}
\label{eq:g_function}
\log g(\mu)\equiv \log \langle {\bm L}_\Sigma \rangle =\frac{c}{3}\log\!\left(\frac{\mu+\sqrt{\mu^2-4}}{2}\right)\,.
\end{equation}
The saddle exists for $\mu>2$ so the defect must be sufficiently ``heavy''.\footnote{Note that the $g$-function here is obtained by analytic continuation of the disk one-point function with normalizable primaries since the identity operator is non-normalizable in the Liouville CFT. Relatedly, the familiar $g$-theorem for defect RG flows in compact CFTs does not apply in this setting. It would be interesting to investigate whether a modified version of such a defect RG monotone exists for defect flows in the Liouville CFT and more general non-compact CFTs.}

Second, we construct finite-temperature saddles on the torus by gluing hyperbolic cylinders with the defect.
The corresponding solutions determine the vacuum energy of the \emph{defect Hilbert space} and allow us to
probe information transmission via the replica construction. For a single defect, we find that the vacuum energy is given by
\begin{equation}
    E_1=-\frac{c}{12\pi^2}\left(\sin^{-1}(\frac{\mu}{2})\right)^2\,.
\end{equation}
The torus saddles exist in a complementary regime ($\mu<2$) to the sphere. In addition, we observe that for $n$ equally spaced defects, the vacuum energy is
$E_n=nE_1$, implying that the \emph{effective central charge} governing entanglement transmission across the
defect vanishes:
\begin{equation}
\label{eq:ceff_vanish}
E_n=nE_1\qquad \Longrightarrow\qquad c_{\rm eff}=0\,.
\end{equation}
This suggests that the defect is perfectly reflecting (in both energy and information \cite{Karch:2024udk}). This is also consistent with the observation that there is no correlation between the stress tensor across the defect since the stress tensor is a light operator and doesn't backreact on the geometry. On the other hand, for heavy operators which backreact on the geometry, we can efficiently compute their correlations across the defect as discussed in (\ref{matstrongsumm}).

Third, the geometric description makes the \emph{fusion of defects} particularly transparent.
Gluing saddles for two defects placed in sequence yields an explicit composition rule for the effective
coupling $\mu_{\text{eff}}$ of the fused defect,
\begin{equation} \label{fus}
    \mu_{\text{eff}}=\mu_1+\mu_2\,.
\end{equation}

In addition, we show that in the limit the defects fuse (in the vacuum state), they exchange an operator of dimension,
\begin{equation}
    \Delta=\frac{c}{3}\frac{\mu_1 \mu_2}{(\mu_1+\mu_2)^2-4}\,.
\end{equation}
Interestingly, this expression shows that if the difference in defect couplings is greater than $2$, then the weight of the exchanged operator is below the threshold of $\frac{c}{12}$. 

We also study fusion in an excited state and at finite temperature and obtain corresponding expressions for the dimension of the exchanged operator, all the while observing that the fusion rule (\ref{fus}) is consistent. Furthermore, we also study fusion of the defect with a conformal boundary by constructing appropriate Liouville saddles. We observe that the fusion with a ZZ (Dirichlet) boundary is singular and has an associated Casimir energy,
\begin{equation}
    E_{\text{fus}}=-\frac{c}{12}\left(\cos^{-1}(1-\mu)\right)^2\,.
\end{equation}
This can be compared to the corresponding expression for the Casimir energy computed at weak coupling in (\ref{Casenergy}). The fusion of the defect with the FZZT boundary is nonsingular and consecutive fusions generate additive shifts in the boundary cosmological constant which is consistent with \eqref{fus}.

Having summarized the key results, the rest of the paper is organized as follows. In Section \ref{sec:weakcoupling}, we describe the defects (\ref{eq:defect_def}) and (\ref{eq:exactmarginal}) at weak defect coupling using perturbation theory. In Section \ref{Secexp}, we describe the defects at strong coupling in a semiclassical limit using hyperbolic geometries. In Section \ref{Seclen}, we describe the decohered FZZT interface (\ref{decoheredefect}) and its relation to the defects (\ref{eq:defect_def}), (\ref{eq:exactmarginal}) at both weak and strong coupling. In Section \ref{sec:GMC}, we review the probabilistic framework called Gaussian Multiplicative Chaos (GMC) introduced by mathematicians to make the Liouville path integral mathematically rigorous and provide a non-perturbative realization of the defect within this framework. In Section \ref{sec:othermodels}, we interpret the defect in other models. Finally, in Section \ref{sec:disc}, we conclude and discuss some future directions related to this work. There are three Appendices containing more details of some of the calculations done in the paper.

\section{Weak defect coupling: Perturbative CFT data} \label{sec:weakcoupling}

Liouville theory is described by the following path integral over a single real field $\phi$,
\begin{equation}
    Z=\int [D\phi] e^{-S[\phi]} \quad{\rm with}\quad S=\frac{1}{\pi}\int d^2 z\ (\partial \phi \overline{\partial}\phi +\pi \mu_{\text{bulk}} e^{2b\phi})\,.
\end{equation}
Here, $\phi$ is called the Liouville field and the parameter $b$ is the Liouville coupling. We restrict $b\in (0,1)$. The bulk cosmological constant is denoted $\mu_{\text{bulk}}$. We have written down the action for a flat background metric. More generally, the action can be written in a covariant form,
\begin{equation}\label{eq:covariant_lv_action}
   S = \frac{1}{4\pi}\int d^2z\sqrt{\Hat{g}}\ (\Hat{g}^{ab}\partial_a\phi\partial_b\phi+Q\Hat{R}\phi+4\pi\mu_{\text{bulk}} e^{2b\phi})\,,
\end{equation}
where $\Hat{g}$ is the fiducial background metric and $\Hat{R}$ is its Ricci scalar. The path integral describes a $2d$ CFT with Virasoro symmetry and central charge given by
\begin{equation}
    c=1+6Q^2 \quad{\rm with}\quad Q=b+\frac{1}{b}\,.
\end{equation}
Often $Q$ is referred to as the background charge. The local Virasoro primary operators are the vertex operators
\begin{equation}
    V_\alpha =e^{2\alpha \phi}\,,
\end{equation}
which are scalars with total scaling dimension $\Delta_\alpha =2\alpha(Q-\alpha)$. The spectrum of Liouville CFT is non-compact and consists of scalar primaries (and their Virasoro descendents) with 
\begin{equation}
    \alpha=\frac{Q}{2}+iP \quad{\rm and}\quad \Delta_\alpha=2\left(\frac{Q^2}{4}+P^2\right)\,,
\end{equation}
given $P \in \mathbb{R}^+$. Vertex operators are normalized such that the two-point function
\begin{align} \label{DOZZnorm}
    \langle V_{\frac{Q}{2}+iP}(\infty) V_{\frac{Q}{2}+iP'}(0)\rangle &= \lim_{\epsilon \to 0} C_{\rm DOZZ}\left(\frac{Q}{2}+iP, \frac{Q}{2}+iP', \epsilon\right) \,,\nn\\
    &= 2\pi \delta(P-P') \quad{\rm given}\quad P,P'\in \mathbb{R}^+\,,
\end{align}
is given by a suitable limit of the DOZZ formula \cite{Dorn:1994xn, Zamolodchikov:1995aa}.

In this section, we will be studying the dynamics of the line operator described by a `localized cosmological constant' (since it deforms the action by the term $\mu_D\int_\Sigma e^{b\phi}$) on a closed curve $\Sigma$, 
\begin{equation} \label{expdefect}
    {\bm L}_{\Sigma}=\exp\left[\mu_D\int_{\Sigma}ds\  e^{b\phi(s)}\right]\,,
\end{equation}
inserted into the Liouville path integral. $s$ is a local coordinate along the curve $\Sigma$. We refer to $\mu_D$ as the defect cosmological constant. The Liouville action with the defect term written in a covariant form is given
\begin{equation} \label{exactact}
     S = \frac{1}{4\pi}\int_{\mathcal{M}} d^2z\sqrt{\Hat{g}}\ (\Hat{g}^{ab}\partial_a\phi\partial_b\phi+Q\Hat{R}\phi+4\pi\mu_{\text{bulk}} e^{2b\phi})-\mu_D\int_{\Sigma} ds\sqrt{\Hat{h}}\ e^{b\phi}\,,
\end{equation}
where $\Hat{h}$ is the pull-back of the fiducial metric $\Hat{g}$ onto the curve $\Sigma$. Notice that the action (\ref{exactact}) has a form which is very similar to the Liouville action with an FZZT boundary (a Neumann boundary condition for the Liouville field) \cite{Fateev:2000ik}. The important distinction is that in the present case, the curve $\Sigma$ is not the boundary of the domain $\mathcal{M}$, i.e. $\Sigma \neq \partial \mathcal{M}$.
The vertex operator $V_{b/2}=e^{b\phi}$ has total scaling dimension
\begin{equation} \label{deltabounds}
    \Delta_{b/2}=1+\frac{b^2}{2} \in \left(1,\frac{3}{2}\right)\,.
\end{equation}
The limits arise due to the restriction to $b\in (0,1)$.
Since $\Delta_{b/2}>1$, the defect is not conformal away from the $b\to 0$ limit. Equivalently, the insertion of ${\bm L}_\Sigma$ into the Liouville path integral introduces an irrelevant deformation that triggers an RG flow along the defect. To study the flow, we can introduce a dimensionless coupling $g_D$ related to the defect cosmological constant 
\begin{equation}
    \mu_D=\frac{g_D}{\Lambda^{\Delta_{b/2}-1}}=\frac{g_D}{\Lambda^{\frac{b^2}{2}}}\,.
\end{equation}
Here, $\Lambda$ is a renormalization energy scale. The above relation implies that the $\beta$-function that governs the defect-RG flow is given to leading order in $g_D$ by
\begin{equation}
    \beta_{g_D}\equiv \frac{\partial g_D}{\partial \log \Lambda}=\frac{b^2}{2}g_D \,.
\end{equation}
It turns out that the above $\beta$-function is perturbatively exact (in $\mu_D$) since the repeated fusion of $V_{b/2}$ with itself, $V_{b/2}\otimes V_{b/2} \otimes V_{b/2}\otimes \dots$ does not generate $V_{b/2}$.\footnote{See for example \cite{Harlow:2011ny, Collier:2018exn, Post:2024itb} for discussions on fusion rules for Liouville vertex operators. Here we assume $b$ to be generic such that $b^2\notin \mathbb{Q}$.} One way to argue for this is that the lightest operator appearing in the repeated fusion is $V_{b}$ which has scaling dimension $\Delta_b=2$ which is greater than $\Delta_{b/2}$. This argument suggests that in the $b\to 0$ limit where $\Delta_{b/2}\to 1$, the deformation on the line becomes \emph{exactly marginal}.  

Now, consider a closely related defect,\footnote{We thank Nina Holden for an interesting discussion on this defect.}
\begin{equation} \label{alphac}
    {\bm L}_\Sigma(\alpha_c):=\exp\left[\mu_D\int_{\Sigma}ds\  V_{\alpha_c}\right]\,, \qquad \Delta_{\alpha_c}=1 \,.
\end{equation}
The vertex operator used to construct this defect has scaling dimension $1$, so the defect sources a marginal deformation on the line for any $b$. The value of the Liouville momentum $\alpha_c$ is given by
\begin{equation} 
    \Delta_{\alpha_c}=1 \implies \alpha_c=\frac{Q}{2}-\frac{1}{2}\sqrt{Q^2-2}=\frac{1}{2}\left(b+\frac{1}{b}\right)-\frac{1}{2}\sqrt{b^2+\frac{1}{b^2}}\,.
\end{equation}
Notice that as $b\to 0$, $\alpha_c \to \frac{b}{2}$ so in this limit, this defect reduces to the defect in (\ref{expdefect}).\footnote{It is possible to distinguish the two defects in the $b\to 0$ limit by computing observables which can distinguish between a marginal and an almost marginal deformation. We compute one such observable in Section \ref{Seccuspbound}.} Using the fusion rules for Liouville vertex operators \cite{Harlow:2011ny, Collier:2018exn, Post:2024itb}, the lightest operator that appears in the repeated fusion $V_{\alpha_c}\otimes V_{\alpha_c}\otimes V_{\alpha_c}\otimes \dots$ has scaling dimension,
\begin{equation}
    \Delta_{\rm min}=2\sqrt{Q^2-2}(Q-\sqrt{Q^2-2}) =2 \sqrt{b^2+\frac{1}{b^2}}\left(b+\frac{1}{b}-\sqrt{b^2+\frac{1}{b^2}}\right) > 1 \,.
\end{equation}
Since $\Delta_{\text{min}}>1$ for any $b$, it means that no relevant or marginal operators are generated by repeated fusion of $V_{\alpha_c}$. This also means that observables involving ${\bm L}_\Sigma(\alpha_c)$ do not exhibit UV divergences at any order in perturbation theory in coupling $\mu_D$. This argument implies that for any $b$, the defect ${\bm L}_\Sigma(\alpha_c)$ sources an exactly marginal deformation on the line. In the subsequent analysis, we refer to (\ref{alphac}) as a \emph{manifestly conformal defect}. 

In this section, we will compute various observables involving the defects (\ref{expdefect}) and (\ref{alphac}) at weak coupling, to leading non-trivial order in perturbation theory in $\mu_D$. In Section~\ref{Secexp}, we will compute the corresponding observables at strong coupling in a semiclassical limit. This allows us to compare the properties of the defect at weak and strong defect coupling. The observables that we will compute in the rest of this section are: (i) correlations of local primary operators across the defect; (ii) stress tensor correlations across the defect using which we quantify the fraction of incident energy reflected by the defect, which is a standard way to characterize a conformal defect in $2d$ CFTs; (iii) the leading correction to cylinder transition amplitudes between conformal boundaries due to the defect and hence to the open string spectrum; (iv) the Casimir energies of fusion of the defect with conformal boundaries which determines whether the defect is attracted or repelled by the boundary; (v) finally, we compute the effect of introducing a cusp on the defect on these observables. 

\subsection{Correlations of local operators across the defect} \label{seccorrpert}

In this section, we want to understand quantitatively how the defect induces correlations between local operators in the presence of the line defect.
To this end, we compute the 2-point functions of vertex operators inserted on either side of the defect,
\begin{equation}
    \langle V_{\alpha}(z_1) {\bm L}_{\Sigma} V_{\alpha'}(z_2)\rangle=\sum_{n=0}^{\infty} \frac{\mu_D^n}{n !}\langle V_{\alpha}(z_1)\left(\int_{\Sigma}V_{b/2}\right )^n  V_{\alpha'}(z_2)\rangle\,.
\end{equation}
In the above expression, $\left(\int_{\Sigma}V_{b/2}\right )^n$ needs to be evaluated with a suitable $i\epsilon$-prescription to regulate the coincident divergences. Since the observables are sensitive to the curve $\Sigma$ on which the defect is placed, we consider two canonical choices of $\Sigma$ in the subsequent analysis: the real axis corresponding to an infinite line defect and the unit circle corresponding to a circular defect.

\begin{figure}
    \centering
    \begin{subfigure}{0.45\textwidth}
        \centering
        \includegraphics[width=0.55\textwidth]{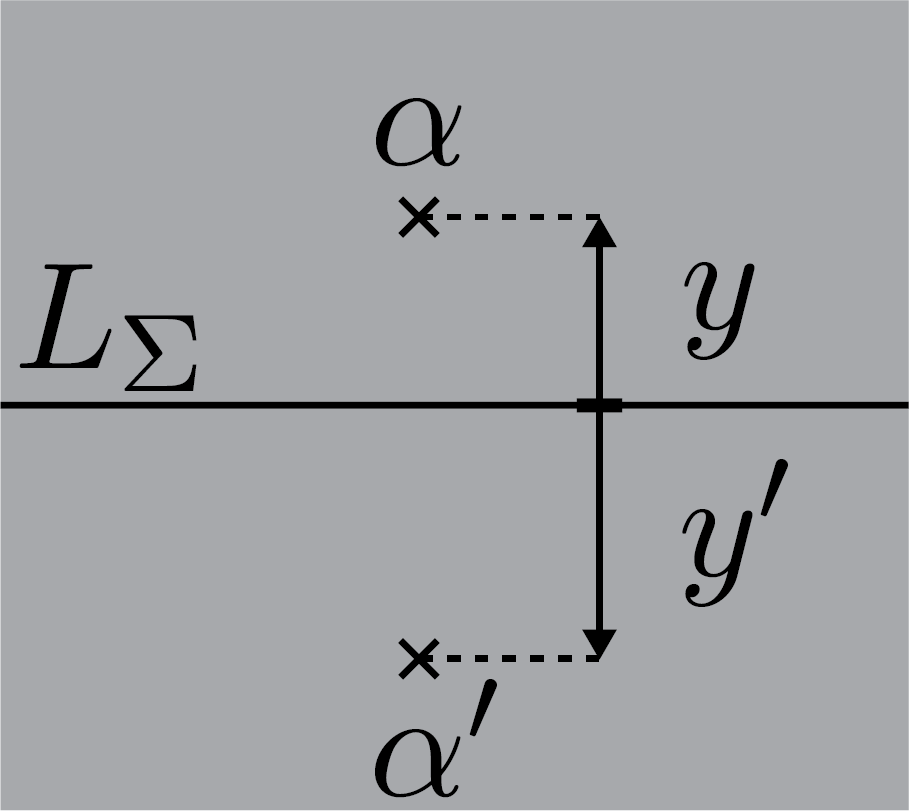}
        \caption{Infinite Line Defect}
    \end{subfigure}
    \begin{subfigure}{0.45\textwidth}
        \centering
        \includegraphics[width=0.4\textwidth]{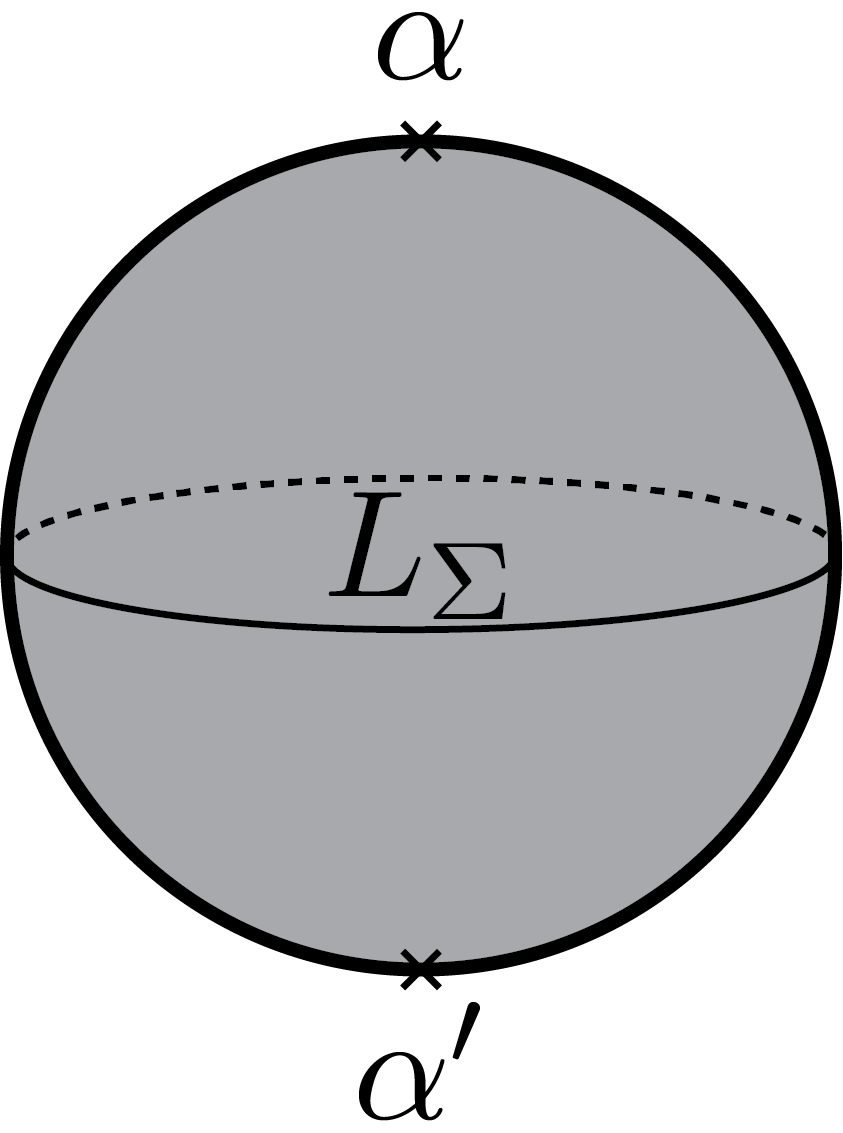}
        \caption{Circular Defect}
    \end{subfigure}
    \caption{These figures describe the kinematics of the setup used to compute correlations of local operators across the defect with two canonical choices for the curve $\Sigma$ on which the defect is placed. Figure (a) shows an infinite line defect placed along the real axis, with a pair of local vertex operator insertions with Liouville momenta $\alpha$ and $\alpha'$ on the imaginary axis. Figure (b) shows a circular defect around the equator of the Riemann sphere or equivalently around the unit circle on the plane, with the pair of local operators inserted at the poles. }
    \label{fig:canonical_sigma}
\end{figure}

Consider first the scenario described in part (a) of Figure~\ref{fig:canonical_sigma} where the defect is placed along the real axis and two vertex operators are placed on the imaginary axis at $z=\pm iy$. So, the 3-point function of two vertex operators placed on either side of the line defect is given by
\begin{equation} \label{vertexcorr}
\begin{split}
    \langle V_{\alpha}(iy) {\bm L}_\Sigma V_{\alpha'}(-iy)\rangle = & \mu_D\int_{-\infty}^{\infty}dx \langle V_{\alpha}(iy)V_{b/2}(x) V_{\alpha'}(-iy) \rangle +O(\mu_D^2)\,, \\
    =& \mu_D\int_{-\infty}^{\infty} dx\frac{C_{\text{DOZZ}}\left(\alpha,\alpha',\frac{b}{2}\right)}{(x^2+y^2)^{\Delta_{b/2}}(2y)^{\Delta_\alpha+\Delta_{\alpha'}-\Delta_{b/2}}}+O(\mu_D^2)\,,\\
    =& \mu_D\frac{C_{\text{DOZZ}}\left(\alpha,\alpha',\frac{b}{2}\right)}{2^{\Delta_\alpha+\Delta_{\alpha'}-\Delta_{b/2}}} \sqrt{\pi} \frac{\Gamma(\Delta_{b/2}-\frac{1}{2})}{\Gamma(\Delta_{b/2})}y^{1-\Delta_\alpha-\Delta_{\alpha'}-\Delta_{b/2}}+O(\mu_D^2)\,.
    \end{split} 
\end{equation}
Here $\Delta_\alpha=2\alpha(Q-\alpha)$ is the scaling dimension. $C_{\text{DOZZ}}$ are the DOZZ structure constants \cite{Dorn:1994xn, Zamolodchikov:1995aa}. In (\ref{dozb}) of Appendix \ref{AppDOZZ}, we have derived an expression for $C_{\text{DOZZ}}(\alpha,\alpha',\frac{b}{2})$. In (\ref{vertexcorr}), we have assumed that the vertex operators have different scaling dimensions just to avoid writing the $\delta$-function at $O(\mu_D^0)$. This also means that in the absence of the defect, there is no correlation between the two operators. Any non-zero correlation is induced by the defect. 

Notice that even though the defect has infinite length, the integral does not exhibit any divergence. For vertex operators in the physical spectrum of Liouville, the sum over scaling dimensions is greater than 1 which means there is a singularity in the $y\to 0$ limit i.e., as the operators approach the defect symmetrically. From the expression (\ref{dozb}), it is clear that the magnitude of the 3-point function in (\ref{vertexcorr}) depends on the bulk and defect cosmological constants only in terms of the scale-invariant ratio of $\mu_D/\sqrt{\mu_{\text{bulk}}}$. This is true to any order in perturbation theory, so we also organize the perturbation series in terms of this scale invariant ratio. In the $b\to 0$ limit when the vertex operator $V_{b/2}$ becomes marginal on the line, the above correlator (\ref{vertexcorr}) becomes,\footnote{For the manifestly conformal defect (\ref{alphac}), the scaling in (\ref{VCVb0}) applies for any $b$,
\begin{equation} 
    \langle V_{\alpha}(iy) {\bm L}_\Sigma(\alpha_c) V_{\alpha'}(-iy)\rangle=2\pi \mu_D\times \frac{C_{\text{DOZZ}}\left(\alpha,\alpha',\alpha_c\right)}{(2y)^{\Delta_\alpha+\Delta_{\alpha'}}}+O(\mu_D^2)\,.
\end{equation}}
\begin{equation} \label{VCVb0}
    \langle V_{\alpha}(iy) {\bm L}_\Sigma V_{\alpha'}(-iy)\rangle=2\pi \mu_D\times \lim_{b\to 0} \frac{C_{\text{DOZZ}}\left(\alpha,\alpha',\frac{b}{2}\right)}{(2y)^{\Delta_\alpha+\Delta_{\alpha'}}}+O(\mu_D^2)\,.
\end{equation}
We analyze the $b\to 0$ limit of the above DOZZ structure constant in Appendix \ref{AppDOZZ}. When the momenta of the vertex operators are of the form $\alpha=\frac{Q}{2}+ibk$ and $\alpha'=\frac{Q}{2}+ibk'$ so that in the $b\to 0$ limit, the operators are at the edge of the normalizable spectrum,\footnote{Note that this is the limit that gives matrix elements of operators in Schwarzian Quantum Mechanics from the DOZZ structure constants in Liouville CFT \cite{Mertens:2017mtv}. So, we could refer to this limit as the Schwarzian limit.} the limit takes a nice form,
\begin{equation} \label{DOZZLim}
    \left|C_{\text{DOZZ}}\left(\frac{Q}{2}+ibk,\frac{Q}{2}+ibk',\frac{b}{2}\right)\right |\overset{b \to 0}{=}\left(\frac{4\pi}{\mu_{\text{bulk}}}\frac{kk' \sinh(2\pi k)\sinh(2\pi k')}{\cosh^2(\pi(k+k'))\cosh^2(\pi(k-k'))}\right)^{\frac{1}{2}}\,.
\end{equation}
We have taken the absolute value to remove the phases in the expression since the expression is not real. In Appendix \ref{AppDOZZlight}, we have also computed the phases in the above expression. In Appendix \ref{AppDOZZheavy}, we also computed the DOZZ structure constants in the $b\to 0$ limit with the pair of vertex operators taken to be very heavy,
\begin{align}
    \log \bigg | C_{\rm DOZZ}\bigg(\frac{Q}{2}+ & \frac{ik_1}{b},\frac{Q}{2}+\frac{ik_2}{b},\frac{b}{2}\bigg) \bigg | = \nn\\
    &-\frac{1}{b^2}\left(F(ik_1+ik_2)+F(-ik_1-ik_2)+F(ik_1-ik_2)+F(ik_2-ik_1)\right)+\nn\\
    &\frac{1}{2b^2}\left(4F(0)+F(-2ik_1)+F(2ik_1)+F(-2ik_2)+F(2ik_2)\right)+O(\log(b)) \,,
\end{align}
where $F(x)\equiv \int_{\frac{1}{2}}^x dt \log(\gamma(t))$. See Appendix \ref{AppDOZZheavy} for more details and fine-print. The terms in the second line factorize between the momenta $k_{1,2}$ and hence can be absorbed into a normalisation of the vertex operators whereas the terms in the first line don't factorize and hence govern the correlations between the vertex operators induced by the defect in this limit.

We can also compute the three-point function with the defect placed along a different curve. For instance, another canonical setup described in part (b) of Figure~\ref{fig:canonical_sigma} is when the defect is placed along the unit circle with the vertex operators at $0, \infty$. For this circular defect, the three-point function evaluates to
\begin{equation}
    \langle V_\alpha(\infty) {\bm L}_\Sigma V_{\alpha'}(0)\rangle =2\pi \mu_D  C_{\text{DOZZ}}\left(\alpha,\alpha',\frac{b}{2}\right)+O(\mu_D^2)\,.
\end{equation}
The integrand in this case is uniform over the unit circle and hence we just get the prefactor of $2\pi$. In the $b\to 0$ limit where the defect is effectively conformal and when the vertex operators are close to the edge of the normalizable spectrum, using (\ref{DOZZLim}), the magnitude of the matrix elements of the circular defect are given by
\begin{equation} \label{matweak}
   \left |\langle V_{\frac{Q}{2}+ibk}(\infty) {\bm L}_\Sigma V_{\frac{Q}{2}+ibk'}(0)\rangle \right | \overset{b \to 0}{=} \frac{\mu_D}{\sqrt{\mu_{\text{bulk}}}}\left(\frac{16\pi^3 kk' \sinh(2\pi k)\sinh(2\pi k')}{\cosh^2(\pi(k+k'))\cosh^2(\pi(k-k'))}\right)^{\frac{1}{2}} + O(\mu_D^2)\,.
\end{equation}
In summary, we have quantified the correlations between the vertex operators induced by the defect working perturbatively in the defect cosmological constant. In principle, we could compute the 2-point function of vertex operators across the defect to any order in the perturbation theory since the correlation functions of any number of vertex operators can be expressed by integrating DOZZ structure constants against Virasoro blocks. In practice, this is hard since we will have to integrate the Virasoro blocks which are not known in closed form along the curve $\Sigma$. In addition, the terms at higher orders in perturbation theory would be plagued by divergences coming from OPE singularities. In Section~\ref{Secexp}, we will show how to compute these correlations for the circular defect non-perturbatively at large $\mu_D$ in a semiclassical limit using hyperbolic geometry.

\subsection{Energy transmission across the defect} \label{etrans}

A natural way to quantify how ``permeable'' a line defect is to energy flow is to study stress-tensor correlators with insertions on opposite sides of the defect \cite{Bachas:2001vj,Quella:2006de}. For a general conformal defect $L_\Sigma$ separating identical CFTs, conformal symmetry fixes the mixed two-point function, 
\begin{equation}
    \langle T(z_1)L_\Sigma T(z_2) \rangle =\frac{c_{LR}}{2(z_1-z_2)^4} \,,
\end{equation}
and allows one to define a dimensionless transmission coefficient $c_{LR}/c$ (together with a complementary reflection coefficient). 
The coefficient $c_{LR}$ is bounded $0\leq c_{LR}\leq c$ and is a measure of the fraction of incident energy transmitted across the defect \cite{Quella:2006de}. If $c_{LR}=c$, then the defect is perfectly transmitting whereas if $c_{LR}=0$, then the defect is perfectly reflecting. We now compute the stress tensor two-point function across $L_\Sigma$ to analyze the energy transmission across it. We place the stress tensor insertions on the imaginary axis at $z=\pm iy$ and the defect is along the real axis. We wish to compute the normalised two-point function defined below,
\begin{equation}
    \langle T(iy) T(-iy) \rangle_\Sigma \equiv \frac{\langle T(iy) T(-iy) L_\Sigma \rangle}{\langle L_\Sigma \rangle} \,.
\end{equation}
We will now evaluate the above two-point function for the pinning line defect
\begin{equation} \label{defcompact}
    L_\Sigma=\exp\left[\lambda \int_\Sigma \mathcal{O}\right]\,,
\end{equation}
in a compact $2d$ CFT constructed by integrating a scalar primary operator $\mathcal{O}$ of dimension $\Delta$. We will work perturbatively in the coupling $\lambda$. 
At the end, we will comment on applicability of this calculation to the Liouville defect and the subtleties due to IR divergences.

The leading contribution to the two-point function just comes from the TT-OPE and evaluates to $\frac{c}{32y^4}$. The contribution at $O(\mu_D)$ vanishes since the 3-point function $\langle T(iy) \mathcal{O}(x) T(-iy)\rangle =0$. This is because the operator $\mathcal{O}$ does not belong to the vacuum module. Therefore, expanding the normalised two-point function using a perturbative expansion in $\lambda$, we get
\begin{align}
     &\langle T(iy) T(-iy) \rangle_\Sigma = \nn\\
     &\qquad \langle T(iy) T(-iy)\rangle +\frac{\lambda^2}{2}\int_{-\infty}^{\infty} dx_1 \int_{-\infty}^{\infty} dx_2\ \langle T(iy)  \mathcal{O}(x_1)  \mathcal{O}(x_2) T(-iy)\rangle_{\text{conn}}+O(\lambda^3)\,.
\end{align}
The effect of dividing by $\langle { L}_\Sigma \rangle$ is to cancel the disconnected contribution to the 4-point function, 
\begin{align}
    &\langle T(iy)  \mathcal{O}(x_1)  \mathcal{O}(x_2) T(-iy)\rangle_{\text{conn}} \equiv \nn\\
    &\qquad\qquad \langle T(iy)  \mathcal{O}(x_1)  \mathcal{O}(x_2) T(-iy)\rangle- \langle T(iy) T(-iy) \rangle \langle  \mathcal{O}(x_1) \mathcal{O}(x_2) \rangle \,.
\end{align}
In Appendix \ref{AppTT}, we used the conformal Ward identities to derive an expression for the above connected 4-point function using which we get
\begin{equation}
    \frac{\langle T(iy)T(-iy) \mathcal{O}(x_1) \mathcal{O}(x_2)\rangle_{\text{conn}}}{\langle  \mathcal{O}(x_1) \mathcal{O}(x_2) \rangle}=\left [\frac{\Delta^2}{4}\frac{(x_1-x_2)^4}{(y^2+x_1^2)^2(y^2+x_2^2)^2}-\frac{\Delta}{4}\frac{(x_1-x_2)^2}{y^2(y^2+x_1^2)(y^2+x_2^2)}\right]\,.
\end{equation}
Substituting the form of the 2-point function, we evaluate the following two integrals,
\begin{equation}
    \begin{split}
        I_1(\Delta)=& \int_{-\infty}^{\infty}dx_1\int_{-\infty}^{\infty}dx_2\ \frac{|x_1-x_2|^{4-2\Delta}}{(y^2+x_1^2)^2(y^2+x_2^2)^2}=\frac{2^{1-2\Delta}\Delta(2\Delta-3)\pi^2}{\cos(\pi \Delta)y^{2+2\Delta}}\,,\\
        I_2(\Delta)=& \int_{-\infty}^{\infty}dx_1\int_{-\infty}^{\infty}dx_2\ \frac{|x_1-x_2|^{2-2\Delta}}{y^2(y^2+x_1^2)(y^2+x_2^2)}=-\frac{4^{1-\Delta}\pi^2}{\cos(\pi \Delta)y^{2+2\Delta}}\,.
    \end{split}
\end{equation}
The integral $I_1$ converges for $\frac{1}{2}<\Delta<\frac{5}{2}$ and the integral $I_2$ converges for $\frac{1}{2}<\Delta<\frac{3}{2}$. Hence, we have the final expression for the stress-tensor 2-point function across the defect (\ref{defcompact}), 
\begin{equation} \label{TexpT}
      \frac{\langle T(iy)\exp\left(\lambda \int_\Sigma \mathcal{O}\right) T(-iy) \rangle}{\langle \exp\left(\lambda \int_\Sigma \mathcal{O}\right)\rangle} =\frac{c/2}{(2y)^4}+\pi^2\lambda^2\frac{\Delta(\Delta^2(2\Delta-3)+2)}{\cos(\pi \Delta)}\frac{1}{(2y)^{2+2\Delta}}+O(\lambda^3)\,,
\end{equation}
and the result applies for $\frac{1}{2}<\Delta <\frac{3}{2}$ for which the $O(\lambda^2)$ correction is negative.
When $\Delta=1$, the 2-point function takes the same form as that of a conformal defect so we can compute the quantity $c_{LR}$ which governs the energy transmission,
\begin{equation} \label{cLR}
    c_{LR}=c-2\pi^2 \lambda^2 \,.
\end{equation}
The reflection coefficient that quantifies the fraction of incident energy reflected by the defect is given by\footnote{See also \cite{Brehm:2020agd, Brunner:2015vva} for calculations of reflection coefficient in defect perturbation theory. For the $\Delta=1$ marginal case, the result (\ref{cLR}) matches with the expression written in \cite{Brehm:2020agd} where the calculation was done for a circular defect. }
\begin{equation} \label{refstline}
    \mathcal{R}\equiv 1-\frac{c_{LR}}{c}=\frac{2\pi^2 \lambda^2}{c} \,.
\end{equation}

Now, we make a few comments about the result (\ref{TexpT}). The first term is the stress-tensor 2-point function when there is no defect. The second term is the leading correction in $\mu_D$ due to the defect. We see that this term is negative. Intuitively, we expect this because the defect reflects some part of the energy incident on it and hence screens the stress tensor correlations. The magnitude of the correction quantifies the amount of energy reflected off the defect.

Coming back to the defect \eqref{eq:defect_def} we study in the Liouville CFT and using the above results, we can show that the stress-tensor 2-point function across the Liouville defect (\ref{expdefect}) takes the following form,
\begin{align} \label{TexpTL}
    &\langle T(iy)  T(-iy) \rangle_\Sigma = \nn\\
    &\qquad\qquad \frac{c}{32y^4}-\frac{\mu_D^2}{128 y^{4+b^2}}2^{-b^2}(b^2+2)(b^6+3b^4+4)\pi^2 \sec\left(\frac{\pi b^2}{2}\right)\frac{\langle V_{b/2}(\infty)V_{b/2}(0)\rangle}{\braket{0}}+O(\mu_D^3)\,,
\end{align}
with the central charge $c=1+6(b+\frac{1}{b})^2$. The factor of $\frac{\langle V_{b/2}(\infty)V_{b/2}(0)\rangle}{\braket{0}}$ 
must be treated with care: it appears in the second term because neither the Liouville vacuum nor the vertex operator $V_{b/2}$ belongs to the normalizable spectrum. Moreover, both the numerator and the denominator are ill-defined due to a mixture of IR and UV divergences.\footnote{Note that the continuum normalization of Liouville vertex operators imply that the two-point function $\langle V_{b/2}(\infty)V_{b/2}(0)\rangle$ has a $\delta(0)$ IR divergence attributed to a zero-mode integral over a non-compact subgroup of PSL(2,$\mathbb{C}$) as explained in \cite{Harlow:2011ny} (see also \cite{Zamolodchikov:2005jb,Giribet:2022cvw}). In addition, the coefficient of $\delta(0)$ has a pole at $\alpha=\frac{b}{2}$ which comes from a non-perturbative UV divergence (more generally due to the pole in the DOZZ formula at $Q-\sum_i \alpha_i={1\over b}$) \cite{Chatterjee:2024phq}. The norm of the vacuum state which corresponds to the ${\rm S}^2$ partition function of Liouville CFT is also IR divergent due to non-compact zero modes (the divergence is proportional to the volume of ${\rm PSL}(2,\mathbb{C})$). In discussions of Liouville quantum gravity where conformal symmetry is gauged, we divide by the volume of the non-compact stabilizer subgroup which automatically gives a finite result for these divergent quantities in Liouville CFT on a fixed background \cite{Zamolodchikov:2005jb,Giribet:2022cvw}.} Later in Section~\ref{energytrans}, we will regulate these divergences by turning on a cusp angle on the defect and focus on the angle dependent part of the stress-tensor two-point function (and then taking the no cusp limit). Nonetheless, it would be desirable to find a more direct way to make sense of this ratio.

For the manifestly conformal defect (\ref{alphac}), the stress-tensor 2-point function takes the following form,
\begin{equation} \label{TmargTL}
    \langle T(iy)  T(-iy) \rangle_\Sigma =  \frac{c}{32y^4}-\frac{\pi^2 \mu_D^2}{16y^4}\frac{\langle V_{\alpha_c}(\infty)V_{\alpha_c}(0)\rangle}{\braket{0}} +O(\mu_D^3)\,.
\end{equation}
We see that (\ref{TexpTL}) and (\ref{TmargTL}) are ill-defined due to the normalization factors. In Section \ref{energytrans}, we will generalize the above calculation to introduce a cusp on the defect locus $\Sigma$ as a result of which the second term in (\ref{TmargTL}) will be a function of the cusp angle multiplying the ill-defined normalization factor. In this case, we can extract the physics governing the energy transmission across the defect by isolating the angle-dependent piece.

\subsection{Corrections to open string channel spectrum from the defect}

\begin{figure}
    \centering
    \begin{subfigure}{0.45\textwidth}
        \centering
        \includegraphics[width=0.7\textwidth]{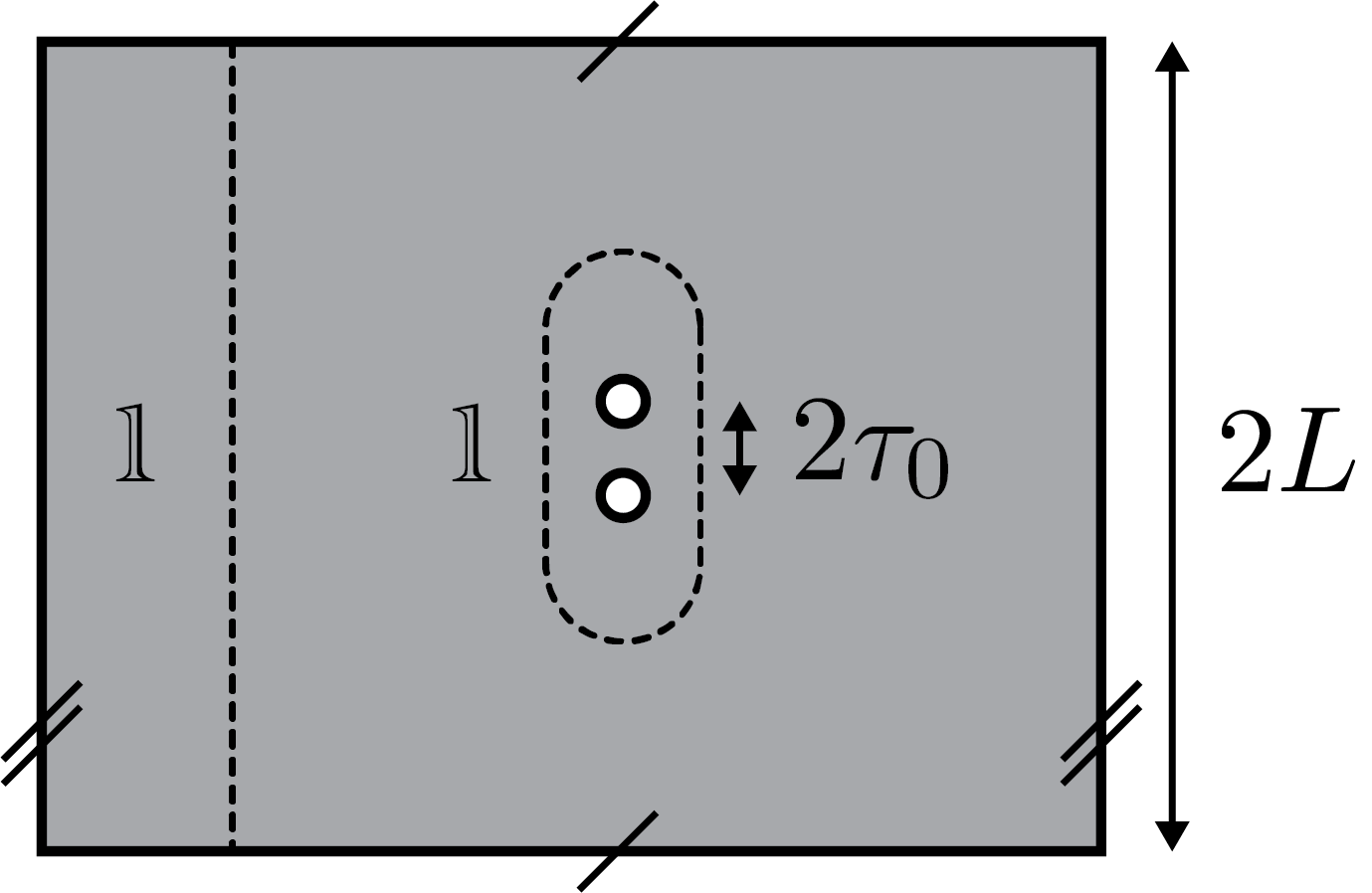}
        \caption{}
    \end{subfigure}
    \begin{subfigure}{0.45\textwidth}
        \centering
        \includegraphics[width=0.7\textwidth]{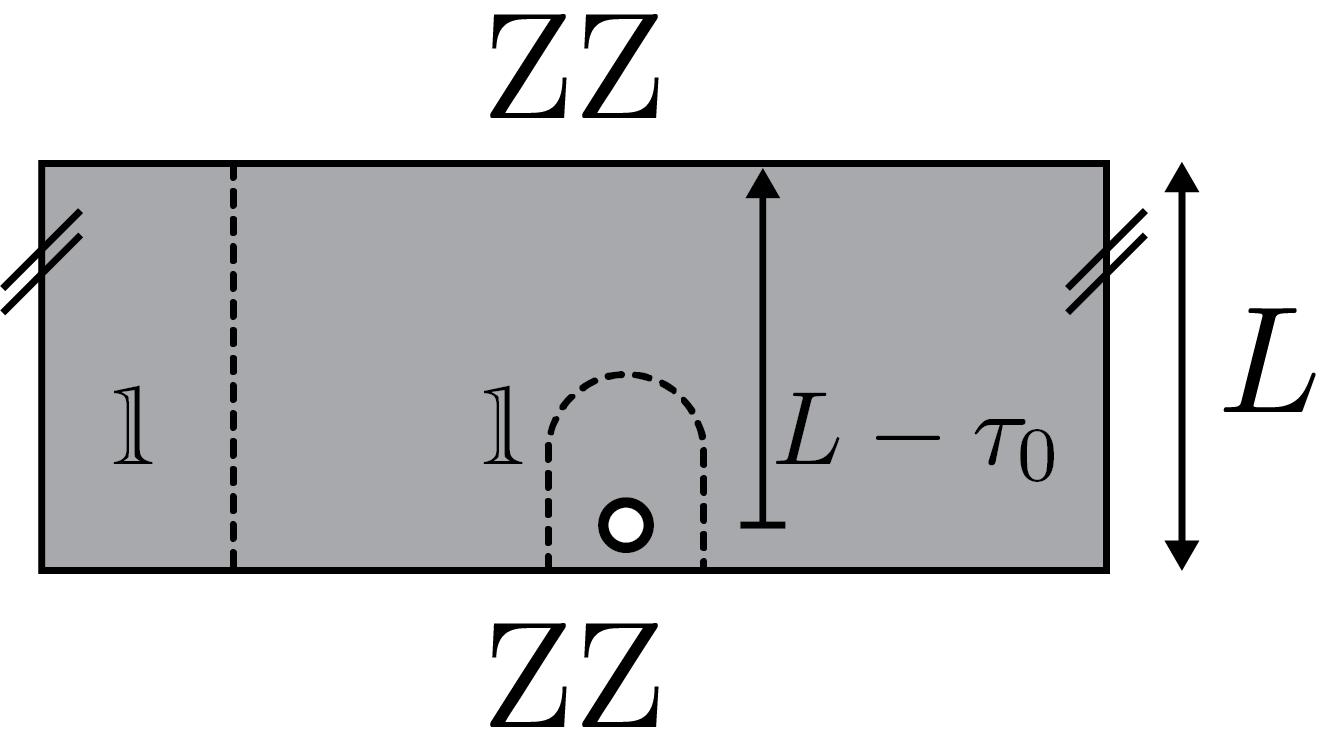}
        \caption{}
    \end{subfigure}
    \caption{This figure illustrates the `doubling trick' that relates boundary Liouville correlators to Virasoro blocks. The left figure defines a torus two-point Virasoro identity block by cutting the path integral of a general compact CFT and projecting to the vacuum module along the pair of non-contractible closed curves denoted by dotted lines. The figure on the right is the corresponding Liouville correlator on a cylinder with ZZ boundaries on either end which computes this block. The small white disks are local operator insertions. Only the identity and its Virasoro descendents propagate along the dotted line segments joining the ZZ boundaries.}
    \label{fig:torus_vacuum_block}
\end{figure}

Conformal boundary states in Liouville CFT have been classified into two families: ZZ \cite{Zamolodchikov:2001ah} and FZZT \cite{Fateev:2000ik}. The consistency of transition amplitudes between these states on a cylinder help determine the open string spectra on line segments stretched between these states. In this section, we compute the leading correction to the transition amplitudes coming from the insertion of the line defect which in turn also corrects the open string spectrum. For concreteness, we focus on the transition amplitudes between two ZZ states which evaluates to the vacuum character in the open string channel,
\begin{equation}
    \bra{\text{ZZ}}e^{-LH}\ket{\text{ZZ}}=\chi_{\id}\left(\frac{i\pi}{L}\right) \,,
\end{equation}
where $L$ is the height of the cylinder and $H=L_0+\overline{L}_0-\frac{c}{12}$ is the closed string Hamiltonian. This shows that that the open string spectrum between two ZZ states consists only of the identity and its Virasoro descendents. Now, we place the defect ${\bm L}_\Sigma$ parallel to the boundaries at a distance of $\tau_0$ from one of the boundaries. The transition amplitude then gets corrected as shown below,
\begin{equation}
\begin{split}
    \bra{\text{ZZ}}e^{-\tau_0 H}{\bm L}_\Sigma e^{-(L-\tau_0)H}\ket{\text{ZZ}}=&\chi_{\id}\left(\frac{i\pi}{L}\right)+2\pi \mu_D \bra{\text{ZZ}}e^{-\tau_0 H}V_{b/2} e^{-(L-\tau_0)H}\ket{\text{ZZ}}+O(\mu_D^2) \,,\\
    =& \chi_{\id}\left(\frac{i\pi}{L}\right)+2\pi \mu_D  \mathcal{F}_{\id,\id}(\tau_0,L) +O(\mu_D^2) \,.
    \end{split}
\end{equation}
Here we have used the fact that the ZZ-transition amplitude with a local operator evaluates to the chiral Virasoro identity block on a twice-punctured torus with moduli $\tau_0,L$,
\begin{equation}
    \bra{\text{ZZ}}e^{-\tau_0 H}V_{b/2} e^{-(L-\tau_0)H}\ket{\text{ZZ}}=\mathcal{F}_{\id,\id}(\tau_0,L)\equiv \inlinefig{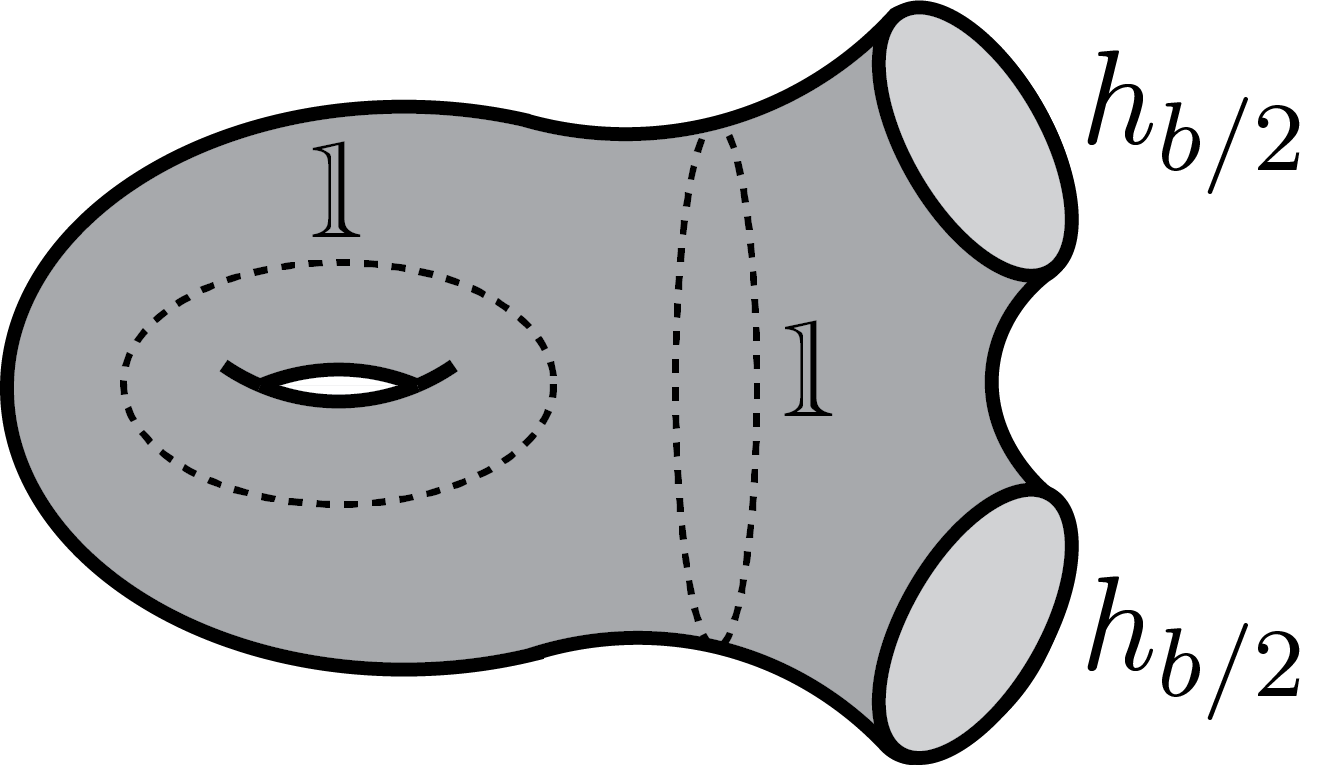} \,,
\end{equation}
where $h_{b/2}=\frac{1}{2}+\frac{b^2}{4}$ is the chiral weight of the vertex operator.
The identity block is defined by cutting open the path integral for the thermal 2-point function of the vertex operator $\langle V_{b/2}(\tau_0)V_{b/2}(-\tau_0)\rangle_{2L}$ at inverse temperature $\beta=2L$ and inserting the identity along the closed curves as shown in the Figure~\ref{fig:torus_vacuum_block}. Interpreting the cylinder amplitude in the open string channel,
\begin{equation}
    \text{Tr}_{\mathcal{H}_L}\left(e^{-2\pi H_{\text{open}}}\right)=\chi_{\id}\left(\frac{i\pi}{L}\right)+2\pi \mu_D  \mathcal{F}_{\id, \id}(\tau_0,L) +O(\mu_D^2) \,,
\end{equation}
where $H_{\text{open}}$ is the Hamiltonian in the open string channel and $\mathcal{H}_L$ is the open string Hilbert space on a segment of length $L$ in the presence of the defect. To leading order, the open string spectrum consists of the identity and its Virasoro descendents which gets corrected by the torus 2-point identity block.

\begin{figure}
    \centering
    \begin{subfigure}{0.49\textwidth}
        \centering
        \includegraphics[width=0.7\textwidth]{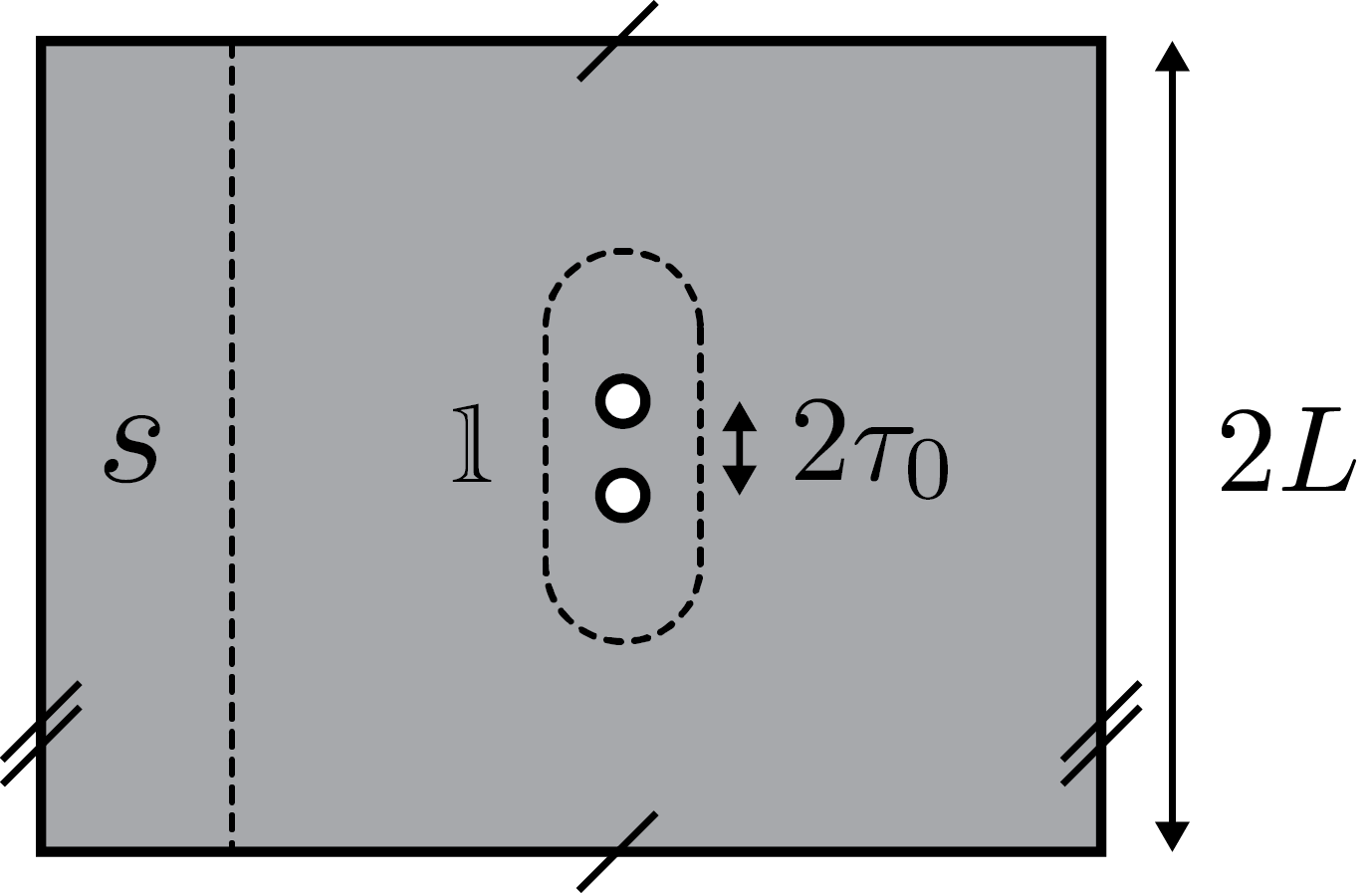}
        \caption{}
    \end{subfigure}
    \begin{subfigure}{0.49\textwidth}
        \centering
        \includegraphics[width=0.7\textwidth]{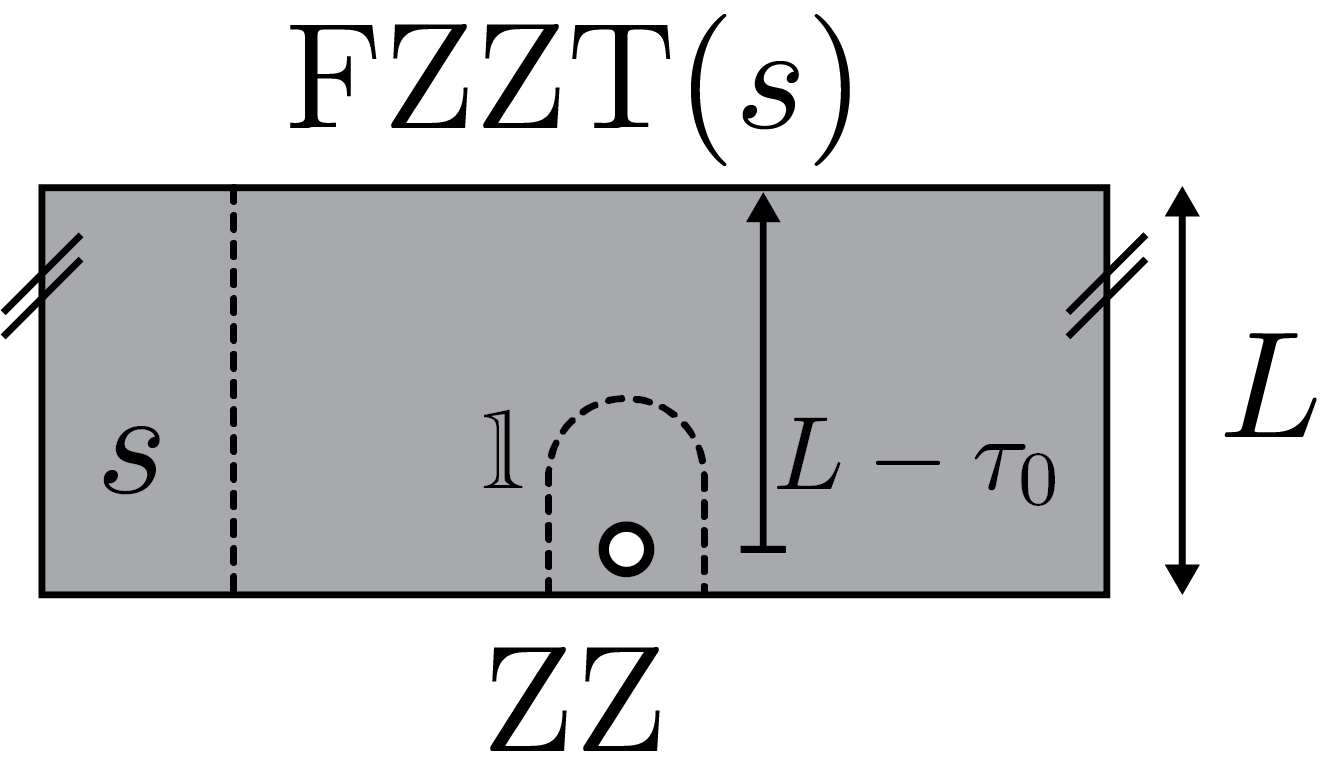}
        \caption{}
    \end{subfigure}
    \caption{This is another illustration of the `doubling trick' analogous to Figure~\ref{fig:torus_vacuum_block}. The left figure defines a Virasoro conformal block on the twice-punctured torus and the right figure is the Liouville correlator on the cylinder with ZZ and FZZT boundaries inserted on the two ends that computes this block. In the right figure, the primary $s$ with conformal weight $h_s=\frac{c-1}{24}+s^2$ and its Virasoro descendents propagate along the line segment joining the ZZ and FZZT boundaries, while the identity and its descendents propagate along the line segment joining the ZZ boundary.}
    \label{fig:torus_block}
\end{figure}

As another example, we will compute the correction to the transition amplitude between a ZZ and an FZZT state in the presence of the defect. Without the defect, it is knwon that the transition amplitude evaluates to give a non-vacuum Virasoro character,
\begin{equation}
    \bra{\text{ZZ}}e^{-LH}\ket{\text{FZZT}(s)}=\chi_{s}\left(\frac{i\pi}{L}\right) \,,
\end{equation}
where the character is associated with the conformal weight $h=\frac{c-1}{24}+s^2$. The leading correction in the presence of the defect is given by
\begin{align}
    &\bra{\text{ZZ}}e^{-\tau_0 H}{\bm L}_\Sigma e^{-(L-\tau_0)H}\ket{\text{FZZT}(s)} \nn\\
    &\qquad\qquad =\chi_{s}\left(\frac{i\pi}{L}\right)+2\pi \mu_D \bra{\text{ZZ}}e^{-\tau_0 H}V_{b/2} e^{-(L-\tau_0)H}\ket{\text{FZZT}(s)}+O(\mu_D^2) \,,\\
    &\qquad\qquad = \chi_{s}\left(\frac{i\pi}{L}\right)+2\pi \mu_D  \mathcal{F}_{\id,s}(\tau_0,L) +O(\mu_D^2) \,,
\end{align}
where we have used the fact that the transition amplitude between a ZZ and an FZZT state in the presence of a local operator evaluates to the torus 2-point conformal block by the doubling trick illustrated in Figure~\ref{fig:torus_block}, 
\begin{equation}
    \bra{\text{ZZ}}e^{-\tau_0 H}V_{b/2} e^{-(L-\tau_0)H}\ket{\text{FZZT}(s)}=\mathcal{F}_{\id,s}(\tau_0,L)\equiv\inlinefig{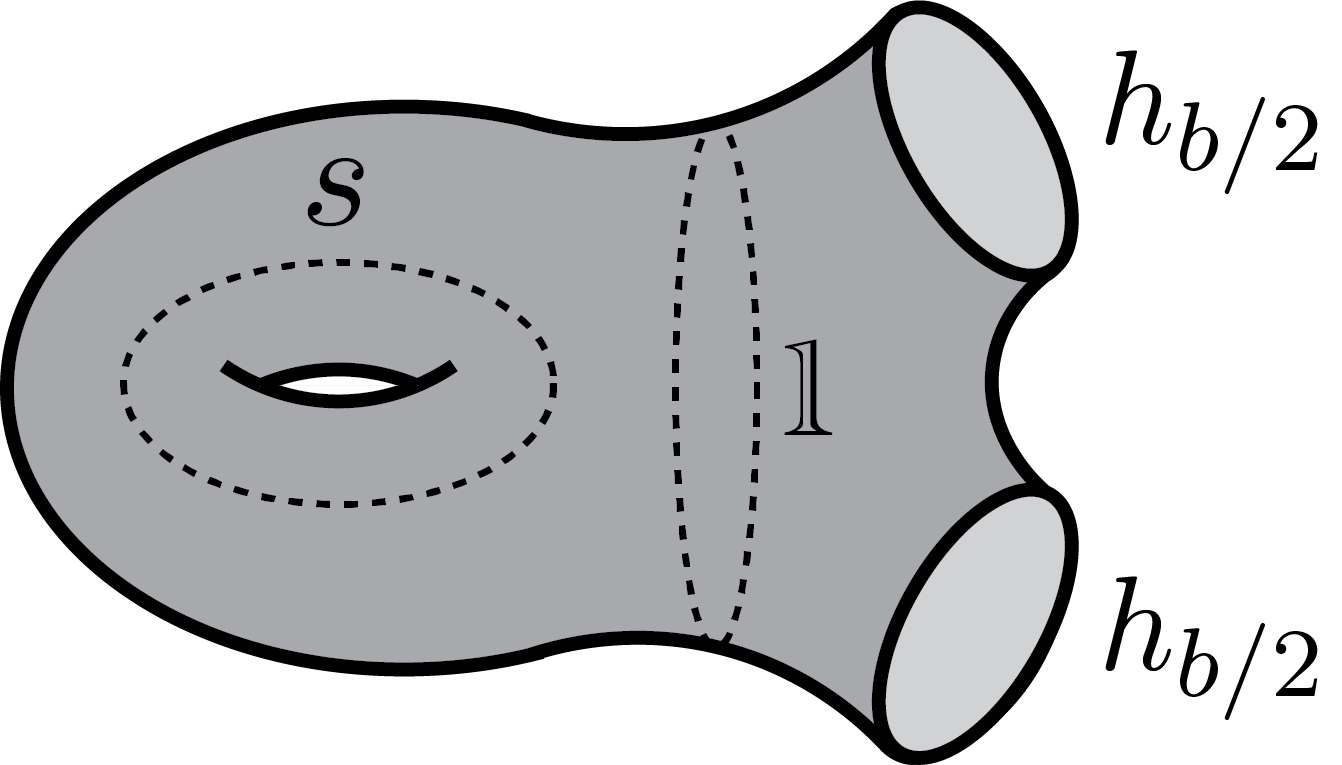}\,.
\end{equation}
In summary, we have shown in this section that the leading corrections to the ZZ-ZZ and ZZ-FZZT cylinder amplitudes and correspondingly, the leading corrections to the open string channel spectrum, due to the defect is captured by a Virasoro conformal block on the twice-punctured torus. The analysis readily generalizes to the manifestly conformal defect ${\bm L}_\Sigma(\alpha_c)$ (\ref{alphac}) with the holomorphic dimension of the external operator in the blocks replaced as $h_{b/2} \to h_{\alpha_c}=\frac{1}{2}$. We can also compute the corrections to the open string channel spectrum at higher orders in perturbation theory using the fact that the cylinder amplitudes evaluate to integrals (over the defect locus) of higher-point torus conformal blocks by the doubling trick. In Section \ref{sec:vacopen}, we compute the corresponding corrections to the open-string channel spectrum at large defect coupling in a semiclassical limit.

\subsection{Fusion of the defect with a conformal boundary}

We study the fusion of the defect with conformal boundaries perturbatively in $\mu_D$ and compute the Casimir energy of fusion at this order.\footnote{In general, when a conformal defect $\mathcal{D}$ is brought close to another defect or to a conformal boundary condition $B$, the fusion $\mathcal{D}\circ B$ can be defined from the strip (or annulus) partition function with $\mathcal{D}$ and $B$ inserted, and is dominated in the small–separation limit by the ground–state (Casimir) energy in the open channel. More precisely, the amplitude behaves as $Z_{\mathcal{D}B}(\ell)\sim e^{- {E^{(\mathcal{D},B)}_0}{/\ell}}$ for separation $\ell\to 0$, and the coefficient $E^{(\mathcal{D},B)}_0$, sometimes called the \emph{fusion Casimir energy}, is a universal, scheme–independent defect observable that controls the leading divergence in defect/boundary fusion and generalizes the familiar Casimir energy between parallel plates. See for instance \cite{Bachas:2007td,Konechny:2015wla,Diatlyk:2024defect,Brax:2024Casimir} for discussions of defect fusion and associated Casimir energies in two and higher dimensions.} To this end, we place the conformal boundary along the real axis and consider the limit in which the defect placed parallel to the boundary approaches it. The leading singularity comes at $O(\mu_D)$ where the defect can be replaced by the local operator. So, at this order, studying the fusion of the defect with a conformal boundary is equivalent to studying the fusion of a local operator with a conformal boundary.  Therefore, the 1-point function of the defect with a ZZ boundary is given by
\begin{equation} \label{LZZ}
    \langle {\bm L}_\Sigma \rangle_{\text{ZZ}}=1+\mu_D L \frac{U_{\text{ZZ}}(b/2)}{(2y)^{\Delta_{b/2}}}+O(\mu_D^2) \,.
\end{equation}
Here, $y$ is the transverse distance of the defect to the ZZ boundary, $L$ is an IR cutoff since the defect is along an infinite line, and $U_{\text{ZZ}}(b/2)$ is the one-point function of the vertex operator $V_{b/2}$ with the ZZ boundary. Using the expression for $U_{\text{ZZ}}(\alpha)$ derived in \cite{Zamolodchikov:2001ah},
\begin{equation}
    U_{\rm ZZ}(\alpha)=\frac{(\pi \mu_{\rm bulk}\gamma(b^2))^{-\alpha/b}\Gamma(Qb)\Gamma(Qb^{-1})Q}{\Gamma(b(Q-2\alpha))\Gamma(b^{-1}(Q-2\alpha))(Q-2\alpha)} \,,
\end{equation}
where $\gamma(x)=\Gamma(x)/\Gamma(1-x)$. This one-point function is normalized such that the cluster decomposition is satisfied, i.e., $U_{\text{ZZ}}(\alpha=0)=1$. Plugging in our defect gives
\begin{equation} \label{UZZb2}
    U_{\text{ZZ}}(b/2)=(\pi \mu_{\rm bulk}\gamma(b^2))^{-\frac{1}{2}}\left(1+b^2\right)\Gamma(b^2) \,.
\end{equation}
From this expression, we see that at this order, $ \langle {\bm L}_\Sigma \rangle_{\text{ZZ}}$ depends on the defect and bulk cosmological constants in terms of the ratio $\mu_D/\sqrt{\mu_{\rm bulk}}$, just like the observables with local operators. In the $b\to 0$ limit, the above expression has the following asymptotics,
\begin{equation}
     U_{\text{ZZ}}(b/2)\overset{b\to 0}{=} \frac{1}{\sqrt{\pi \mu_{\rm bulk}b^2}} \,.
\end{equation}
In this limit, since the defect becomes conformal, the above expression governs the Casimir energy of fusion of the defect with the ZZ boundary,
\begin{equation} \label{Casenergy}
    E_{\rm cas}\sim -\frac{\mu_D}{2\sqrt{\pi \mu_{\rm bulk}b^2}} \,.
\end{equation}
We see that the Casimir energy is negative, so the defect is attracted to the ZZ boundary. In Section~\ref{Seccasbound}, we compute this Casimir energy at large $\mu_D$ using semiclassical geometry. 

We can similarly study the fusion with other members of the ZZ family denoted using a pair of integers $\text{ZZ}_{(m,n)}$. Using the expression for 1-point function of a vertex operator with a $\text{ZZ}_{(m,n)}$ boundary derived in \cite{Zamolodchikov:2001ah},
\begin{equation}
     U_{(m,n)}(\alpha)=\frac{\sin(\pi b^{-1}Q)\sin(\pi bQ)\sin(\pi m b^{-1}(2\alpha-Q))\sin(\pi n b(2\alpha-Q))}{\sin(\pi m b^{-1}Q)\sin(\pi n b Q)\sin(\pi  b^{-1}(2\alpha-Q))\sin(\pi  b(2\alpha-Q))}U_{\rm ZZ}(\alpha) \,,
\end{equation}
we find that when $\alpha=\frac{b}{2}$,
\begin{align}
    U_{(m,n)}(b/2) &= (-1)^{m+1}n\frac{\sin(\pi b^2)}{\sin(\pi n b^2)}U_{\rm ZZ}(b/2) \,,\nn\\
    &=(-1)^{m+1}n \frac{\sin(\pi b^2)}{\sin(\pi n b^2)}(\pi \mu_{\rm bulk}\gamma(b^2))^{-\frac{1}{2}}\left(1+b^2\right)\Gamma(b^2) \,.
\end{align}
In the $b\to 0$ limit, 
\begin{equation} \label{casmn}
     U_{(m,n)}(b/2)\overset{b \to 0}{=} (-1)^{m+1} \frac{1}{\sqrt{\pi \mu_{\rm bulk}b^2}} \,.
\end{equation}
Interestingly, the result is independent of $n$.
Since this governs the Casimir energy of fusion, we see that for odd values of $m$, the Casimir energy is negative, so the defect is attracted to the boundary while for even values of $m$, the Casimir energy is positive which means the defect is repelled by the boundary. We have not discussed fusion with FZZT boundary due to the presence of a pole in the FZZT 1-point function $U_s(\alpha)$ \cite{Fateev:2000ik} at $\alpha=b/2$ and additional subtleties associated with the $b\to 0$ limit. However, for the manifestly conformal defect ${\bm L}_\Sigma(\alpha_c)$ (\ref{alphac}), both $U_{(m,n)}(\alpha_c)$ and $U_s(\alpha_c)$ are well-defined and respectively govern the Casimir energy of fusion with ZZ and FZZT boundaries for small $\mu_D$ at any value of the Liouville coupling, and in the $b\to 0$ limit, the Casimir energy of fusion with ZZ boundary reduces to (\ref{casmn}). In Section~\ref{Seccasbound}, we compute the corresponding Casimir energies for these defects at strong coupling using semiclassics.

\subsection{Effect of introducing a cusp on the defect locus} \label{Seccusp}

\begin{figure}
    \centering
    \begin{subfigure}{0.45\textwidth}
        \centering
        \includegraphics[width=0.55\textwidth]{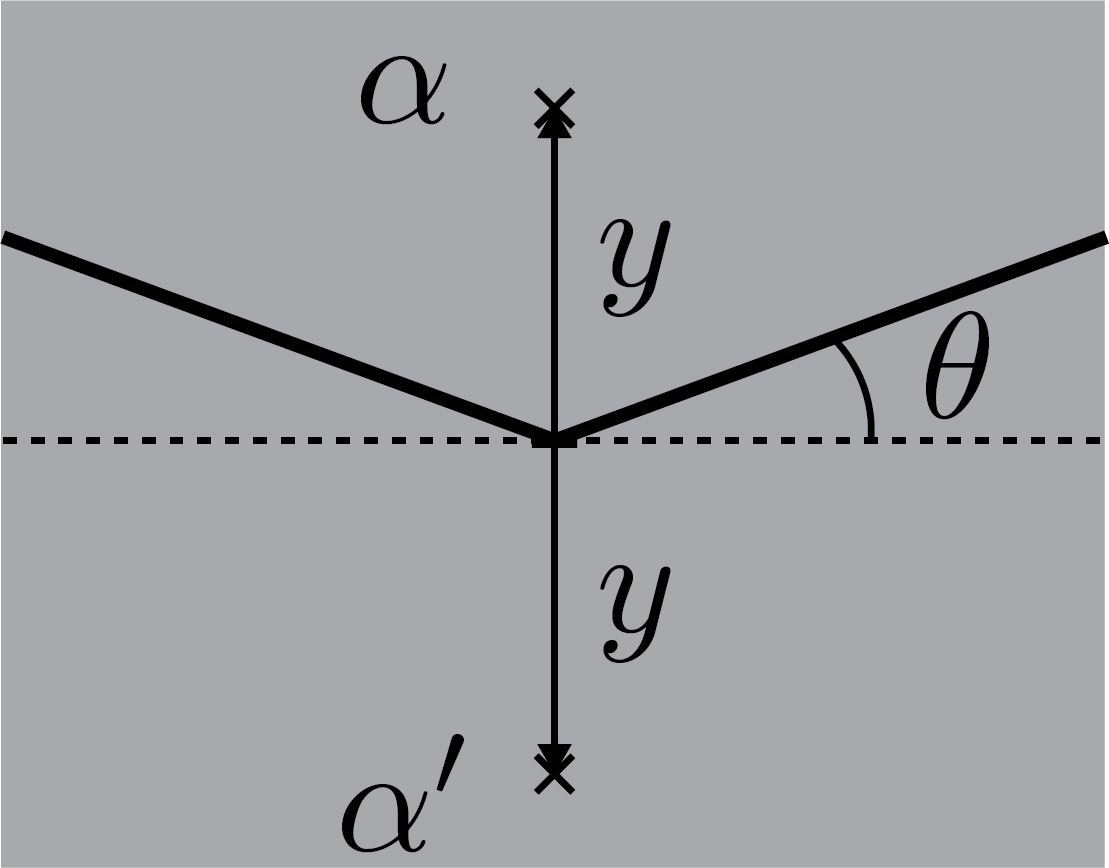}
        \caption{}
    \end{subfigure}
    \begin{subfigure}{0.45\textwidth}
        \centering
        \includegraphics[width=0.55\textwidth]{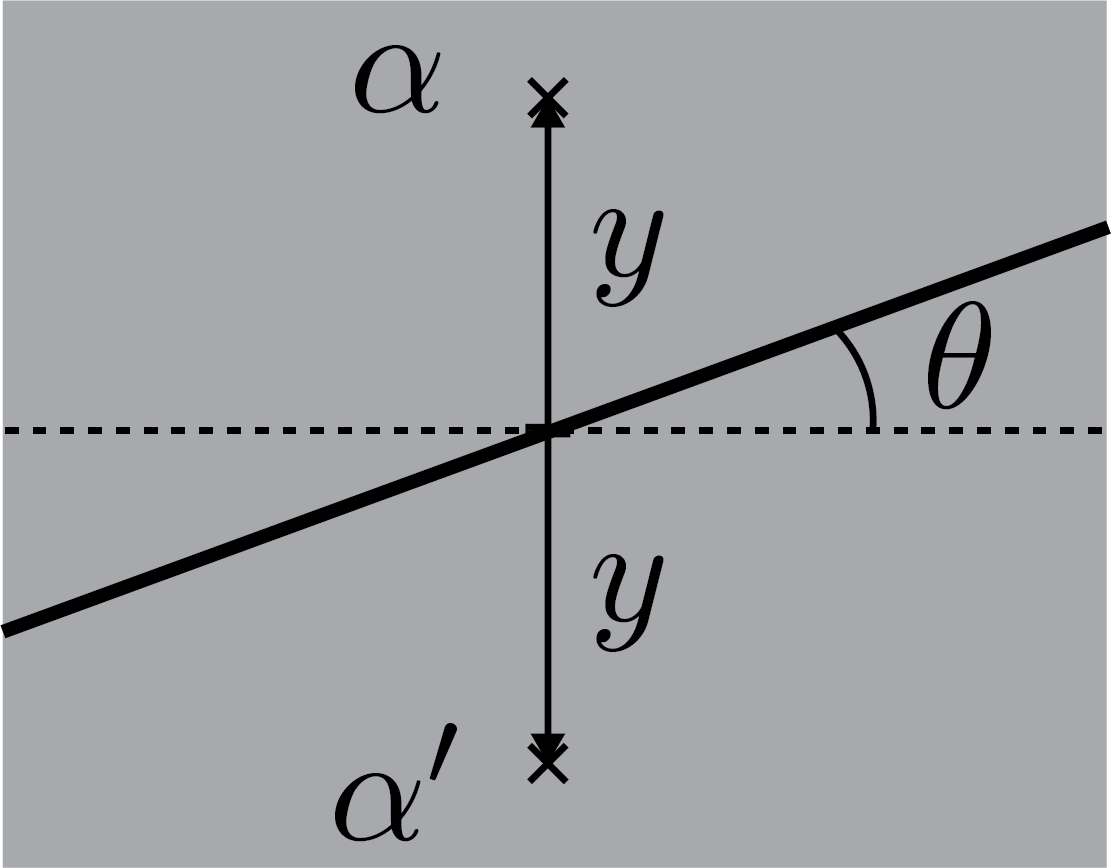}
        \caption{}
    \end{subfigure}
    \caption{The figure on the left describes the kinematics used to study the effect of introducing a cusp on the defect, on the correlation between the local vertex operators of Liouville momenta $\alpha$ and $\alpha'$ placed on the imaginary axis at $z=\pm iy$ respectively. The cusp has an opening angle $\phi=\pi-2\theta$. The figure on the right describes a similar setup but instead of introducing a cusp, we rotate the defect by the same angle $\theta$.}
    \label{fig:thetaloc}
\end{figure}

So far, we have computed observables placing the defect on an infinite straight line or a circle. So, it was assumed that the curve $\Sigma$ on which the defect is placed is smooth. Now, we introduce a cusp in $\Sigma$ and compute its effect on observables. In particular, we study the effect of introducing a cusp on the defect: (i) on fusion of the defect with a conformal boundary; (ii) on correlations of local operators across the defect; (iii) on energy transmission across the defect; and (iv) on information transmission across the defect. In the first two situations, there is a non-trivial effect of the cusp at linear order in perturbation theory while in the third and fourth situations, the effect starts at quadratic order. See Figures \ref{fig:theta}, \ref{fig:thetaloc}, \ref{fig:cuspreflection} and \ref{fig:cusp_replicas} for a sketch of the respective setups. In the absence of a boundary or local operator insertions, the leading contribution from the cusp occurs at quadratic order in perturbation theory. This setup is used to compute the cusp anomalous dimension defined as the coefficient of the logarithmically divergent piece in the free energy, and has been computed for various models of conformal line defects in the literature (See for example \cite{Polyakov:1980ca, Korchemsky:1987wg, Correa:2012at, Bianchi:2018scb, Cuomo:2024cusp, Giombi:2025cuspFermions}). In \cite{Correa:2012at}, the cusp anomalous dimension of a Wilson line was shown to capture the energy radiated by a quark in $\mathcal{N}=4$ SYM. It was shown in \cite{Cuomo:2024cusp} that the cusp anomalous dimension is a concave function of the cusp angle. Working at leading non-trivial order in perturbation theory, we will describe novel monotonicity properties of observables as a function of the cusp angle in each of the 4 setups described above. The final result in the presence of conformal boundaries is given in equation (\ref{cuspbound}), in the presence of local operators is given in equation (\ref{cusploc}), for the reflection coefficient is given in equation (\ref{refcusp}) and for the effective central charge is given in table \ref{tab:E2-cusp-6}.

\subsubsection{On fusion of the defect with a conformal boundary} \label{Seccuspbound}

\begin{figure}
    \centering
    \includegraphics[width=0.25\linewidth]{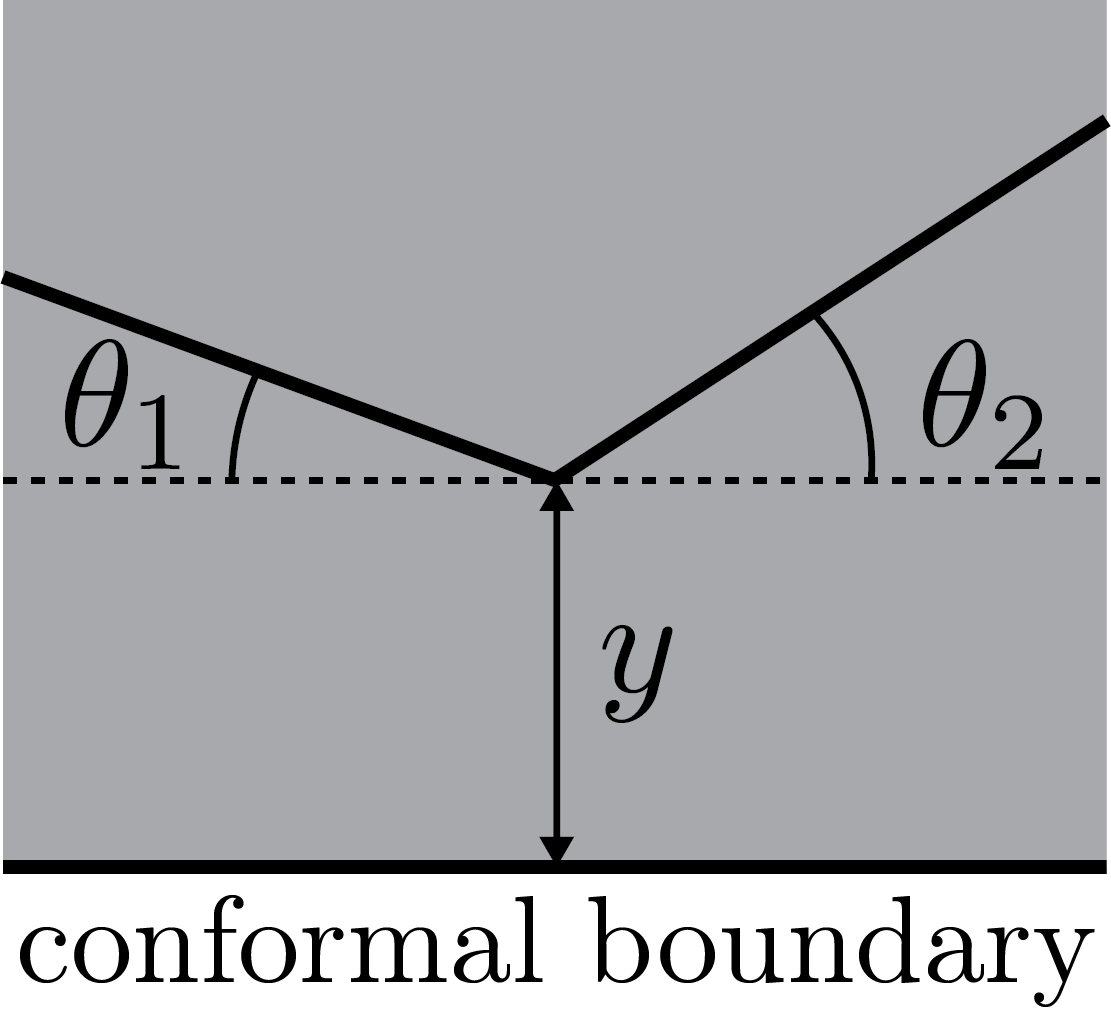}
    \caption{This figure describes the kinematics used to compute the 1-point function of the defect on the upper half-plane with a conformal boundary condition along the real axis. A cusp making angles $\theta_{1,2}$ with the real axis is introduced on the defect and the cusp is at a distance $y$ from the conformal boundary.}
    \label{fig:theta}
\end{figure}

First, we use the setup described in the previous subsection where we computed the 1-point function of the defect with a conformal boundary. Previously, $\Sigma$ was chosen to be parallel to the real axis but now we introduce a cusp such that the portions of $\Sigma$ on either side of the cusp make angles $\theta_{1,2}$ with the real axis as shown in Figure~\ref{fig:theta}. So, the 1-point function of the defect with a ZZ boundary in the presence of a cusp is given by 
\begin{equation} \label{cuspboundint}
     \langle {\bm L}_{\text{cusp}} \rangle_{\text{ZZ}}=1+\mu_D U_{\text{ZZ}}(b/2)\int_0^{\infty}dx \left( \frac{\sec(\theta_1)}{\left(2y+2x \tan(\theta_1)\right)^{\Delta_{b/2}}}+\frac{\sec(\theta_2)}{\left(2y+2x \tan(\theta_2)\right)^{\Delta_{b/2}}}\right )+O(\mu_D^2)\,.
\end{equation}
Here, $y$ is the distance of the cusp to the ZZ boundary.
The integral converges for $b>0$ for any non-zero cusp angles $\theta_{1,2}$. This shows that the IR divergence in (\ref{LZZ}) gets resolved by a cusp of small opening angle. Evaluating the integral, we find
\begin{equation}
\begin{split}
    \langle {\bm L}_{\text{cusp}} \rangle_{\text{ZZ}}=& 1+\frac{\mu_D}{2} \frac{U_{\text{ZZ}}(b/2)}{(\Delta_{b/2}-1)}\frac{1}{(2y)^{\Delta_{b/2}-1}}\left(\csc(\theta_1)+\csc(\theta_2)\right)+O(\mu_D^2) \,,\\
    = & 1+\frac{\mu_D}{\sqrt{\pi \mu_{\rm bulk}\gamma(b^2)}}\left(1+\frac{1}{b^2}\right)\Gamma(b^2)\frac{1}{(2y)^{\frac{b^2}{2}}}\left(\csc(\theta_1)+\csc(\theta_2)\right)+O(\mu_D^2) \,.
    \end{split}
\end{equation}
We used (\ref{UZZb2}) to arrive at the second line.
We see that the divergence as $y\to 0$ has also been softened due to the addition of the cusp. In the $b\to 0$ limit when the defect approaches conformality, the expression becomes independent of the distance of the cusp from the ZZ boundary,
\begin{equation} \label{cuspZZ}
    \langle {\bm L}_{\text{cusp}} \rangle_{\text{ZZ}} \overset{b \to 0}{=} 1+\frac{\mu_D}{\sqrt{\pi \mu_{\rm bulk}b^2}}\left(1+\frac{1}{b^2}\right)\left(\csc(\theta_1)+\csc(\theta_2)\right)+O(\mu_D^2)\,.
\end{equation}
Note that in the $b\to 0$ limit, $\Delta_{b/2}=1^+$. If instead the scaling dimension was exactly equal to $1$ which is the case for the manifestly conformal defect (\ref{alphac}), then the integral in (\ref{cuspboundint}) is logarithmically divergent, 
\begin{equation} \label{cuspalphaZZ}
    \begin{split}
        \langle {\bm L}_{\text{cusp}}(\alpha_c) \rangle_{\text{ZZ}} & = 1+\frac{\mu_D}{2}U_{\rm ZZ}(\alpha_c)\left(\csc(\theta_1)+\csc(\theta_2)\right)\log\left(\frac{L}{y}\right)+O(\mu_D^2)\\
        &   \overset{b \to 0}{=} 1+\frac{\mu_D}{2\sqrt{\pi \mu_{\rm bulk}b^2}}\left(\csc(\theta_1)+\csc(\theta_2)\right)\log\left(\frac{L}{y}\right)+O(\mu_D^2)\,.
    \end{split}
\end{equation}
Here, the length $L$ of the defect serves as an IR cutoff and the distance $y$ from the boundary serves as a UV cutoff. We see that the coefficient of the logarithmically divergent piece is governed by the Casimir energy of fusion of the defect with the ZZ boundary. For small angles $\theta \to 0$, we expect the coefficient to be $E_{\rm cas}/\theta$ upto numerical factors. For the simple angular dependence in (\ref{cuspalphaZZ}), this is obviously true. Comparing (\ref{cuspalphaZZ}) and (\ref{cuspZZ}), we see that the 1-point function of a cusped defect with a conformal boundary is sensitive enough to distinguish the two defects in the $b\to 0$ limit.

Although we presented the results for the Liouville line defect, the perturbative analysis is applicable to pinning defects in a compact $2d$ CFT constructed similarly by integrating scalar primary operators of scaling dimension $\Delta$. The logarithmic divergence of the cusped defect in the presence of a conformal boundary is a signature of marginal deformations. For a defect sourcing a relevant deformation, introducing a cusp would give rise to a power-law divergence in the perturbative analysis above. This shows that the cusp is a sensitive probe of the nature of the deformation sourced by the defect: power-law divergence in the IR cutoff scale for relevant deformation ($\Delta <1$), logarithmic divergence in the scale for a marginal deformation ($\Delta=1$), and independent of the IR cutoff for an irrelevant deformation ($\Delta>1$). We will now rewrite the result (\ref{cuspZZ}) in such a way that it applies more generally to conformal pinning defects in $2d$ CFTs constructed by integrating a scalar primary of dimension $1$ denoted $\mathcal{O}$, and a conformal boundary condition denoted $B$,
\begin{equation} \label{cuspbound}
    \log\left[\frac{\left\langle \exp\left(\lambda \int_{\text{cusp}(\theta_1,\theta_2)}\mathcal{O}\right)\right\rangle_B}{\langle \id \rangle_B}\right]=\frac{\lambda}{2} \frac{\langle \mathcal{O}\rangle_B }{\langle \id \rangle_B}\left(\csc(\theta_1)+\csc(\theta_2)\right)\log\left(\frac{L}{y}\right) + O(\lambda^2) \,.
\end{equation}
Here, the cusp is at a distance $y$ from the conformal boundary and $\theta_{1,2}$ are the two half-angles made by segments of the defect on either side of the cusp with the direction of the boundary. $\langle \mathcal{O}\rangle_B$ and $\langle \id \rangle_B$ are respectively the 1-point functions of the local operator $\mathcal{O}$ and identity in the presence of the conformal boundary $B$. A special limit of this expression is $\theta_{1,2}\to \frac{\pi}{2}$ in which case the defect is straddling the imaginary axis. In this limit, the coefficient of the logarithmically divergent piece becomes $\lambda  \frac{\langle \mathcal{O}\rangle_B }{\langle \id \rangle_B} $. This is the minimum value of the expression as a function of the two cusp angles. Another interesting feature of the above result is that the angular dependence is given by a sum of two \emph{completely monotonic} functions when expressed in terms of non-compact angular coordinates $x_{1,2}\equiv \tan^2(\theta_{1,2})$ so that $x_{1,2}\in (0,\infty)$.

The form of the above expression (\ref{cuspbound}) is suggestive of interpreting the coefficient of the logarithmically divergent piece as a generalization of the notion of cusp anomalous dimension to include the effect of conformal boundaries. In the absence of boundaries, the cusp anomalous dimension is usually expressed in terms of the cusp opening angle, but in the presence of boundaries, it is better to express it in terms of the two half-angles since the result does not just depend on the cusp opening angle but also on the orientation of the setup relative to the boundary. For the sake of comparison, we would like to mention that the cusp anomalous dimension (in the absence of a boundary) calculated in perturbation theory with two defects of coupling $\lambda$ meeting at an angle $\phi$ has a leading contribution at $O(\lambda^2)$ given by
\begin{equation}
    \Gamma_{\lambda \lambda}(\phi)=\lambda^2\left(1-\frac{\pi -\phi}{\sin(\phi)}\right) \,.
\end{equation}
See for example \cite{Cuomo:2024cusp} where the above result was derived for the pinning field defects in free field theory in $4d$ but the result is applicable to the present case of conformal pinning defects in $2d$ CFTs since the calculation only involves integrating the 2-point function. Notably, the above function is concave but not completely monotonic as a function of the cusp opening angle $\phi$. As explained in \cite{Cuomo:2024cusp}, the behaviour of the cusp anomalous dimension at small angles can be used to extract information about the dynamics of fusion of two defects at weak coupling. It turns out that defects with couplings $\lambda_1$ and $\lambda_2$ fuse into a defect with coupling $\lambda_{\text{eff}}=\lambda_1+\lambda_2$ and the Casimir energy associated with the fusion is $-\pi \lambda_1 \lambda_2$. The sign of the Casimir energy determines whether the defects attract or repel: Defects with the same sign of the coupling attract while defects with opposite signs of the coupling repel.

\subsubsection{On correlations of local operators across the defect} \label{Secloccusp}

We can similarly compute the effect of introducing a cusp on the correlation of local operators across the defect. We will again observe monotonicity in the cusp angle. Consider the setup discussed in Section~\ref{seccorrpert} with the vertex operators placed at $z=\pm iy$ and the defect along the real axis. Now, we introduce a cusp at the origin such that the two halves of the line each make an angle $\theta$ with the real axis. The three-point function takes the form,
\begin{align}
    \langle V_{\alpha}(iy) & {\bm L}_{\text{cusp}}  V_{\alpha'}(-iy)\rangle  \nn\\
    &= \frac{2\mu_D C_{\text{DOZZ}}(\alpha,\alpha',\frac{b}{2})}{(2y)^{\Delta_\alpha+\Delta_{\alpha'}-\Delta_{b/2}}}\int_0^{\infty}\frac{dx}{\cos(\theta)} \frac{(x^2\sec^2(\theta)+2xy\tan(\theta)+y^2)^{\frac{1}{2}(\Delta_{\alpha}-\Delta_{\alpha'}-\Delta_{b/2})}}{(x^2\sec^2(\theta)-2xy\tan(\theta)+y^2)^{\frac{1}{2}(\Delta_\alpha-\Delta_{\alpha'}+\Delta_{b/2})}}+O(\mu_D^2) \,.
\end{align}
The above integral converges since for large $x$ since the integrand decays as $\frac{1}{x^{2\Delta_{b/2}}}$ and $\Delta_{b/2}>1$.
This needs to be contrasted with the three-point function with the defect rotated about the origin by the same angle $\theta$. We denote the rotated defect by ${\bm L}_{\Sigma_{\theta}}$,
\begin{align}
    \langle V_{\alpha}(iy) & {\bm L}_{\Sigma_\theta}  V_{\alpha'}(-iy)\rangle  \nn\\
    &=\frac{\mu_D C_{\text{DOZZ}}(\alpha,\alpha',\frac{b}{2})}{(2y)^{\Delta_\alpha+\Delta_{\alpha'}-\Delta_{b/2}}}\int_{-\infty}^{\infty}\frac{dx}{\cos(\theta)} \frac{(x^2\sec^2(\theta)+2xy\tan(\theta)+y^2)^{\frac{1}{2}(\Delta_{\alpha}-\Delta_{\alpha'}-\Delta_{b/2})}}{(x^2\sec^2(\theta)-2xy\tan(\theta)+y^2)^{\frac{1}{2}(\Delta_\alpha-\Delta_{\alpha'}+\Delta_{b/2})}}+O(\mu_D^2) \,.
\end{align}
We see that when the vertex operators are identical, the three-point functions with the rotated defect and with the cusp on the defect give the same answer as expected from the kinematics. 

To extract some physics from these expressions, let us work with the special case of $\Delta_\alpha=\Delta_{\alpha'}$. Since we are interested in computing the correlation induced by the defect, we ignore the $\delta$-function term at $O(\mu_D^0)$. In this limit, the integral can be rewritten after a change of variables as 
\begin{equation} \label{VLcuspV}
    \langle V_{\alpha}(iy){\bm L}_{\text{cusp}}V_{\alpha}(-iy)\rangle =\frac{2^{2\Delta_{b/2}}\mu_D C_{\text{DOZZ}}(\alpha,\alpha,\frac{b}{2})}{(2y)^{2\Delta_\alpha+\Delta_{b/2}-1}} I(\Delta_{b/2},\theta) +O(\mu_D^2)\,,
\end{equation}
where $I(\Delta,\theta)$ is the following integral which can be evaluated in closed form,
\begin{equation} \label{cuspcoor}
\begin{split}
    I(\Delta,\theta)\equiv &\int_0^{\infty}\frac{dt}{(t^4+2\cos(2\theta)t^2+1)^{\Delta/2}} \,,\\
    =& \pi 2^{-\Delta}\frac{\Gamma(\Delta-\frac{1}{2})}{\Gamma(\frac{\Delta}{2})\Gamma(\frac{\Delta+1}{2})}{}_2F_1\left[\frac{1}{2},\Delta-\frac{1}{2};\frac{\Delta+1}{2};\sin^2(\theta)\right] \,.
    \end{split}
\end{equation}
If we further take the $\Delta \to 1$ limit of the above expression, the integral can be expressed in terms of the Elliptic-K function (complete Elliptic integral of the first kind\footnote{The Elliptic-K has an integral representation given by
\begin{equation}
    K(m)=\int_0^{\frac{\pi}{2}}\frac{d\phi}{\sqrt{1-m\sin^2(\phi)}} \,,
\end{equation}
This matches with the integral in the first line of (\ref{cuspcoor}) when $\Delta=1$ with $m=\sin^2(\theta)$. The easiest way to see this is to do the change of variables, $t=\tan(\frac{\phi}{2})$.
}),
\begin{equation} \label{IDelta1}
    I(\Delta=1,\theta)=K(\sin^2(\theta)) \,.
\end{equation}
This is a monotonically increasing function of the angle $\theta$ which means that the correlation induced by the defect increases monotonically as the opening angle of the cusp decreases (i.e., the cusp becomes sharper). The result (\ref{IDelta1}) implies that for the manifestly conformal defect (\ref{alphac}), the correlation across the cusp is given by
\begin{equation}
     \langle V_{\alpha}(iy){\bm L}_{\text{cusp}}(\alpha_c)V_{\alpha}(-iy)\rangle =\frac{4\mu_D C_{\text{DOZZ}}(\alpha,\alpha,\alpha_c)}{(2y)^{2\Delta_\alpha}} K(\sin^2(\theta)) +O(\mu_D^2)\,.
\end{equation}
In the $\theta \to 0$ limit, we recover the results of Section \ref{seccorrpert}.

More generally, we can state the above result for a conformal pinning defect constructed by integrating a scalar primary $\mathcal{O}$ of scaling dimension 1 in a $2d$ CFT, with a cusp of angles $\theta_{1,2}$ made by the two segments with the real axis, 
\begin{align} 
    \langle \mathcal{O}_i(iy) \exp\left(\lambda \int_{\text{cusp}(\theta_1,\theta_2)} \mathcal{O}\right) & \mathcal{O}_i(-iy) \rangle - \langle \mathcal{O}_i(iy)\mathcal{O}_i(-iy)\rangle = \nn\\
    &\frac{2\lambda}{(2y)^{2\Delta_i}}C_{ii\mathcal{O}}\left(K\left(\sin^2(\theta_1)\right)+K\left(\sin^2(\theta_2)\right)\right)+O(\lambda^2) \,.
\end{align}
Here $\mathcal{O}_i$ is a scalar primary with scaling dimension $\Delta_i$, $C_{ii\mathcal{O}}$ is the OPE coefficient between the external operators $\mathcal{O}_i$ and the operator $\mathcal{O}$. We have subtracted the two-point function of the external operators so that we can interpret the RHS as a measure of the correlation between the operators induced by turning on the defect perturbatively. Another way to extract the $O(\lambda)$ piece and get rid of the position dependence is
\begin{equation} \label{cusploc}
    \log\left[\frac{ \langle \mathcal{O}_i(iy) \exp\left(\lambda \int_{\text{cusp}(\theta_1,\theta_2)}\mathcal{O}\right)\mathcal{O}_i(-iy) \rangle}{\langle \mathcal{O}_i(iy)\mathcal{O}_i(-iy)\rangle}\right]=2\lambda C_{ii\mathcal{O}}\left(K\left(\sin^2(\theta_1)\right)+K\left(\sin^2(\theta_2)\right)\right)+O(\lambda^2) \,.
\end{equation}

\subsubsection{On energy transmission across the defect} \label{energytrans}

We now turn to computing the effect of a cusp on the energy transmitted across the defect. 
As we showed in Section~\ref{etrans}, the energy reflected by the defect at leading order in perturbation theory is calculated by the connected part of the 4-point function, $\langle T(iy) T(-iy)\int \mathcal{O}\int \mathcal{O}\rangle$, where $\mathcal{O}$ is a scalar primary with scaling dimension $1$. The subsequent analysis applies to the Liouville defect (\ref{expdefect}) in the $b\to 0$ limit and its manifestly conformal cousin (\ref{alphac}) for any $b$, and more generally to conformal pinning defects in $2d$ CFTs. The integration contours for the operator $\mathcal{O}$ are a pair of semi-infinite lines denoted $\Sigma_{1,2}$ originating at the cusp (placed at the origin) and making angles $\theta_{1,2}$ respectively with the positive real axis as shown in Figure~\ref{fig:cuspreflection}. In Appendix \ref{cuspTTOO}, we have computed this 4-point function and shown that it takes the form\footnote{Note that if we had introduced the cusp at any other location apart from the origin, then we would be introducing a new distance scale in addition to the distance scale $y$ between the stress tensor insertions. As a consequence, the 4-point function in (\ref{conn}) would also depend on the new distance scale and so it would not have a $\frac{1}{y^4}$ scaling.}
\begin{equation} \label{conn}
     \langle T(iy) T(-iy)\int_{\Sigma_1} \mathcal{O}\int_{\Sigma_2} \mathcal{O}\rangle_{\text{conn}}= -\frac{1}{4y^4}I(\theta_1,\theta_2) \,.
\end{equation}
Due to the symmetry of the setup, we assume $0< \theta_2 <\frac{\pi}{2}$, then the function $I(\theta_1,\theta_2)$ has the following piecewise definition depending on whether $\theta_1$ is greater or smaller than $\frac{\pi}{2}$ (since the function is ill-defined at $\theta_1=\frac{\pi}{2}$),
\begin{equation}
    I(\theta_1,\theta_2)=\begin{cases}
        \int_0^{\infty}dt\ \left(\frac{2e^{i(\theta_1-\theta_2)} t \log(t)+2it(\theta_2-\theta_1)e^{i(\theta_1-\theta_2)}+t^2-e^{2i(\theta_1-\theta_2)}}{(t+e^{i(\theta_1-\theta_2)})^3(t-e^{i(\theta_2-\theta_1)})}\right) \quad &{\rm if}\  \theta_2<\theta_1<\frac{\pi}{2}\,, \\
         \int_0^{\infty}dt\ \left(\frac{2e^{i(\theta_1-\theta_2)} t \log(t)+2it(\theta_2-\theta_1+\pi)e^{i(\theta_1-\theta_2)}+t^2-e^{2i(\theta_1-\theta_2)}}{(t+e^{i(\theta_1-\theta_2)})^3(t-e^{i(\theta_2-\theta_1)})}\right) \quad &{\rm if}\ \frac{\pi}{2}<\theta_1\leq \pi\,.
    \end{cases}
\end{equation}
In either patch, the function $I$ only depends on the difference in angles and we have checked numerically that the function is real and positive\footnote{Note that $I(\theta_1,\theta_2)$ can be made negative by allowing $\theta_2<0$ with $\theta_1<\frac{\pi}{2}$. In fact, when $\theta_1=\frac{\pi}{2}^-$ and $\theta_2=-\frac{\pi}{2}^+$ i.e., when $\Sigma_{1,2}$ are straddling the imaginary axis on the same side of the stress tensor insertions, $I(\theta_1,\theta_2)$ can be made arbitrarily negative due to a divergence in the $t$ integral.} for any value of the difference in angles and furthermore, it is monotonically decreasing as we increase $\theta_1$ holding $\theta_2$ fixed in each patch.
The function jumps across $\theta_1=\frac{\pi}{2}$ with the magnitude of jump being a function of $\theta_2$,
\begin{equation} \label{jumpI}
    I\left(\frac{\pi}{2}^+,\theta_2\right)-I\left(\frac{\pi}{2}^-,\theta_2\right)=\frac{\pi}{4}\left(\frac{2\theta_2-\sin(2\theta_2)}{\sin^3(\theta_2)}\right) \,.
\end{equation}
The above jump is always positive and increases monotonically with $\theta_2$ and ranges between $(\frac{\pi}{3},\frac{\pi^2}{4})$ with the minimum value occurring at $\theta_2=0$ and maximum at $\theta_2= \frac{\pi}{2}^-$.

\begin{figure}
    \centering
    \includegraphics[width=0.25\linewidth]{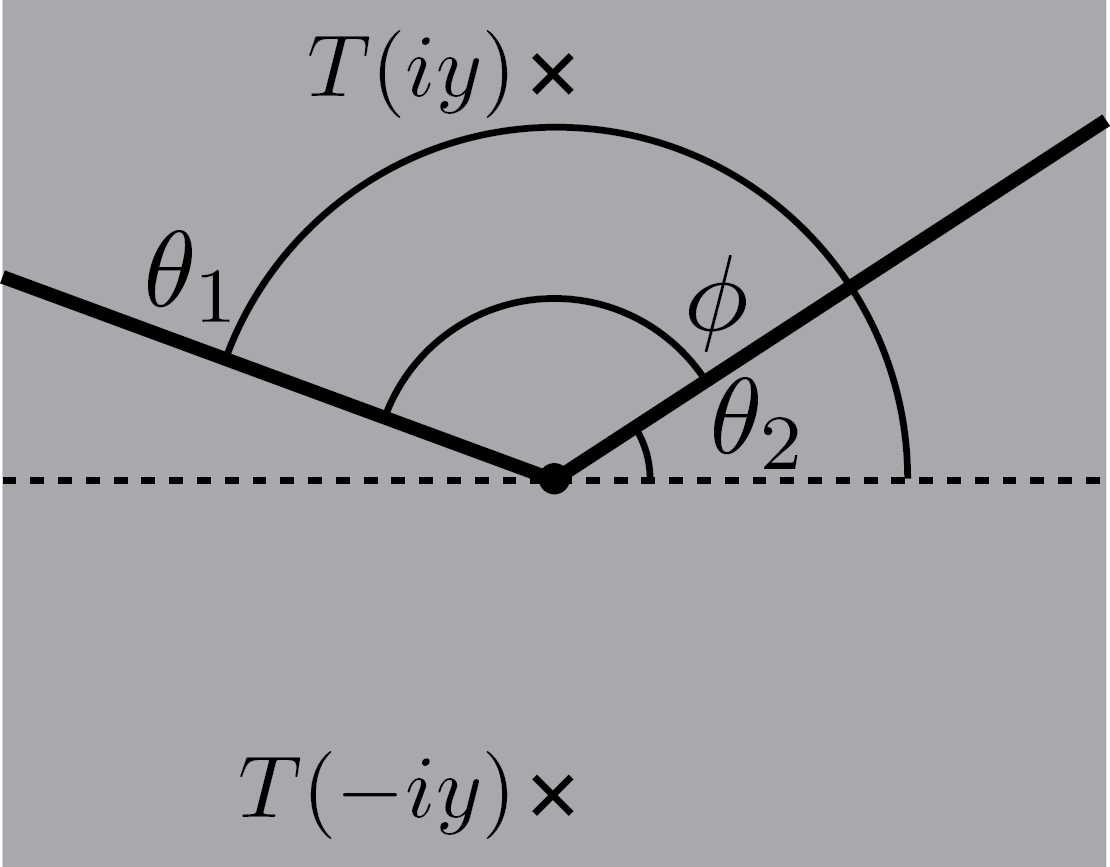}
    \caption{This figure describes the setup used to compute the reflection coefficient in the presence of a cusp. The holomorphic stress tensors are inserted at $z=\pm iy$. The angles made by the two semi-infinite defect lines is $\theta_{1,2}$ with the real axis. The angle $\phi$ is the opening angle at the cusp. }
    \label{fig:cuspreflection}
\end{figure}

The integrated 4-point function measures the energy reflected by the defect due to `cross-correlations' between $\Sigma_1$ and $\Sigma_2$. To compute the total energy reflected, we also need to compute the `self-correlations' along $\Sigma_1$ and $\Sigma_2$. To this end, note that the $\theta_1 \to \theta_2$ limit computes the contribution of self-correlations of the defect to the reflected energy since $\Sigma_1$ and $\Sigma_2$ coincide in this limit. 
It is sensible to define the reflection coefficient which measures the fraction of incident energy reflected by the defect only when $\theta_1>\frac{\pi}{2}$ and $\theta_2<\frac{\pi}{2}$ so that the rays $\Sigma_{1,2}$ are on either side of the stress tensor insertion. In this case, the reflection coefficient depends on the cusp opening angle. 
Therefore, for a defect with coupling $\lambda$, the reflection coefficient is given by 
\begin{equation} \label{refcusp}
    \mathcal{R}(\phi)=\frac{\lambda^2}{c}\left(\pi^2+4+8I(\phi)\right)+O(\lambda^3) \,,
\end{equation}
where $I(\phi)$ follows from the second line of $I(\theta_1,\theta_2)$ by defining $\phi \equiv \theta_1-\theta_2$,
\begin{equation}
    I(\phi)\equiv \int_0^{\infty}dt\  \left(\frac{2e^{i\phi} t \log(t)+2ite^{i\phi}(\pi-\phi)+t^2-e^{2i\phi}}{(t+e^{i\phi})^3(t-e^{-i\phi})}\right) \,.
\end{equation}
For the extreme values of $\phi$, the above integral can be evaluated to give a closed-form expression
\begin{equation}
    I(\phi)=\begin{cases}
         \frac{1}{8}(\pi^2 -4) \quad  &{\rm if}\ \phi =\pi\,, \\
         \frac{1}{8}(3\pi^2 +4) \quad  &{\rm if}\ \phi=0^+ \,,
    \end{cases}
\end{equation}
and we have verified numerically that $I(\phi)$ is monotonically decreasing as $\theta$ is increased.
The angular dependence in $\mathcal{R}(\phi)$ comes only from the cross-correlations since the self-correlations are independent of the angles. As a consistency check for the expression (\ref{refcusp}), when $\theta_1=\pi, \theta_2=0$, we recover the expression (\ref{refstline}) for the infinite straight line defect extending along the real axis. The minimum reflection occurs when the cusp angle is $\pi$ since the cross-correlations are minimal in this configuration. The maximum reflection occurs when $\phi\to 0$ ($\theta_2=\frac{\pi}{2}^-$ and $\theta_1=\frac{\pi}{2}^+$) i.e, when $\Sigma_{1,2}$ straddle the imaginary axis on either side. In summary, the reflection coefficient decreases monotonically as the cusp angle is increased,
\begin{equation} \label{Rmonotone}
    \partial_\phi \log(\mathcal{R}(\phi))<0\,,
\end{equation}
and is bounded in between
\begin{equation}
   \frac{2\pi^2 \lambda^2}{c} \leq \mathcal{R}(\phi)< \frac{4\lambda^2}{c}(\pi^2+2) \,.
\end{equation}
Despite the presence of divergent normalization factors in the reflection coefficient for the Liouville defects \eqref{TexpTL} and (\ref{TmargTL}), since those factors are independent of the cusp angle, the monotonicity result as stated in (\ref{Rmonotone}) is also applicable to the Liouville defect.

So far, we have assumed that the defect couplings along $\Sigma_{1,2}$ are same. More generally, if the defect couplings along $\Sigma_1$ and $\Sigma_2$ are $\lambda_1$ and $\lambda_2$ respectively (if $\lambda_1 \neq \lambda_2$, then an appropriate defect changing operator needs to be inserted at the cusp), then the reflection coefficient is given by 
\begin{equation} \label{refcusp2}
    \mathcal{R}(\phi)=\frac{1}{c}\left(\frac{1}{2}(\lambda_1^2+\lambda_2^2)(\pi^2+4)+8\lambda_1 \lambda_2 I(\phi)\right) \,.
\end{equation}
Since for small cusp angles\footnote{Numerically, we checked that for $\phi <0.8,\, 8I(\phi)>\pi^2+4$.}, the cross-correlation given by $8I(\phi)$ can exceed the self-correlation of $(\pi^2+4)$, when the defect couplings have the opposite signs, the reflection coefficient can become negative. This is the case when defects repel each other. This suggests that the defect setup could transmit more energy than what is incident on it when the defects meeting at a cusp repel each other. 

\subsubsection{On information transmission across the defect} \label{secperinf}

As a final illustration of the effect of introducing a cusp on the defect, we compute the information transmitted across the defect with a cusp. 
In $2d$ CFTs, the ground-state entanglement entropy of an interval of length \(L\) across a conformal interface separating two identical CFTs of central charge $c$ scales as \cite{Peschel:2005InterfaceEE}
\begin{equation}
  S_A = \frac{c_{\text{eff}}}{6}\,\log\frac{L}{\epsilon}\,,
\end{equation}
where \(c_{\text{eff}} \in [0,c]\) is the \emph{effective central charge} of the interface and \(\epsilon\) is a UV cutoff.\footnote{See also \cite{Brehm:2017cew,Quella:2006de,Sakai:2008tt,Calabrese:2004eu,Azeyanagi:2007qj}.} 
The quantity \(c_{\text{eff}}\) provides a quantitative measure of how much entanglement (or information) is transmitted across the defect: it reduces to the bulk central charge \(c\) for a topological, perfectly transmitting interface and vanishes for a totally reflecting one. 

\begin{figure}
    \centering
    \includegraphics[width=0.75\linewidth]{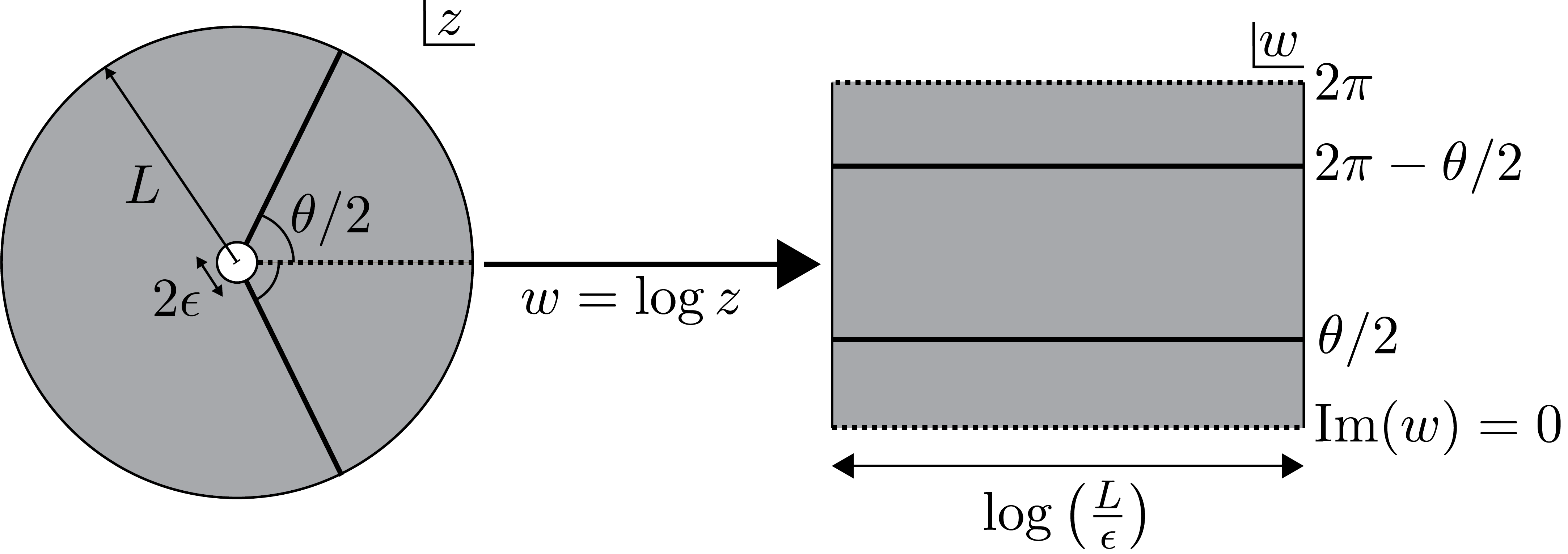}
    \caption{The figure on the left describes the setup used to compute the entanglement entropy of a semi-infinite line (drawn using a dashed brown line) across a defect with a cusp of angle $\theta$. $\epsilon$ and $L$ are UV and IR cutoffs. The figure on the right is obtained by a conformal map. The $n^{\rm th}$ replica saddle used to compute the entropy is obtained by cyclically gluing $n$ copies of this figure across the cuts.}
    \label{fig:cusp_replicas}
\end{figure}

We will describe the calculation for a conformal pinning defect $\exp\left[\lambda \int_\Sigma \mathcal{O}\right]$, where $\mathcal{O}$ is a scalar primary of dimension $1$ in a compact $2d$ CFT. At the end, we will comment on the applicability of the result to the Liouville line defects (\ref{expdefect}) and (\ref{alphac}).
For the setup described in Figure~\ref{fig:cusp_replicas}, we will use the replica trick to compute the Renyi entropies and analytically continue in the Renyi index to extract the entanglement entropy and hence the effective central charge. The leading non-trivial contribution to $c_{\text{eff}}$ comes at quadratic order in defect coupling. In the absence of the cusp i.e., when the cusp angle is $\pi$, this computation was done in \cite{Brehm:2020agd} and the result was shown to be
\begin{equation}
    c_{\text{eff}}(\theta=\pi)=c-\frac{3\pi^2}{2}\lambda^2+O(\lambda^3) \,,
\end{equation}
We will generalise this result to any cusp angle. Mapping the setup from the plane to cylinder via the conformal map $w=\log z$, the defect rays get mapped to the infinite lines $\text{Im}(w)=\frac{\theta}{2}, 2\pi-\frac{\theta}{2}$ and the cuts along the entangling interval map to $\text{Im}(w)=0,2\pi$. The length of the cylinder is $\log(L/\epsilon)$ where $L$ is the length of the entangling interval (an IR-cutoff) and $\epsilon$ is a UV-cutoff. The $n^{\text{th}}$-Renyi entropy is computed by replicating around the cuts to get a cylinder of cirumference $2n\pi$ with $2n$-defects placed symmetrically in $n$-pairs with a pairwise angular separation of $\theta$ and the angular separation between adjacent pairs being $2\pi-\theta$. To quadratic order in the defect coupling, the partition function of this $n$-sheeted cylinder is given by
\begin{equation} \label{replican}
    Z_n=\exp\left[\frac{c}{12n}\log\left(\frac{L}{\epsilon}\right)\right]\left(1+\frac{\lambda^2}{2}\sum_{\alpha_n,\alpha'_n}\int_{\log(\epsilon)}^{\log(L)}dxdx'\ \langle \mathcal{O}(x+i\alpha_n)\mathcal{O}(x'+i\alpha'_n)\rangle_{\text{cyl}_n} \right) \,,
\end{equation}
where $\alpha_n, \alpha_n'$ are angular positions of a pair of defects on the cylinder chosen from the set $\{2\pi m-\frac{\theta}{2}, 2\pi m+\frac{\theta}{2}\}$ where $m$ is an integer from $0$ to $n-1$. The 2-point function on the cylinder of circumference $2\pi n$ denoted $\text{cyl}_n$ can be evaluated using the plane-to-cylinder conformal map,
\begin{equation}
     \langle \mathcal{O}(x+i\alpha_n)\mathcal{O}(x'+i\alpha'_n)\rangle_{\text{cyl}_n} =\frac{1}{2n^2}\left(\frac{1}{\cosh((x-x')/n)-\cos((\alpha_n-\alpha_n')/n)}\right)\,.
\end{equation}
Due to the translational invariance of the 2-point function, the integrated 2-point function is extensive in the length of the cylinder\footnote{The summand in (\ref{doublesumreplica}) is divergent when $\alpha_n=\alpha_n'$. We discard the divergent piece and retain the finite piece in the evaluation of the replica partition function, i.e.,
\begin{equation}
    \frac{\pi-|\alpha_n-\alpha'_n|/n}{\sin\left(|\alpha_n-\alpha_n'|/n\right)}\to -1\,.
\end{equation}},
\begin{equation} \label{doublesumreplica}
     Z_n=\exp\left[\frac{c}{12n}\log\left(\frac{L}{\epsilon}\right)\right]\left(1+\frac{\lambda^2}{2n}\log\left(\frac{L}{\epsilon}\right)\sum_{\alpha_n,\alpha'_n} \left(\frac{\pi-|\alpha_n-\alpha'_n|/n}{\sin\left(|\alpha_n-\alpha_n'|/n\right)}\right) \right) \,.
\end{equation}
The entanglement entropy is then given by
\begin{equation}\label{eq:ee_replicas}
    \frac{c_{\text{eff}}}{6}\log\left(\frac{L}{\epsilon}\right)=\lim_{n\to 1}\frac{1}{1-n}\log\left(\frac{Z_n}{Z_1^n}\right) \,,
\end{equation}
Performing the sum in $Z_n$ numerically for various values of $n$ and interpolating to $n\to 1$, we find the value of the effective central charge for various cusp angles to be \ref{tab:E2-cusp-6}
\begin{table}[t]
\centering
\begin{tabular}{cc}
\hline
Cusp angle ($\theta/\pi$) & $(c_{\text{eff}}-c)/\lambda^2$ \\
\hline
1.00 & -14.8 \\
0.90 & -15.0\\
0.80 & -15.2 \\
0.70 & -15.5 \\
0.60 & -15.9 \\
0.50 & -16.4 \\
0.40 & -17.1 \\
0.30 & -18.0 \\
\hline
\end{tabular}
\caption{This table gives numerical estimates of the effective central charge $c_{\text{eff}}$ expressed as a deviation from the central charge $c$ for various cusp angles computed to quadratic order in the defect coupling $\lambda$. We extracted $c_{\rm eff}$ by numerically computing replica entropies $Z_n$ for $n\leq 100$ and a fixed $\theta$. Entanglement entropies were subsequently computed through (\ref{eq:ee_replicas}) by fitting a degree 25 spline to the replica entropies.}
\label{tab:E2-cusp-6}
\end{table}
As the cusp angle is decreased, the effective central charge decreases so it appears that more information is `reflected' by the defect with smaller cusp opening angle. This is consistent with the observation made in Section~\ref{energytrans} where we saw that the defect reflects more energy as the cusp angle is lowered.

Note that the above calculation was done perturbatively around the vacuum state.
Since Liouville CFT does not have a normalizable vacuum state, the above calculation of entanglement entropy is not well-defined. However, we could define a vacuum-subtracted entropy by stripping off the prefactor in (\ref{replican}) in the replica analysis. For the Liouville defect (\ref{expdefect}) in the $b\to 0$ limit or its manifestly conformal cousin (\ref{alphac}) for any $b$, the vacuum-subtracted entropy computed to $O(\mu_D^2)$ would give
\begin{equation}
    S-S_{\text{vac}}=\mu_D^2 \mathcal{N} f(\theta)\log\left(\frac{L}{\epsilon}\right) \,,
\end{equation}
where $\mathcal{N}$ is the normalization factor for the 2-point function of the respective vertex operators and $f(\theta)$ is the function governing the dependence on cusp angle whose numerical values are the same as the ones tabulated in \ref{tab:E2-cusp-6}. In Section~\ref{Secceff}, we will compute the effective central charge for these defects at large $\mu_D$ semiclassically and see that it is better-defined since we won't be doing perturbation theory around the vacuum.

\section{Strong defect coupling: Semiclassical hyperbolic geometries} \label{Secexp}

In the previous section, we analyzed various properties of the defects (\ref{expdefect}) and (\ref{alphac}) working perturbatively at small defect coupling $\mu_D$. At large $\mu_D$, the perturbative expansion breaks down and we would have to resum the perturbation series which is hard to do in practice despite the fact that the defect observables can be written down at any order in perturbation theory. In this section, we will instead use hyperbolic geometry to efficiently compute observables involving the defect in a strongly coupled double scaling limit where 
\begin{equation}
    \mu_D\to \infty\,,\quad b\to 0\,, \quad{\rm with}\ \mu_D b^2\ \text{fixed} \,.
\end{equation}
We refer to the above limit as a semiclassical limit where Liouville becomes a theory of hyperbolic metrics on Riemann surfaces. We implement this limit using the scaling 
\begin{equation}
    b\to 0\,, \quad{\rm with}\ 2b\phi=\Phi \ {\rm and}\ \pi \mu_{\text{bulk}}b^2=\frac{1}{4}\ {\rm fixed}\,.
\end{equation}
We refer to $\Phi$ as the classical Liouville field.
The semiclassical Liouville action evaluated on the closed domain $\mathcal{M}$,
\begin{align}
S =\frac{c}{6}S_L \quad{\rm with}\quad
S_L = \frac{1}{4\pi}\int_{\mathcal{M}} d^2z\left( \partial \Phi \bar{\partial}\Phi +e^{\Phi}\right)\,.
\end{align}
Varying the action gives the Liouville equation,
\begin{equation}
    \partial\overline{\partial}\Phi=\frac{e^{\Phi}}{2} \xleftrightarrow{} R[e^{\Phi}|dz|^2]=-2\,.
\end{equation}
To study the dynamics of the line defect ${\bm L}_{\Sigma}$\footnote{The defect (\ref{expdefect}) or its manifestly conformal cousin (\ref{alphac}) are equivalent in the semiclassical limit. Therefore, the results in this section apply to both the defects and we refer to both of them by the same notation ${\bm L}_\Sigma$.} in the semiclassical limit, we scale the defect cosmological constant as 
\begin{equation}
    \mu_D=\frac{\mu}{2\pi b^2}\,, \quad{\rm where}\ \mu>0\,.
\end{equation}
Inserting the line into the path integral deforms the Liouville action to include a source term,
\begin{align}
S_L = \frac{1}{4\pi}\int_{\mathcal{M}} d^2z\left( \partial \Phi \bar{\partial}\Phi +e^{\Phi}\right) - \frac{\mu}{2\pi} \int_\Sigma dx\ e^{\frac{\Phi}{2}} \,.
\end{align}
Varying this action, we get the Liouville equation with a $\delta$-function source on the line, 
\begin{equation}
    \partial\overline{\partial}\Phi=\frac{e^\Phi}{2}-\frac{\mu}{2}e^{\frac{\Phi}{2}}\delta(y)\,, \quad{\rm where}\ y=\text{Im}z \,.
\end{equation}
Integrating the equation in a collar around the defect (chosen to be placed at $y=0$ for convenience) yields the following jump condition in the normal derivative of the Liouville field,
\begin{equation} \label{jumpcond}
    \partial_y \Phi |_+-\partial_y \Phi |_-= -2\mu e^{\frac{\Phi}{2}} \,.
\end{equation}
The jump condition can also be written covariantly in terms of a jump in the extrinsic curvature (calculated with respect to the hyperbolic metric) as
\begin{equation} \label{jumpK}
    K_+-K_-=-\mu \,,
\end{equation}
where $K_+$ and $K_-$ are the extrinsic curvatures above and below the defect. 

In this section, we will construct various Liouville saddles by gluing hyperbolic geometries across the defect, and provide a physical interpretation for these saddles. The two basic building blocks needed to construct the saddles are the hyperbolic disk and the cylinder whose metric in a suitable coordinate system is given by
\begin{equation}
    \begin{split}
        ds^2_{\text{Disk}}=\frac{1}{\sinh^2(y)}(dx^2+dy^2)\,, &\qquad{\rm where}\ 0<y<\infty\,, \quad x\sim x+2\pi \,,\\
        ds^2_{\text{Cyl}}=\frac{r_H^2}{\cos^2(r_H y)}(dx^2+dy^2)\,, &\qquad{\rm where}\ -\frac{\pi}{2r_H}<y<\frac{\pi}{2r_H}\,, \quad x\sim x+2\pi \,.
    \end{split}
\end{equation}
\begin{figure}
    \centering
    \begin{subfigure}{0.45\textwidth}
        \centering
        \includegraphics[]{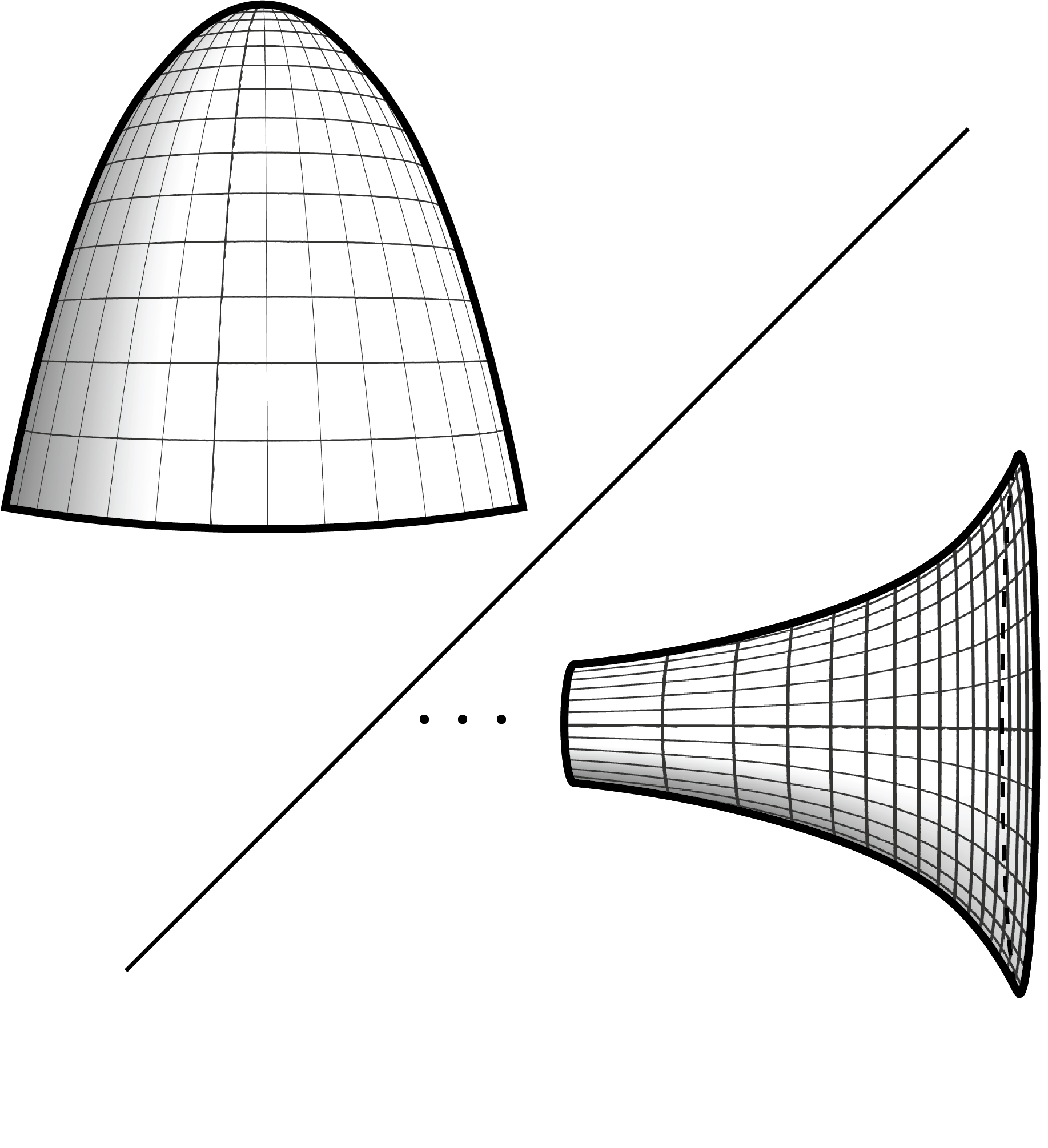}
        \caption{Disk}
    \end{subfigure}
    \begin{subfigure}{0.45\textwidth}
        \centering
        \includegraphics[]{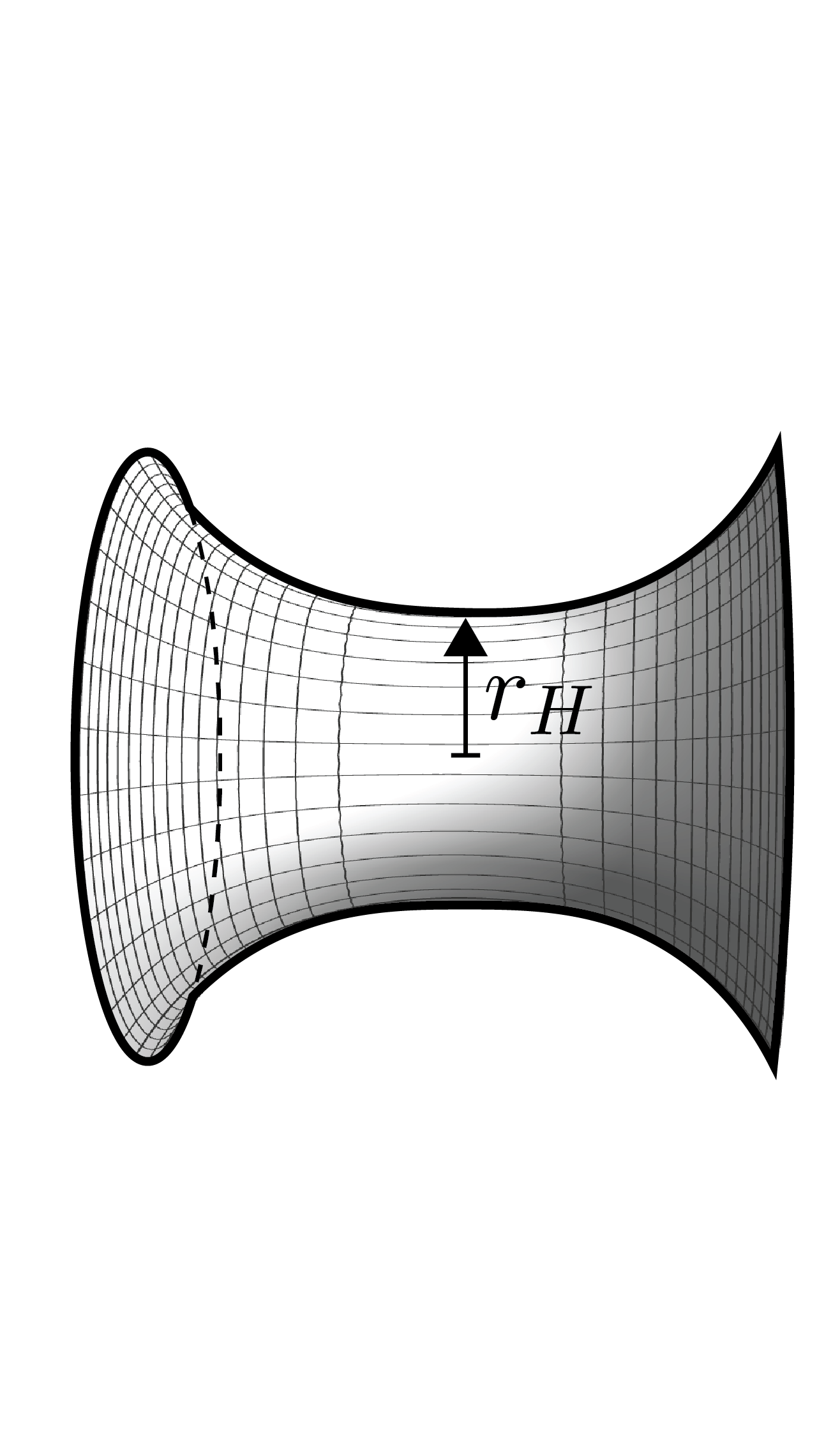}
        \caption{Cylinder}
    \end{subfigure}
    \caption{The figure shows the two building blocks used to construct Liouville saddles with the line defect drawn in the hyperbolic metric. Figure $(a)$ shows the hyperbolic disk drawn to scale in the disk and cylinder conformal frames. For the calculations in this section, we will be using the cylinder conformal frame. Figure $(b)$ shows the hyperbolic cylinder drawn to scale with the radius of the waist specified to be $r_H$.}
    \label{fig:metrics}
\end{figure}
We will use the following notation for a couple of geometric quantities that show up in the construction of Liouville saddles in the rest of the section: We denote the proper length of the defect (in the hyperbolic metric) as $2\pi r_0$ and the length of the minimal geodesic around the waist of the cylinder as $2\pi r_H$. The stress energy localized on the defect can be expressed in terms of the defect cosmological constant and the proper length as
\begin{equation} \label{stressdef}
    T(z) \bigg|_{\text{defect}}=\frac{\mu r_0}{2}\delta (\Sigma) \,,
\end{equation}
where $\delta(\Sigma)$ denotes a delta function localised on the defect. The length parameter $r_0$ is dynamically determined by solving the Liouville equation and is affected by the presence of other defects. This means the stress energy on the defect is not universal and is affected by the presence of other defects.

\subsection{Conformality in the semiclassical limit} \label{Secconformality}

In this section, we verify that the defect is conformal by checking that the Liouville stress tensor obeys the Cardy gluing condition across the defect. We will exploit some features of conformality of the defect in the subsequent calculations in this section.
First, notice that the defect is scale invariant in the $b\to 0 $ limit, because under a scale transformation, the Liouville field shifts by a logarithm of the Jacobian of the transformation which cancels the Jacobian arising from the integration measure.
We also verify that the defect satisfies the local conformal gluing condition at the classical level using its equations of motion (\ref{jumpcond}). To this end, we need to verify the continuity of the difference between the left-moving and right-moving stress tensors across the defect,
\begin{equation}
    (T-\overline{T})_+=(T-\overline{T})_- \,.
\end{equation}
This is the condition that there is no net energy flux across the defect. This condition was first formulated in the context of surface critical phenomena and boundary CFT by Cardy \cite{Cardy:1984bb} and has since become the standard defining property of conformal defects/ interfaces.
We now show that this is true when the jump condition for the Liouville field is satisfied. Let us choose coordinates $z=x+iy$ and $\overline{z}=x-iy$ where $x$ is tangential to the defect and $y$ is normal to the defect. Expressing the stress tensors in this coordinate system, we see that
\begin{equation}
    T=\frac{1}{2}\partial^2\Phi-\frac{1}{4}(\partial\Phi)^2\implies T-\overline{T}=-\frac{i}{2}\partial_x\partial_y\Phi+\frac{i}{4}\partial_x\Phi \partial_y \Phi \,.
\end{equation}
Now consider the difference in the energy fluxes across the defect,
\begin{equation}
\begin{split}
    (T-\overline{T})_+- (T-\overline{T})_-=-\frac{i}{2}\partial_x(\Delta (\partial_y \Phi))+\frac{i}{4}\partial_x\Phi\Delta (\partial_y\Phi)=\frac{i}{2}\mu e^{\frac{\Phi}{2}}\partial_x\Phi-\frac{i}{2}\mu e^{\frac{\Phi}{2}}\partial_x\Phi=0 \,,
    \end{split}
\end{equation}
Here, we have used the jump condition across the defect (\ref{jumpcond}).
Therefore, we have shown that the defect obeys the conformal gluing condition.

\subsection{Vacuum expectation value} \label{Secvev}

We compute the vacuum expectation value of the defect, $\langle {\bm L}_\Sigma \rangle$ in the semiclassical limit by constructing a saddle on the sphere with the defect placed along its equator.
The Liouville action with a single line defect on the sphere is given by 
\begin{align}
S_L = \frac{1}{4\pi}\int_{\Gamma} d^2z\left( \partial \Phi \bar{\partial}\Phi +e^{\Phi}\right) - \frac{\mu}{2\pi} \int_\Sigma dz\ e^{\frac{\Phi}{2}}+ \frac{1}{4\pi}\int_{\Gamma_+\sqcup \Gamma_-} \!\! dz\ \Phi + T+2(1-\log2) \,,
\end{align}
where $\Gamma$ is the domain $\text{Im}z\in (-T,T)$ with the boundaries denoted by $\Gamma_\pm$. The other terms are added to ensure that the action has a good variational principle and is finite \cite{Chandra:2024vhm}.
The Liouville solution takes the form
\begin{align}
\Phi = \begin{cases}
-2\log \sinh\left(y+ A\right) &\qquad {\rm if}\ y > 0 \,,\\
-2\log \sinh\left( A- y \right) &\qquad {\rm if}\ y< 0 \,,
\end{cases}
\end{align}
where the junction condition reads
\begin{equation}
    2\sqrt{1+r_0^2}=\mu r_0 \quad{\rm and}\quad A=\sinh^{-1}(\frac{1}{r_0}) =\cosh^{-1}(\frac{\mu}{2})\,.
\end{equation}
A classical solution exists only if $\mu> 2$. This should be interpreted as a Gauss-Bonnet type constraint which requires the defect to be sufficiently `heavy' to backreact on the sphere to produce a hyperbolic metric. The stress tensor for this solution takes the form
\begin{equation}
    T^{\Phi}(z)=\frac{1}{2}\partial^2 \Phi-\frac{1}{4}(\partial\Phi)^2=\frac{1}{4}+\frac{\mu r_0}{2}\delta(y)=\frac{1}{4}+\frac{\mu}{\sqrt{\mu^2-4}}\delta(y) \,.
\end{equation}
The on-shell Liouville action evaluates to give
\begin{equation}
    S_L=-2A=-2\cosh^{-1}\left(\frac{\mu}{2}\right)=-2\log\left (\frac{\mu+\sqrt{\mu^2-4}}{2} \right ) \,.
\end{equation}
This means that the vacuum expectation value sometimes referred to as the  $g$-function \cite{Affleck:1991tk, Cuomo:2021rkm} of the line defect is given by
\begin{equation}
    \log g \equiv \log \langle {\bm L}_\Sigma \rangle= \frac{c}{3}\log\left (\frac{\mu+\sqrt{\mu^2-4}}{2} \right ) \,.
\end{equation}
Note that for the $\mu=2$ line, the $g$-function is unity because in this limit, $r_0 \to \infty$ so $A=0$.

\begin{figure}
    \centering
    \includegraphics[width=0.65\linewidth]{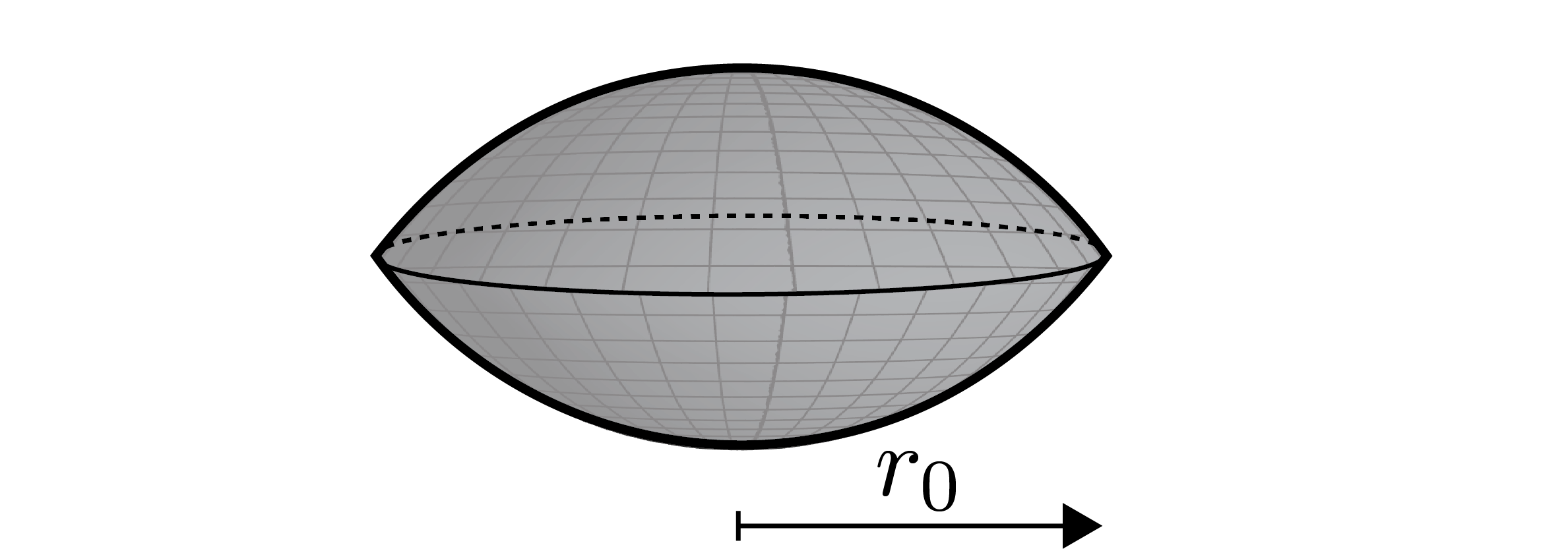}
    \caption{The figure shows the Liouville saddle used to compute the vacuum expectation value of the defect. The saddle is constructed by gluing two hyperbolic disk across the defect, which is along the equator of the squashed sphere shown above. The proper length of the defect is $2\pi r_0=\frac{4\pi}{\sqrt{\mu^2-4}}$.}
    \label{fig:vacuum_expectation}
\end{figure}

It is instructive to relate the above Liouville solution to uniformization of the two patches of the cylinder above and below the defect,
\begin{equation}
    e^{\Phi(z,\overline{z})}=\frac{4|w'(z)|^2}{(1-|w(z)|^2)^2} \,.
\end{equation}
Here $w(z)$ is the holomorphic uniformization map from a patch of the cylinder into the unit Poincare disk. For the present case, the two uniformization maps respectively into the interior and exterior of the disk are
\begin{equation}
    \begin{split}
        w_+(z)=e^{-A}e^{iz} &\qquad {\rm for}\ y>0 \,,\\
        w_-(z)=e^{+A}e^{iz} &\qquad {\rm for}\ y<0 \,.
    \end{split}
\end{equation}
On the defect ($y=0$), the two maps are related by a rescaling, 
\begin{equation}
    w_+(x)=e^{-2A}w_-(x) \quad{\rm with}\quad G=\begin{bmatrix}
      e^{-A}& 0\\
      0 & e^{A}
    \end{bmatrix}\,,
\end{equation}
where we have expressed the Mobius transformation in terms of a gluing matrix $G$. Observe that $G$ belongs to the hyperbolic conjugacy class of PSL($2,\mathbb{R}$).

\subsection{Vacuum energy of the defect Hilbert space} \label{Sectorus1point}

We compute the vacuum energy of the defect Hilbert space by constructing the saddle for the thermal 1-point function of the defect $\text{Tr}(e^{-\beta H}{\bm L}_{\Sigma})$ on the torus and use modular invariance to interpret the partition function using the defect Hilbert space.
The Liouville solution used to compute the one-point function of the line defect on the torus (whose fundamental domain is chosen to be $z\sim z+2\pi \sim z+i\beta$) is constructed by gluing the ends of a finite hyperbolic cylinder of height $\beta$ and proper length of the waist $2\pi r_H$, and is given by
\begin{equation}
    \Phi=-2\log \left (\frac{1}{r_H}\cos(r_H y)\right) \,.
\end{equation}
Here, $y\in (-\frac{\beta}{2},\frac{\beta}{2})$ with $y=\pm \frac{\beta}{2}$ identified and $r_H$ is determined by solving the junction conditions across the defect placed at $y=\frac{\beta}{2}$,
\begin{equation}
    2\sqrt{r_0^2-r_H^2}=\mu r_0 \implies r_0=\frac{2r_H}{\sqrt{4-\mu^2}}=\frac{r_H}{\cos(r_H\frac{\beta}{2})} \,,
\end{equation}
meaning $r_H$ is thus given by
\begin{equation}
    r_H=\frac{2}{\beta}\sin^{-1}(\frac{\mu}{2}) \,.
\end{equation}
For the solution to be real, we require $0<\mu<2$. This is complementary to the allowed range on the sphere. Notice from the obtained value of $r_H$ that the argument of the cosine in the expression for the Liouville field is always in $(-\frac{\pi}{2},\frac{\pi}{2})$ so we have a real classical saddle.
We now compute the on-shell Liouville action for this solution,
\begin{equation}
    S_L = \frac{1}{4\pi} \int_{\Gamma}d^2z(\partial \Phi \bar{\partial}\Phi + e^{\Phi})  - \frac{\mu}{2\pi } \int_\Sigma dz\ e^{\frac{\Phi}{2}} \,,
\end{equation}
to get
\begin{equation}
    S_L = -\frac{\beta r_H^2}{2} = -\frac{2}{\beta}\left(\sin^{-1}(\frac{\mu}{2})\right)^2 \,.
\end{equation}
In the $\mu \to 2$ limit, we see that the action gives
\begin{equation}
    S_L \to -\frac{\pi^2}{2\beta} \quad{\rm and}\quad \mu \to 2 \,,
\end{equation}
Note that in the absence of the defect, the computation is ill-defined since there is no classical Liouville solution on a torus. In the presence of the defect, we can define a Hilbert space of Liouville CFT on the circle intersected by the defect. Since we have computed above the partition function at any temperature, the reslicing of the path integral implies that the ground state energy of the defect Hilbert space in the $b\to 0$ limit is given by
\begin{equation}
   e^{-\frac{c}{6}S_L}=e^{-\frac{4\pi^2}{\beta}E_{\text{vac}}}\implies E_{\text{vac}}=-\frac{c}{12\pi^2}\left(\sin^{-1}(\frac{\mu}{2})\right)^2 \,.
\end{equation}
In the above argument, we have also made use of the fact that the defect is scale invariant to rescale the lengths of the two sides of the fundamental domain of the torus. 
In the $\mu \to 2$ limit, the vacuum energy becomes $E_{\text{vac}}=-\frac{c}{48}$. 

Now, we would like to uniformize the Liouville solution near the defect and compute the gluing matrix that relates the uniformization maps on the defect. Expressing the Liouville solution as
\begin{equation}
    e^{\Phi(z,\overline{z})}=\frac{4 w'(z)\overline{w}'(\overline{z})}{(1-w(z)\overline{w}(\overline{z}))^2} \,,
\end{equation}
for independent $w$ and $\overline{w}$ functions, we see that near the defect placed at $y=0$ in these coordinates,
\begin{equation}
    \begin{split}
        w_+(z)= ie^{\frac{i\beta r_H}{2}}e^{r_Hz},\quad \overline{w}_+(\overline{z})=ie^{\frac{i\beta r_H}{2}}e^{-r_H\overline{z}} &\qquad{\rm if}\ y>0 \,,\\
        w_-(z)= ie^{\frac{-i\beta r_H}{2}}e^{r_Hz},\quad \overline{w}_-(\overline{z})=ie^{-\frac{i\beta r_H}{2}}e^{-r_H\overline{z}} &\qquad{\rm if}\ y<0 \,.
    \end{split}
\end{equation}
On the defect, we see that the uniformization maps are related by the gluing matrix,
\begin{equation}
    w_+(x)=e^{i\beta r_H}w_-(x) \quad{\rm with}\quad G=\begin{bmatrix}
       e^{\frac{i\beta r_H}{2}} & 0\\
       0 & e^{-\frac{i\beta r_H}{2}}
    \end{bmatrix}\,,
\end{equation}
Notice that since $\frac{\beta r_H}{2}=\sin^{-1}(\frac{\mu}{2})$, the gluing matrix is independent of temperature. Observe that $G$ is an element of the elliptic conjugacy class of PSU$(1,1)$. This is to be contrasted with the Liouville solution on the sphere where the gluing matrix was in the hyperbolic conjugacy class.

Using the Liouville solution for the defect, we can readily compute thermal correlators of probe operators in the defect-background. For example, the 1-point function of a probe vertex operator $V_{\alpha}=e^{2\alpha \phi}$ with $\alpha=\frac{\eta}{b}$ and $\eta \ll 1$ on the torus with the line defect is 
\begin{equation}
    \langle e^{2\alpha \phi}\rangle = \left (\frac{r_H}{\cos(r_H(\frac{\beta}{2}- y_0))}\right )^{\frac{c\eta}{3}} \qquad{\rm where}\ r_H=\frac{2}{\beta}\sin^{-1}(\frac{\mu}{2})\,.
\end{equation}
Here, $y_0$ is the separation between the defect and the probe operator.

\subsection{Vanishing effective central charge} \label{Secceff}

The effective central charge that governs the entanglement entropy across a conformal defect can be expressed in terms of the ground state energies of the defect Hilbert space \cite{Wen:2017smb},
\begin{equation} \label{ceff}
    c_{\text{eff}}=\frac{12n}{1-n^2}(nE_1-E_n) \,.
\end{equation}
Here, $E_n$ is the vacuum energy of the defect Hilbert space with $2n$ defects evenly spaced with a separation $\pi$ between the neighbouring ones. This expression was derived for compact $2d$ CFTs. Even though Liouville CFT has a non-compact spectrum, in the presence of the defect, we can still define and compute the ground state energies of the defect Hilbert space as we did in the previous section. We now compute this energy by constructing a saddle for the $2n$-point function on the torus by gluing $2n$ hyperbolic cylinders each of height $\frac{\beta}{2n}$ end-to-end, thereby generalizing the construction in Section~\ref{Sectorus1point}. Since we are interested in the ground state energy of the defect Hilbert space, we take the high temperature limit $\beta \to 0$ in the closed string channel. In this symmetric situation, the jump conditions across each defect are solved by setting the waist lengths of the cylinders and thereby the saddlepoint weights to be equal and given by

\begin{figure}
    \centering
    \includegraphics[width=0.5\linewidth]{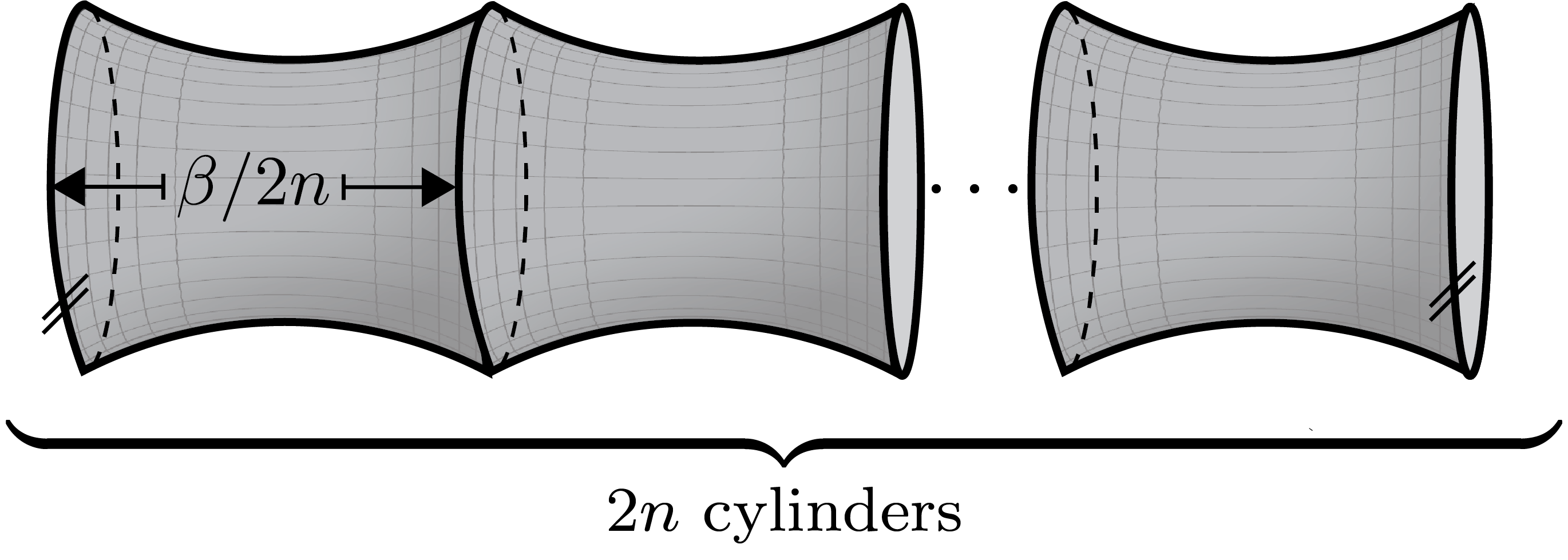}
    \caption{The figure shows the Liouville saddle used to compute the effective central charge. It is a torus with $2n$ evenly spaced defects. The saddle is constructed by gluing $2n$ identical hyperbolic cylinders each of height $\frac{\beta}{2n}$ end-to-end in a cyclic fashion.}
    \label{fig:torus_replicas}
\end{figure}

\begin{equation}
    r_H=\frac{4n}{\beta}\sin^{-1}(\frac{\mu}{2}) \,.
\end{equation}
The on-shell Liouville action evaluates to
\begin{equation}
    S_L=-\frac{\beta r_H^2}{2}=-\frac{8n^2}{\beta}\left(\sin^{-1}(\frac{\mu}{2})\right)^2 \,,
\end{equation}
By reslicing the path integral, we compute the vacuum energy of the defect Hilbert space on a circle of circumference $2n\pi$ so that the defects are evenly spaced with a separation of $\pi$,
\begin{equation}
    \text{Tr}_{\mathcal{H}_n} e^{-\frac{4\pi^2 n}{\beta}H_n}=e^{-\frac{4\pi^2 n}{\beta}E_n}+\dots=e^{\frac{4cn^2}{3\beta}\left(\sin^{-1}(\frac{\mu}{2})\right)^2} \,.
\end{equation}
The $\dots$ denote contributions from excited states in the defect Hilbert space which are suppressed in the $\beta \to 0$ limit. Therefore, the vacuum energy of the $2n$-defect Hilbert space is given by 
\begin{equation}
    E_n=-\frac{cn}{3\pi^2}\left(\sin^{-1}(\frac{\mu}{2})\right)^2=nE_1 \,.
\end{equation}
For the energies we have obtained, we see using (\ref{ceff}) that $c_{\text{eff}}=0$ indicating that the defect does not transmit information and is perfectly reflecting. To reinforce this observation, we can also check that the energy transmission across the defect measured using the stress tensor 2-point function across the defect, also vanishes. This is because the stress tensor is a light operator and does not backreact on the geometry, and hence there are no cross-correlations of the stress tensor in this semiclassical limit. However, once we include heavy enough operators which beackreact on the geometry, we can reliably compute their correlations across the defect as we will show in Section~\ref{secmatrix}.

In Section~\ref{secperinf}, we computed the effective central charge perturbatively in the defect coupling. We observed that the effect of introducing a cusp on the defect can be computed by shifting the locations of $n$ non-adjacent defects on the $n$-sheeted cylinder by an amount specified by the cusp angle. We could similarly study the effect of introducing a cusp in the present semiclassical analysis by changing the lengths of the non-adjacent hyperbolic cylinders in Figure~\ref{fig:torus_replicas} by an amount specified by the cusp angle. Still, by symmetry, the effective central charge computed using these replica saddles vanishes.

\subsection{Fusion of two defects}

In this section, we analyze the fusion of two defects by constructing the Liouville solution in the limit where the separation between the defects goes to zero. We will analyze fusion both in the vacuum state and at finite temperature by constructing the Liouville saddles on the sphere and torus respectively. In the subsequent analysis, we will show that defects with cosmological constants $\mu_1$ and $\mu_2$ fuse into a defect with cosmological constant $\mu_1+\mu_2$. We will show that the Liouville solution is non-singular in the fusion limit and derive the saddlepoint energy of the defect 2-point function in the fusion limit. This corresponds to the weight of a local operator exchanged by the defects in the limit where they fuse together.
 
First, we construct the Liouville saddle on the sphere for two parallel defects with cosmological constants $\mu_1$ and $\mu_2$. To this end, we work with the following general ansatz for the Liouville field,
\begin{align}
\Phi = \begin{cases}
-2\log \sinh(y-\tau_0 + A_1) & {\rm if}\ y>\tau_0\,, \\
-2\log \left(\frac{1}{r_H}\cos(r_H (y-y_0)) \right) & {\rm if}\ |y| < \tau_0\,, \\
-2\log \sinh(A_2-y-\tau_0 ) & {\rm if}\ y < -\tau_0 \,.
\end{cases}
\end{align}
The four unknowns in the ansatz: $A_1,A_2,y_0,r_H$ are determined by continuity and jump conditions across the two defects.
Since we are mainly interested in the $\tau_0 \to 0$ limit, we work with the specific branch of solutions where the defects are on the same side of the equator of the sphere. Assuming $\mu_1>\mu_2$ without loss of generality, we see that the jump conditions take the form,
\begin{equation}
    \mu_1 R_1=\sqrt{1+R_1^2}+\sqrt{R_1^2-r_H^2} \quad{\rm and}\quad  \mu_2 R_2=\sqrt{1+R_2^2}-\sqrt{R_2^2-r_H^2}\,,
\end{equation}
with
\begin{equation}
    R_1=\frac{1}{\sinh(A_1)}=\frac{r_H}{\cos(r_H(\tau_0-y_0))} \quad{\rm and}\quad R_2=\frac{1}{\sinh(A_2)}=\frac{r_H}{\cos(r_H(\tau_0+y_0))} \,.
\end{equation}
\begin{figure}
    \centering
    \begin{subfigure}{0.49\textwidth}
        \centering
        \includegraphics[height=0.4\textwidth]{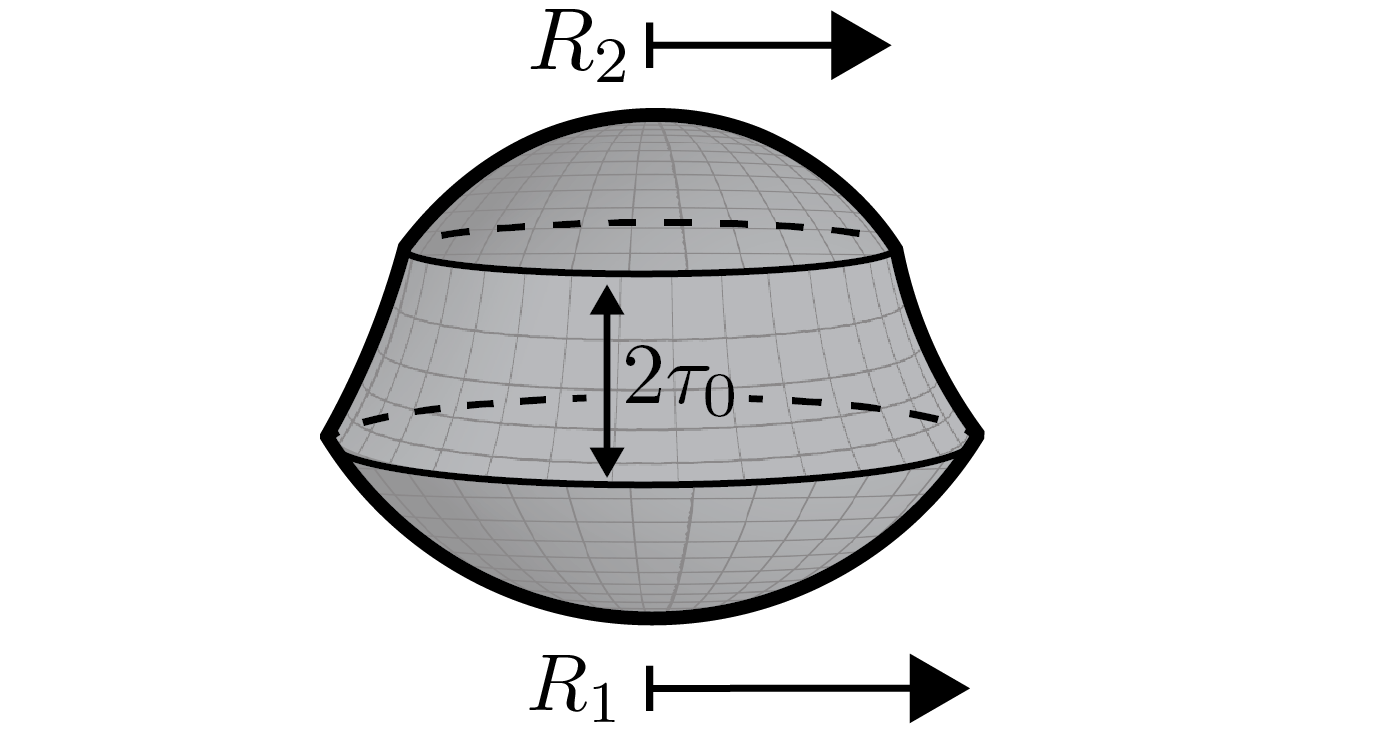}
        \caption{Sphere}
    \end{subfigure}
    \begin{subfigure}{0.49\textwidth}
        \centering
        \includegraphics[height=0.4\textwidth]{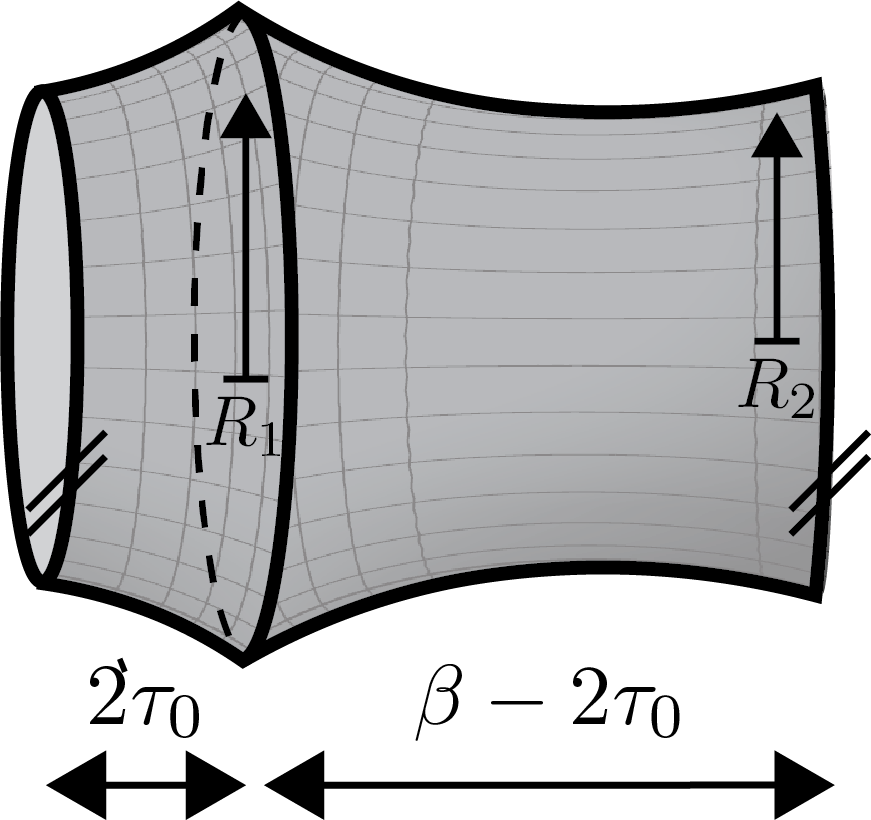}
        \caption{Torus}
    \end{subfigure}
    \caption{The figure shows the Liouville saddles used to study the fusion of two defects of different cosmological constants on the sphere and on the torus. Here $\tau_0$ measures the spacing between the two defects and the fusion limit corresponds to taking $\tau_0 \to 0$. The proper lengths of the two defects in the backreacted geometry are given by $2\pi R_1$ and $2\pi R_2$.}
    \label{fig:defect_fusion}
\end{figure}
In the limit $\tau_0\to 0$, we see that $R_1=R_2\equiv R$ so adding and subtracting the two jump conditions gives the following pair of equations,
\begin{equation}
    (\mu_1+\mu_2)R=2\sqrt{1+R^2} \quad{\rm and}\quad r_H=R\sqrt{1-\left(\frac{\mu_1-\mu_2}{2}\right)^2}\,.
\end{equation}
From the first equation above, we see that the defects fuse into a new defect with $\mu_{\text{eff}}=\mu_1+\mu_2$. Solving the above equations, we get the following expression for the saddlepoint weight in the fusion limit,
\begin{equation}
   r_H=\sqrt{\frac{4-(\mu_1-\mu_2)^2}{(\mu_1+\mu_2)^2-4}} \implies \Delta=\frac{c}{12}(1+r_H^2)=\frac{c}{3}\frac{\mu_1\mu_2}{(\mu_1+\mu_2)^2-4} \,.
\end{equation}
From the expression for $r_H$, we see that only if $|\mu_1-\mu_2|<2$, the saddle-point weight is above the threshold i.e $\Delta>\frac{c}{12}$. Conversely, if $|\mu_1-\mu_2|>2$, $r_H$ is imaginary so $\Delta<\frac{c}{12}$ and in the extreme case when $\mu_1\gg \mu_2$, $\Delta \to 0$. The existence of a threshold in the difference in cosmological constants above which the saddlepoint weight goes below $\frac{c}{12}$ can be explained geometrically: As the difference in cosmological constants is increased, the hyperbolic cylinder needs to get thinner and eventually, it pinches off at the waist which is when the saddlepoint weight is at the threshold. We see that if we hold $\mu_1+\mu_2$ fixed, the weight of the operator exchanged in the fusion limit decreases monotonically with the difference $|\mu_1-\mu_2|$. The Liouville action in the fusion limit evaluates to 
\begin{equation}
    S_L=-2\cosh^{-1}\left(\frac{\mu_1+\mu_2}{2}\right) \,,
\end{equation}
consistent with the observation that $\mu_{\text{eff}}=\mu_1+\mu_2$.

In the above analysis, we have studied fusion of defects in the vacuum state. But we can readily generalize the analysis to study fusion in any energy eigenstate. The conclusion that the cosmological constants add under defect fusion does not change. The energy of the exchanged operator in the fusion limit changes. If the energy of the excited state is $E$, then the scaling dimension of the exchanged operator is given by
\begin{equation}
    \Delta=\frac{c}{12}(1+r_H^2)\,, \qquad {\rm where}\ r_H=\sqrt{\frac{4-(\mu_1-\mu_2)^2}{(\mu_1+\mu_2)^2+\frac{48 E}{c}}} \,.
\end{equation}
From this expression, we see that the energy of the exchanged operator decreases monotonically with the energy of the excited state. 

In order to study the fusion of defects with different cosmological constants at finite temperature, we work with the following general ansatz for the Liouville solution on the torus,
\begin{align}
\Phi = \begin{cases}
-2 \log \left( \frac{1}{r_H'} \cos(r_H' (y-y_1))\right) &{\rm if}\ |y| < \tau_0 \,,\\
-2\log\left( \frac{1}{r_H}\cos(r_H(\frac{\beta}{2}-|y-y_2|))\right) &{\rm if}\  \tau_0 < |y| < \frac{\beta}{2} \,.
\end{cases}
\end{align}
The defect with cosmological constant $\mu_1$ is placed at $y=\tau_0$ while the defect with cosmological constant $\mu_2$ is placed at $y=-\tau_0$.
Assuming $\mu_1>\mu_2$, the jump conditions across the two defects give 
\begin{equation}
    \mu_1 R_1=\sqrt{R_1^2-(r_H')^2}+\sqrt{R_1^2-r_H^2} \quad{\rm and}\quad \mu_2 R_2=\sqrt{R_2^2-r_H^2}-\sqrt{R_2^2-(r_H')^2} \,.
\end{equation}
where the parameters are related to the moduli by the continuity of the Liouville field,
\begin{align}
    R_1 &= \frac{r_H'}{\cos(r_H'(\tau_0-y_1))} = \frac{r_H}{\cos(r_H(\frac{\beta}{2}-|\tau_0-y_2|))}\,,\nn\\  
    R_2 &= \frac{r_H'}{\cos(r_H'(\tau_0+y_1))} =\frac{r_H}{\cos(r_H(\frac{\beta}{2}-|\tau_0+y_2|))}\,.
\end{align}
Altogether, we have six equations for six unknowns: $R_1,R_2,r_H,r_H',y_1,y_2$. In the $\tau_0\to 0$ limit, the analysis simplifies considerably and we get $y_2=0, R_1=R_2\equiv R$. Adding the two jump equations, we get the jump condition derived previously for the torus one-point function with $\mu_{\text{eff}}=\mu_1+\mu_2$ and subtracting them shows that $|\mu_1-\mu_2|<2$ which is the same result that we observed on the sphere. The complete solution in the fusion limit is given by
\begin{equation}
   r_H=\frac{2}{\beta}\sin^{-1}\left(\frac{\mu_1+\mu_2}{2}\right) \quad{\rm and}\quad r_H'=\frac{2}{\beta}\sin^{-1}\left(\frac{\mu_1+\mu_2}{2}\right)\sqrt{\frac{4-(\mu_1-\mu_2)^2}{4-(\mu_1+\mu_2)^2}} \,.
\end{equation}
The conformal weight of the operator exchanged by the defects in the fusion limit is $\Delta=\frac{c}{12}(1+(r_H')^2)$. The Liouville action in the fusion limit at finite temperature evaluates to 
\begin{equation}
    S_L  = -\frac{2}{\beta}\left(\sin^{-1}\left(\frac{\mu_1+\mu_2}{2}\right)\right)^2 \,,
\end{equation}
reinforcing that when the defects fuse, their cosmological constants simply add.

\subsection{Fusion of the defect with a conformal boundary} \label{Seccasbound}

In this section, we study the fusion of the defect with a conformal boundary. Conformal boundary conditions in Liouville have been classified into ZZ and FZZT depending on whether we impose Dirichlet or Neumann boundary conditions for the Liouville field at the boundary. We will study the fusion by constructing a Liouville saddle for the defect on the disc with the appropriate boundary condition imposed on the Liouville field. We will observe that the fusion with a ZZ boundary is singular and has an associated Casimir energy which we compute while the fusion with an FZZT boundary is non-singular.

We start with the disk 1-pt function of the defect with ZZ boundary conditions. The relevant Liouville solution for this case is
\begin{equation}
    \Phi = 
    \begin{cases}
        -2\log \left( \sinh(y-\tau_0+A) \right) &{\rm if}\ y>\tau_0 \,,\\
        -2\log \left( \frac{1}{r_H}\sin(r_H y) \right) &{\rm if}\ 0<y<\tau_0 \,.
    \end{cases}
\end{equation}
\begin{figure}
    \centering
    \includegraphics[width=0.25\linewidth]{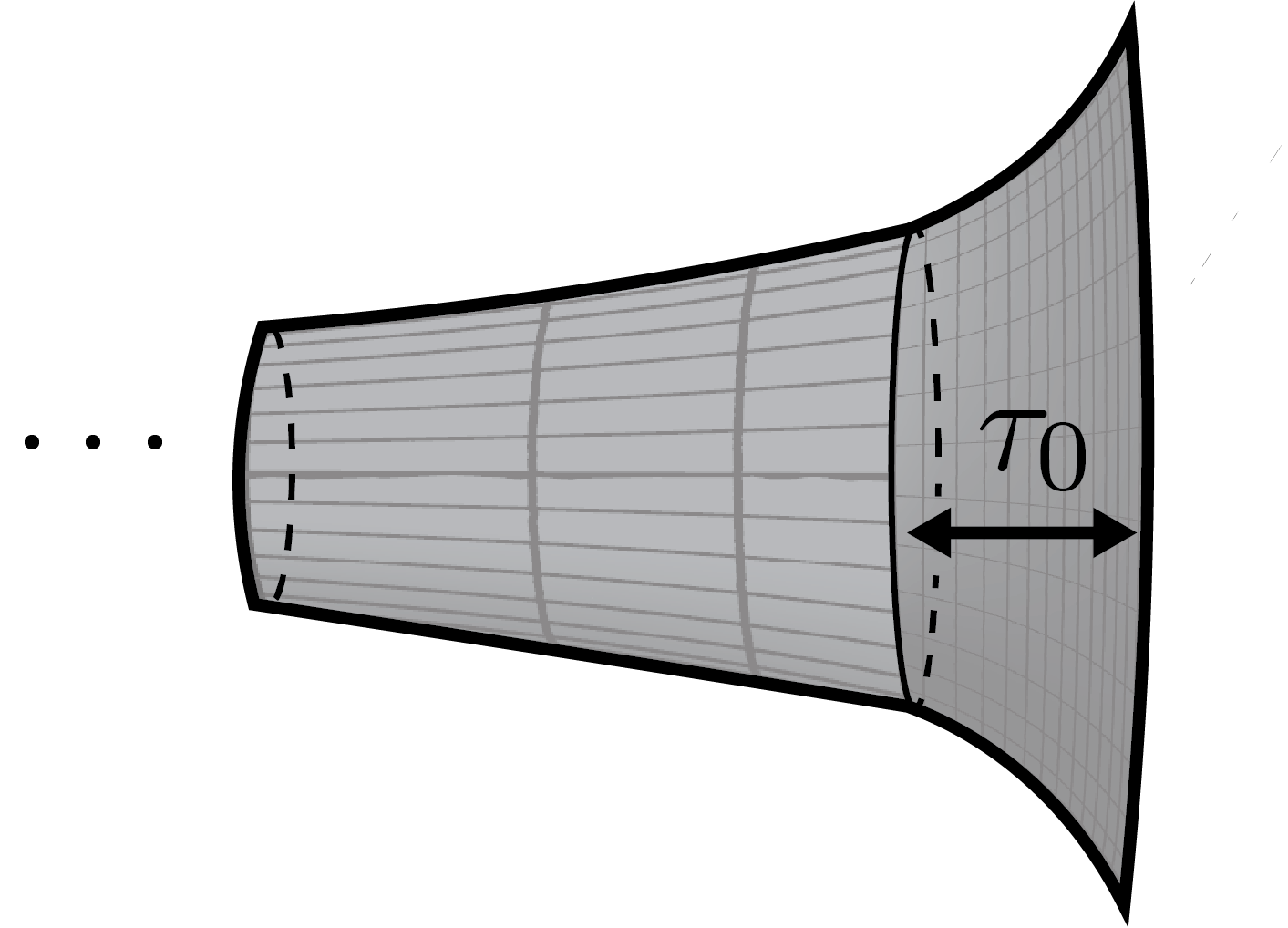}
    \caption{The figure shows the Liouville saddle on the disk (drawn in the cylinder conformal frame) used to study the fusion of the defect with a conformal boundary where a Dirichlet or Neumann boundary condition has been imposed on the Liouville field. $\tau_0$ is the separation between the defect and the boundary of the disk. The fusion limit corresponds to taking $\tau_0 \to 0$.}
    \label{fig:zz_sphere_fusion}
\end{figure}
Here, we have placed the ZZ boundary at $y=0$ and the defect at $y=\tau_0$. The boundary condition that we impose at the ZZ boundary is $\Phi \to -2\log y$. The continuity of the Liouville solution gives
\begin{equation}
    \sinh A = \frac{\sin(r_H \tau_0)}{r_H} = \frac{1}{r_0} \,,
\end{equation}
and the jump condition gives
\begin{equation}
     \mu r_0 = \sqrt{r_0^2 + 1} - \sqrt{r_0^2 - r_H^2} \,.
\end{equation}
The Liouville action with a defect and a ZZ boundary with appropriate boundary and counter-terms added is given by
\begin{equation}
    S_L = \frac{1}{4\pi} \int_{\Gamma} d^2 z \left( \p \Phi \bar{\p} \Phi + e^{\Phi}\right) - \frac{\mu}{2\pi} \int_\Sigma dz \, e^{\frac{\Phi}{2}} +\frac{1}{4\pi }\int_{\Gamma_+}dz\ \Phi- \frac{T}{2} - \frac{1}{\epsilon} + 1 -\log 2 \,.
\end{equation}
Here, $\epsilon \to 0$ is a regulator added to cure the divergences near the ZZ boundary. $\Gamma$ is the region $y \in (\epsilon, T)$ with $\Gamma_+$ being the boundary at $y=T$. The constant term of $(1-\log 2)$ ensures the normalization of the ZZ state $\langle 0\ket{\rm ZZ}=1$.
For the Liouville solution constructed above, we have
\begin{equation}
    S_L = -(r_H^2 - 1)\frac{\tau_0}{2} \,.
\end{equation}
Now, we would like to study the fusion limit of the above solution i.e $\tau_0\to 0$ to understand how the defect fuses with the ZZ boundary. In this limit, we expect $r_0\to \infty$ since the defect is close to the ZZ boundary where the Liouville solution diverges. The jump condition in this limit gives,
\begin{equation}
    \mu r_0=\sqrt{1+r_0^2}-\sqrt{r_0^2-r_H^2} \implies r_H=r_0\sqrt{1-(1-\mu)^2} \,.
\end{equation}
A solution exists only if $\mu<1$.
Plugging this into the continuity equation gives
\begin{equation}
    r_0=\frac{r_H}{\sin(r_H\tau_0)}\implies r_H=\frac{1}{\tau_0}\cos^{-1}(1-\mu) \,,
\end{equation}
so a solution exists only if $\mu \leq 1$. In this limit, the Liouville action evaluates to
\begin{equation}
    S_L=-\frac{1}{2}\tau_0(r_H^2-1)\to -\frac{(\cos^{-1}(1-\mu))^2}{2\tau_0} \,.
\end{equation}
This shows that the fusion of the defect with the ZZ boundary is singular and it incurs a large energy cost. We interpret the coefficient of the divergence as the Casimir energy associated with fusion of the defect with a ZZ boundary,
\begin{equation}
    E_{\text{fus}}=-\frac{c}{12}(\cos^{-1}(1-\mu))^2 \,.
\end{equation}

Now, we study the fusion of the defect with an FZZT boundary described classically by a Neumann boundary condition for the Liouville field. The action on the disk with a defect and an FZZT boundary is given by
\begin{align} 
S_L  = \frac{1}{4\pi} \int_{\Gamma} d^2 z \left( \p \Phi \bar{\p} \Phi + e^{\Phi}\right)
 - \frac{\mu}{2\pi} \int_\Sigma dz\ e^{\frac{\Phi}{2}}-\frac{\mu_B}{2\pi}\int_{ \Gamma_-}dz\ e^{\frac{\Phi}{2}}+\frac{1}{4\pi}\int_{\Gamma_+}dz\ \Phi-\frac{T}{2}+1-\log 2 \,,
\end{align}
where $\Gamma$ is the region $y \in (0,T)$ with $\Gamma_-$ being the boundary at $y=0$ and $\Gamma_+$ being the boundary at $y=T$.
To construct the Liouville solution, we use a general ansatz,
\begin{equation}
     \Phi = 
    \begin{cases}
        -2\log \left( \sinh(y-\tau_0+A) \right) &{\rm if}\ y>\tau_0 \,,\\
        -2\log \left( \frac{1}{r_H}\sin(r_H (y+y_0)) \right) &{\rm if}\ 0<y < \tau_0 \,.
    \end{cases}
\end{equation}
The FZZT boundary is at $y=0$, and the defect is placed at $y=\tau_0$. The ansatz is non-singular only if $y_0>0$ and $A>0$. There are three equations: Continuity and jump across the defect and Neumann at the FZZT boundary to determine the three unknowns $y_0,r_H,A$,
\begin{equation}
    \cos(r_Hy_0)=\mu_B\,, \quad r_0\equiv \frac{1}{\sinh(A)}=\frac{r_H}{\sin(r_H(\tau_0+y_0))}\,, \quad{\rm and}\quad \sqrt{1+r_0^2}-\sqrt{r_0^2-r_H^2}=\mu r_0
\end{equation}
The action takes the same parametric form,
\begin{equation}
     S_L = -(r_H^2 - 1)\frac{\tau_0}{2} \,,
\end{equation}
but $r_H$ is a different function of the modulus. In the fusion limit i.e, $\tau_0\to 0$, we have 
\begin{equation}
    r_0=\frac{r_H}{\sqrt{1-\mu_B^2}}\implies r_0=\frac{1}{\sqrt{(\mu+\mu_B)^2-1}} \quad{\rm and}\quad r_H=\frac{\sqrt{1-\mu_B^2}}{\sqrt{(\mu+\mu_B)^2-1}}\,.
\end{equation}
A solution exists only if $\mu_B<1$ and $\mu+\mu_B>1$. In this regime, in the fusion limit, $S_L\to 0$ indicating that the fusion is regular and there is no Casimir energy associated with fusion of the defect with an FZZT boundary.

\subsection{Vacuum energy in open string channel with the defect} \label{sec:vacopen}

In this section, we evaluate the cylinder transition amplitudes in the presence of the defect. These amplitudes help determine the open string spectrum on a line segment with a marked point where the defect intersects the line segment. For concreteness, we compute the 1-point function on the cylinder of height $L$ with ZZ boundary conditions imposed on either ends. The calculation can readily be generalized to include FZZT boundary conditions as well but we will omit the details here. The Liouville solution on the cylinder of height $L$ takes the form
\begin{equation} \label{Liouann}
\Phi=
\begin{cases}
  -2\log \left (\frac{1}{r_H}\sin(r_Hy) \right ) &{\rm if}\ 0<y<\tau_0 \,,\\
   -2\log \left (\frac{1}{r_H'}\sin(r_H'(L-y)) \right ) &{\rm if}\  \tau_0<y<L \,.
\end{cases}
\end{equation}
\begin{figure}
    \centering
    \includegraphics[width=0.5\linewidth]{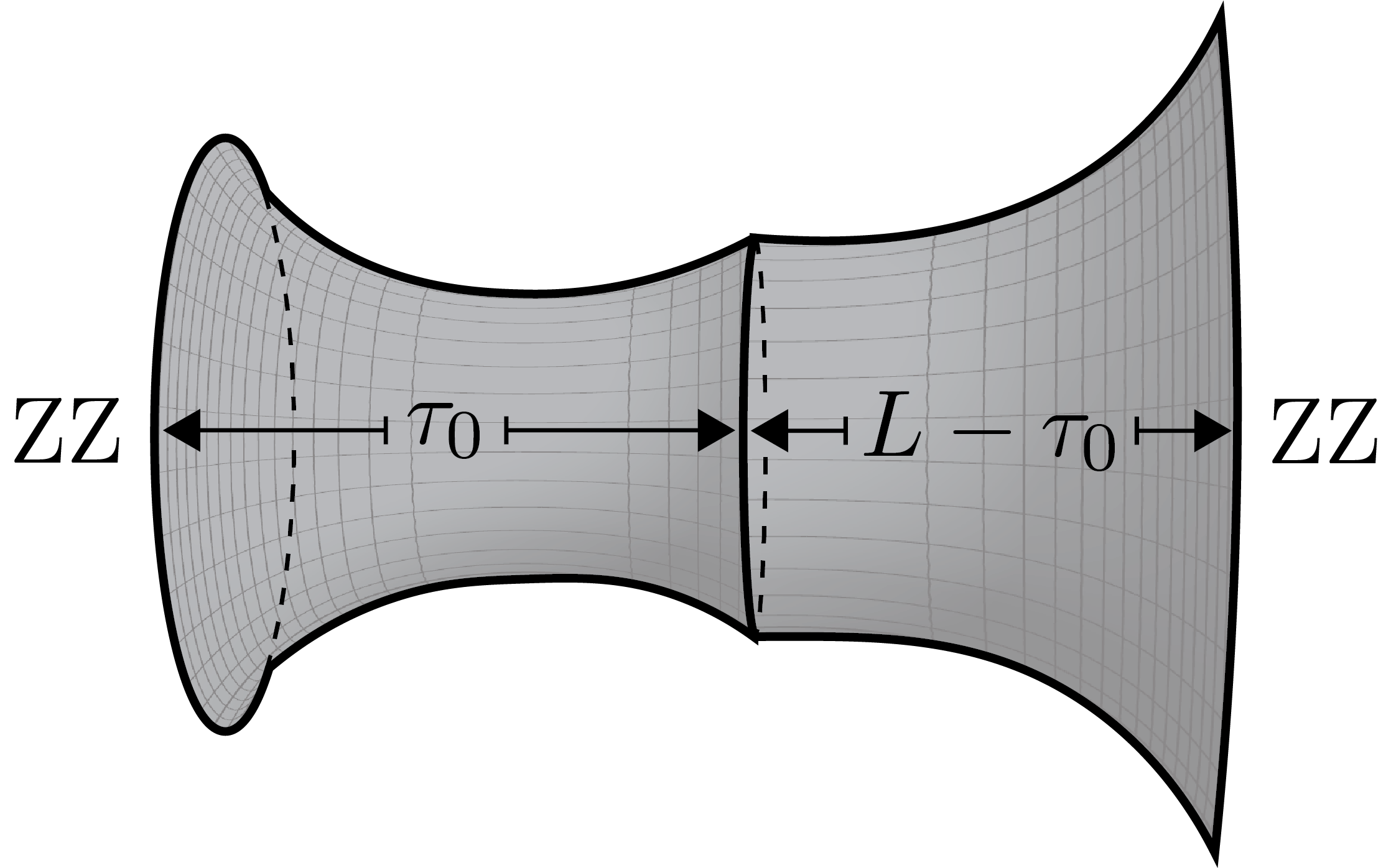}
    \caption{The figure shows the Liouville saddle used to compute the cylinder transition amplitude between two ZZ states with the defect placed asymmetrically between the two boundaries. The saddle is constructed by gluing two hyperbolic cylinders across the defect.}
    \label{fig:zz_L_zz}
\end{figure}
The defect is placed at $y=\tau_0$.
The classical Liouville action with a single line defect is given by
\begin{align} \label{LiouvannZZ}
S_L  = \frac{1}{4\pi} \int_{\Gamma} d^2 z \left( \p \Phi \bar{\p} \Phi + e^{\Phi}\right)
 - \frac{\mu}{2\pi} \int_\Sigma dz e^{\frac{\Phi}{2}} - \frac{2}{\epsilon} \,,
\end{align}
where $\Gamma$ is the region $y \in (\epsilon,L-\epsilon)$ with $\epsilon \to 0$. The on-shell action evaluates to
\begin{equation} \label{SLann}
    S_L=-\tau_0 \frac{r_H^2}{2}-(L-\tau_0)\frac{(r_H')^2}{2} \,,
\end{equation}
where $r_0$ can be expressed in terms of $\tau_0,L$ and $m$ by solving the junction conditions for the Liouville field,
\begin{equation} \label{juncannulus}
    r_0=\frac{r_H}{\sin(r_H \tau_0)}=\frac{r_H'}{\sin(r_H'(L-\tau_0))} \quad{\rm and}\quad -r_H'\cot(r_H'(L-\tau_0))-r_H\cot(r_H \tau_0)=\mu r_0 \,.
\end{equation}
As a special case, when the defect is symmetrically placed between the ZZ boundaries i.e $\tau_0=\frac{L}{2}$, then 
\begin{equation}
    r_H=\frac{2}{L}\cos^{-1}(\frac{\mu}{2})\implies S_L=-\frac{2}{L}\left(\cos^{-1}(\frac{\mu}{2})\right)^2 \,.
\end{equation}
In the $\mu\to 0$ limit when there is no defect, we recover the semiclassical vacuum character as expected. In the $\mu \to 2$ limit, interestingly, the action vanishes. This suggests that the $\mu=2$ defect `screens' the transition between the two ZZ states. We can also use the above expression to read off the vacuum energy of the open-string Hilbert space on a segment of length $\pi$ ending on the two ZZ boundaries with the defect intersecting the line segment midway between the ZZ boundaries,
\begin{equation}
    \text{Tr}_{\mathcal{H}_{\text{open}}}e^{-\frac{2\pi^2}{L}H_{\text{open}}}=e^{\frac{c}{3L}(\cos^{-1}(\frac{\mu}{2}))^2} \,.
\end{equation}
Since we have evaluated the partition function at any $\beta$, this shows that in the semiclassical limit, only the vacuum state in the open channel contributes and it has energy given by
\begin{equation}
    E_{\text{vac}}=-\frac{c}{6\pi^2}\left(\cos^{-1}(\frac{\mu}{2})\right)^2 \,.
\end{equation}
We see that in the $\mu\to 0$ limit, we recover the familar expression of $E_{\text{vac}}=-\frac{c}{24}$ for the open string vacuum energy between two ZZ boundaries.

\subsection{Correlations of local operators across the defect} \label{secmatrix}

In this section, we compute the semiclassical matrix elements of the defect between primary states by computing the thermal 2-point function of the defect and comparing the result with its spectral decomposition. This is a measure of correlations of local operators across the defect. We computed the matrix elements at weak defect coupling in Section~\ref{seccorrpert}, where we showed that they are proportional to the DOZZ structure constants (\ref{matweak}). This should be contrasted with the expression for the matrix elements at strong defect coupling computed in this section.
The Liouville solution on the torus with two defects placed at $y=\pm \tau_0$ can be parametrised as
\begin{align}
\Phi = \begin{cases}
-2 \log \left( \frac{1}{r_H} \cos(r_H y)\right) &{\rm if}\ |y| < \tau_0 \,,\\
-2\log\left( \frac{1}{r_H'}\cos(r_H' (\frac{\beta}{2}-|y|))\right) &{\rm if}\  \tau_0 < |y| < \frac{\beta}{2} \,.
\end{cases}
\end{align}
\begin{figure}
    \centering
    \includegraphics[width=0.4\linewidth]{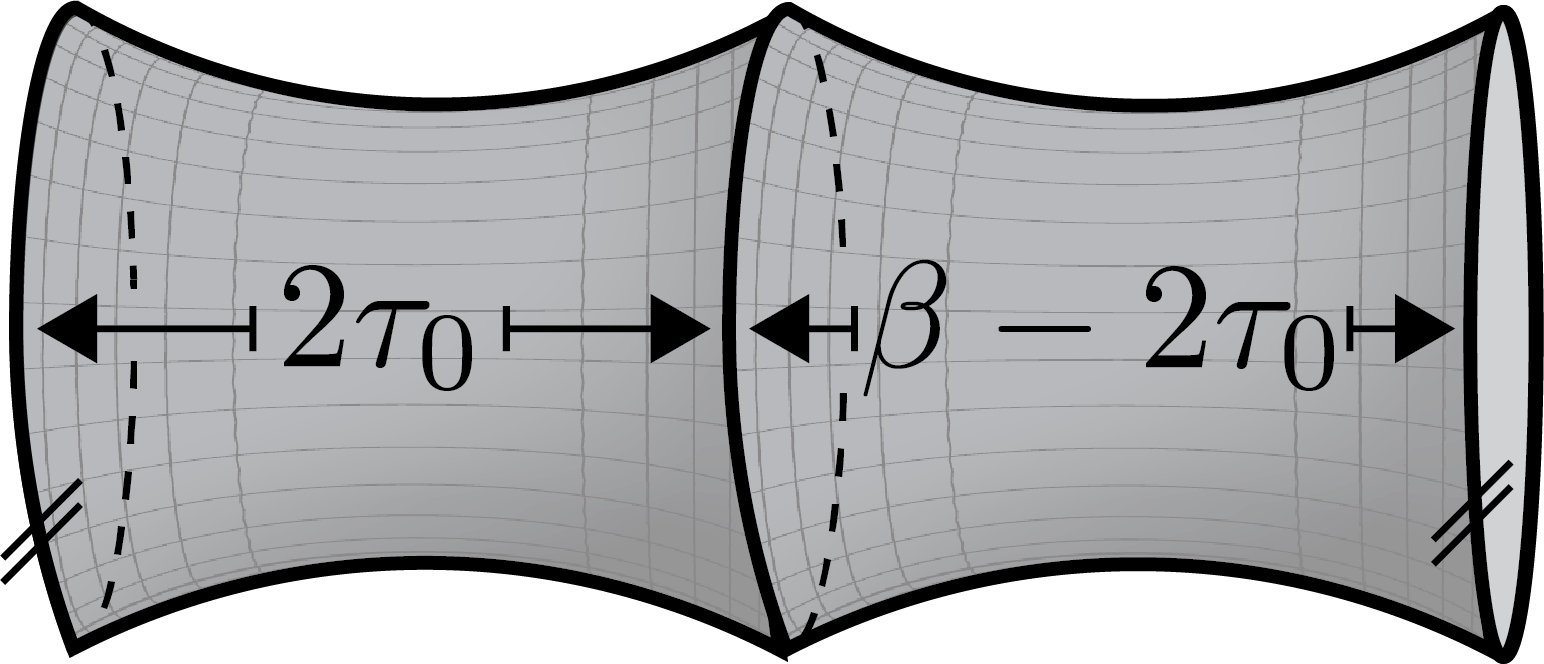}
    \caption{The figure shows the Liouville saddle used to compute the semiclassical matrix elements of the defect. It is a torus with two defects constructed by gluing two hyperbolic cylinders across the defects.}
    \label{fig:torus_two}
\end{figure}
The junction conditions determine $(r_H, r_H')$ in terms of $(\tau_0, \beta)$,
\begin{align}
r_0 &= \frac{r_H}{\cos(\tau_0 r_H)} = \frac{r_H'}{\cos((\frac{\beta}{2}-\tau_0)r_H')} \,,\\
\mu r_0 &= r_H \tan(\tau_0 r_H)  + r_H' \tan(r_H' (\frac{\beta}{2}-\tau_0)) \,.
\end{align}
There is a real solution to the above equation only if $\mu<2$.
The classical action is
\begin{align}
S_L = \frac{1}{4\pi} \int_{\Gamma}(\partial \Phi \bar{\partial}\Phi + e^{\Phi})  - \frac{\mu}{2\pi } \int_\Sigma dz\ e^{\frac{\Phi}{2}} - \frac{\mu}{2\pi} \int_{\Sigma'} dz\ e^{\frac{\Phi}{2}} \,,
\end{align}
with $\Gamma$ the fundamental domain of the torus. Evaluating this on the solution above gives
\begin{align}
S_L = 2\left( - \frac{r_H^2}{2}  \tau_0 -\frac{ r_H'^2}{2} (\frac{\beta}{2}-\tau_0) \right) \,.
\end{align}
Comparing with the spectral decomposition of the thermal two-point function, 
\begin{align}\label{spectorus}
\Tr \left (e^{-(\beta-2\tau_0)H} {\bm L}_\Sigma e^{-2\tau_0 H} {\bm L}_\Sigma \right ) \nn\\
= \frac{1}{4}e^{c \beta/12} & \int_{\mathbb{R}} dP \int_{\mathbb{R}} dP' |\langle \alpha|{\bm L}_\Sigma|\alpha'\rangle|^2 \left(e^{-4h_P \tau_0 - 4h_{P'} (\frac{\beta}{2}-\tau_0)}  +\mbox{descendants} \right) \,,
\end{align}
and ignoring the descendant contributions since they are subleading in the semiclassical limit,
we get the general semiclassical structure constants defined in terms of the matrix elements as
\begin{align}
|\langle \alpha|{\bm L}_\Sigma | \alpha'\rangle|^2 &= \exp\left[ \frac{c}{3}c_L(\gamma,\gamma') + \cdots \right] \,,
\end{align}
to give 
\begin{equation} \label{semimat}
    c_L(\gamma,\gamma')=\gamma\cos^{-1}\left(\frac{\gamma}{r_0(\gamma,\gamma')}\right)+\gamma'\cos^{-1}\left(\frac{\gamma'}{r_0(\gamma,\gamma')}\right) \,,
\end{equation}
where $r_0$ is determined by solving the junction condition, $\sqrt{r_0^2-\gamma^2}+ \sqrt{r_0^2-\gamma'^2}=\mu r_0$. In the above expression, we have used the semiclassical momentum labels for the conformal weights of the primaries,
\begin{equation}
    h=\frac{c}{24}(1+\gamma^2) \quad{\rm and}\quad h'=\frac{c}{24}(1+\gamma'^2) \,,
\end{equation}
and we have used the fact that the geometric parameters $r_H, r_H'$ that measure the geodesic lengths (see Figure~\ref{fig:scsl_limit}) match with these semiclassical saddlepoint momenta, $r_H=\gamma$ and $r_H'=\gamma'$ \cite{Chandra:2024vhm}.
Here, we have written the structure constants assuming $h$ and $h'$ are above the threshold of $\frac{c}{24}$. The structure constants evaluated at $\gamma=\gamma'$ have a simple closed form expression,
\begin{equation}
    c_L(\gamma,\gamma)=2\gamma \sin^{-1}(\frac{\mu}{2})\,.
\end{equation}

Now, we make a couple of comments about the structure constants (\ref{semimat}). The first comment is that, since $r_0$ is a non-trivial function of the conformal weights of the external operators, the matrix elements of the defect don't factorize into functions of either of the external conformal weights. This is indicative of non-trivial correlation across the defect. Since $2\pi r_0$ is the proper length of the defect, it can be treated as a geometric measure of the correlation across the defect. In Section~\ref{Sechypdef}, we will reproduce the above expression (\ref{semimat}) for the structure constants in terms of the matrix elements of another line operator obtained by ``decohering'' the FZZT boundary state thereby providing a more quantum mechanical interpretation of this correlation. The second comment is that, since we have defined the structure constants in terms of the absolute values of the matrix elements, we have ignored the phases. We can fix these for example using the cylinder transition amplitude between ZZ boundaries computed in \ref{sec:vacopen} since it is sensitive to the phases in the matrix element unlike the thermal 2-point function which only depends on the absolute value of the matrix elements. Since the phases are not very important for the subsequent discussion, we will omit writing them explicitly.

\section{The line defect from decohering FZZT boundaries} \label{Seclen}

In this section, we introduce the decohered FZZT interface,
\begin{equation}
    L(\Tilde{\mu}_D)=2\sqrt{2}\pi b\int_0^{\infty}\frac{d\ell}{\ell}\ e^{\Tilde{\mu}_D \ell}\ketbra{\ell}\,,
    \label{decoheredinterface}
\end{equation}
and describe its connection to the line defects (\ref{expdefect}) and (\ref{alphac}) at both weak and strong coupling.
The geometric boundary state $\ket{\ell}$ was first introduced in \cite{Fateev:2000ik} and we review its relation to the FZZT boundary in Section~\ref{SeclenFT}. We have also introduced an interface cosmological constant $\Tilde{\mu}_D$ in (\ref{decoheredinterface}) that is distinct from the defect cosmological constant $\mu_D$ of the line defect ${\bm L}_\Sigma$ (\ref{expdefect}). We compute the exact matrix elements of the decohered FZZT interface (\ref{decoheredinterface}) in Section~\ref{Secmat}. We then evaluate these matrix elements at small and large values of the cosmological constant $\Tilde{\mu}_D$ in Sections \ref{sec:deco_weak} and \ref{Sechypdef} respectively, before matching these with the matrix elements of the line defect ${\bm L}_\Sigma$ computed at weak and strong coupling in Sections \ref{seccorrpert} and \ref{secmatrix} respectively. This helps establish that the correlations across the defect ${\bm L}_\Sigma$ can be realized by decohering FZZT boundaries in a way described by (\ref{decoheredinterface}). 
Finally, in section \ref{sec:cyldeco}, we compute the effect of decohering FZZT boundaries on cylinder transition amplitudes between Liouville boundary states. 


\subsection{The fixed-length wavefunction of an FZZT boundary}  \label{SeclenFT}

In this section, we review the wavefunctions of FZZT boundary states computed in \cite{Fateev:2000ik}. Since we will be working with a different normalisation for the wavefunctions from that of \cite{Fateev:2000ik}, we will first set up some conventions. Conformal boundary states like the FZZT state can be expanded using a basis of Ishibashi states \cite{Ishibashi:1988kg} which we normalize such that they satisfy the following transition amplitude,
\begin{equation} \label{ishicyl}
    \begin{split}
        \langle\bra{P}e^{-\pi tH}\ket{P'}\rangle=\frac{\delta(P-P')}{\rho_0(P)}\chi_P(it) \quad{\rm given}\quad P,P'\in \mathbb{R}^+\,,
    \end{split}
\end{equation}
where $\rho_0(P)$ is the Plancherel measure
\begin{equation}
    \rho_0(P) = 4\sqrt{2}\sinh(2\pi bP)\sinh(2\pi b^{-1}P)
    \label{rho0} \,,
\end{equation}
and $\chi_P(\tau)\equiv \frac{e^{2\pi i\tau P^2}}{\eta(\tau)}$ is the torus character. We will use the notation $\ket{P}$ to denote the primary state associated to the Ishibashi state $\kett{P}$ which obeys the normalization,\footnote{
    The normalizations (\ref{ishicyl}) and (\ref{Pnormalization}) for the Ishibashi and primary states do not fix the phases. The phases can be fixed by relating the boundary wavefunctions (\ref{eq:liouville_bdy_wvfcns}) to the corresponding ones in the original literature \cite{Fateev:2000ik, Zamolodchikov:2001ah}.
}
\ie 
\langle P |P'\rangle={\delta(P-P')\over \rho_0(P)} \quad{\rm given}\quad P,P'\in \mathbb{R}^+\,.
\label{Pnormalization}
\fe
The normalization of the primary states in (\ref{Pnormalization}) should be contrasted  with the DOZZ normalization of the corresponding vertex operators derived by taking the identity limit of the DOZZ structure constants (\ref{DOZZnorm}). It is important to note that primary states are $\delta$-function normalized whereas the Ishibashi states are not normalizable, which is evidenced from the divergence in the $t\to 0$ limit of the transition amplitude (\ref{ishicyl}). 

The ZZ and FZZT families of boundary states in Liouville CFT can be expanded using the basis of Ishibashi states as follows,\footnote{
    Our conventions for the boundary wavefunctions agree with those used in \cite{Collier:2023cyw}
}
\begin{alignat}{2}\label{eq:liouville_bdy_wvfcns} 
    &\ket{\text{ZZ}_{(m,n)}}=\int_0^{\infty} \rho_0(P)dP\ \psi_{(m,n)}(P)\ket{P}\rangle &&\quad{\rm with}\ \psi_{(m,n)}(P)=\frac{4\sqrt{2}\sinh(2\pi m bP)\sinh(2\pi n b^{-1}P)}{\rho_0(P)} \,,\nn\\
    &\ket{\text{FZZT}(s)}=\int_0^{\infty} \rho_0(P)dP\ \psi_s(P)\ket{P}\rangle &&\quad{\rm with}\ \psi_s(P)=\frac{2\sqrt{2}\cos(4\pi sP)}{\rho_0(P)} \,.
\end{alignat}
Note that, with the above conventions, $\psi_{(1,1)}(P)=1$. We will often refer to $\ket{\text{ZZ}_{(1,1)}}$ as $\ket{\rm ZZ}$ and to $\ket{\text{FZZT}(s)}$ as $\ket{s}$.

Having established the conventions, we now go on to compute the wavefunction of the FZZT boundary state in the fixed-length basis.
FZZT boundaries can be labelled by the boundary cosmological constant $\mu_B$ or equivalently the $s$ parameter related as
\begin{equation}
    \cosh^2(2\pi bs)=\frac{\mu_B^2}{\mu_{\text{bulk}}}\sin(\pi b^2)\,.
\end{equation}
A more geometric representation of FZZT wavefunctions was studied in \cite{Fateev:2000ik} in terms of the total boundary length
\begin{equation}
    \ell = \int e^{b\phi}dx\,,
\end{equation}
which is conjugate to the boundary cosmological constant. It follows that we can define the wavefunctions $\psi_\ell(P)$ of this operator through Laplace conjugation\footnote{
    The $\ell$-measure is chosen such that the measure for the boundary Liouville-zero mode
    \begin{equation}
        \phi_0 = \int \phi(x) dx\,,
    \end{equation}
    is flat in the mini-superspace approximation.
}
\begin{equation}
    \psi_s(P) = \int_0^{\infty}\frac{d\ell}{\ell}\ e^{-\mu_B\ell}\psi_{\ell}(P) \,.
\end{equation}
With the normalization chosen in (\ref{eq:liouville_bdy_wvfcns}) for $\psi_s(P)$, the fixed-length wavefunction of the FZZT boundary state is given by
\begin{equation}
    \psi_\ell(P) =\frac{P}{\pi b \sinh(2\pi Pb)} K_{-2iP/b}(\kappa\ell) \,,
    \label{flwavefunction}
\end{equation}
where the parameter $\kappa$ is such that
\begin{equation} \label{eq:kappa_def}
    \cosh(2\pi bs) = \frac{\mu_B}{\kappa} \quad{\rm or}\quad  \kappa^2 = \frac{\mu_{\text{bulk}}}{\sin(\pi b^2)} \,.
\end{equation}
We temporarily pause to justify (\ref{flwavefunction}) since the expression
\begin{equation}\label{eq:s_from_ell_integral}
    \psi_{s}(P) \overset{?}{=} \frac{P}{\pi b \sinh(2\pi Pb)}\int_0^\infty \frac{d\ell}{\ell}e^{-\mu_B\ell}\ K_{-2iP/b}(\kappa\ell) \,,
\end{equation}
contains an integrand which is singular at $\ell=0$. We can make sense of this equation through the limit
\begin{equation}
    \psi_{s}(P) = \lim_{\delta\to 0^+}\frac{P}{\pi b \sinh(2\pi Pb)}\int_0^\infty \ell^{\delta-1}d\ell e^{-\mu_B\ell}K_{-2iP/b}(\kappa\ell) \,.
\end{equation}
The integral evaluates to
\begin{align}
    \lim_{\delta\to 0^+}\int_0^\infty \ell^{\delta-1}d\ell\ e^{-\mu_B\ell}K_{-2iP/b}(\kappa\ell) &= 
    \frac{\pi\csch(2\pi Pb^{-1})}{4Pb^{-1}}\times \nn\\
    &\qquad \left[\left(\frac{\mu_B}{\kappa}+\sqrt{\frac{\mu_B^2}{\kappa^2}-1}\right)^{2iP/b}+\left(\frac{\mu_B}{\kappa}+\sqrt{\frac{\mu_B^2}{\kappa^2}-1}\right)^{-2iP/b}\right] \,, \nn\\
    &= \frac{\pi\csch(2\pi Pb^{-1})}{2Pb^{-1}}\cos(4\pi sP) \,,
\end{align}
where in the second line we substituted (\ref{eq:kappa_def}). Plugging this back into (\ref{eq:s_from_ell_integral}) recovers the FZZT wavefunction introduced in (\ref{eq:liouville_bdy_wvfcns}). We note that one utility of the parameter $s$ over $\ell$ is that the $\psi_s(P)$ wavefunctions manifestly exhibit the $b\to b^{-1}$ symmetry while the $\psi_\ell(P)$ wavefunctions do not.

Given our exact expression for the fixed-length wavefunction, we can now define a purely geometric boundary state denoted as $\ket{\ell}$ which admits the following expansion in a basis of Ishibashi states,
\begin{equation}
    \ket{\ell} = \int_0^{\infty} \rho_0(P)dP\ \psi_\ell(P)\ket{P}\rangle \,.
\end{equation}
Just like other boundary states, states of fixed boundary length are not normalizable.

\subsection{Matrix elements of the decohered FZZT interface} \label{Secmat}

Having defined the fixed-length boundary states in our conventions, we now move to describing the decohered FZZT interface which we study for the remainder of this section. This operator is obtained by a diagonal gluing of two copies of the fixed length states $\ket{\ell}$ defined as follows  
\footnote{
    We refer to this as a ``decohered'' state because this resembles einselection in an open quantum system. 
}
\begin{equation}
   L(\tilde{\mu}_D) \equiv 2\sqrt{2}\pi b\int_0^{\infty}\frac{d\ell}{\ell}\ e^{\tilde{\mu}_D\ell}\ketbra{\ell} \,.
\end{equation}
The integration measure including the prefactor is chosen such that when $\Tilde{\mu}_D=0$, the decohered FZZT interface furnishes a resolution of the identity on the Schwarzian subspace with the DOZZ normalisation (\ref{DOZZnorm}) for the states in the subspace. These are states spanned by primaries of the form $\ket{V_{\frac{Q}{2}+ibk}}$ in the $b\to 0$ limit (i.e. by the Liouville primary states near the edge of the normalizable spectrum). We discuss the normalization in more detail in Section \ref{sec:deco_weak}.

The matrix elements of this decohered FZZT interface between the primary states are given by
\begin{align} \label{matdecFZZT}
    \mel{P}{L(\tilde{\mu}_D)}{P'}=\frac{2\sqrt{2}PP'}{\pi b \sinh(2\pi Pb)\sinh(2\pi P'b)}\int_0^{\infty}\frac{d\ell}{\ell}\ e^{\tilde{\mu}_D\ell}K_{-2iP/b}(\kappa\ell)K_{-2iP'/b}^*(\kappa\ell) \,.
\end{align}
Using the large-$z$ asymptotics of the Bessel function,
\begin{equation}
    K_\nu(z)\sim e^{-z}\sqrt{\frac{\pi}{2z}} \,,
\end{equation}
we see that there is no divergence in the integral at large $\ell$ for $\Tilde{\mu}_D<2\kappa$. However, the final expression can be analytically continued to outside this range as we will discuss shortly. We can treat the logarithmic divergence near $\ell=0$ by regulating the integral and analytically continuing the answer in exactly the same way as we showed for the FZZT 1-point function in Section~\ref{SeclenFT}. The result is 
\begin{align}\label{eq:L_mel_exact}
    &\mel{P}{L(\tilde{\mu}_D)}{P'} =\frac{\pi PP'}{\sqrt{2}b\sinh(2\pi Pb)\sinh(2\pi P'b)}\times \nn\\
    &\qquad\left(\frac{\tilde{\mu}_D^2(P^2-P'^2)}{b^2\kappa^2}\frac{_4F_3\left(1\pm\frac{iP}{b}\pm\frac{iP'}{b};\frac{3}{2},\frac{3}{2},2;\frac{\tilde{\mu}_D^2}{4\kappa^2}\right)}{\sinh\left(\frac{\pi(P-P')}{b}\right)\sinh\left(\frac{\pi(P+P')}{b}\right)}+\frac{\tilde{\mu}_D}{\kappa}\frac{_4F_3\left(\frac{1}{2}\pm\frac{iP}{b}\pm\frac{iP'}{b};\frac{1}{2},1,\frac{3}{2};\frac{\tilde{\mu}_D^2}{4\kappa^2}\right)}{\cosh\left(\frac{\pi(P-P')}{b}\right)\cosh\left(\frac{\pi(P+P')}{b}\right)}\right) \,,
\end{align}
where we have introduced the generalized hypergeometric function $_4F_3$. The expression simplifies slightly when considering diagonal matrix elements. The limit as $P\to P'$ of equation \ref{eq:L_mel_exact} gives 
\begin{align}
    \mel{P}{L(\tilde{\mu}_D)}{P} &= \frac{\pi P^2}{\sqrt{2}b\sinh^2(2\pi Pb)}\times  \nn\\
    &\qquad \Bigg(\frac{2\tilde{\mu}_D^2 P {_4F_3}\left(1,1,1\pm \frac{2iP}{b};\frac{3}{2},\frac{3}{2},2;\frac{\tilde{\mu}_D^2}{4\kappa^2}\right)}{\pi b \kappa^2\sinh(2\pi Pb^{-1})}+\frac{\tilde{\mu}_D{_3F_2}\left(\frac{1}{2},\frac{1}{2}\pm\frac{2iP}{b};1,\frac{3}{2};\frac{\tilde{\mu}_D^2}{4\kappa^2}\right)}{\kappa\cosh(2\pi Pb^{-1})}\Bigg) \,,
\end{align}
These diagonal matrix elements will later prove useful in Section ~\ref{secJT} where we compute observables in JT gravity.

We pause to make some comments about the properties of these matrix elements. The expression in parenthesis does not factorize into independent functions of $P$ and $P'$. We take this to mean that this expression contains dynamical information about the correlation induced by decohering the FZZT boundary. The expression depends on the defect cosmological constant $\Tilde{\mu}_D$ in terms of a scale invariant ratio of the defect and bulk cosmological constants $\Tilde{\mu}_D/\sqrt{\mu_{\text{bulk}}}$, written using $\Tilde{\mu}_D/\kappa$. This can be easily seen by rescaling the integration variable in (\ref{matdecFZZT}). Also, notice that the expression is reflection symmetric in both the momenta. In other words, it is independently invariant under $P\to -P$ and under $P'\to -P'$. It is also symmetric under the exchange $P\leftrightarrow P'$, which means that $L(\tilde{\mu}_D)$ corresponds to a Hermitian operator.  

We will treat these matrix elements as complex valued function of the two momenta $P$ and $P'$. The hypergeometric functions are regular in these momenta but the denominators $\cosh \left(\frac{2 \pi  P_1}{b}\right)\pm\cosh \left(\frac{2 \pi  P_2}{b}\right)$ introduce 4 infinite lines of simple poles wherever
\begin{equation}
    \pm P_1\pm P_2=\frac{ib}{2}m \quad{\rm for}\quad m \in \mathbb{Z}-\{0\} \,.
\end{equation}
Notice that the locations of these poles is independent of $\Tilde{\mu}_D$. Liouville observables computed in the literature typically exhibit a $2d$ lattice of simple poles characteristic of fusion with degenerate operators and exhibit a $b\leftrightarrow\frac{1}{b}$ exchange duality. Interestingly, these poles do not manifest this symmetry. We reviewed in Section~\ref{SeclenFT} that the bulk 1-point function with an FZZT boundary does exhibit this property. This implies that decohering the FZZT boundary is what breaks the self-duality property of Liouville CFT. 

\begin{figure}
    \centering
    \includegraphics[width=0.5\linewidth]{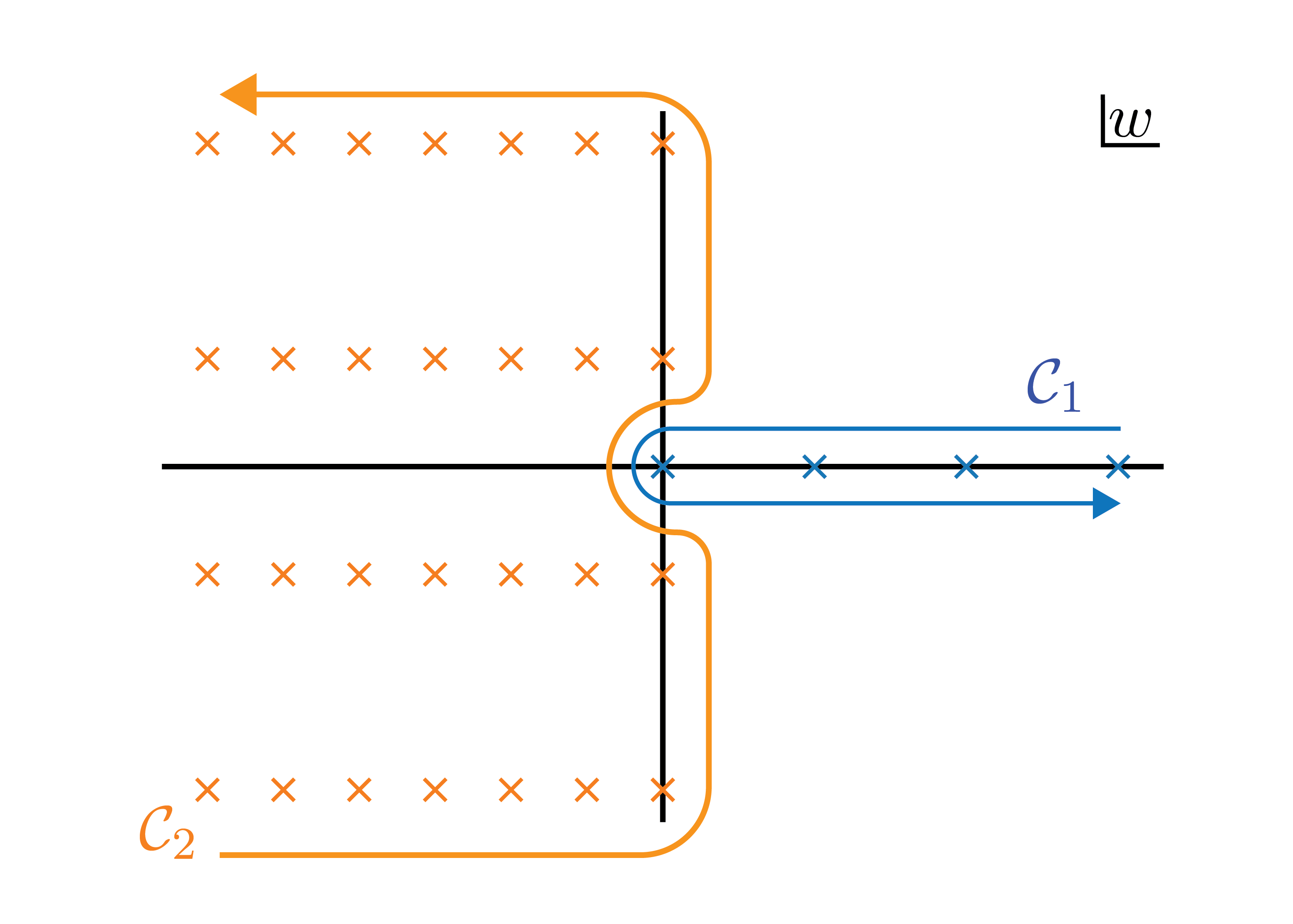}
    \caption{The poles of the integrand of (\ref{eq:L_mel_continue}) are plotted in the $w$-plane. Integrating along $\cC_1$ picks up the blue poles (c.f. (\ref{eq:c1_poles})) and recovers the usual series representation of the hypergeometric function which converges when $|z|\leq 1$. This is the standard interpretation of (\ref{eq:L_mel_exact}). Outside of the unit disk, the unique analytic continuation is given by integrating along $\cC_2$ and picking up the orange poles (c.f. (\ref{eq:c2_poles})). }
    \label{fig:mel_contour}
\end{figure}

We have expressed the matrix elements of $L(\Tilde{\mu}_D)$ in terms of generalized hypergeometric functions in the complex argument 
\begin{equation}
    z\equiv \frac{\Tilde{\mu}_D}{2\kappa} \,.
\end{equation}
Standard practice is to define hypergeometric functions on the unit disk $|z|\leq 1$ where the series representation is convergent. We can uniquely continue these functions away from this domain using the Meijer G-Function
\begin{equation}
    G^{(m,n)}_{(s,t)}(a_1,\cdots,a_s;b_1,\cdots,b_t;z) = \frac{1}{2\pi i}\int_\cC \frac{\prod_{j=1}^m\Gamma(b_j-w)\prod_{j=1}^n\Gamma(1-a_j+w)}{\prod_{j=m+1}^t\Gamma(1-b_j+w)\prod_{j=n+1}^s\Gamma(a_j-w)}z^wdw \,,
\end{equation}
since
\begin{equation}
    _sF_t[a_1,\cdots,a_s;b_1,\cdots,b_t;z] = \frac{\prod\Gamma(b_j)}{\prod\Gamma(a_k)}G^{1,s}_{s,t+1}(1-a_1,\cdots,1-a_s;0,1-b_1,\cdots,1-b_t;-z) \,.
\end{equation}
The prefactors cancel nicely with the coefficients to the $_4F_3$'s in (\ref{eq:L_mel_exact}) giving
\begin{align}\label{eq:L_mel_continue}
    &2\pi i\mel{P}{L(z)}{P'} = \frac{\pi PP'}{\sqrt{2}b \sinh(2\pi Pb)\sinh(2\pi P'b)} \times \nn\\
    & \left(\frac{z^2}{2} \int_\cC\frac{\Gamma(-w)\Gamma\left(w+1\pm\frac{iP}{b}\pm\frac{iP'}{b}\right)}{\Gamma\left(w+\frac{3}{2}\right)^2\Gamma(w+2)}(-z^2)^wdw+z\int_\cC\frac{\Gamma(-w)\Gamma\left(w+\frac{1}{2}\pm\frac{iP}{b}\pm\frac{iP'}{b}\right)}{\Gamma\left(w+\frac{1}{2}\right)\Gamma\left(w+1\right)\Gamma(w+\frac{3}{2})}(-z^2)^wdw\right)\,.
\end{align}
Once again, these matrix elements do not exhibit the familiar $b\leftrightarrow\frac{1}{b}$ duality of Liouville CFT. 

The analytic continuation is specified by the choice of $\cC$ contour. A graphical representation of two possible contour choices is provided in Fig. \ref{fig:mel_contour}. A contour which exclusively encircles the poles of $\Gamma(b_j-w)$ recovers the usual series definition of the hypergeometric function. These poles are spaced along the real $w$-axis on non-negative integers
\begin{equation}\label{eq:c1_poles}
    \Re w \in \Z^{\geq 0}\quad{\rm and}\quad \Im w = 0 \,.
\end{equation}
This contour choice, labeled $\cC_1$, only yields a convergent integral if we take $|z|\leq 1$. There is another choice for $\cC$ which generically agrees with $\cC_1$ in the intersection of their domains of convergence. When $|z|>1$, we use a contour which starts and ends at $-\infty$ and wraps all poles of the numerator to the left of the imaginary axis. We label this contour $\cC_2$. These poles are spaced in half-integer steps along four semi-finite lines where
\begin{equation}\label{eq:c2_poles}
    \Re w\in \Z^{\leq 0}/2\quad{\rm and}\quad \Im w = b^{-1}(\pm P\pm P')\,.
\end{equation}
Poles with $\Re w\in\Z^{\leq 0}$ come from the first integral, while poles with $\Re w\in\Z^{\leq 0}-1/2$ come from the second.

\subsection{Relation to matrix elements of the weakly coupled line defect}\label{sec:deco_weak}

Now, we relate the matrix elements of the decohered FZZT interface to the matrix elements of the line defect ${\bm L}_\Sigma$ (\ref{expdefect}) computed at weak coupling  in Section~\ref{seccorrpert}. Specifically, we will show that the matrix elements of the decohered FZZT state in the Schwarzian limit (accessed by taking $b\to 0$ while zooming in on states near the edge of the normalizable spectrum) match with (\ref{matweak}) for small $\Tilde{\mu}_D$. Using an orthogonality relation for Bessel functions of imaginary order \cite{Yakubovich:2006KL},
\begin{equation}\label{eq:KK_overlap}
    \int_0^{\infty}\frac{dx}{x}\ K_{-i\nu}(x)K_{-i\nu'}(x) = \frac{\pi^2}{2\nu\sinh(\pi \nu)}\delta(\nu-\nu') \,,
\end{equation}
we get that the leading contribution to the matrix element is given by a $\delta$-function in the difference in momenta, 
\begin{equation} \label{PlP'}
    \mel{P}{L(\Tilde{\mu}_D=0)}{P'} = \frac{\pi Pb}{2\sqrt{2}\sinh^2(2\pi P b)\sinh(2\pi P b^{-1})}\delta(P-P') \,.
\end{equation}
This shows that, when $\Tilde{\mu}_D=0$, the decohered FZZT interface acts as an identity operator on primary states. In other words, the decohered FZZT interface furnishes a resolution of the identity on the Schwarzian subspace of near-threshold primaries in a way that is consistent with the normalization of states given in (\ref{Pnormalization}). 

To compute the $O(\Tilde{\mu}_D)$ term, note the following integral identity,
\begin{equation} \label{decoweak}
    \int_0^{\infty}dx\ K_{-i\nu}(x)K_{-i\nu'}(x) =\frac{\pi^2}{4\cosh\left(\frac{\pi(\nu+\nu')}{2}\right)\cosh\left(\frac{\pi(\nu-\nu')}{2}\right)} \,,
\end{equation}
From this, we get that the matrix elements of the decohered FZZT interface at this order are given by
\begin{equation}
     \mel{P}{L(\Tilde{\mu}_D)}{P'}=\frac{\Tilde{\mu}_D}{\kappa} \frac{\pi PP'}{\sqrt{2}b\cosh\left(\frac{\pi(P+P')}{b}\right)\cosh\left(\frac{\pi(P-P')}{b}\right)\sinh(2\pi P b)\sinh(2\pi P'b)} \,.
\end{equation}
This expression can also be extracted from the exact result (\ref{eq:L_mel_exact}). In the $b\to 0$ limit with $k=\frac{P}{b}$ and $k'=\frac{P'}{b}$ held fixed, the above expression reduces to 
\begin{equation} \label{weakdec}
     \mel{P=bk}{L(\Tilde{\mu}_D)}{P'=bk'}\overset{b\to 0}{=}\frac{\Tilde{\mu}_D}{\sqrt{\mu_{\rm bulk}}}\frac{1}{\sqrt{32\pi} b^2 \cosh(\pi(k+k'))\cosh(\pi(k-k'))} \,.
\end{equation}
Compare this with (\ref{matweak}) where we computed the matrix elements of ${\bm L}_\Sigma$ to linear order in $\mu_D$ and in the $b\to 0$ limit. We get a match by equating the two  defect cosmological constants $\mu_D=\Tilde{\mu}_D$ and working near the edge of normalizable spectrum. We must also be careful to account for the change in normalization between the vertex operators (\ref{DOZZnorm}) used in (\ref{matweak}) and the normalization 
\begin{equation}
    \langle V_\alpha(\infty)V_{\alpha'}(0)\rangle =2\pi \rho_0(P) \langle P \ket{P'}
\end{equation}
of the primary states (\ref{Pnormalization}) used in (\ref{weakdec}). After accounting for these technical details, we get
\begin{align} 
    \langle V_{\frac{Q}{2}+ibk}(\infty)\, L(\Tilde{\mu}_D=\mu_D)\, V_{\frac{Q}{2}+ibk'}(0)\rangle &\overset{b\to 0}{=} \nn\\
    & \frac{\mu_D}{\sqrt{\mu_{\text{bulk}}}}\left(\frac{16\pi^3 kk' \sinh(2\pi k)\sinh(2\pi k')}{\cosh^2(\pi(k+k'))\cosh^2(\pi(k-k'))}\right)^{\frac{1}{2}} + O(\mu_D^2) \,.
\end{align} 
Thus, we have shown that, at weak coupling, the matrix elements of the decohered FZZT interface between states in the Schwarzian subspace match with the corresponding matrix elements of the defect ${\bm L}_\Sigma$. 

\subsubsection*{Beyond the Schwarzian limit:}

Beyond the Schwarzian limit, we can reproduce the matrix elements of the defect ${\bm L}_\Sigma$ at weak coupling (up to linear order in $\mu_D$) by generalizing the decohering ansatz to a non-diagonal identification of the fixed-length states: 
\begin{equation} \label{nondiag}
    L(\Tilde{\mu}_D) \to 2\sqrt{2}\pi b\int_0^{\infty}\frac{d\ell}{\ell} \frac{d\ell'}{\ell'} f(\ell,\ell';\Tilde{\mu}_D)\ketbra{\ell}{\ell'}\,.
\end{equation}
Using the completeness relation for the modified Bessel functions of imaginary order
\begin{equation}
    \int_0^\infty d\nu \nu\sinh(\pi\nu) K_{-i\nu}(x)K_{-i\nu}(x') = \frac{\pi^2}{2}x\delta(x-x')\,,
\end{equation}
we find, to linear order in $\Tilde{\mu}_D$, the kernel $f(\ell,\ell';\Tilde{\mu}_D)$ is given by
\begin{align} \label{nondiagkernel}
    f&(\ell,\ell';\Tilde{\mu}_D)=  \frac{1}{\sqrt{2}\pi^3 b^3}\int_0^{\infty}\rho_0(P)dP\ K_{-2iP/b}(\kappa \ell)K_{-2iP/b}(\kappa \ell') \qquad \\ \notag
    +& \Tilde{\mu}_D\frac{\sqrt{2}}{\pi^2 b^3 }\int_0^{\infty}\rho_0(P)dP\rho_0(P')dP'\ \Hat{C}_{\rm DOZZ}\left(\frac{Q}{2}+iP,\frac{Q}{2}+iP',\frac{b}{2}\right) K_{-2iP/b}(\kappa \ell)K_{-2iP'/b}(\kappa \ell')\,.
\end{align}
The hatted DOZZ structure constant in the above expression is the matrix element of $V_{b/2}$ between two primary states normalized as in (\ref{Pnormalization})
\begin{equation}
    \Hat{C}_{\rm DOZZ}\left(\frac{Q}{2}+iP,\frac{Q}{2}+iP',\frac{b}{2}\right)=\bra{P}V_{b/2}\ket{P'}\,.
\end{equation}
The kernel (\ref{nondiagkernel}) ensures that at weak coupling, the matrix elements of the defect ${\bm L}_\Sigma$ agree with those of the non-diagonal interface (\ref{nondiag}),
\begin{equation}
    \bra{P}{\bm L}_\Sigma \ket{P'}=2\sqrt{2}\pi b \int_0^{\infty}\frac{d\ell}{\ell}\frac{d\ell'}{\ell'} f(\ell,\ell';\mu_D)\braket{P}{\ell}\braket{\ell'}{P'}\,.
\end{equation}
Note also that in the Schwarzian limit, the kernel decoheres,
\begin{equation}
    f(\ell,\ell';\Tilde{\mu}_D)\underset{\rm schw}{=}(1+\Tilde{\mu}_D \ell)\ell\delta(\ell-\ell')\approx e^{\Tilde{\mu}_D\ell} \ell\delta(\ell-\ell')\,,
\end{equation}
thereby recovering the earlier results in this subsection. To compute the gluing kernel $f(\ell,\ell';\Tilde{\mu}_D)$ at higher orders in $\Tilde{\mu}_D$, we need to use the matrix elements of ${\bm L}_\Sigma$ at higher orders in $\mu_D$. However, as noted in Section \ref{sec:weakcoupling}, the defect generates UV divergences at higher orders in perturbation theory, so these matrix elements require UV cutoffs. But, for the manifestly conformal defect (\ref{alphac}), we could use this procedure to extract the gluing kernel $f$ to any order in perturbation theory since the defect does not generate any UV divergences perturbatively. This provides a representation of the matrix elements of the defect between primaries in terms of gluing a pair of 1-punctured disks with FZZT boundaries according to (\ref{nondiag}).

\subsection{Relation to matrix elements of the strongly coupled line defect} \label{Sechypdef}

We now relate the matrix elements of the decohered FZZT interface to the semiclassical matrix elements of the strongly coupled line defect ${\bm L}_\Sigma$ ((\ref{expdefect}) or (\ref{alphac})) computed in Section~\ref{secmatrix}. This will help establish the connection between the hyperbolic geometries with the line defect ${\bm L}_\Sigma$ constructed in Section~\ref{Secexp} with the decohered FZZT interface, thereby motivating use of the phrase `decohered hyperbolic geometries' in the title of the paper. To this end, we start with the integral form of the matrix elements written down in (\ref{matdecFZZT}) which we rewrite here,
\begin{equation}
    C_L(\alpha,\alpha'; \Tilde{\mu}_D)=\mathcal{N}\int_0^{\infty}\frac{d\ell}{\ell}e^{\Tilde{\mu}_D \ell} K_{(Q-2\alpha)/b}(\kappa l) K_{(Q-2\alpha')/b}(\kappa l) \,.
\end{equation}
In the subsequent semiclassical analysis, the normalization factor $\mathcal{N}$ won't play any role.
We evaluate the above integral in the semiclassical limit $b\to 0$ using the saddlepoint approximation. We take the $b\to 0$ limit holding $\eta, \eta', \mu$ fixed,
\begin{equation}
    \alpha = \frac{\eta}{b}\,, \quad \alpha'=\frac{\eta'}{b}\,, \quad \Tilde{\mu}_D = \frac{\mu}{2\pi b^2},\quad {\rm and}\ \quad\mu_{\text{bulk}}=\frac{1}{4\pi b^2}\,.
\end{equation}
The Bessel function in the $b\to 0$ limit has the following asymptotics,
\begin{equation}
    K_{\frac{1-2\eta}{b^2}}\left( \frac{r}{b^2} \right) \sim \exp(\frac{1}{b^2}\left[-\sqrt{(1-2\eta)^2 + r^2} + (1-2\eta)\sinh^{-1}\left(\frac{1-2\eta}{r}\right)\right]) \,.
\end{equation}
For real $1-2\eta>0$, we can derive these asymptotics from the saddlepoint of the following integral representation valid for $\text{Re}(x)>0$,
\begin{equation} \label{BesselIntegralrep}
    K_{\alpha}(x)=\int_0^{\infty}dt \cosh(\alpha t)e^{-x\cosh(t)} \,.
\end{equation}
For purely imaginary $\alpha=i\beta$ with $x>|\alpha|$, the asymptotics are given by 
\begin{equation}
    K_{i\beta}(x)\sim \sqrt{\frac{\pi}{2\sqrt{x^2-\beta^2}}}\exp\left(-\sqrt{x^2-\beta^2}-\beta\sin^{-1}(\frac{\beta}{x})\right) \,.
\end{equation}
Using these aymptotics, we can express the integrand in terms of an effective action which is a function of the length parameter $r\equiv\frac{\ell}{2\pi}$,
\begin{equation}
    C_L(\eta,\eta'; \mu) \sim  \int_0^{\infty} dr \, e^{I(r)/b^2} \,,
\end{equation}
where the effective action is given by
\begin{equation}
    \begin{split}
        I(r) = \mu r &- \sqrt{(1-2\eta)^2 + r^2} + (1-2\eta)\sinh^{-1}(\frac{1-2\eta}{r}) \\
        &- \sqrt{(1-2\eta')^2 + r^2} + (1-2\eta')\sinh^{-1}(\frac{1-2\eta'}{r})\,.
    \end{split}
\end{equation}
Since we are working in the limit $b\rightarrow0$, we can evaluate the integral using a saddlepoint approximation with the saddlepoint equation taking the form,
\begin{equation}
    I'(r_0) = 0 \implies \mu - \frac{\sqrt{(1-2\eta)^2 + r_0^2}}{r_0} - \frac{\sqrt{(1-2\eta')^2 + r_0^2}}{r_0} = 0\,.
\end{equation}
We denoted the location of the saddle as $r_0$.
We are interested in the regime where the weights are above the threshold, so we parametrise them as
\begin{equation}
    \eta = \frac{1}{2}(1 + ir_H) \quad{\rm and}\quad \eta' = \frac{1}{2}(1 + ir_H') \,.
\end{equation}
With this parametrisation, the saddlepoint equation reduces to the jump condition for the normal derivative of the Liouville field across the exponential line defect derived in the previous section,
\begin{equation}
    \mu r_0 = \sqrt{r_0^2 - r_H^2} + \sqrt{r_0^2 - r_H'^2} \,.
\end{equation}
When the weights are above the threshold, the saddle in the $r$ integral, at $r=r_0$ is stable because
\begin{equation}
    I''(r_0)=-\frac{r_H^2}{r_0^2\sqrt{r_0^2-r_H^2}}-\frac{r_H'^2}{r_0^2\sqrt{r_0^2-r_H'^2}}<0 \,.
\end{equation}
Furthermore, evaluating the effective action using the above saddlepoint equation, we get
\begin{equation}
    \begin{split}
        b^2 \log |C_L(\gamma,\gamma')| \sim \gamma\cos^{-1}\left(\frac{\gamma}{r_0(\gamma,\gamma')}\right) + \gamma'\cos^{-1}\left(\frac{\gamma'}{r_0(\gamma,\gamma')}\right) \,, 
    \end{split}
\end{equation}
where we have replaced $r_H$ and $r_H'$ by $\gamma$ and $\gamma'$ respectively to match the notation of Section~\ref{secmatrix}.
Since we define the semiclassical structure constants in terms of the absolute value of the exact matrix elements, $|C_L| \sim \exp(\frac{c}{6} c_L)$, we get the expression
\begin{equation}
     c_L(\gamma,\gamma')=\gamma\cos^{-1}\left(\frac{\gamma}{r_0(\gamma,\gamma')}\right)+\gamma'\cos^{-1}\left(\frac{\gamma'}{r_0(\gamma,\gamma')}\right) \,,
\end{equation}
which matches with (\ref{semimat}) computed using the on-shell Liouville action. We thus establish a match between the matrix elements of the decohered FZZT interface and the matrix elements of the line defect ${\bm L}_\Sigma$ in the strongly coupled semiclassical limit.

\subsection{Effect of decohering FZZT boundaries on cylinder amplitudes} \label{sec:cyldeco}

\begin{figure}
    \centering
    \includegraphics[width=0.6\linewidth]{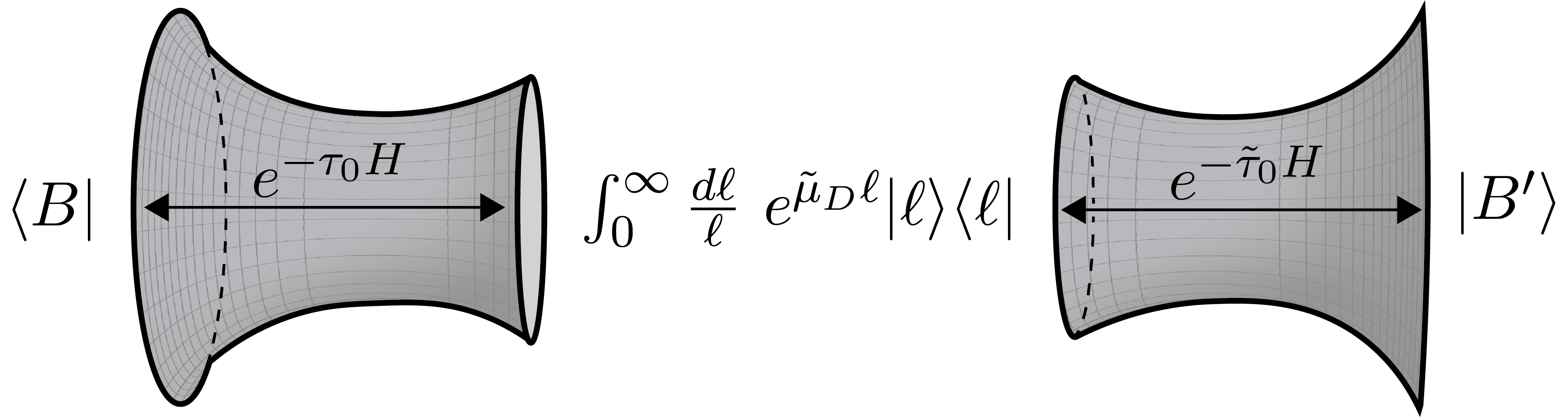}
    \caption{The figure shows the cylinder transition amplitude between two conformal boundary states denoted $\ket{B}$ and $\ket{B'}$ with the decohered FZZT interface inserted at distances $\tau_0$ and $\Tilde{\tau}_0$ from the boundaries.} 
    \label{fig:P_L_P}
\end{figure}

In this section, we compute cylinder transition amplitudes between Liouville boundary states in the presence of the decohered FZZT interface. This will help us quantify the effect of decohering the FZZT boundary on the correlation induced between the two portions of the cylinder as described pictorially in Figure~\ref{fig:P_L_P}.

In terms of the matrix elements that we computed earlier, we can expand the decohered FZZT interface using a tensor product of Ishibashi states,
\begin{equation}
    L(\Tilde{\mu}_D)=\int_0^{\infty} dPdQ\rho_0(P)\rho_0(Q)  C_L(P,Q;\Tilde{\mu}_D) \ket{P}\rangle \langle \bra{Q} \,,
\end{equation}
where $C_L(P,Q;\Tilde{\mu}_D)$ is the matrix element computed in (\ref{matdecFZZT}),
\begin{equation}
    C_L(P,Q;\Tilde{\mu}_D)=\bra{P}L(\Tilde{\mu}_D)\ket{Q} \,. 
\end{equation}
Let us compute the cylinder transition amplitudes with the decohered FZZT interface placed parallel to the boundaries at a separation of $\tau_0$ and $\Tilde{\tau}_0$ from either boundary. The transition amplitudes between ZZ states is given by
\begin{equation}
\begin{split}
    \bra{\text{ZZ}}e^{-\tau_0 H}L(\Tilde{\mu}_D) e^{-\Tilde{\tau}_0 H} & \ket{\text{ZZ}} = \nn\\
    &\int_0^{\infty} dP dQ \rho_0(P)\rho_0(Q)\Psi_{\rm ZZ}(P)\Psi_{\rm ZZ}(Q)C_L(P,Q;\Tilde{\mu}_D)\chi_P\left(\frac{i\tau_0}{\pi}\right)\chi_Q\left(\frac{i\Tilde{\tau}_0}{\pi}\right) \,.
    \end{split}
\end{equation}
Note that with the normalization of the boundary states used in (\ref{eq:liouville_bdy_wvfcns}), $\psi_{\rm ZZ}=1$. 
\begin{align} \label{ZZtrans}
    \bra{\text{ZZ}}e^{-\tau_0 H}L(\Tilde{\mu}_D)  e^{-\Tilde{\tau}_0H}\ket{\text{ZZ}} &= \int_0^{\infty} dP dQ \rho_0(P)\rho_0(Q)C_L(P,Q;\Tilde{\mu}_D)\chi_P\left(\frac{i\tau_0}{\pi}\right)\chi_Q\left(\frac{i\Tilde{\tau}_0}{\pi}\right) \,,\nn\\
    &= \frac{64\sqrt{2}}{\pi b \eta(\frac{i\tau_0}{\pi}) \eta(\frac{i\Tilde{\tau}_0}{\pi})}\int_0^{\infty}\frac{d\ell}{\ell}e^{\Tilde{\mu}_D\ell}\int_0^{\infty} dPP\sinh(\frac{2\pi P}{b})K_{-\frac{2iP}{b}}(\kappa \ell)e^{-2\tau_0 P^2} \nn\\
    &\qquad \qquad \qquad \qquad \qquad \times\int_0^{\infty} dQQ\sinh(\frac{2\pi Q}{b})K_{-\frac{2iQ}{b}}(\kappa \ell)  e^{-2\Tilde{\tau}_0 Q^2} \,,\nn\\
    &= \frac{\sqrt{2}b^3 \pi^3}{\eta(\frac{i\tau_0}{\pi}) \eta(\frac{i\Tilde{\tau}_0}{\pi})}\int_0^{\infty}d\ell \ell e^{\frac{\Tilde{\mu}_D}{\kappa}\ell}F_{b^2\tau_0}(\ell)F_{b^2\Tilde{\tau}_0}(\ell) \,,
\end{align}
where we noted the appearance of the Yor integral denoted $F_t(r)$ expressed in terms of the inverse Kontorovich-Lebedev (KL) transform\footnote{The Kontorovich-Lebedev transform is an integral transform of a function with the kernel being a Bessel function of imaginary order. See for example \cite{Yakubovich:1996IndexTransforms, Lebedev:SpecialFunctions} for definition and properties.} of the Gaussian (See for example \cite{yakubovich2012yorintegralpolynomialsrelated} where it is also mentioned that the Yor integral plays an important role in certain models of mathematical finance),
\begin{equation}
    F_t(r)=\frac{2}{r\pi^2}\int_0^{\infty}d\nu e^{-\frac{t \nu^2}{2}}\nu \sinh(\pi \nu)K_{i\nu}(r) \implies \int_0^{\infty}dr K_{i\nu}(r)F_t(r)=e^{-\frac{t\nu^2}{2}} \,.
\end{equation}
Hence, the ZZ-ZZ transition amplitude with the decohered FZZT interface is given by
\begin{equation} \label{ZZZZdec}
    \bra{\text{ZZ}}e^{-\tau_0 H}L(\Tilde{\mu}_D)  e^{-\Tilde{\tau}_0H}\ket{\text{ZZ}}= \frac{\sqrt{2}b^3 \pi^3}{\eta(\frac{i\tau_0}{\pi}) \eta(\frac{i\Tilde{\tau}_0}{\pi})}\int_0^{\infty}d\ell \ell e^{\frac{\Tilde{\mu}_D}{\kappa}\ell}F_{b^2\tau_0}(\ell)F_{b^2\Tilde{\tau}_0}(\ell) \,.
\end{equation}
If we didn't decohere the FZZT boundaries, then the above transition amplitude would factorize into a product of ZZ-FZZT transition amplitudes. The integral over $\ell$ in (\ref{ZZZZdec}) doesn't factorize and hence captures the correlation between the two halves of the cylinder induced by decohering the FZZT boundaries. The $\eta$-functions in the denominator of (\ref{ZZZZdec}) capture the contribution from Virasoro descendants. Since they factorize, it suggests that the decoherence only correlates Virasoro primaries across the interface but doesn't correlate the descendants.

We can evaluate the $\ell$-integral in (\ref{ZZZZdec}) analytically in the $\Tilde{\mu}_D \to 0$ limit. To this end, we use the Parseval identity for the KL-transform which states that overlap between two square-integrable functions is invariant under the KL-transform,
\begin{equation}
    \int_0^{\infty} dr r f(r) g(r)=\frac{2}{\pi^2}\int_0^{\infty}d\nu \nu \sinh(\pi \nu) \Hat{f}(\nu)\Hat{g}(\nu) \,.
\end{equation}
In our case, the functions are the two Yor integrals whose KL-transforms are Gaussians and hence the integral is simple to evaluate,
\begin{equation}
    \int _0^{\infty}dr rF_t(r) F_s(r)=\frac{2}{\pi^2}\int_0^{\infty}d\nu \nu \sinh(\pi \nu)e^{-\frac{s\nu^2}{2}}e^{-\frac{t\nu^2}{2}}=\sqrt{\frac{2}{\pi}}\frac{e^{\frac{\pi^2}{2(t+s)}}}{(t+s)^{\frac{3}{2}}} \,,
\end{equation}
Thus, in this limit, the ZZ transition amplitude in the presence of the decohered FZZT state is given by
\begin{equation}
     \bra{\rm ZZ}e^{-\tau_0 H}L(\Tilde{\mu}_D = 0)  e^{-\Tilde{\tau}_0H}\ket{\rm ZZ}=2\pi^{\frac{5}{2}}\frac{e^{\frac{\pi^2}{2b^2(\tau_0+\Tilde{\tau}_0)}}}{(\tau_0+\Tilde{\tau}_0)^{\frac{3}{2}} \eta(\frac{i\tau_0}{\pi}) \eta(\frac{i\Tilde{\tau}_0}{\pi})} \,.
\end{equation}
This result can also be derived directly using (\ref{PlP'}).

\section{From Gaussian Multiplicative Chaos to Geometric Decoherence} \label{sec:GMC}

As mentioned in the Introduction, the Liouville CFT plays an important role in the formulation of $2d$ quantum gravity, which, in the Euclidean setting, is conjectured to emerge as the scaling limit of random planar maps. This connection between Liouville CFT and discretized random surfaces has led to many important developments at the interface of mathematics and physics, including probability theory, random geometry, Teichm\"uller theory, statistical physics, conformal field theory, and string theory (see \cite{Vargas:2017swx,Chatterjee:2024phq,Guillarmou:2024lqk,Rhodes:2025fti,Holden:2025mmu} for recent reviews).

More recently, a rigorous definition of Liouville CFT as a theory of random surfaces has been put forward within the probabilistic approach. This framework places the path integral formulation of the theory on a solid mathematical footing and has established several previously conjectured nontrivial results in connection with the conformal bootstrap formulation of Liouville CFT \cite{David:2014aha,David:2015kea,Huang:2018ncv,Guillarmou:2020wbo,Guillarmou:2021frk,Guillarmou:2024mwf}.

Although much of the existing literature has focused on conformal observables in Liouville CFT, the probabilistic approach has the advantage of also addressing observables away from criticality, such as the ``localized cosmological constant'' defect  we consider in \eqref{eq:defect_def}. The key ingredient is a rigorous definition of a regularized and renormalized local exponential interaction of the form $e^{\gamma \phi(x)}$ as a random measure on $\mathbb{C}$, known as Gaussian multiplicative chaos (GMC), which we will briefly review below.
We will see that GMC can be used to define our line defect \eqref{eq:defect_def} non-perturbatively, and we will sketch how, within this approach, one may find further evidence for the emergence of a decohered FZZT interface describing effectively \eqref{eq:defect_def} at least at strong defect coupling.

\subsection{Primer on Probability Approach to Liouville CFT and Beyond}

There are many excellent reviews on the probability approach to Liouville CFT (see e.g. \cite{Vargas:2017swx,Chatterjee:2024phq,Guillarmou:2024lqk,Rhodes:2025fti,Holden:2025mmu}). Here we gather the main ingredients necessary for the application in Section~\ref{sec:GMCtoDecoherence}. We also provide a quick summary of the probability and CFT concepts together with their relations in Table~\ref{tab:LCFTprob}. We will mostly follow the reviews \cite{Vargas:2017swx,Chatterjee:2024phq}
below.\footnote{We adopt the physics convention for Liouville vertex operators $e^{2\alpha \phi}$ and the Liouville momenta here differ from those in \cite{Vargas:2017swx} by a factor of 2.}

In the path integral formulation of the Liouville CFT on the flat spacetime, expectation values of observables $F(\phi)$ build from the Liouville field, are calculated by the following formal integral over the field space,
\ie 
\langle F(\phi)\rangle =\int F(\phi) e^{-S_{\rm Liouville}(\phi)}  D\phi \,,
\fe
where $S_{\rm Liouville}$ is given in \eqref{eq:covariant_lv_action} in the covariant form. An obvious problem with the above expression is that $D\phi$ does not exist as a Lebesgue measure on the infinite dimensional field space. However, it is natural to consider the combination together with the free field action $e^{-S_{\rm free}(\phi)}  D\phi$ which can be rigorously defined as the infinite dimensional Gaussian probability measure, thanks to the exponential damping and translation symmetry breaking in field space. 

\paragraph{Gaussian Free Field (GFF)}
Following the probabilistic approach, it's convenient to decompose the Liouville field as,
\ie 
\phi_g(x)=c+X(x)\,,
\label{separation}
\fe 
into a real constant $c$ and the Gaussian Free Field $X(x)$.  The choice of metric $ds^2=g(x) |dx|^2$  enters in the definition of $X(x)$ as a random distribution and the associated measures. Recall that the flat spacetime Liouville field can be obtained from the Weyl transformation,
\ie 
\phi(x)=\phi_g(x)+{Q\over 2} \log g(x)\,.
\label{flatphi}
\fe

Here we follow \cite{Vargas:2017swx} and choose the metric on the sphere ${\rm S}^2$ to be that from gluing two flat disks by inversion, and thus the conformal factor is
\ie g(x)={1\over |x|_+^4}\,,\quad  |x|_+\equiv \max (1,|x|)\,,
\label{metricchoice}
\fe
which turns out to simplify various integral formulae for Liouville observables. In particular, note that the curvature for \eqref{metricchoice} is concentrated at $|x|=1$. The decomposition in \eqref{separation} is achieved by enforcing
\ie 
\int_0^{2\pi} d\theta X(e^{2\pi i\theta}) =0\,.
\fe
The covariance of the Gaussian free field is given by
\ie 
\mathbb{E}[X(x)X(y)]=G(x,y)=\log {|x|_+|y|_+\over |x-y|}\,.
\label{GFFcov}
\fe
Clearly $X(x)$ is not a function but should be regarded as a distribution that lives in a negative-index Sobolev space.\footnote{Suppose the spectral decomposition of the Laplacian $-\Delta$ on a domain $D$ is given in terms of non-negative eigenvalues $\lambda_0=0<\lambda_1<\lambda_2<\dots$ and corresponding $L^2$-normalized eigenfunctions $e_n(x)$. The Sobolev space $H^{s}(D)$ with index $s$ is defined with respect to the norm $|\cdot|_s$ such that for a sequence $f=\sum_{n\geq 1} f_n e_n$, $|f|_s^2=\sum_n{f_n^2  \lambda_n^{s}}$. Consequently, the sequence $\sum_{n\geq 1} X_n {e_n\over \sqrt{\lambda_n}}$ converges in $H^{s}(D)$ as long as $s<0$ and this defines the GFF $X(x)$. The Gaussian distribution of $X_n$ then naturally induces the infinite measure $d\mathbb{P}[X]$.} 
The standard infinite measure on GFF is denoted as $d\mathbb{P}[x]$ with the corresponding expectation values denoted as $\mathbb{E}[\cdot]$. The GFF enjoys manifest conformal invariance (after including the zero mode $c$) and Markov property which are essential in the construction of a local CFT in the probabilistic approach.

\paragraph{Gaussian Multiplicative Chaos (GMC)}

The exponential interaction $e^{2b\phi}$ in the Liouville action  requires regularization and renormalization due to short distance singularities that are common in QFT. In an interesting way, this same problem also arises in probability theory and a different physical context, where random measures of the form $e^{\gamma X(x)}d^d x$ for log-correlated GFF $X(x)$ on domains in $\mathbb{R}^d$ were considered to model energy dissipation in the Kolmogorov-Obukhov model of turbulence (see \cite{Rhodes:2013iua} for a comprehensive review of GMC and its applications). The regularized and renormalized random measure on $\mathbb{R}^d$ is known as the Gaussian Multiplicative Chaos, which is known to exist for $0<\gamma\leq \sqrt{2d}$. For our purpose $d=2$ and $\gamma=2b$, and so this condition is obeyed for $0<b\leq 1$ by the bulk Liouville interaction. 

To define the  exponential measure properly requires first replacing $X(x)$, which is not a random variable  by its regularized version $X_\epsilon(x)$ which has finite variance, and defines a random variable. Then $e^{\gamma X_\ep(x)}$ is a non-negative random variable and thus $e^{\gamma X_\ep(x)} d^2 x$ defines a random Radon measure. The GMC measure is defined by the limit of these random Radon measures
\ie 
M_\gamma(d^2 x)\equiv \lim_{\ep \to 0 } e^{\gamma X_\ep(x)-{
		\gamma^2\over 2} \mathbb{E}[X_\ep(x)^2]}g(x) d^2 x\,,
\label{GMCrr}
\fe
which converges (weakly) to a positive random measure for $0<\gamma<2$. The GMC measure for the borderline case $\gamma=2$ is more subtle but can be dealt with by an additional logarithmic renormalization. It has been shown that the limit does not depend on the choice of the regularization. Here we will take a spacetime (soft) cutoff regulator, known as mollification in probability theory, given by
\ie 
X_\ep(x)\equiv {1\over 2\pi }\int_0^{2\pi}d\theta X(x+\ep e^{i\theta})\,,
\label{Xep}
\fe 
via smearing on a unit circle of radius $\ep$ centered at $x$.
One could also implement a momentum space type cutoff, known as Martingale regularization. The convergence property is mostly transparent in the latter regularization, since the limit involves a sequence of Martingales (see \cite{Chatterjee:2024phq} for a recent review). On the other hand, the spacetime cutoff regulator is more convenient when calculating correlation functions.

The regulated GFF \eqref{Xep} is a random variable. Indeed, $\mathbb{E}[X_\ep(x) X_\ep(y)]$ is bounded everywhere for fixed $\ep$ and the variance is
\ie 
\mathbb{E}[X_\ep(x)^2]=-\log \ep +2\log |x|_+ \,.
\fe
Therefore  \eqref{GMCrr} can be written as
\ie 
M_\gamma(d^2 x)=e^{-\gamma c}\lim_{\ep \to 0 } \ep^{\gamma^2\over 2} e^{\gamma  \phi_\ep(x)}  d^2 x\,,
\fe 
where $\phi_\ep$ is the regularized Liouville field with flat metric from \eqref{flatphi},
\ie 
\phi_\ep = c+X_\ep(x)-2Q\log |x|_+
\,,
\fe
where the last time implements the Weyl transformation from \eqref{metricchoice} to flat metric.

\paragraph{Integrability of GMC} An important property of the GMC measure is its integrability against power-law singularities. Namely the expectation values of
\ie 
I(\alpha)\equiv \int_{|x-z|\leq 1}  {M_{2b}(d^2 x)\over |x-z|^{4b \alpha}}  \,,
\label{Ialpha}\fe
satisfy 
\ie 
\mathbb{E}\left[ I(\alpha)^p\right]<+\infty\,,\quad \text{if and only if}~ p<\min\left({1\over b^2}, {Q-2\alpha\over b}\right)\,.
\label{Ialphamom}
\fe
which is a consequence of the multifractal nature of the GMC random measure. Such a singularity in the integrand of \eqref{Ialpha} is induced by a local vertex operator insertion $V_\alpha(z)$ and the Seiberg bound $\alpha<{Q\over 2}$ ensures that $I(\alpha)$ defines a positive random variable on the probability space of GFF.  In particular, the mass of the GMC measure $M_\gamma({\rm S}^2)\equiv \int_{{\rm S}^2} M_\gamma(d^2 x)$ is a positive random variable.

\paragraph{Liouville Path Integral} In the probability approach, the Liouville path integral is defined with respect to the following measure
\ie 
e^{-S_{\rm Liouville}(\phi)}D\phi \equiv 2e^{-2Qc} e^{-\m_{\rm bulk} e^{2b c}  M_{2b}({\rm S}^2)} dc\otimes d\mathbb{P}[X]
\label{Lmeasure}
\fe
where the LHS is a multiplicative modification of the GFF measure with the zero mode Lebesgue measure $dc$ by a random positive variable built from the mass of the GMC on ${\rm S}^2$. Correlation functions in the Liouville CFT are then defined by expectation values with respect to this probability measure. 

\paragraph{Local Correlation Functions} 
The correlation functions of local vertex operators $V_{\alpha_i}=e^{2\alpha_i\phi}$ are defined by regulated and renormalized insertions in the path integral,\footnote{As is the case of the field $\phi$ (and the GFF $X$), the bare operator $e^{2\alpha_i\phi}$ is not a random variable but its regularized version $\ep^{2\alpha_i^2}e^{2\alpha_i\phi_\ep}$ is a random variable and positive for $\alpha_i\in \mathbb{R}$.}\ie
\langle \prod_i V_{\alpha_i}(z_i)\rangle
=\lim_{\ep\to 0}\int \prod_{i} \ep^{2\alpha_i^2} e^{2\alpha_i \phi_\ep(z_i)} e^{-S_{\rm Liouville}(\phi)}D\phi\,.
\fe 
It is convenient to use the probability expression of the measure in \eqref{Lmeasure} and rewrite the RHS above as an expectation value in the GFF, where the factor of 2 is inherited from \eqref{Lmeasure} which ensures the correlators obey the usual CFT normalizations \cite{Vargas:2017swx} (e.g. the DOZZ formula \cite{Nakayama:2004vk}),
\ie 
\langle \prod_i V_{\alpha_i}(z_i)\rangle=2\lim_{\ep \to 0}\int_\mathbb{R} dc \,e^{-2Qc} \mathbb{E}
\left[ e^{-\m_{\rm bulk} e^{2b c}  M_{2b}({\rm S}^2)} \prod_i\ep^{2\alpha_i^2}
e^{2\alpha_i \phi_\ep (z_i)}
\right]\,.
\fe
Performing the integral over $c$ (after applying the Fubini's theorem), we obtain 
\ie 
\langle \prod_i V_{\alpha_i}(z_i)\rangle
=
b^{-1}\m_{\rm bulk}^{-s} \Gamma(s) \lim_{\ep \to 0}\mathbb{E}\left[ 
M_{2b}({\rm S}^2)^{-s}\prod_i\ep^{2\alpha_i^2 }
e^{2\alpha_i \phi_\ep (z_i)}
\right]~{\rm with}~s\equiv {\sum_i \alpha_i -Q\over b}\,,
\fe
which converges if the first Seiberg bound $\sum_i \alpha_i>Q$ is satisfied.\footnote{In fact, the following weaker bound is sufficient
	\ie 
	-s <\min \left({1\over b^2},{Q-2\alpha_i\over b} \right)~\forall i\,,
	\fe
	which follows from the integrability of the GMC \eqref{Ialphamom} and the expression \eqref{probcorrelator}.} 
Then by applying the Girsanov theorem for the GFF,\footnote{This is also commonly known as the ``complete-the-square'' trick used in manipulating QFT path integral. The Girsanov theorem makes this formal manipulation rigorous in the GFF.} we have after taking the $\ep \to 0$ limit,
\ie 
\langle \prod_i V_{\alpha_i}(z_i)\rangle
=b^{-1}\m_{\rm bulk}^{-s} \Gamma(s)\prod_{i<j}|z_i-z_j|^{-4\alpha_i \alpha_j}\mathbb{E}\left[ 
\left(\int_{{\rm S}^2} F(x,\bm{z})M_{2b}(d^2 x)\right)^{-s}
\right]\,,
\label{probcorrelator}
\fe 
where the local operator insertions modify the random GMC measure by a Radon-Nikodym derivative,
\ie 
F(x,\bm{z})\equiv \prod_{i} \left({|z_i|_+\over |x-z_i|}\right)^{4b\alpha_i}\,.
\fe
As explained around \eqref{Ialpha}, this modified random measure integrates to a positive random variable if the second Seiberg bound $\alpha_i<Q/2$ is obeyed, and thus the GFF expectation value in  
\eqref{probcorrelator} is nontrivial and well-defined, and the remaining integrals have been shown to reproduce the known correlation functions in the Liouville CFT obtained from conformal bootstrap when the Seiberg bounds are obeyed. The correlation functions for more general $\alpha_i$ are then obtained from analytic continuation (assuming analyticity in $\alpha_i$). For example, the DOZZ formula for the three-point function is derived this way explicitly \cite{Kupiainen:2017eaw}, without assumptions on the $b\xleftrightarrow{} 1/b$ duality which were important  in the original argument of \cite{Teschner:2001rv}.
Furthermore the generalizations of the correlation functions on ${\rm S}^2$ to the case of higher genus Riemann surfaces takes a very similar form to \cite{Guillarmou:2016ked} and consistency with the conformal bootstrap axioms have been established in \cite{Guillarmou:2020wbo}.

\begin{table}[!htb]
	\centering
	\renewcommand{\arraystretch}{1.2}
	\begin{tabular}{|cc|}\hline
		Probability  & Liouville CFT \\\hline
		Gaussian free field (GFF)  $X(x)$    & \multicolumn{1}{c|}{\multirow{2}{*}{Liouville field $\phi(x)=c+X(x)$}} \\
		Lebesgue variable $c \in \mathbb{R}$
		& \multicolumn{1}{c|}{} 
		\\\hline
		Covariance of GFF $\mathbb{E}[X(x)X(y)]$& Free scalar propagator $G(x,y)$
		\\
		Dirichlet energy $\int |\nabla X|^2$ & Free scalar action $S_{\rm free}(\phi)$
		\\
		Law of GFF (with $c$) $dc\otimes d\mathbb{P}[X]$ & Free path integral measure $e^{-S_{\rm free}(\phi)} D\phi
		$
		\\\hline 
		Mollification of GFF $X_\epsilon(x)$ &  Smooth short-distance regularization
		\\
		Martingale approximation of GFF $X_t(x)$ & Sharp spectral cutoff regularization 
		\\\hline
		Gaussian Multiplicative Chaos  $M_\gamma(d^2 x)$   & Renormalized exponential interaction $ e^{\gamma X(x)}d^2x$
		\\
		Random surface measure  from $e^{2bc}M_{2b}(d^2x)   $ & Renormalized Liouville interaction 
		$e^{2b\phi}d^2 x$
		\\
		Random boundary measure  $e^{bc}M_{b}^{\partial}(ds) $ & Renormalized FZZT boundary interaction $e^{b\phi_\partial}ds$
		\\
		Random length measure  $e^{bc}M_{b}(ds)$ & Renormalized defect interaction $e^{b\phi}ds$
		\\
		Law of Liouville field  & Liouville path integral measure $e^{-S_{\rm Liouville}(\phi)}D\phi$
		\\
		Random variables from GFF & Local operator insertions (regularized)
		\\\hline
		Moments of random variables from GMC  & Correlation functions (and conformal blocks) 
		\\
		Moments of random area & DOZZ bulk structure constants
		\\
		Moments of random area and boundary lengths & FZZT bulk-boundary structure constants
		\\
		Finite moments & Seiberg bound  I: $\sum_i\alpha_i>(1-g)Q$
		\\
		Non-negative moments & Seiberg bound II: $\alpha_i\leq  Q/2$
		\\
		Law of random area and boundary lengths & Fixed area and fixed length correlators
		\\\hline 
		\multirow{3}{6.5cm}{\centering Gaussian integration by parts identities  
			(Malliavin calculus)}   & KPZ identities (higher equations of motion)
		\\
		& BPZ identities (null state equations)
		\\
		& Conformal Ward identities
		\\\hline
		\multirow{3}{8cm}{\centering Markov property of GFF  \\[5pt] Positive measure (and positive masses from GMC)}  
		&
		Locality 
		\\
		&Reflection Positivty \\
		
		& Analyticity in Liouville momenta 
		\\\hline 
	\end{tabular}
	\caption{A concise dictionary matching concepts in probability theory with their counterparts in Liouville CFT. In several cases, a single structure or property on one side corresponds to multiple structures on the other side. The relative positioning of the entries is intended to make these correspondences explicit.}
	\label{tab:LCFTprob}
\end{table}

\paragraph{Correlation functions with boundaries}
Conformal boundaries can also be incorporated in the probabilistic approach to the Liouville theory. Let us consider the simplest nontrivial setting where the theory is defined on the upper half plane $\mathbb{H}$ with $\Im z>0$ where we impose FZZT boundary conditions labeled by boundary cosmological constant $\m_B$ at $\Im z=0$.  
The probabilistic formula for general bulk-boundary correlation functions can be derived in much the same way as above, after noting that the GFF on $\mathbb{H}$ with Neumann boundary condition, which is the relevant case here, has the following covariance,
\ie 
\mathbb{E}[X(x)X(y)]=\log{|x|_+^2 |y|_+^2\over |x-y||x-\bar y|} \,,
\fe
from the method of images.
With the same background metric in  \eqref{metricchoice}, we implement the regularization of the GFF as before at interior points of $\mathbb{H}$ as in 
\eqref{Xep}, and a similar average over a half-circle for points on the boundary. This allow us to define, in additional to the bulk GMC, the following boundary GMC for the Neumann GFF $X(x)$ and the corresponding Liouville field $\phi(x)$,
\ie 
M_b^\partial(dx)
\equiv \lim_{\ep \to 0 } e^{ b X_\ep(x)- 
	{b^2\over 2} \mathbb{E}[X_\ep(x)^2]} {d  x\over |x|_+^2}
=
e^{-b c}\lim_{\ep \to 0 } \ep^{b^2} e^{b\phi_\ep(x)}  d x\,,
\label{bdyGMC}
\fe 
where $x\in \mathbb{R}$ is restricted to the boundary of $\mathbb{H}$. The correlation functions with bulk  and boundary operator insertions are then computed by a similar probability formula,
\ie 
\langle
&\prod_i V_{\alpha_i}(z_i) \prod_m U_{\beta_m}(t_m)
\rangle 
\\
=&2\lim_{\ep \to 0}\int_\mathbb{R} dc \,e^{-2Qc} \mathbb{E}
\left[ e^{-\m_{\rm bulk} e^{2b c}  M_{2b}(\mathbb{H})-\m_B e^{ b c}  M^\partial_{ b}(\mathbb{R})} \prod_i\ep^{2\alpha_i^2}
e^{2\alpha_i \phi_\ep (z_i)}
\prod_m\ep^{ 
	\beta_m^2}
e^{\beta_m \phi_\ep (t_m)
}
\right]\,,
\label{bulkboundaryprob}
\fe
where $U_\beta(t)\equiv e^{\beta \phi(t)}$ denotes the boundary operator of dimension $\Delta=\beta(Q-\beta)$ inserted at $t\in \mathbb{R}$. For the basic mixed bulk and boundary structure constants, the above expression has been shown to reproduce results from (boundary) conformal bootstrap in the Liouville CFT.

\paragraph{Beyond Liouville}
Beyond Liouville, the probabilistic framework built on the Gaussian free field and Gaussian multiplicative chaos has been extended to a broader class of non-rational CFTs and QFTs, providing a genuinely nonperturbative definition of renormalized exponential interactions both in the bulk and at the boundary. In Toda theories \cite{Cercle:2023xfz}, multi-component exponential potentials associated with simple Lie algebras are constructed rigorously. In the sinh-Gordon model \cite{Guillarmou:2024uou}, the two-sided interaction $e^{\pm \gamma \phi}$ is treated simultaneously in the probabilistic approach, establishing the mass gap and a discrete spectrum rigorously. 
The cases of exponential interactions with imaginary exponents are more intricate and have been recently dealt with in the boundary sine-Gordon theory via the probability approach \cite{lacoin2020probabilisticapproachultravioletrenormalisation}. 
The approach has further been extended to more complicated theories, including time-like Liouville \cite{Guillarmou:2023exh,Usciati:2025cdn} with $c\leq 1$ (see also \cite{Ribault:2015sxa}), and to the non-compact $\mathbb{H} _3$ WZNW model \cite{Guillarmou:2025vrq}, where affine symmetry is encoded directly in the probabilistic formulation, providing a structural explanation of the Ribault–Teschner correspondence with Liouville CFT observables.

\subsection{From GMC Defect to Geometric Decoherence}
\label{sec:GMCtoDecoherence}

As reviewed above, much of the success of the probabilistic  approach to Liouville CFT is due to the existence of non-perturbative renormalization of exponential interactions via the GMC random measure that manifestly respects conformal symmetry. Nonetheless, the same framework can also be used to study more general observables, such as extended defects, in the same theory, which may not have manifest conformal symmetry. This should not come up as a surprise because the probabilistic approach should be thought of as a rigorous way to perform path integral calculations which have been practiced in many contexts.

Our interest here concerns the ``localized cosmological constant'' line defect introduced in \eqref{eq:defect_def}. As mentioned there, since the perturbing operator $V_{b/ 2}=e^{b\phi}$ is irrelevant on the line, a short-distance cutoff $\Lambda$ is necessary to define the defect. In the probabilistic approach, this defect can be defined non-perturbatively where the cutoff can be safely removed. This is achieved via the Gaussian multiplicative chaos on the defect locus $\Sigma$,\footnote{Here, to avoid further notations, we have abused the notation (comparing to \eqref{eq:defect_def})  to denote the renormalized coupling simply by $\m_D$.}
\ie 
{\bm L}_\Sigma(\mu_D) \to  e^{\m_D e^{bc}  M_b(\Sigma)}~{\rm with}~ M_b(\Sigma)\equiv \int_\Sigma M_b(ds)\,,
\label{GMCdefect}
\fe
interpreted as an insertion in the Liouville path integral. We emphasize that the curve GMC is built from the bulk GFF which satisfies covariance \eqref{GFFcov} and that the regularization and renormalization procedure discussed around \eqref{GMCrr} carries over to this setting. In particular, the insertion \eqref{GMCdefect} defines a positive random variable which we will refer to as the GMC defect.\footnote{More generally, 1d GMC $M_\gamma(ds)$ as defined in a similar way as in \eqref{GMCrr} converges to a positive random measure for $0\leq \gamma <\sqrt{2}$. Here since we have set $b\in (0,1]$, $M_b(ds)$ always exists as a positive random measure.} In the following, we will take the GMC defect locus $\Sigma$ to be  the unit circle at $|x|=1$.

After having identified the defect insertion as a random variable, it is straightforward to write down the probabilistic expressions for correlation functions in the presence of the GMC defect. For example, the correlation function with local operators inserted away from the defect is
\ie 
\langle \prod_i V_{\alpha_i}(z_i) {\bm L}_\Sigma(\m_D)\rangle =2\lim_{\ep \to 0}\int_\mathbb{R} dc \,e^{-2Qc} \mathbb{E}
\left[ e^{-\m_{\rm bulk} e^{2b c}  M_{2b}({\rm S}^2) + \m_D e^{bc} M_{b}(\Sigma)} \prod_i\ep^{2\alpha_i^2}
e^{2\alpha_i \phi_\ep (z_i)}
\right]\,,
\label{defectcorr}
\fe
which is very similar to \eqref{bulkboundaryprob} and well-defined for real $\alpha_i$ obeying the Seiberg bounds. However, conformal invariance is clearly broken here for $\m_D\neq 0$. Concretely, under a general ${\rm SL}(2,\mathbb{R})$ M\"obius transformation $x\to \psi(x)$ preserving the defect locus $\Sigma$, 
\ie 
\phi(x) \to \phi(\psi(x)) + Q \log |\psi'(x)|\,,\quad M_b(ds) \to |\psi'(x(s))|^{b^2 \over 2}M_b(ds) \,,
\fe 
whereas the bulk GMC and the Neumann boundary GMC are invariant which were key to the conformally invariant bulk-boundary correlation functions found by the probability approach in \cite{Remy:2017phd,Remy:2020suk,Ang:2021ird,Ang:2023fky}, confirming the earlier bootstrap results \cite{Fateev:2000ik,Teschner:2000md,Zamolodchikov:2001ah,Hosomichi:2001xc}. 

Since the defect coupling (localized cosmological constant) $\m_D$ is irrelevant near the IR fixed point (which describes the trivial defect), one could wonder if there is a UV fixed point (or potential run-away behavior). This can be diagnosed by studying the correlation function \eqref{defectcorr} as the insertions $V_{\alpha_i}(z_i)$ approach the defect, namely the limit $|z_i|\to 1$.\footnote{The naive UV limit by sending $\m_D\to \infty$ (with $z_i$ and $\m$ fixed) would be too restrictive.} 

Based on the explicit results at strong defect cosmological constant  in this limit, we expect the Liouville field measure on ${\rm S}^2$ factorize into a product of two independent Liouville field measures on two disks $D_1,D_2$ with a shared boundary $\Sigma=\partial D_1 =\partial D_2$ with Neumann boundary conditions and conditioning on a fixed quantum length defined by $\ell=\int_\Sigma e^{bc}M_b^{\partial}(ds)$ for either disk.  

Concretely, in the case of two bulk insertions at $z_1,z_2$, on the two sides of $\Sigma$ (i.e. $|z_1|>1>|z_2|$ thus belonging to disks $D_1$ and $D_2$ respectively), the correlation function with the GMC defect in the UV at strong defect coupling is proposed to induce decoherence in the quantum length, and given below up to a normalization constant $C$, 
\ie 
\langle  V_{\alpha_1}(z_1)  V_{\alpha_2}(z_2) {\bm L}_\Sigma(\m_D)\rangle_{\rm UV}
=
C\int_0^\infty {d\ell \over \ell}\,  e^{\tilde\m_D \ell}  \cM_{1,1}^{\rm disk}(2\alpha_1;\ell)[e^{-\m_{\rm bulk} A}] \cM_{1,1}^{\rm disk}(2\alpha_2;\ell)[e^{-\m_{\rm bulk} A}] 
\label{decohereGMC}
\fe
where $\cM_{1,1}^{\rm disk}(2\alpha_1;\ell)$ is the  probability measure defined in \cite{Ang:2021ird}\footnote{More specifically, the quantum disk measure with bulk marked insertion $V_{2\alpha}$ and an exactly marginal boundary insertion $U_{b}$ defines $\cM_{1,1}^{\rm disk}(2\alpha)$ in Definition 4.2 of \cite{Ang:2021ird}. The disintegration of $\cM_{1,1}^{\rm disk}(2\alpha)$ over the boundary length $\ell$ then defines the finite measure $\cM_{1,1}^{\rm disk}(2\alpha;\ell)$.} for quantum disk with one bulk insertion of weight $2\alpha_1$, one boundary marking (of weight $b$), and a fixed quantum boundary length $\ell$. The quantum area of the quantum disk is denoted as $A=\int_{D_1} e^{2bc} M_{2b}(d^2 x)$ (similarly for $D_2$). The expectation $\cM_{1,1}^{\rm disk}(2\alpha_1;\ell)[e^{-\m_{\rm bulk} A}]=\ell\cM_1^{\rm disk}(2\alpha_1;\ell)[e^{-\m_{\rm bulk} A}]$ precisely yields the fixed length FZZT boundary wavefunction in the original normalization of \cite{Fateev:2000ik} (denoted as $W_\alpha(\ell)$ in \cite{Fateev:2000ik}) and $C$ is $\alpha_i$ independent in that convention.\footnote{The relative factor of $\ell$ relating the expectations in the two measures $\cM_{1,1}^{\rm disk}(2\alpha)$ and $\cM_{1}^{\rm disk}(2\alpha)$ come from forgetting the boundary marked point in the first quantum disk measure.} For dimensional reasons, the emergent coupling $\tilde\mu_D$ is determined in terms of bulk and defect cosmological constants by
\ie
\tilde \m_D =\m_D f(\m_D/\sqrt{\m_{\rm bulk}},b)\,.
\label{IRUVmapping}
\fe
Our results at strong  coupling indicate that $\lim_{b\to 0} f(\m_D/\sqrt{\m_{\rm bulk}},b)=1$ with $\m_D b^2$ and $\m_{\rm bulk}b^2$ fixed. The probability approach outlined here provides a potential way to test this proposal at finite $b$ and also to identify the precise mapping in \eqref{IRUVmapping}. Furthermore, the probability approach that gives the  non-perturbative definition of the localized cosmological constant defect \eqref{eq:defect_def} presented here will also help us understanding other pinning defects discussed in the Section \ref{sec:diskRG} and their RG flows. We hope to come back to these questions in the future.

\section{An interpretation of the line defect in other models} \label{sec:othermodels}

In this section, we provide an interpretation for the Liouville line defect as a brane/ interface/ domain wall in other models. In Section~\ref{secJT}, we relate the defect to a tensionless end-of-the-world (EOW) brane in Jackiw-Teitelboim (JT) gravity by reducing the thermal 1-point function of the decohered FZZT interface in the Schwarzian limit to the partition function of an EOW brane in JT gravity. In Section~\ref{WZW}, we interpret the defect as an interface in the WZW model by deriving the off-shell Liouville action with the defect from the WZW model with a domain wall interaction term added. In Section~\ref{shells}, we relate the Liouville solutions with the defect that we constructed in Section~\ref{Secexp} to wormholes in $3d$ gravity sourced by thin shells of dust particles. In Section~\ref{sec:AGT}, we interpret the decohered FZZT interface in $4d$ $\mathcal{N}=2$ supersymmetric gauge theories using the AGT correspondence.

\subsection{Relation to end-of-the-world branes in JT gravity}\label{secJT}

Our goal in this section is to relate the decohered FZZT interface to an EOW brane in JT gravity. Specifically, we will show that, in a certain limit, the thermal 1-point function of the decohered FZZT state reduces to the  partition function in the presence of tensionless EOW brane in JT gravity. JT gravity \cite{Jackiw:1984je,teitelboim_gravitation_1983} is a $2d$ theory of gravity coupled to a dilaton $\phi$ with action
\begin{equation}
    S_{\rm JT} = \frac{1}{16\pi G_N}\int_{M}\sqrt{-g}\ \phi(R+2)+\frac{1}{8\pi G_N}\int_{\partial M}\sqrt{-\gamma}\ \phi(K-1)\,.
\end{equation}
This theory has garnered much attention in recent years as a solvable toy model for the near-extremal dynamics of charged black holes \cite{Mertens:2022irh}. Calculations are especially tractable because the genus expansion can be rewritten in terms of a double-scaled matrix integral \cite{Saad:2019lba}. The duality between JT gravity and the matrix integral implies that two-boundary amplitudes fail to factorize into products of single-boundary amplitudes \cite{Harlow:2018tqv}. It is believed that this factorization puzzle is ameliorated in a UV theory which admits JT as an effective long-range description \cite{Blommaert:2021fob}. Such a completion is expected to have additional non-perturbative dynamical objects like spacetime branes and defects \cite{Turiaci:2020fjj,Witten:2020wvy,Maxfield:2020ale}. This motivates the study of JT coupled to a dynamical end-of-the-world (EOW) brane:
\begin{equation}
    S_{\rm JT} \mapsto S_{\rm JT} + \frac{1}{8\pi G_N}\int \sqrt{-h}\ (\phi K-\mu_{\rm JT})\,.
\end{equation}
Here $\mu_{\rm JT}$ is the brane tension and $h_{ij}$ is the induced metric on the brane.

The theory dictates the dynamics of a Schwarzian boundary particle whose trajectory defines a UV cutoff for the bulk geometry. After fixing
\begin{equation}
    \gamma_{tt} = \frac{1}{\epsilon^2} \quad{\rm and}\quad \phi|_{\partial M}=\frac{\phi_b}{\epsilon} \,,
\end{equation}
the Hamiltonian for the renormalized geodesic length $L$ separating the two AdS boundaries is
\begin{equation}
    H_{\rm JT} = -\frac{1}{2}\partial_L^2+2\mu_{\rm JT} e^{-L}+2e^{-2L} \,.
\end{equation}
In the tension-less $\mu_{\rm JT}=0$ limit, energy eigenstates obeying $H_{\rm JT}\ket{k}=2k^2\ket{k}$ have wavefunctions
\begin{equation}
    \varphi_k(L) \equiv \braket{k}{L} = K_{-2ik}(2e^{-L}) \,.
\end{equation}
The normalization of these states reveals that the measure for a resolution of the identity in the energy basis is given by the Schwarzian density of states
\begin{equation}
    \rho_{\rm schw}(k) = 8\pi^{-2}k\sinh(2\pi k) \,.
\end{equation}
The partition function $Z_{\rm EOW}(\beta)$ for the theory can be written as an integral transform of the trumpet partition function \cite{Gao:2021uro,Okuyama:2021eju}\footnote{
    We thank J. Boruch for bringing these works to our attention.
}
\begin{equation}\label{eq:jt_eow_divergent}
    Z_{\rm EOW}(\beta) = \int_0^\infty \frac{d\lambda}{2\sinh(\lambda/2)}\ Z_{\rm trumpet}(\beta, \lambda) \quad{\rm where}\quad Z_{\rm trumpet}(\beta,\lambda)=\frac{1}{2\pi}\int_0^\infty dk\ e^{-\beta k^2}\cos(\lambda k) \,.
\end{equation}
This integral has a clear operational meaning. The end-of-the-world brane cuts off the spacetime at a boundary of length $\lambda$. The partition function sums up contributions for each choice of $\lambda$. One difficulty with this expression is the lack of a saddle in the integral over boundaries. The action favors vanishingly small brane lengths and diverges as $\lambda\to 0$. We now return to Liouville theory and explain how our defect introduces a saddle at finite $\lambda$.

\begin{figure}
    \centering
    \includegraphics[width=0.65\linewidth]{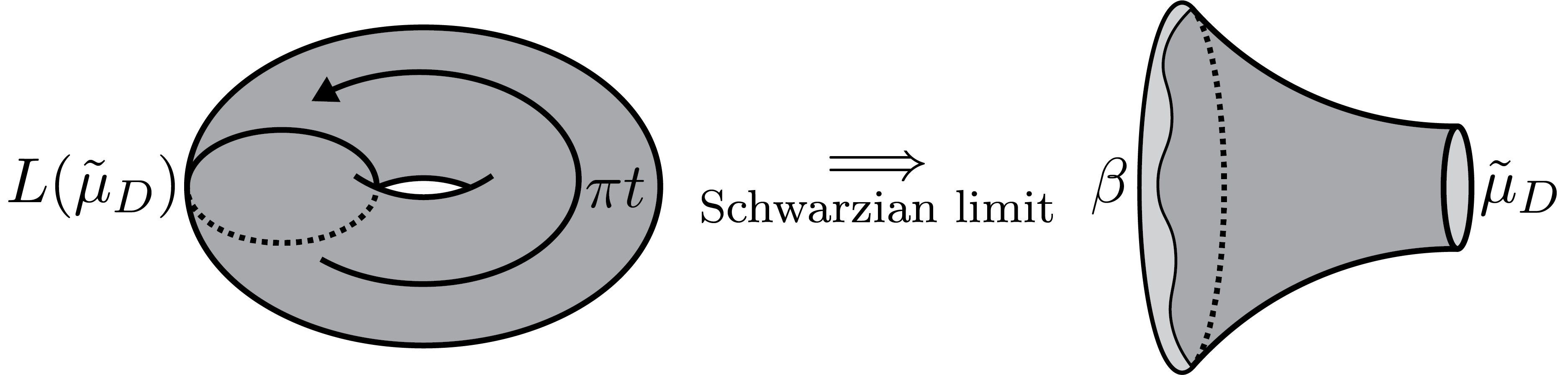}
    \caption{One the left we have drawn the geometry which computes the thermal one-point function of the decohered FZZT interface $L(\tilde{\mu}_D)$. For small $\tilde{\mu}_D$, the Schwarzian sector computes a partition function in JT gravity in the presence of a dynamical EOW brane as shown on the right. The length $t$ of the thermal circle on the torus is identified with the length $\beta$ of the asymptotic boundary.}
    \label{fig:schwarzian}
\end{figure}

Amplitudes in JT gravity can be bootstrapped using the fact that near-threshold states in a large-$c$ CFT$_2$ admit a Schwarzian description \cite{Mertens:2018fds,Mertens:2017mtv,Collier:2019weq,Ghosh:2019rcj}. We access such states by fixing the ratio $k=Pb^{-1}$ between momenta and the Liouville coupling and taking the semiclassical limit $b\to 0$. We refer to this as the Schwarzian limit. In this regime, disk partition functions are computed by ZZ-ZZ cylinder transition amplitudes in Liouville CFT. In order to introduce a dynamical EOW brane, we will study the thermal one-point function of the decohered FZZT interface (\ref{decoheredinterface}) 
\begin{align}
    \Tr[e^{-\pi t H}L(\tilde{\mu}_D)] &= 2\sqrt{2}\pi b\int_0^\infty \rho_0(P)dP\ \chi_P(it)\int_{0}^\infty \frac{d\ell}{\ell}\ e^{\tilde{\mu}_D\ell}\psi_P(\ell)\psi_P^*(\ell) \,,\nn\\
    &= \frac{16 b^2}{\pi}\int_0^\infty  \frac{k^2\sinh(2\pi k)}{\sinh(2\pi kb^2)}dk\ \chi_{bk}(it)\int_0^\infty \frac{d\ell}{\ell}\ e^{\tilde{\mu}_D\ell}K_{-2ik}(\kappa\ell)K_{-2ik}(\kappa\ell) \,.
\end{align}
Taking the Schwarzian limit gives
\begin{equation}
    \Tr[e^{-\pi t H}L(\tilde{\mu}_D)] \underset{\rm schw}{=} \int_0^\infty dk\left(\frac{8}{\pi^2}k\sinh(2\pi k)\right)\ e^{-\beta(t)\times(2k)^2}\int_0^\infty\frac{d\ell}{\ell}\ e^{\tilde{\mu}_D\ell}K_{-2ik}(\kappa\ell)K_{-2ik}(\kappa\ell) \,.
\end{equation}
where, in accordance with the JT nomenclature, we define an asymptotic boundary length $\beta(t)=\frac{\pi b^2 t}{2}$ in terms of the torus modulus $t$. We let the modulus scale as $t \sim b^{-2}$ so that $\beta =O(1)$ in the $b\to 0$ limit.  We note the appearance of the Schwarzian density of states $\rho_{\rm schw}$ and the fixed length states $\varphi_k(L=-\log(\kappa\ell/2))$ now that we have isolated the Schwarzian sector of the theory. The subsequent details of the calculation closely follow those in Section 2.2 of \cite{Gao:2021uro}. Using eq. 6.647.2 and 6.621.3 of \cite{Gradshteyn:1943cpj}, we find
\begin{align} 
    \Tr[e^{-\pi t H}L(\tilde{\mu}_D)] \underset{\rm schw}{=}& \frac{2}{\pi }\int_0^\infty dk\ e^{-\beta\times(2k)^2}\int_0^1\frac{dw}{1-w^2}\times \nn\\
    &\quad \cos\left(8k{\rm arcsinh}\left(\sqrt{\frac{1}{2}\left(\frac{1}{2}(w+w^{-1}+\frac{\tilde{\mu}_D}{2\kappa}(w-w^{-1}))-1\right)}\right)\right) \,,
\end{align}
where the variable $w$ is as in eq. 2.45 of \cite{Gao:2021uro}. 

Thus far we have made no claim as to how the interface cosmological constant $\tilde{\mu}_D$ should scale in the Schwarzian limit. There are two natural choices. In \ref{eq:kappa_def}, FZZT boundary wavefunctions were expressed in terms of the parameter $s$. The trumpet in JT can be derived from a Schwarzian limit of the ZZ-FZZT cylinder transition amplitude where $\lambda=2bs$ \cite{Mertens:2019tcm,Blommaert:2018iqz}. As $b\to 0$, this amounts to the strongly coupled semiclassical limit studied in Section~\ref{Secexp}. We will see that the variable $w$ fulfills the role of the trumpet throat length $\lambda$ so there is no a priori reason to make this choice here. Though it would be interesting to consider the near-extremal implications of this regime, we will instead focus on the weak defect coupling limit studied in Sections \ref{seccorrpert} and \ref{sec:deco_weak}. In this case, we can substitute
\begin{equation}
    1+\frac{\tilde{\mu}_D}{2\kappa} \approx e^{\frac{\tilde{\mu}_D}{2\kappa}} \quad{\rm and}\quad  1-\frac{\tilde{\mu}_D}{2\kappa} \approx e^{-\frac{\tilde{\mu}_D}{2\kappa}} \,,
\end{equation}
which greatly simplifies the argument of the cosine above. Setting $w=\exp(-\frac{\lambda}{2}-\frac{\Tilde{\mu}_D}{2\kappa})$ gives
\begin{align}
    \Tr[e^{-\pi t H}L(\tilde{\mu}_D)] &\underset{\rm schw}{=} \frac{1}{2\pi}\int_0^\infty dk\ e^{-\beta\times(2k)^2}\int_{-\frac{\tilde{\mu}_D}{\kappa}}^\infty \frac{d\lambda}{\sinh(\frac{\lambda}{2}+\frac{\tilde{\mu}_D}{2\kappa})}\cos(2k\lambda) \,,\nn\\
    &\underset{\rm schw}{=} \int_{0}^\infty\frac{d\lambda}{2\sinh(\lambda/2)}\ Z_{\rm trumpet}(\beta, \lambda-\tilde{\mu}_D/\kappa) \,.
\end{align}
When $\tilde{\mu}_D=0$, the interface one-point torus amplitude exactly computes a partition function in the presence of a tensionless EOW brane. For small $\tilde{\mu}_D\neq 0$, the trumpet length is offset relative to the measure factor
\begin{equation}
    \cM(\lambda) = \frac{d\lambda}{2\sinh(\lambda/2)} \,.
\end{equation}
This introduces a saddle in the integral over boundary lengths whenever the equation
\begin{equation}
    \lambda + \beta\coth\left(\frac{\lambda}{2}\right) = \frac{\tilde{\mu}_D}{\kappa} \,,
\end{equation}
admits a solution. In the limit of infinite temperature ($\beta=0$) we find that the defect cosmological constant exactly sets the saddle-point value of the brane length.  More generally, the ratio $\tilde{\mu}_D/\kappa$ sets the characteristic length scale of the EOW brane. This gives an a posteriori reason for taking $\tilde{\mu}_D\ll\kappa$ --- this is precisely the limit where the defect acts as a good UV regulator.

\subsection{Relation to interfaces in the WZW model} \label{WZW}

In this section, we interpret the line defect ${\bm L}_\Sigma$ as as an interface in the $H_3^+$ model which is a gauged WZW model with target space being the coset $SL(2,\mathbb{C})/SU(2)$. The concrete goal is to reproduce the Liouville action with the exponential interaction on the line from the action for the WZW model with a suitable interface action added. We place the WZW model on the plane with the interface along the real axis $z=\overline{z}$. We will closely follow the calculation done in \cite{Fateev:2007wk} where they reduce the WZW action on a half-plane to the Liouville action with an FZZT boundary term, and generalize their calculation for the case of the defect which has an exponential interaction term similar to the FZZT boundary. In the Appendix \ref{AppDS}, we review the Drinfeld-Sokolov reduction of the closely related non-chiral SL$(2,\mathbb{R})$ WZW model using the WZW action written as a sigma model in terms of the SL$(2,\mathbb{R})$ group-valued field.

Following \cite{Fateev:2007wk}, we work with the WZW action written in first order variables in terms of fields $\phi, \gamma, \overline{\gamma}, \beta, \overline{\beta}$,
\begin{equation}
    S_{\text{bulk}}=\frac{1}{\pi}\int d^2 z \left (\partial \phi \overline{\partial}\phi+\beta \overline{\partial}\gamma+\overline{\beta}\partial \overline{\gamma}-\lambda b^2 \beta \overline{\beta} e^{2b\phi} \right ) \,.
\end{equation}
The field $\phi$ will turn out to be the Liouville field once we impose suitable constraints on the fields as explained below. The parameter $b$ will turn into the Liouville coupling. It is related to the level $k$ by
\begin{equation}
    b^2=\frac{1}{k-2} \,.
\end{equation}
The affine $sl(2)$ currents expressed in terms of these fields take the form, 
\begin{equation}
    J^-=\beta, \quad J^3=\beta \gamma+b^{-1}\partial \phi, \quad{\rm and}\quad J^+=\beta \gamma^2+2b^{-1}\gamma \partial \phi- (k-2)\partial\phi \,.
\end{equation}
There are corresponding expressions for the anti-holomorphic currents $\overline{J}$ in terms of the barred fields.
Now, we have two copies of this theory on either side of the real axis. We distinguish the fields in the two copies by subscripts $\pm$. Now, we add the line interaction term that glues the two copies,
\begin{equation}
    S_{\text{interface}}=-\frac{i}{2\pi}\int_{y=0}dx \left(\beta_+(\gamma_++\overline{\gamma}_--\alpha e^{b\phi_d})+\beta_-(\gamma_-+\overline{\gamma}_+-\alpha e^{b\phi_d})\right) \,,
\end{equation}
$\alpha$ is a free real parameter. Here, we are imposing that the field $\phi$ is continuous across the interface with the value on the wall given by $\phi_d(x)$. The total action of the theory is therefore,
\begin{equation}
    S=S_{\text{bulk}+}+S_{\text{bulk}-}+S_{\text{interface}} \,.
\end{equation}
Now, we impose the following pair of constraints on the $\beta$ fields on the interface, 
\begin{equation} \label{betacons}
    (\beta_++\overline{\beta}_-)|_{y=0}=0 \quad{\rm and}\quad (\beta_-+\overline{\beta}_+)|_{y=0}=0 \,.
\end{equation}
Varying the action gives the following gluing conditions for the fields across the interface,
\begin{equation}
    \gamma_++\overline{\gamma}_-=\alpha e^{b\phi_d}\,, \quad \gamma_-+\overline{\gamma}_+=\alpha e^{b\phi_d}\,, \quad{\rm and}\quad \partial_y\phi_+- \partial_y\phi_-=i\alpha b(\beta_++\beta_-)e^{b\phi_d} \,.
\end{equation}
The latter condition resembles the jump condition across the Liouville defect. Now, to reduce the WZW action to the Liouville action, we impose the Drinfeld-Sokolov constraints in the bulk which set the $J^-$ and $\overline{J}^-$ components of the currents on the two sides to be constants,
\begin{equation}
    J^{-}_{\pm}=\beta_{\pm}=1 \quad{\rm and}\quad \overline{J}^-_\pm=\overline{\beta}_\pm=-1 \,.
\end{equation}
These constraints are consistent with the constraints (\ref{betacons}) on the interface. With these constraints, the total action reduces to
\begin{equation}
    S=\frac{1}{\pi}\int d^2z \left(\partial \phi\overline{\partial}\phi+\lambda b^2 e^{2b\phi}\right)+\frac{i\alpha}{\pi}\int_{y=0} dx e^{b\phi} \,.
\end{equation}
This is exactly of the form of the Liouville action with an exponential defect interaction term with $\lambda b^2$ identified with the bulk cosmological constant and 
\begin{equation}
    \mu_D=-\frac{i \alpha}{\pi} \,,
\end{equation}
being the defect cosmological constant. For the action to be real we need to choose $\alpha$ imaginary which results in a real defect cosmological constant.

\subsection{Relation to dust shell wormholes in $3d$ gravity} \label{shells}

\begin{figure}
    \centering
    \includegraphics[width=0.7\linewidth]{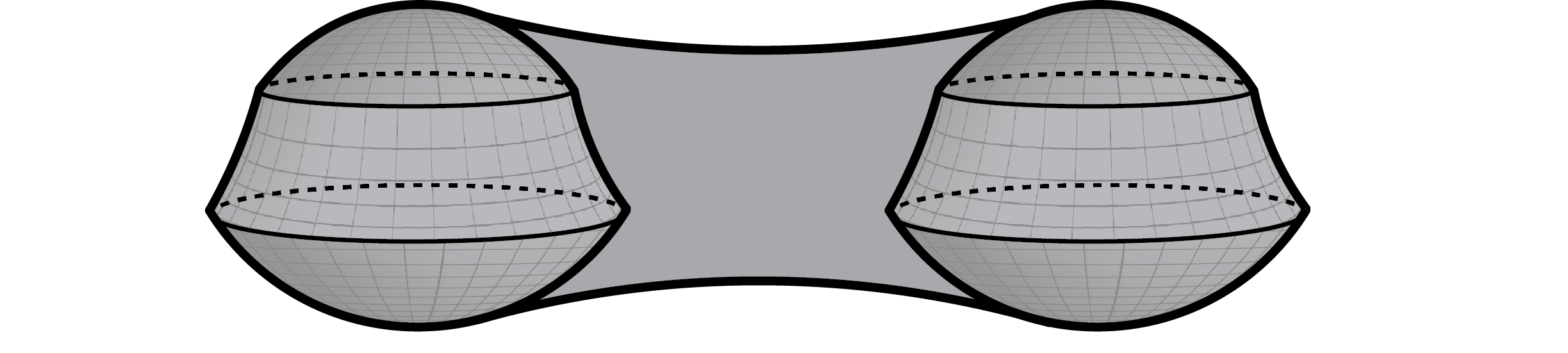}
    \caption{A two-boundary wormhole sourced by two dust shells. If we tune the mass of the shells in such a way that it matches the energy density on the Liouville line defect, then the geometry of each transverse slice in this figure is governed by the Liouville solution with two defects on the sphere.}
    \label{fig:wormhole}
\end{figure}

In this section, we relate the Liouville solutions with the defect constructed in \ref{Secexp} to two-boundary Euclidean wormholes in $3d$ gravity sourced by dust shells \cite{Chandra:2022fwi, Chandra:2024vhm, Sasieta:2022ksu}. As part of the boundary conditions for these wormholes, one has to specify the mass of the shell. In order to relate to the Liouville solutions with the exponential line defect, we specify the mass of the shell such that it matches the stress energy localized on the defect (\ref{stressdef}). Since we noted in Section~\ref{Secexp} that the stress energy on the defect is also affected by the presence of other defects, it would mean that the mass of the shell is similarly affected by the presence of other shells. So, the mass of the shell is not a free parameter in this setup and is determined in terms of other parameters in the boundary conditions by solving the Liouville equation sourced by exponential line defects placed at the same locations as the shells on the boundary Riemann surface. This `dynamical' boundary condition on the mass of the shell is what distinguishes the analysis in this section with that of \cite{Chandra:2024vhm} where the thin-shell wormholes were interpreted using a different class of Liouville line defects constructed by taking the continuum limit of a product of local operators.

The gravitational action for the theory coupled to dust shells is given by
\begin{equation}
    I=-\frac{1}{16\pi G}\int_{\mathcal{M}}d^3 x\sqrt{g}(R+2)-\frac{1}{8\pi G}\int_{\partial{\mathcal{M}}} \sqrt{h}(K-1)+\int_{\Sigma } d^2x \sqrt{h}\sigma(x)(h_{ij}u^i u^j-1) \,.
\end{equation}
Here, $\Sigma $ denotes the worldvolume of the shell. The action on the worldvolume is expressed in terms of a field $\sigma(x)$ that parametrises the energy density on the shell. Varying the action with respect to this field enforces the normalisation of the 4-velocity $h_{ij}u^iu^j=1$. We therefore view the last term in the action,
\begin{equation}
    I_{\text{shell}}\equiv \int_{\Sigma } d^2x \sqrt{h}\sigma(x)(h_{ij}u^i u^j-1) \,,
\end{equation}
as an effective action for dust shells. It is better to use this effective action as opposed to the worldline action of a massive point particle as it isolates the gravitational solutions sourced by dust shells.

We now construct a two-boundary wormhole solution in this theory, compute its renormalized gravitational action, and relate it to the corresponding Liouville action governing the hyperbolic metric on a transverse slice of the wormhole.
The wormhole metric in hyperbolic slicing coordinates \cite{Maldacena:2004rf, Chandra:2022bqq} is given by
\begin{equation}
    ds^2=d\rho^2+\cosh^2(\rho)e^{\Phi(y)}(dy^2+d\theta^2) \,.
\end{equation}
The scalar curvature associated with this metric is
\begin{equation}
    R=-6+2\sech^2(\rho)-e^{-\Phi(y)}\sech^2(\rho)\Phi''(y) \,.
\end{equation}
Away from the location of the shell, the vacuum Einstein equations should be satisfied, hence we require $\Phi''(y)=2e^{\Phi(y)}$. To solve the Israel junction conditions, we first determine the surface energy density by requiring that we are fixing the mass density of the shell in terms of $\mu$. For the shell described by the trajectory $y=y_0$, the junction conditions takes the form,
\begin{equation}
     \Delta K_{ij}-\Delta K h_{ij}=\sigma(x) u_i u_j \implies -\frac{1}{2}\frac{\Delta \Phi'(y_0)}{e^{\frac{\Phi(y_0)}{2}}\cosh(\rho)}=\sigma(\rho) \,.
\end{equation}
This gives the jump in the normal derivative of the function $\Phi(y)$ across the shell,
\begin{equation}
    \Delta \Phi'(y_0)=-2\sigma(\rho)\cosh(\rho)e^{\frac{\Phi(y_0)}{2}} \,.
\end{equation}
Note that this equation also fixes the density function profile to be $\sigma(\rho)\sim \frac{1}{\cosh(\rho)}$ with the proportionality factor fixed by the asymptotic boundary condition for $\sigma$ in terms of a function that governs the stress energy localized on the Liouville defect $T^\Phi(z)\supset \sigma_L \delta(\text{defect})$. The Liouville defect is placed at the same location where the shell reaches the boundary and the function $\sigma_L$ which depends on the defect cosmological constant $\mu$ and the location of the defect as computed in (\ref{stressdef}),
\begin{equation}
    \sigma |_{\partial \mathcal{M}}=2\epsilon \sigma_L \,.
\end{equation}
This boundary condition has been designed such that the function $\Phi(y)$ entering the metric satisfies the following second order differential equation, 
\begin{equation}
    \Phi''(y)=2e^{\Phi(y)}-2\mu e^{\frac{\Phi(y)}{2}}\delta(y-y_0) \,,
\end{equation}
which matches with the Liouville equation sourced by the exponential line defect. Now, we turn to computing the on-shell action with a wiggly cutoff $\rho_c=\pm (\log(\frac{2}{\epsilon})-\frac{\Phi}{2})$ so the induced metric on the boundary is flat (to leading order in the cutoff $\epsilon$). On shell, the action of the shell term vanishes due to the normalization of 4-velocity. The other terms evaluate to give,
\begin{equation}
    \begin{split}
        & -\frac{1}{16\pi G}\int_{\mathcal{M}}d^3 x\sqrt{g}(R+2)=\frac{2\pi}{8\pi G}\int dy \left(\frac{2}{\epsilon^2}+\Phi''(y)\left(\log(\frac{2}{\epsilon})-\frac{\Phi(y)}{2}\right)\right) \,,\\
        &  -\frac{1}{8\pi G}\int_{\partial{\mathcal{M}}}d^2x \sqrt{h}(K-1)=-\frac{2\pi}{8\pi G}\int dy\left(\frac{2}{\epsilon^2}-e^{\Phi(y)}+\frac{1}{4}(\Phi'(y))^2+\Phi''(y)\right) \,.
    \end{split}
\end{equation}
Combining all the terms, the renormalized on-shell action takes the form,
\begin{equation}
    I_{\text{grav}}=\frac{c}{6}\int dy \left(\frac{1}{4}\Phi''(y)+e^{\Phi(y)}\right) \,.
\end{equation}
In the above calculation, we have already included the Hayward term contribution into the GHY term,
\begin{equation}
    I_{\text{Hay}}\equiv -\frac{1}{8\pi G}\int_{\Sigma} \sqrt{\gamma}(\pi -\Theta)=\frac{\mu}{8\pi G} \int_{\Sigma}dz e^{\frac{\Phi}{2}} \,.
\end{equation}
We see that the Hayward term evaluates to the source term corresponding to the Liouville defect.
The gravitational action on the wormhole and the Liouville action on the hyperbolic surface at the boundary differ by the Hayward term and are related by
\begin{equation}
    I_{\text{grav}}-I_{\text{Hay}}=\frac{c}{3}S_L \,.
\end{equation}
We have thus illustrated that the Liouville solutions constructed in Section \ref{Secexp} govern the geometries of the transverse hyperbolic slices of two-boundary wormholes in $3d$ gravity sourced by dust shells, and hence determine the partition functions of wormholes in the semiclassical limit. More generally, by tuning the energy density function on the shell $\sigma(x)$ to match the stress energy on the defect, we can relate the dust shell wormholes to Liouville saddles with the corresponding Liouville line defect.

\subsection{Relation to interfaces in $4d$ $\cN=2$ gauge theories} \label{sec:AGT}

As mentioned in the Introduction, Liouville CFT plays a foundational role in the AGT conjecture \cite{Alday:2009aq} which relates general $4d$ $\cN=2$ supersymmetric field theories to $2d$ Toda CFTs. This conjecture
is deeply rooted in the mysterious 6d $\cN=(2,0)$ superconformal field theories (SCFT) labelled by an ADE Lie algebra $\mathfrak{g}$, whose existence is implied by M-theory and type IIB string theory \cite{Witten:1997sc}. Indeed, via the Class S construction \cite{Gaiotto:2009we}, where the $4d$ $\cN=2$ theory  is obtained by a twisted compactification of the 6d $\cN=(2,0)$ theory of type $\mathfrak{g}$ on a Riemann surface $\cC$ (with prescribed punctures), the AGT dual is given by the $2d$ $\mathfrak{g}$ Toda CFT on $\cC$ (with local operator insertions corresponding to the punctures). 
The AGT correspondence 
predicts precise $4d$-$2d$ relations between $4d$ \textit{supersymmetric observables} (that can be assessed by supersymmetric localization on the squashed sphere ${\rm S}^4$ as in \cite{Pestun:2007rz}) and \textit{conformal correlators} in the $2d$ Toda CFTs on $\cC$. Such observables include both local point-like operator insertions on the ${\rm S}^4$, as well as lines, surfaces and interfaces, as long as the localizing supercharge is preserved \cite{Drukker:2010jp} (see also related works \cite{Alday:2009fs,Drukker:2009id,Hosomichi:2010vh,
Gomis:2010kv,Teschner:2012em,Gomis:2014eya,LeFloch:2015bto,LeFloch:2017lbt,Bawane:2017gjf} and recent review \cite{LeFloch:2020uop}). 
 
The most well-understood instance of AGT conjecture is when $\mathfrak{g}=\mathfrak{sl}_2$, also known as the rank-one case. Here the $2d$ side is described by the Liouville CFT and $4d$ side usually by gauge theories with gauge group SU$(2)$. The Liouville $b$ parameter is identified with the squashing parameter of the squashed sphere  S$^4_b$ which can be embedded in ${\mathbb{R}}^5$ with coordinates $X_i$ as follows,
\ie 
b^2 (X_1^2+X_2^2)+ {1\over b^2}  (X_3^2+X_4^2)+X_5^2=1\,.
\label{S4b}
\fe
For a fixed $4d$ $\cN=2$ theory from Class S construction on the punctured Riemann surface $\cC$, its supersymmetric observables, denoted schematically as $\cO$, on  ${\rm S}_b^4$ are then determined by corresponding observables $\cF(\cO)$ in the Liouville CFT on $\cC$ in the presence of certain ``background'' local operator insertions corresponding to the punctures \cite{Alday:2009aq}. For example, for the $\cN=4$ super-Yang-Mills (SYM) theory with $\cN=2$ preserving mass deformation by the mass parameter $m$, also known as the $\cN=2^*$ theory, we have the following equation satisfied by the $4d$ and $2d$ expectation values,
\ie 
\langle \cO  \rangle_{{\rm S}_b^4}^{\cN=2^*} (m,\tau)=\langle \cF(\cO) V_{{Q\over 2}+im}(0)\rangle_{T^2_\tau}^{\rm Liouville}\,,
\fe
where the $4d$ complexified gauge coupling $\tau$ is identified with the complex moduli of the torus where the Liouville CFT lives. As another familiar example, for the Seiberg-Witten theory \cite{Seiberg:1994aj}, namely the $\cN=2$ SU$(2)$ SYM with four fundamental flavor hypermultiplets, the AGT conjecture states that,
\ie 
\langle \cO  \rangle_{{\rm S}_b^4}^{{\rm SW}} (m_i,\tau)=\langle \cF(\cO) V_{{Q\over 2}+im_1}(0)V_{{Q\over 2}+im_2}(1)V_{{Q\over 2}+im_3}(z(\tau))V_{{Q\over 2}+im_4}(\infty)\rangle_{{\rm S}^2}^{\rm Liouville}\,,
\fe
where $m_i$ with $i=1,2,3,4$ denotes the four independent mass parameters for the hypermultiplets and the complexified gauge coupling here determines the locations of the punctures (up to conformal transformations). We emphasize that, in general, the map $\cF(\cO)$ assigns specific placement for the insertion on the Riemann surface $\cC$ relative to the ``background'' on the Liouville CFT side, according to the choice of the $4d$ observable $\cO$ (see \cite{Drukker:2010jp} for details).

Having reviewed the basic ingredients behind the AGT conjecture, we are now ready to give an interpretation of our decohered FZZT interface in the Liouville CFT in general $4d$ $\cN=2$ SU$(2)$ gauge theories (and in fact general rank-one theories which could be non-Lagrangian). 
Recall that this interface is built from the fixed-length boundaries $|\ell\rangle$, which is related to the usual FZZT boundaries by a Laplace transform in the boundary cosmological constant. In the following, we will identify the fixed-length boundary in terms of a novel boundary condition for the $4d$ $\cN=2$ theory. More specifically, we will make use of the existing results for the localization of $4d$ and $3d$ supersymmetric field theories to identify a boundary condition for the $4d$ gauge theory that involve novel couplings on its $3d$ boundary, such that under the AGT dictionary, reproduce the fixed-length boundary state in the Liouville CFT (see \cite{Pestun:2016zxk} for an extensive review on the localization technique). It would be interesting to derive this correspondence more directly, from the $6d$ or M-theory origin of the AGT correspondence mentioned earlier, for example, by generalizing the argument in \cite{Cordova:2016cmu}.

As explained in \cite{LeFloch:2017lbt,Bawane:2017gjf},
the familiar FZZT boundaries $|{\rm FZZT}(s)\rangle$ correspond to a symmetry breaking boundary condition of the gauge theory on the equator ${\rm S}^3_b$ (from SU$(2)$ to the Cartan U$(1)$) with $s$ identified as the $3d$ Fayet–Iliopoulos (FI) parameter for the remaining U$(1)$ gauge field on the boundary. We thus have the following relation between the   Liouville correlation function 
and the hemisphere partition function, 
\ie 
\langle {\rm FZZT}(s)| {\rm HS}^4_b[\cT_\cC]\rangle_\cC^{\rm Liouville} = {1\over 2}\int_{\mathbb{R}} d P 
 \underbrace{2\sqrt{2}{e^{4\pi i sP}\over \rho_0(P)} }_{\substack{{\rm sym~breaking}\\{\rm 
FI~term}}} \rho_0(P)
\underbrace{\vphantom{{e^{4\pi i sP}\over \rho_0(P)}}Z_{{\rm HS}^4_b} [\cT_\cC]_{\rm Dir}(P)}_{{\rm HS}^4_b~{\rm wavefunction}}\,.
\label{AGTfzzt}\fe
Here $| {\rm HS}^4_b[\cT_\cC]\rangle$ denotes an abstract state in the Liouville CFT that keeps track of the insertions on $\cC$ due to the punctures in the class S construction of $\cT_\cC$.\footnote{For $\mathbb{Z}_2$ quotient of $\cC$ which has a fixed locus with multiple disconnected components (i.e. disconnected S$^1$'s), the state $| {\rm HS}^4_b[\cT_\cC]\rangle$ also includes assignment of boundary conditions for the other circles (see also relevant discussions in \cite{LeFloch:2017lbt,Bawane:2017gjf}). This is because the 6d origin of the AGT correspondence for hemisphere partition functions involves studying the $(2,0)$ SCFT on orbifold geometry $(\cC\times {\rm S}^4_b)/\mathbb{Z}_2$. The choices that go into specifying such an orbifold in the 6d theory is reflected in the state $|{\rm HS}^4_b[\cT_\cC]\rangle$. It would be interesting to understanding this better, possibly by generalizing the work of \cite{Cordova:2016cmu}.} The Liouville momenta $P$ is identified with the vev of the scalar in the Cartan vector multiplet that is integrated over in the localization formula \cite{Pestun:2007rz} for the $4d$ gauge theory. The supersymmetric Dirichlet boundary condition fixes its value on the boundary.
The hemisphere wavefunction in the $|P\rangle\rangle$ basis is 
\ie 
Z_{{\rm HS}^4_b} [\cT_\cC]_{\rm Dir}(P)\equiv \langle\langle P |{\rm HS}^4_b[\cT_\cC]\rangle\,,
\fe
and the RHS of \eqref{AGTfzzt} follows from \eqref{eq:liouville_bdy_wvfcns}.\footnote{There are also implicitly prescribed boundary conditions for the hypermultiplets which will not be important for the discussion here (see \cite{LeFloch:2017lbt,Bawane:2017gjf} for relevant discussions).}

Note that the Plancherel measure $\rho_0(P)$ is the natural measure for gluing the hemisphere partition functions, which coincides, up to a normalization constant, with the one-loop determinant of $3d$ $\cN=2$ ${\rm SU}(2)$ vector multiplets on ${\rm S}^3_b$ \cite{Willett:2016adv,Dedushenko:2018tgx}.

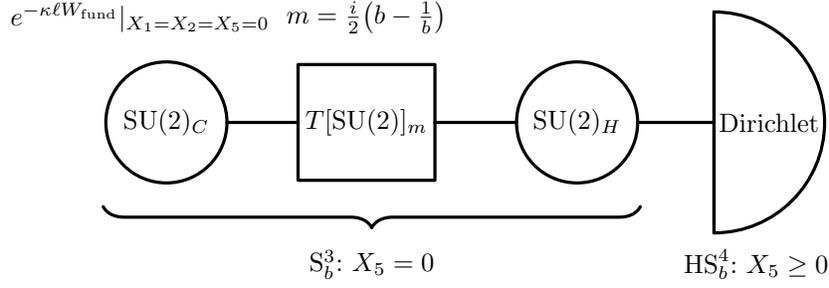
\begin{figure}
\centering
\usetikzlibrary{decorations.pathreplacing}

\begin{tikzpicture}[line cap=round,line join=round,scale=.7]

\tikzset{
  every node/.style={black},
  every path/.style={black}
}

\def\r{1.15}

\draw[very thick] (-6,0) circle (\r);
\node at (-6,0) {${\rm SU}(2)_C$};

\draw[very thick] (-3.5,-1.1) rectangle (-.9,1.1);
\node at (-2.2,0) {$T[{\rm SU}(2)]_m$};

\draw[very thick] (1.8,0) circle (\r);
\node at (1.8,0) {${\rm SU}(2)_H$};

\draw[very thick] (4.4,-2.1) -- (4.4,2.1);
\draw[very thick]
  (4.4,2.1)
  arc[start angle=90,end angle=-90,
      x radius=2.1,y radius=2.1]
  -- cycle;

\node at (5.45,0) {Dirichlet};

\draw[very thick] (-4.85,0) -- (-3.5,0);
\draw[very thick] (-.9,0) -- (0.65,0);
\draw[very thick] (2.95,0) -- (4.4,0);

\node at (-2.2,2)
{$m=\frac{i}{2}\!\left(b-\frac{1}{b}\right)$};

\node at (-6.5,2)
{$e^{-\kappa\ell W_{\rm fund}}|_{X_1=X_2=X_5=0}$};

\draw[
  very thick,
  decorate,
  decoration={brace,mirror,amplitude=8pt}
]
  (-7.2,-1.6) -- (3,-1.6)
  node[midway,below=10pt] {};

\node at (-2.1,-2.7) {${\rm S}^{3}_{b}$:~$X_5= 0$};
\node at (5.2,-2.7) {${\rm HS}^{4}_{b}$:~$X_5\geq 0$};

\end{tikzpicture}
\caption{The $4d$ boundary condition dual to the fixed-length brane from the Dirichlet boundary condition  on ${\rm HS}^4_b$ coupled to $3d$ modes on ${\rm S}^3_b$ described by the mass-deformed theory $T[{\rm SU}(2)]_m$. The coupling is realized by gauging the ${\rm SU}(2)_C \times {\rm SU}(2)_H$ flavor symmetry of the $T[{\rm SU}(2)]_m$ theory.
There is also a proliferation of Wilson loops $W_{\rm fund}$ for the ${\rm SU}(2)_C$ gauge fields.} 
    \label{fig:AGT}
\end{figure}

Given the explicit expression of the wavefunction $\psi_\ell(P)$    for the fixed-length FZZT boundary in \eqref{flwavefunction}, we now describe its AGT dual, in terms of the Dirichlet boundary condition for the $4d$ ${\rm SU}(2)$ gauge fields on a squashed hemisphere ${\rm HS}^4_b$, coupled to a strongly interacting $3d$ $\cN=2$ supersymmetric theory on the boundary ${\rm S}^3_b$ which we refer to as $T[{\rm SU}(2)]_m$, together with the proliferation of certain Wilson loop operators on the ${\rm S}^3_b$.
This is summarized in Figure~\ref{fig:AGT}. The corresponding relation between the Liouville correlation function and the hemisphere partition function reads,
\ie 
&\langle \ell| {\rm HS}^4_b[\cT_\cC]\rangle_\cC^{\rm Liouville} 
\\
=&{b\kappa\ell\over 8\sqrt{2}}\int_{\mathbb{R}} d P \int_{\mathbb{R}} ds 
\,\underbrace{\vphantom{Z_{{\rm HS}^4_b}}e^{-\kappa\ell \langle W_{\rm fund} \rangle(s)}}_{\rm WL ~proliferation}
 \rho_0(s) 
\underbrace{ Z_{{\rm S}^3_b}\big[T[{\rm SU}(2)]_m\big](s,P)}_{\rm duality~kernel} \rho_0(P) 
\underbrace{Z_{{\rm HS}^4_b} [\cT_\cC]_{\rm Dir}(P)}_{{\rm HS}^4_b~{\rm wavefunction}}\,.
\label{fixedlengthagt}
\fe
Equivalently, the fixed-length brane wavefunction \eqref{flwavefunction} can be written as
\ie 
\psi_\ell(P)
=
{b\kappa\ell\over 4\sqrt{2} } \int_{\mathbb{R}} ds \,e^{-\kappa\ell \cosh(2\pi b s)}
 \rho_0(s) 
 Z_{{\rm S}^3_b}\big[T[{\rm SU}(2)]_m\big](s,P)\,,
\label{fixedlength4d}
\fe
In what follows, let us unpack the ingredients on the RHS of the above equations, using results from localization.

The $T[{\rm SU}(2)]_m$ theory is a mass deformation of the famous $3d$ $\cN=4$ SCFT $T[{\rm SU}(2)]$ that plays an important role in AGT as the S-duality wall for the $\cN=4$ SYM \cite{Gaiotto:2008ak,Hosomichi:2010vh}. The $T[{\rm SU}(2)]$ SCFT is the IR fixed point of $\cN=4$ super-QED with two hypermultiplets of unit gauge charge.
This SCFT  has ${\rm SU}(2)_{\rm 1}\times {\rm SU}(2)_{\rm 2}$ R-symmetry and a commutant global symmetry ${\rm SU}(2)_{\rm H}\times {\rm SU}(2)_{\rm C}$. 
In the super-QED description, the ${\rm SU}(2)_{\rm H}$ is the usual 
flavor symmetry for the gauged hypermultiplets, whereas the ${\rm SU}(2)_{\rm C}$ symmetry is an enhancement of the topological ${\rm U}(1)$ symmetry under which monopoles are charged. The mass-deformation relevant here is along a diagonal Cartan direction of the ${\rm SU}(2)_{\rm 1}\times {\rm SU}(2)_{\rm 2}$ R-symmetry, breaking the $\cN=4$ supersymmetry to $\cN=2$ along the way. 

As is common for supersymmetric theories, a wealth of observables is accessible via supersymmetric localization \cite{Pestun:2016zxk}.
In particular, 
the supersymmetric partition function of $T[{\rm SU}(2)]_m$ on ${\rm S}^3_b$ follows from its super-QED description,
\ie 
Z_{{\rm S}^3_b}\big[T[{\rm SU}(2)]_m\big] (s,P)
=
{1\over s_b(m)}
\int_{\mathbb R} d\sigma \,e^{4\pi i s \sigma}
\prod_{\pm}
\frac{
s_b\!\left(
P \pm \sigma + \frac{m}{2} + \frac{iQ}{4}
\right)
}{
s_b\!\left(
P \pm \sigma - \frac{m}{2} - \frac{iQ}{4}
\right)
}\,,
\fe 
where $s$ and $P$ are the mass parameters associated with the Cartans of the ${\rm SU}(2)_C$ and ${\rm SU}(2)_H$ global symmetries respectively.
The double-sine function $s_b(x)$ is defined by the regularized product (see Appendix A.2 of \cite{Teschner:2012em} for its properties) 
\ie 
s_b(x)=\left.\prod_{m,n\geq 0}{\left( m+{1\over 2}\right)b+\left( n+{1\over 2}\right){1\over b}-i x\over \left( m+{1\over 2}\right)b+\left( n+{1\over 2}\right){1\over b}+i x}\right|_{\rm reg} \,,
\fe 
and keeps track of the contributions of the $\cN=2$ chiral multiplets from decomposing the $\cN=4$ vector multiplet and hypermultiplets. The following identity for double sine functions will be useful
\ie 
\frac{s_b\!\left(x+\tfrac{i b}{2}\right)}
     {s_b\!\left(x-\tfrac{i b}{2}\right)}
=
\frac{1}{2\cosh(\pi b x)}\,. 
\label{sbid}
\fe

Indeed after specializing the mass of the $3d$ hypermultiplets to\footnote{See also recent works \cite{Chester:2021gdw,Minahan:2021pfv} where this special mass was considered. } 
\ie 
m={i\over 2}\left (b -{1\over b}\right)\,,
\label{specialmass}\fe
the partition function of $T[{\rm SU}(2)]_m$ simplifies using the identity \eqref{sbid} to\footnote{At the special mass \eqref{specialmass},
the $T[{\rm SU}(2)]_m$ partition function coincides with the modular $S$-kernel for torus one-point blocks with an external operator of dimension $h=\bar h = 1$ (i.e. the dual Liouville potential $e^{2\phi/b}$).
It is well-known that the torus one-point block with an external operator of dimension $h=1$ coincides with the Virasoro character when normalized such that the leading term is 1 in the $q$-expansion \cite{Hadasz:2009db,Kraus:2016nwo}, so one may be confused why the expression in \eqref{agtkernel} differs from the usual $S$-kernel $\cos(4\pi s P)$ for the characters. This is because in AGT, the convenient normalization of conformal blocks (which makes clear the connection to representations of the modular double of the quantum group ${\cal U}_q(\mathfrak{sl}_2(\mathbb{R}))$) differs from the usual ones used in conformal bootstrap by products of $\sqrt{C_{\rm DOZZ}}$ \cite{Ponsot:1999uf,Ponsot:2000mt} (see also \cite{LeFloch:2020uop} for a review). In particular, for torus one-point blocks, the two conventions are related by $\cF_{\rm AGT}(\alpha_{\rm ext},\alpha)=N(\alpha,\alpha_{\rm ext},\alpha)^{-1}\cF_{\rm bootstrap}(\alpha_{\rm ext},\alpha)$ with $N(\alpha,\alpha_{\rm ext},\alpha)$ defined in \cite{Ponsot:1999uf}. In the special case of $\alpha_{\rm ext}={1\over b}$ and $\alpha={Q\over 2}+iP$, one finds $N(\alpha,\alpha_{\rm ext},\alpha)^{-1}\propto P(\sinh 2\pi b^{-1} P)^{-1}$ up to a $P$ independent constant and this explains the form of the $S$-kernel in \eqref{agtkernel}.
} 
\ie 
Z_{{\rm S}^3_b}\big[T[{\rm SU}(2)]_m\big](s,P)
=&
{b\over 4}
\int_{\mathbb R} d\sigma\;
\frac{
e^{4\pi i s \sigma}
}{
\prod_\pm \cosh \pi b(\sigma\pm P) 
}
=
 {\sin (4\pi s P)\over 2\sinh (2\pi  b P)\sinh (2\pi s/b)}\,,
\label{agtkernel}
\fe 
where in the first equality we have also used the following special value of the double sine function,
\ie 
s_b\left ({i\over 2}\left(b-{1\over b}\right)\right)={1\over b}\,,
\fe
which can be deduced using properties of the double sine function (see e.g. Appendix A of \cite{Teschner:2012em})
and the second equality in \eqref{agtkernel} comes from a standard contour integral that picks up the residues (e.g. for $s>0$) at the poles  $\sigma=\pm P+{(2n-1)i\over 2b}$ with $n\in \mathbb{Z}_+$ in the upper half plane. This explains the duality kernel in \eqref{fixedlengthagt}.

In the $4d$ description of the fixed-length  boundary state, the ${\rm SU}(2)_C$ global symmetry of the 
$T[{\rm SU}(2)]_m$ theory is gauged, together with a proliferation of  BPS  ${\rm SU}(2)_C$ Wilson loop in the fundamental representation coupled to the vector multiplet scalar $\varphi$ \cite{Willett:2016adv},
\ie 
W_{\rm fund}(\gamma)\equiv {1\over 2} {\rm tr}_{\rm fund}{\rm P}e^{i\oint_\gamma (A-i \varphi d|x|)}\,,
\fe
 on the great circle  at $X_1=X_2=X_5=0$ in ${\rm S}^3_b$ (see \eqref{S4b}). Upon localization of the $3d$ ${\rm SU}(2)$ gauge theory, the fields are projected to the BPS locus $A_\m=0$ and $\varphi=s$ and each such Wilson loop translates to the following insertion \cite{Willett:2016adv}
\ie 
\langle W_{\rm fund} \rangle(s)= \cosh (2\pi b s)\,,
\fe
in the localization formula and its proliferation is characterized by the exponentiated insertion with parameter $\kappa\ell$ in \eqref{fixedlengthagt}.

Together with the factor $\rho_0(s)$ in \eqref{rho0} that captures the one-loop determinant from gauging ${\rm SU}(2)_C$ on the ${\rm S}^3_b$, we can write \eqref{fixedlength4d} as
\ie 
\psi_\ell(P)={b\kappa\ell}\int_0^\infty ds e^{-\kappa\ell \cosh(2\pi b s)} {\sinh (2\pi bs )\sin (4\pi s P)\over  \sinh (2\pi  b P)}\,,
\fe
which evaluates to \eqref{flwavefunction}, after using the integral representation of the Bessel function (\ref{BesselIntegralrep}).

We have thus identified the novel boundary condition of the $4d$ $\cN=2$ gauge theory on the squashed hemisphere that corresponds to the fixed-length FZZT boundary condition, which is unconventional as it involves a proliferation of Wilson loops wrapping circles of size $2\pi b$ on the boundary ${\rm S}^3_b$ (thus breaking the boundary isometry including the $b\to 1/b$ invariance). Interestingly, as explained, these Wilson loops are not the ordinary boundary Wilson loops of the $4d$ theory. Rather they are transformed by a duality kernel implemented by the nontrivial $3d$ $\cN=2$ theory $T[{\rm SU}(2)]_m$. Coming back to the decohered FZZT interface which we discuss in the main body of the paper, the corresponding AGT dual is clear: it describes an interface in the $4d$ theory on ${\rm S}^4_b$ at the equator ${\rm S}^3_b$ where it imposes the boundary conditions $|\ell\rangle$ described above for the two hemispheres which now involve a correlated proliferation of the ${\rm SU}(2)_C$ Wilson loops with the specific distribution in \eqref{decoheredefect}. It would be interesting to investigate whether there is a more direct origin of this interface from the $6d$ $(2,0)$ theory and M-theory.

\section{Discussion} \label{sec:disc}

In summary, we have constructed a Liouville line defect realized as a localized cosmological constant: a line insertion into the Liouville path integral that approaches conformality in a certain limit and is tractable at both weak and strong defect coupling.. At weak coupling, we analyzed the defect perturbatively and characterize it through its correlations with local operators, energy and information transport, the Casimir energies associated with fusion, and corrections to the open string channel spectrum. We also studied the effect of a cuspidal deformation of the defect locus on these observables, and describe interesting monotonicity properties as the cusp angle is tuned. For example, the energy transmitted across the defect increases monotonically with the opening angle at the cusp. These results derived using perturbation theory are more generally applicable to pinning defects constructed by integrating scalar primary operators in compact $2d$ CFTs. At strong coupling, in a semiclassical limit, we showed that the defect admits a geometric interpretation in terms of a discontinuity in the extrinsic curvature of the $1d$ defect locus embedded in $2d$ hyperbolic geometries. We computed the observables characterizing the defect by gluing hyperbolic surfaces across the defect, and compared the results to the corresponding results at weak coupling. For example, we showed that the defect is strongly transmitting at weak coupling but becomes strongly reflecting at strong coupling. We then showed that the correlations across the defect, at weak and strong coupling, can be realized effectively by a decohered FZZT interface constructed by decohering the FZZT boundary state in the length basis. We also comment on how the localized cosmological constant defect can be studied non-perturbatively using the probabilistic approach to Liouville CFT. Finally, we provided an interpretation of the line defect in other models: in terms of end-of-the-world branes in JT gravity; interfaces in the WZW model; dust shells in $3d$ gravity; and symmetry-breaking interfaces with a proliferation of BPS Wilson loops in $4d$ $\mathcal{N}=2$ gauge theories.

Now, we highlight certain interesting observations inspired by this work that could serve as potentially concrete future directions.

\subsection{Renormalisation Group flows on Liouville line defects} \label{sec:diskRG}

We can use Liouville CFT as a model to study RG flows on line defects, especially of the pinning field type. For example, we could consider line defects constructed from vertex operators,
    \begin{equation}
        {\bm L}_\Sigma=\exp\left[\mu_D \int_\Sigma V_{\alpha} \right] \,,
    \end{equation}
    with $\alpha=bk$ where $b$ is the Liouville coupling. $k\in (0,1)$ for the vertex operator to obey the Seiberg bound $0<\text{Re}(\alpha)<\frac{Q}{2}$. The total scaling dimension of these operators is given by
    \begin{equation}
        \Delta_{bk}=2k+2b^2k-2b^2k^2 \,,
    \end{equation}
    In the $b\to 0$ limit where we keep $k$ fixed, the scaling dimension becomes 
    \begin{equation}
        \Delta_{bk}\to 2k \,.
    \end{equation}
    In the strongly coupled semiclassical limit with $\mu_D\sim \frac{\mu}{b^2}$ where the defect can be studied using hyperbolic geometry, we see that if $k<\frac{1}{2}$, then $\Delta_{bk}<1$ which means the defect triggers a relevant RG flow on the line. On the other hand, if $k>\frac{1}{2}$, then $\Delta_{bk}>1$ in the semiclassical limit, so the deformation on the line is irrelevant. It would be interesting to characterize the flow to the defect CFT (DCFT) at the non-trivial fixed point (in the UV or IR depending on whether the flow is triggered by a relevant or irrelevant deformation) using hyperbolic geometry.

    In fact, if $k<1-\frac{1}{\sqrt{2}}$, then the defect triggers a relevant flow even away from the semiclassical limit for any $b\in (0,1)$. In addition, we can show that under repeated fusion, $V_{bk}\otimes V_{bk}\otimes V_{bk}\otimes \dots$, the vertex operators with $\frac{1}{4}<k<\frac{2+\sqrt{2}}{4}$ do not generate any other marginal or relevant operators. Therefore, the vertex operators that do not generate other relevant or marginal operators under fusion have $\frac{1}{4}<k<1-\frac{1}{\sqrt{2}}$. For the defects constructed from these operators, perturbatively, there are no UV divergences, which means we do not have to turn on any additional couplings. In addition, the $\beta$-function for the defect coupling does not receive any perturbative corrections. Hence, we can compute observables involving these defects to any order in perturbation theory, and we can also compute observables non-perturbatively using semiclassical hyperbolic saddles and quantum corrections around these saddles.
    
    \subsection{Line defects in holographic $2d$ CFTs}
    
    It would be interesting to generalize the ideas used in this paper to compact irrational $2d$ CFTs where we could construct analogous line defects by integrating Virasoro primary operators,
    \begin{equation}
        D_{\Sigma}=\exp\left[\lambda \int_\Sigma \mathcal{O}\right] \,.
    \end{equation}
    For small defect coupling $\lambda$, we can use perturbation theory to compute observables involving the defect just as we have done in this paper. However, the important question remains: How do we analyze the defects at strong coupling where perturbation theory breaks down? For holographic $2d$ CFTs, if we scale $\lambda \sim \frac{1}{G_N}$, we could possibly use $3d$ gravity to compute observables involving the defect at strong coupling. To this end, it would be important to construct the domain wall sourced by the defect, in the bulk. An interesting application of constructing such a domain wall would be to compute the effect of introducing a cusp on the defect at strong coupling using semiclassical gravity. In Section~\ref{Seccusp}, we described interesting monotonicity properties of observables with a cusp on the defect at weak coupling using perturbation theory. It would be interesting to compute these observables at strong coupling possibly using a domain wall description in semiclassical gravity, and determine whether the monotonicity properties persist at strong coupling.

 \subsection{Incorporating line defects into the CFT$_2$ ensemble dual to pure $3d$ gravity}
    
     In \cite{Chandra:2024vhm}, a class of line defects was constructed by taking the continuum limit of a product of local operators. These defects were shown to source dust shells in $3d$ gravity. However, such defects may not exist in pure $3d$ gravity as they are constructed using very light operators in the dual $2d$ CFT. Instead we could construct line defects by exponentiating the integrals of black hole operators,
    \begin{equation} \label{DBH}
        D_\Sigma(\lambda)=\exp\left[\lambda \int_\Sigma \mathcal{O}_{BH}\right] \,,
    \end{equation}
    where $\mathcal{O}_{BH}$ is a Virasoro primary with conformal weights above the black hole threshold. Since the putative $2d$ CFT dual to pure $3d$ gravity is considered to have a spectrum of Virasoro primaries above the black hole threshold, such line defects can be constructed. It would be interesting to incorporate line defects of the form (\ref{DBH}) into the recent developments on defining an ensemble of random $2d$ CFTs dual to pure $3d$ gravity (See for example \cite{Belin:2020hea, Chandra:2022bqq, Jafferis:2025vyp} for various attempts to define a random $2d$ CFT by averaging the OPE data of local operators). To address this question, we would need to construct multi-boundary wormholes in $3d$ gravity sourced by domain walls dual to these line defects since these wormholes compute averages of correlators with the line defect. A systematic way to do this would be to incorporate line defects into TQFT frameworks for $3d$ gravity \cite{Collier:2023fwi, Hartman:2025cyj}, analogous to how local operators have been incorporated as Wilson lines. As a simple illustration of the role played by these line defects in the CFT$_2$ ensemble, consider the following 2-copy observable that computes the covariance of the matrix elements of the line defect,
    \begin{equation}
       \overline{\bra{i}D_\Sigma(\lambda_+) \ket{j}\bra{i}D_\Sigma(\lambda_-) \ket{j}}\equiv \overline{\langle \mathcal{O}_i(\infty) e^{\lambda_+ \int_\Sigma \mathcal{O}}\mathcal{O}_j(0)\rangle \langle \mathcal{O}_i(\infty) e^{\lambda_- \int_\Sigma \mathcal{O}}\mathcal{O}_j(0)\rangle} \,.
    \end{equation}
    We can compute this observable perturbatively in the couplings $\lambda_\pm$,
    \begin{align}
          \overline{\bra{i}D_\Sigma(\lambda_+) \ket{j}\bra{i}D_\Sigma(\lambda_-) \ket{j}}=\begin{cases}
             4\pi^2 \lambda_+ \lambda_- |C_0(P_i,P_j,P_{\mathcal{O}})|^2+\dots &\quad{\rm if}\  i\neq j \,,\\
              1+4\pi^2 \lambda_+ \lambda_- |C_0(P_i,P_i,P_{\mathcal{O}})|^2+\dots &\quad{\rm if}\ i =j \,.
         \end{cases}
    \end{align}
    In the first line, we have expanded the off-diagonal matrix elements of the defect as a Taylor series in the pair of couplings $\lambda_{\pm}$ and used the fact that the variance of the OPE coefficients of local operators in the ensemble is set by the smooth function $C_0$ which is proportional to the DOZZ structure constants \cite{Collier:2019weq}. The $\dots$ are higher-order terms in the couplings. In the second line, we have similarly expanded the diagonal matrix elements for which the leading contribution is $1$ (assuming unit-normalized states). In this example, we have considered a circular defect but we can readily generalise the calculation to the infinite line defect case as well. On the gravity side, each term in the perturbation series appears to be captured by a 2-boundary wormhole with a network of Wilson lines whose partition function can be computed using Virasoro TQFT. It would be interesting to resum this perturbation series and check if the result can be captured by the partition function of a single 2-boundary wormhole with a domain wall going across. Resumming the perturbation series would also help translate Virasoro ETH for local operators \cite{Collier:2019weq} into a concrete realization of ETH for these line defects. This is a signature of chaos exhibited by these line defects.

\subsection{A Spectral Form Factor for line defects}
    
It would be interesting to define an observable analogous to the spectral form factor for the line defect by complexifying the defect couplings,
\begin{equation} \label{defSFF}
    \text{SFF}_{D_\Sigma}(x,y):=\overline{\text{Tr}D_\Sigma(x+iy)\text{Tr}D_\Sigma(x-iy)}=\overline{\text{Tr}\left(e^{(x+iy)\int_\Sigma \mathcal{O}}\right)\text{Tr}\left(e^{(x-iy)\int_\Sigma \mathcal{O}}\right)} \,,
\end{equation}
where we have complexified the defect couplings $\lambda_\pm=x\pm iy$.
It would be interesting to check if this observable that captures the level-spacing statistics in the spectrum of the defect exhibits a linear ramp i.e., is proportional to $y$ at large $y$. As written in (\ref{defSFF}), the defect spectral form factor involves taking traces over the full CFT Hilbert space which is infinite-dimensional.
A standard way to make this well-defined is to project the integrated local operator to a microcanonical energy band,
\begin{equation}
    \int_\Sigma \mathcal{O} \to P_{E,\Delta E}\left(\int_\Sigma \mathcal{O}\right )P_{E,\Delta E} \,,
\end{equation}
where $P_{E,\Delta E}$ is a projector onto an energy window centered at $E$ of width $\Delta E$. Then, the windowed operator can be modeled using a finite-dimensional matrix, so its spectral statistics can be compared directly to a
finite-dimensional random-matrix ensemble, as in standard discussions of the spectral form factor and the ramp/plateau phenomenology \cite{CotlerEtAl2017,DyerGurAri2017, Saad:2018bqo}.  This is directly analogous to the ``matter spectral form factor'' introduced by Lin--Maldacena--Rozenberg--Shan \cite{Lin:2022zxd} where they project a local operator in $\mathcal{N}=2$ super-JT gravity onto the finite dimensional protected BPS subspace and study the random matrix statistics of the projected operator. This was later generalized to the non-supersymmetric case in \cite{Antonini:2025rmr}.

\ \\

\section*{Acknowledgments}

We thank Jan Boruch, Scott Collier, Gabriele di Ubaldo, Gaston Giribet, Tom Hartman, Nina Holden, Luca Iliesiu, Yuya Kusuki, Viraj Meruliya, Massimo Porrati, Martin Sasieta and Nico Valdes-Meller for helpful discussions. The work of AIA and JC is supported in part by the Leinweber Institute for Theoretical Physics at UC Berkeley. AIA is supported by the National Science Foundation through Grant No.\ DGE-2146752. JC is supported by the U.S. Department of Energy through GeoFlow DE-SC0019380.
The work of YW was supported in part by the NSF grant PHY-2210420 and by the Simons Junior Faculty Fellows program.

\appendix

\section{More details on energy transmission across the defect}

\subsection{Derivation of $\langle TT\mathcal{O}\mathcal{O} \rangle$ using Ward identities in $2d$ CFTs} \label{AppTT}

In this appendix, we derive the form of the 4-point function of two holomorphic stress tensors and two identical Virasoro primaries in a $2d$ CFT using the Ward identities. First, note that the holomorphic Ward identity takes the form
\begin{equation}
    \langle T(z)\prod_{i=1}^n \mathcal{O}_i(w_i,\overline{w}_i)\rangle =\sum_{i=1}^n \left(\frac{h_i}{(z-w_i)^2}+\frac{1}{z-w_i}\partial_{w_i}\right)\langle \prod_{i=1}^n \mathcal{O}_i(w_i,\overline{w}_i)\rangle \,.
\end{equation}
Let us first compute the three-point function using the Ward identity,
\begin{equation}
    \langle T(z) \mathcal{O}(w_1,\overline{w}_1)\mathcal{O}(w_2.\overline{w}_2)\rangle =\left(\frac{h}{(z-w_1)^2}+\frac{h}{(z-w_2)^2}+\frac{1}{z-w_1}\partial_{w_1}+\frac{1}{z-w_2}\partial_{w_2}\right) \langle  \mathcal{O}(w_1,\overline{w}_1)\mathcal{O}(w_2.\overline{w}_2)\rangle \,.
\end{equation}
Here $\mathcal{O}$ is a Virasoro primary with weights $(h,\overline{h})$. The two-point function takes the form,
\begin{equation}
    \langle\mathcal{O}(w_1,\overline{w}_1)\mathcal{O}(w_2.\overline{w}_2)\rangle =\frac{1}{(w_1-w_2)^{2h}(\overline{w}_1-\overline{w}_2)^{2\overline{h}}} \,.
\end{equation}
Plugging into the Ward identity, we get the form of the 3-point function, 
\begin{equation} \label{3pointApp}
     \langle T(z) \mathcal{O}(w_1,\overline{w}_1)\mathcal{O}(w_2.\overline{w}_2)\rangle = \frac{h(w_1-w_2)^2}{(z-w_1)^2(z-w_2)^2}\langle \mathcal{O}(w_1,\overline{w}_1)\mathcal{O}(w_2,\overline{w}_2)\rangle \,.
\end{equation}
Now, we turn to computing the 4-point function
\begin{equation}
\langle T(z_1)T(z_2)\mathcal{O}(w_1,\overline{w}_1)\mathcal{O}(w_2,\overline{w}_2) \rangle \,.
\end{equation}
Treated as a function of $z_1$, there are OPE singularities (poles) as $z_1 \to z_2$ from the $TT$-OPE and as $z_1 \to w_{1,2}$ coming from the $T\mathcal{O}$ OPEs,
\begin{equation}
    \begin{split}
        & T(z_1) T(z_2) \sim \frac{c}{2(z_1-z_2)^4}+\frac{2T(z_2)}{(z_1-z_2)^2}+\frac{1}{(z_1-z_2)}\partial_{z_2}T(z_2) \,,\\
        & T(z) \mathcal O(w,\overline{w}) \sim \frac{h}{(z-w)^2}\mathcal{O}(w,\overline{w})+\frac{1}{(z-w)}\partial_w \mathcal{O}(w,\overline{w}) \,.
    \end{split}
\end{equation}
Hence the 4-point function is given by the action of the following differential operator on the 3-point function, 
\begin{multline}
    \langle T(z_1)T(z_2)\mathcal{O}(w_1,\overline{w}_1)\mathcal{O}(w_2,\overline{w}_2) \rangle =\frac{c}{2z_{12}^4}\langle \mathcal{O}(w_1,\overline{w}_1)\mathcal{O}(w_2,\overline{w}_2) \rangle\\+\left(\frac{2}{z_{12}^2}+\frac{1}{z_{12}}\partial_{z_2}+\sum_{i=1}^2\left(\frac{h}{(z_1-w_i)^2}+\frac{1}{z_1-w_i}\partial_{w_i}\right)\right)\langle T(z_2)\mathcal{O}(w_1,\overline{w}_1)\mathcal{O}(w_2,\overline{w}_2) \rangle\,.
\end{multline}
Using the form of the 3-point function computed in equation (\ref{3pointApp}), we arrive at the final expression for the 4-point function,
\begin{multline}
    \frac{ \langle T(z_1)T(z_2)\mathcal{O}(w_1,\overline{w}_1)\mathcal{O}(w_2,\overline{w}_2) \rangle}{\langle \mathcal{O}(w_1,\overline{w}_1)\mathcal{O}(w_2,\overline{w}_2) \rangle }=\frac{c}{2(z_1-z_2)^4}+\frac{2h (w_1-w_2)^2}{(z_1-z_2)^2(z_1-w_1)(z_1-w_2)(z_2-w_1)(z_2-w_2)}\\+\frac{h^2(w_1-w_2)^4}{(z_1-w_1)^2(z_1-w_2)^2(z_2-w_1)^2(z_2-w_2)^2} \,.
\end{multline}
The expression has the expected analytic structure and symmetries as dictated by the OPEs. It can also be written in a more compact form as
\begin{equation} \label{TTOO}
     \frac{ \langle T(z_1)T(z_2)\mathcal{O}(w_1,\overline{w}_1)\mathcal{O}(w_2,\overline{w}_2) \rangle}{\langle \mathcal{O}(w_1,\overline{w}_1)\mathcal{O}(w_2,\overline{w}_2) \rangle }=\frac{c}{2 z_{12}^4}+\frac{2h w_{12}^2}{z_{12}^2\prod_{i=1}^2\prod_{j=1}^2(z_i-w_j)}+\frac{h^2 w_{12}^4}{\prod_{i=1}^2\prod_{j=1}^2(z_i-w_j)^2} \,.
\end{equation}

\subsection{Computing reflection coefficient of defect with a cusp} \label{cuspTTOO}

We integrate the connected part of (\ref{TTOO}) to derive the contribution to the reflection coefficient from the cross-correlations of the defect. The kinematics relevant to computing the cross-correlation between the two halves of the defect meeting at the cusp are
\begin{equation}
    z_1=iy, \quad z_2=-iy, \quad w_1=\ell_1 e^{i\theta_1}, \quad w_2=\ell_2 e^{i\theta_2}, \qquad \ell_{1,2}\in (0,\infty) \,.
\end{equation}
Here, $\ell_{1,2}$ are distances measured along the two halves of the defect making angles $\theta_{1,2}$ with the positive real axis. For notational convenience, define 
\begin{equation}
    \alpha_{1,2}\equiv e^{i\theta_{1,2}} \,.
\end{equation}
The products appearing in the denominators of (\ref{TTOO}) are given by
\begin{equation}
    \prod_{i=1}^2\prod_{j=1}^2(z_i-w_j)=(y^2+\alpha_1^2 \ell_1^2)(y^2+\alpha_2^2\ell_2^2) \,.
\end{equation}
So, the two terms in (\ref{TTOO}) evaluate to
\begin{align}
        \frac{2h w_{12}^2}{z_{12}^2\prod_{i=1}^2\prod_{j=1}^2(z_i-w_j)}\expval{\cO(w_1,\bar{w}_1)\cO(w_2,\bar{w}_2)} &= -\frac{1}{4y^2}\left(\frac{\alpha_1 \ell_1-\alpha_2 \ell_2}{\overline{\alpha}_1\ell_1-\overline{\alpha}_2\ell_2}\right)\frac{1}{(\alpha_1^2 \ell_1^2+y^2)(\alpha_2^2 \ell_2^2+y^2)} \,,\\
        \frac{h^2 w_{12}^4}{\prod_{i=1}^2\prod_{j=1}^2(z_i-w_j)^2}\expval{\cO(w_1,\bar{w}_1)\cO(w_2,\bar{w}_2)} &= \frac{1}{4}\frac{(\alpha_1 \ell_1-\alpha_2 \ell_2)^3}{(\overline{\alpha}_1\ell_1-\overline{\alpha}_2 \ell_2)}\frac{1}{(\alpha_1^2 \ell_1^2+y^2)^2(\alpha_2^2 \ell_2^2+y^2)^2} \,,\
\end{align}
where we have used $h=\frac{1}{2}$ since the operator is marginal on the line. $\overline{\alpha}$ is the complex conjugate of $\alpha$. So, the two integrals that need to be evaluated are
\begin{equation}
    \begin{split}
        & I_1=-\frac{1}{4y^2}\int_0^{\infty}d\ell_2\int_0^{\infty}d\ell_1 \left(\frac{\alpha_1 \ell_1-\alpha_2 \ell_2}{\overline{\alpha}_1 \ell_1-\overline{\alpha}_2 \ell_2}\right)\frac{1}{(\alpha_1^2 \ell_1^2+y^2)(\alpha_2^2 \ell_2^2+y^2)} \,,\\
        & I_2=\frac{1}{4}\int_0^{\infty}d\ell_2\int_0^{\infty}d\ell_1 \frac{(\alpha_1 \ell_1-\alpha_2 \ell_2)^3}{(\overline{\alpha}_1\ell_1-\overline{\alpha}_2 \ell_2)}\frac{1}{(\alpha_1^2 \ell_1^2+y^2)^2(\alpha_2^2 \ell_2^2+y^2)^2} \,.
    \end{split}
\end{equation}
To evaluate these integrals, we find it convenient to shift to polar coordinates in the first quadrant of the $\ell_1-\ell_2$ plane,
\begin{equation}
    \ell_1= y r\cos(\phi)\,, \quad \ell_2=y r\sin(\phi)\,, \qquad r\in(0,\infty)\,, \quad \phi\in (0,\frac{\pi}{2}) \,.
\end{equation}
Now, the radial integrals can be evaluated in closed form\footnote{Note that the $r$ integrals appearing in the evaluation of $I_1$ and $I_2$ in (\ref{I1}) and (\ref{I2}) respectively do not converge when $\theta_1$ or $\theta_2$ is $\frac{\pi}{2}$ which means that these expressions are valid only for $\alpha_{1,2}\neq i$. At these values of the angles, the operators are integrated over the positive imaginary axis so they hit the stress tensor insertion.},
\begin{equation} \label{I1}
\begin{split}
    I_1=&-\frac{1}{4y^4 }\int_0^{\frac{\pi}{2}}d\phi \left(\frac{\alpha_1-\alpha_2\tan(\phi)}{\overline{\alpha}_1-\overline{\alpha}_2\tan(\phi)}\right )\int_0^{\infty}dr \frac{r}{(1+\alpha_1^2 r^2\cos^2(\phi))(1+\alpha_2^2 r^2 \sin^2(\phi))} \,,\\
    =& -\frac{1}{8y^4 }\int_0^{\frac{\pi}{2}}d\phi \frac{1}{(\alpha_2\sin(\phi)+\alpha_1 \cos(\phi))(\overline{\alpha}_1 \cos(\phi)-\overline{\alpha}_2 \sin(\phi))}\log\left(\frac{\alpha_1^2 \cos^2(\phi)}{\alpha_2^2 \sin^2(\phi)}\right) \,,
    \end{split} 
\end{equation}
and similarly $I_2$ evaluates to
\begin{equation} \label{I2}
    \begin{split}
        I_2= & \frac{1}{4y^4 }\int_0^{\frac{\pi}{2}}d\phi \left(\frac{(\alpha_1 \cos(\phi)-\alpha_2\sin(\phi))^3}{\overline{\alpha}_1\cos(\phi)-\overline{\alpha}_2\sin(\phi)}\right)\int_0^{\infty} dr \frac{r^3}{(1+\alpha_1^2 r^2 \cos^2(\phi))^2(1+\alpha_2^2 r^2\sin^2(\phi))^2} \,,\\
        =& \frac{1}{8y^4}\int_0^{\frac{\pi}{2}}d\phi \left(\frac{(\alpha_1 \cos(\phi)-\alpha_2 \sin(\phi))^3}{\overline{\alpha}_1\cos(\phi)-\overline{\alpha}_2\sin(\phi)}\right)\bigg[\frac{\alpha_1^2 \cos^2(\phi)+\alpha_2^2\sin^2(\phi)}{(\alpha_1^2 \cos^2(\phi)-\alpha_2^2\sin^2(\phi))^3}\log\left(\frac{\alpha_1^2\cos^2(\phi)}{\alpha_2^2\sin^2(\phi)}\right)\\& \qquad \qquad \qquad \qquad \qquad \qquad \qquad \qquad \qquad \qquad \qquad \qquad -\frac{2}{(\alpha_1^2 \cos^2(\phi)-\alpha_2^2\sin^2(\phi))^2}\bigg] \,.
    \end{split}
\end{equation}
These integrals take a simpler form when we work with the following variable $t=\tan(\phi)$,
\begin{equation}
    \begin{split}
        I_1=& \frac{1}{8y^4 }\int_0^{\infty}dt \frac{\log(t^2/\alpha^2)}{(\alpha+t)(\overline{\alpha}-t)} \,,\\
        I_2=& -\frac{1}{8y^4}\int_0^{\infty}dt\frac{(\alpha-t)^3}{(\overline{\alpha}-t)}\left[\frac{\alpha^2+t^2}{(\alpha^2-t^2)^3}\log(t^2/\alpha^2)+\frac{2}{(\alpha^2-t^2)^2}\right] \,.\\
    \end{split}
\end{equation}
Here $\alpha \equiv \frac{\alpha_1}{\alpha_2}=e^{i(\theta_1-\theta_2)}$.
Adding the two, we arrive at the following expression,
\begin{equation}
    I(\theta_1,\theta_2)\equiv -4y^4(I_1+I_2)=\int_0^{\infty}dt \left(\frac{\alpha t \log(t^2/\alpha^2)+t^2-\alpha^2}{(t+\alpha)^3(t-\overline{\alpha})}\right) \,,
\end{equation}
which we have used in (\ref{conn}) for the integrated 4-point function in the main text. Without loss of generality, we assume $0\leq \theta_2 <\frac{\pi}{2}$, then the function has the following piecewise definition depending on whether $\theta_1$ is greater or smaller than $\frac{\pi}{2}$ (since the function is ill-defined at $\theta_1=\frac{\pi}{2}$), 
\begin{equation}
    I(\theta_1,\theta_2)=\begin{cases}
        \int_0^{\infty}dt \left(\frac{2e^{i(\theta_1-\theta_2)} t \log(t)+2it(\theta_2-\theta_1)e^{i(\theta_1-\theta_2)}+t^2-e^{2i(\theta_1-\theta_2)}}{(t+e^{i(\theta_1-\theta_2)})^3(t-e^{i(\theta_2-\theta_1)})}\right) &\qquad{\rm if}\ \theta_2<\theta_1<\frac{\pi}{2} \,,\\
         \int_0^{\infty}dt \left(\frac{2e^{i(\theta_1-\theta_2)} t \log(t)+2it(\theta_2-\theta_1+\pi)e^{i(\theta_1-\theta_2)}+t^2-e^{2i(\theta_1-\theta_2)}}{(t+e^{i(\theta_1-\theta_2)})^3(t-e^{i(\theta_2-\theta_1)})}\right) &\qquad{\rm if}\ \frac{\pi}{2}<\theta_1\leq \pi\,.
    \end{cases}
\end{equation}

\section{Useful limits of the DOZZ formula} \label{AppDOZZ}

The DOZZ formula for the 3-point functions of vertex operators in Liouville CFT is given by
\begin{equation}
    C(\alpha_1,\alpha_2,\alpha_3)=\left(\pi \mu \gamma(b^2)b^{2-2b^2}\right)^{(Q-\sum_i\alpha_i)/b} \frac{\Upsilon_b'(0)\prod_{i=1}^3 \Upsilon_b(2\alpha_i)}{\Upsilon_b(\sum_{i=1}^3\alpha_i-Q)\prod_{j=1}^3\Upsilon_b(\sum_{i=1}^3 \alpha_i -2\alpha_j)} \,,
\end{equation}
where $\gamma(x)=\frac{\Gamma(x)}{\Gamma(1-x)}$ and $\Upsilon_b(x)$ is the Upsilon function defined by the following integral representation,
\begin{equation} \label{upintegral}
   \log \Upsilon_b(x)=\int_0^{\infty}\frac{dt}{t}\left(\left(\frac{Q}{2}-x\right)^2 e^{-t}-\frac{\sinh^2\left(\frac{t}{2}(\frac{Q}{2}-x)\right)}{\sinh(\frac{bt}{2})\sinh(\frac{t}{2b})}\right)\,, \qquad 0<\text{Re}(x)<Q \,.
\end{equation}
We are interested in the case where $\alpha_3=\frac{b}{2}$. Expressing $\alpha=\frac{Q}{2}+iP$, the DOZZ formula for this case can be written as
\begin{equation} \label{dozb}
    C\left(\frac{Q}{2}+iP_1,\frac{Q}{2}+iP_2,\frac{b}{2}\right)=\left(\pi \mu \gamma(b^2)b^{2-2b^2}\right)^{-\frac{1}{2}-\frac{i}{b}(P_1+P_2)}\frac{(\Upsilon'_b(0))^2\Upsilon_b(-2iP_1)\Upsilon_b(-2iP_2)}{\Upsilon_b(\frac{b}{2}\pm iP_1 \pm iP_2)} \,.
\end{equation}
We have used the $\pm$ notation in the denominator to denote the product of Upsilon functions with the 4 combinations of momenta. To arrive at this expression, we used the fact that $\Upsilon_b(x)=\Upsilon_b(Q-x)$ which is evident from (\ref{upintegral}). We also used the result $\Upsilon_b'(0)=\Upsilon_b(b)$ which can be derived using the shift relation obeyed by the Upsilon function,
\begin{equation}
    \Upsilon_b(x+b)=\gamma(bx)b^{1-2bx}\Upsilon_b(x) \,,
\end{equation}
as follows, 
\begin{equation}
\begin{split}
    \Upsilon_b(b)= \lim_{x\to 0} \Upsilon_b(x+b)=\lim_{x\to 0} \gamma(bx)b^{1-2bx} \Upsilon_b(x)=\Upsilon_b'(0) \,.
    \end{split}
\end{equation}
Now, we evaluate (\ref{dozb}) in the $b\to 0$ limit in two different scalings of the momenta: (i) $P_i=b k_i$ with $k_i$ held fixed corresponding to near-threshold operators; (ii) $P_i=\frac{k_i}{b}$ with $k_i$ held fixed corresponding to very heavy operators. 

\subsection{Near-threshold (Schwarzian) limit} \label{AppDOZZlight}

We evaluate (\ref{dozb}) in the $b\to 0$ limit with $k_i=\frac{P_i}{b}$ held fixed. To this end, we need to evaluate $\Upsilon_b(b\sigma)$ in the $b\to 0$ limit with $\sigma$ held fixed. We use the following $b\to 0$ asymptotics of the Upsilon function,
\begin{equation}
    \Upsilon_b(b\sigma)=\Upsilon_b'(0)\frac{b^{1-\sigma}}{\Gamma(\sigma)}+\dots
\end{equation}
Using this, we immediately find that
\begin{equation} \label{schwarz}
\begin{split}
    \lim_{b\to 0}C\left(\frac{Q}{2}+ibk_1,\frac{Q}{2}+ibk_2,\frac{b}{2}\right)=& (\pi \mu)^{-\frac{1}{2}-i(k_1+k_2)}b^{2i(k_1+k_2)}\frac{\Gamma(\frac{1}{2}\pm ik_1 \pm ik_2)}{\Gamma(-2ik_1)\Gamma(-2ik_2)} \,,\\
    =&  \frac{\pi^2 (\pi \mu)^{-\frac{1}{2}-i(k_1+k_2)}b^{2i(k_1+k_2)}}{\cosh(\pi(k_1+k_2))\cosh(\pi(k_1-k_2))\Gamma(-2ik_1)\Gamma(-2ik_2)} \,.
    \end{split}
\end{equation}
To go the second line, we used a reflection identity,
\begin{equation}
    |\Gamma(\frac{1}{2}+ix)|^2=\frac{\pi}{\cosh(\pi x)} \,.
\end{equation}
The expression (\ref{schwarz}) is not real due to the phases in the numerator. The absolute value of (\ref{schwarz}) takes a simpler form,
\begin{equation}
    \left|\lim_{b\to 0}C\left(\frac{Q}{2}+ibk_1,\frac{Q}{2}+ibk_2,\frac{b}{2}\right)\right |^2=\frac{4\pi}{\mu}\frac{k_1k_2 \sinh(2\pi k_1)\sinh(2\pi k_2)}{\cosh^2(\pi(k_1+k_2))\cosh^2(\pi(k_1-k_2))} \,,
\end{equation}
where we used another reflection identity,
\begin{equation}
    |\Gamma(ix)|^2=\frac{\pi}{x\sinh(\pi x)} \,.
\end{equation}

\subsection{Very heavy limit} \label{AppDOZZheavy}

Now, we evaluate (\ref{dozb}) in the $b\to 0$ limit with $k_i=bP_i$ held fixed. To this end, we need to evaluate $\Upsilon_b(\frac{\sigma}{b})$ in the $b\to 0$ limit with $\sigma$ held fixed. To do this, we use the relation between the Upsilon-function and the double Gamma function $\Gamma_b(x)$,
\begin{equation} \label{upbarn}
    \Upsilon_b(x)=\frac{1}{\Gamma_b(x)\Gamma_b(Q-x)} \,,
\end{equation} 
and use the following asymptotics of $\Gamma_b(\frac{\sigma}{b})$ as $b\to 0$ (See for example \cite{Collier:2018exn}),
\begin{multline} \label{doublasym}
    \log \Gamma_b(\frac{\sigma}{b})=\frac{1}{2}\left(\frac{1}{b^2}\left(\frac{1}{2}-\sigma\right)^2+\left(\frac{1}{2}-\sigma\right)\right)\log(b)+\frac{1}{b^2}\left(\frac{2\sigma-1}{4}\log(2\pi)-\int_\frac{1}{2}^\sigma dt \log (\Gamma(t))\right)\\+\frac{1}{2}\log\Gamma(\sigma)-\frac{1}{4}\log(2\pi)+\dots
\end{multline}
where the $\dots$ are terms which vanish in the $b\to 0$ limit.
Using (\ref{upbarn}) and (\ref{doublasym}), we immediately find the asymptotics of the Upsilon function,
\begin{equation}
    \log \Upsilon_b(\frac{\sigma}{b})=-\frac{\log(b)}{b^2}\left(\frac{1}{2}-\sigma\right)^2+\frac{1}{b^2}\int_\frac{1}{2}^\sigma dt \log(\gamma(t))+(\sigma-\frac{1}{2})\log(b)-\frac{1}{2}\log(\gamma(\sigma))+\dots
\end{equation}
To evaluate the denominators in (\ref{dozb}), we also need the following asymptotics
\begin{equation}
     \log \Upsilon_b(\frac{b}{2}+\frac{\sigma}{b})=-\frac{\log(b)}{b^2}\left(\frac{1}{2}-\sigma\right)^2+\frac{1}{b^2}\int_\frac{1}{2}^\sigma dt \log(\gamma(t))+\dots
\end{equation}
Using these results, we have the asymptotics of the DOZZ formula,
\begin{multline} \label{cheav}
    \log C\left(\frac{Q}{2}+\frac{ik_1}{b},\frac{Q}{2}+\frac{ik_2}{b},\frac{b}{2}\right) =\frac{1}{b^2}\bigg(2F(0)+F(-2ik_1)+F(-2ik_2)-F(\pm ik_1 \pm ik_2)-i(k_1+k_2)\log(\pi \mu)\bigg)\\-\frac{2\log b}{b^2}(i(k_1+k_2)+b^2)+\frac{1}{2}\log\left(\frac{4\pi}{\mu}\right)-\frac{1}{2}\log\left(\gamma(-2ik_1)\gamma(-2ik_2)\right)+2i(k_1+k_2)\gamma_E+\dots
\end{multline}
where $\gamma_E$ is the Euler-Mascheroni constant and we have defined the following function, $ F(\sigma)\equiv \int_{\frac{1}{2}}^{\sigma}dt \log(\gamma(t))$. $F(\pm ik_1 \pm ik_2)$ is shorthand for summing over $F$ evaluated at the 4 choices of $\pm$ signs. To arrive at (\ref{cheav}), we also used
\begin{equation}
    \log \Upsilon_b'(0)=-\frac{\log b}{4b^2}+\frac{F(0)}{b^2}-\frac{1}{2}\log(b)+\frac{1}{2}\log(2\pi)+\dots
\end{equation}
If we focus on just the absolute value $|C|$, the expression simplifies to
\begin{multline}
    \log \bigg | C\left(\frac{Q}{2}+\frac{ik_1}{b},\frac{Q}{2}+\frac{ik_2}{b},\frac{b}{2}\right) \bigg |=\frac{1}{b^2}\text{Re}\left[2F(0)+F(-2ik_1)+F(-2ik_2)-F(\pm ik_1 \pm ik_2)\right]\\-2\log(b)+\frac{1}{2}\log\left(\frac{16\pi k_1 k_2}{\mu}\right)+\dots
\end{multline}
$F(0)$ and the sum $F(\pm ik_1\pm ik_2)$ are real so the real part only acts on $F(-2ik_1)$ and $F(-2ik_2)$. To evaluate the real part, it is useful to note that for $y\in \mathbb{R}$, $F(iy)+F(-iy)\in \mathbb{R}$. To evaluate this, we need to choose branch of the $\log$-function such that $F(iy)+F(-iy)$ is continuous in $y$.

\section{Drinfeld-Sokolov reduction of the non-chiral SL(2,$\mathbb{R}$) WZW action} \label{AppDS}

In this appendix, we will first review the derivation of the Liouville action by Drinfeld-Sokolov reduction \cite{Drinfeld:1984qv, Balog:1990mu} of the non-chiral SL(2,$\mathbb{R}$) WZW action written as a sigma model for the group valued field. See also \cite{Donnay:2016iyk} for a nice modern review in the context of asymptotic dynamics of $3d$ gravity. Then, we will introduce an interface separating two copies of the WZW model and write down a condition relating the WZW fields across the wall that reduces to the Liouville jump condition after Drinfeld-Sokolov reduction. This is complementary to the similar calculation done in Section~\ref{WZW} where we interpreted the defect in the $\mathbb{H}_3^+$ model.

First, let us establish the following convention for the generators of the sl(2) algebra,
\begin{equation}
    L_0=\frac{1}{2}\begin{bmatrix}
        1 & 0\\
        0 & -1
    \end{bmatrix}\,, \qquad L_+=\begin{bmatrix}
        0 & 1\\
        0 & 0
    \end{bmatrix}\,, \qquad L_-=\begin{bmatrix}
        0 & 0\\
        1 & 0
    \end{bmatrix}\,.
\end{equation}
These matrices satisfy the commutation relations for sl(2) algebra,
\begin{equation}
    [L_+,L_- ]=2L_0 \,,\qquad [L_+,L_0]=-L_+ \,,\qquad [L_-,L_0]=L_- \,.
\end{equation}
In Euclidean signature, the action of the SL(2,$\mathbb{R}$) WZW model is given by (See for example \cite{Maldacena:2000hw}),
\begin{equation}
    I_{\text{WZW}}[g]=\frac{k}{4\pi}\int_{\Gamma}d^2 z\text{Tr}\left(g^{-1}\partial g g^{-1}\overline{\partial}g\right)-\frac{i k}{12\pi}\int_{B}d^3 y \epsilon^{\alpha\beta\gamma}\text{Tr}\left(g^{-1}\partial_\alpha g g^{-1}\partial_\beta g g^{-1}\partial_\gamma g \right) \,,
\end{equation}
where the WZW field $g(z,\overline{z})$ is a map from $\Gamma$ into the SL$(2,\mathbb{R})$ group manifold. To reduce this action to Liouville, we perform a Gauss decomposition of $g$,
\begin{equation}
    g(z,\overline{z})=e^{X(z,\overline{z})L_+}e^{(\Phi(z,\overline{z})-\log(4)) L_0}e^{Y(z,\overline{z})L_-} \,,
\end{equation}
where $\Phi(z,\overline{z})$ will turn out to be the Liouville field and the shift of $\log(4)$ is just to match with the conventions used in Section~\ref{Secexp}. $X$ and $Y$ are auxiliary fields which we will eliminate below using the Drinfeld-Sokolov constraints.
The above Gauss decomposition expresses the WZW field as a product of three SL(2,$\mathbb{R}$) matrices, the first being upper triangular, the second diagonal and third lower triangular, because of the conventions we chose for the sl(2) generators. Plugging into the WZW action, we get
\begin{equation}
    I_{\text{WZW}}[\Phi, X, Y]=\frac{k}{2\pi}\int d^2 z \left (\frac{1}{2}\partial\Phi \overline{\partial}\Phi+8e^{-\Phi}\partial X\overline{\partial}Y\right ) \,.
\end{equation}
The derivatives of the auxiliary fields $X$ and $Y$ are related to the Liouville field upon imposing the Drinfeld-Sokolov constraints on the WZW currents which we now describe. The WZW equations of motion,
\begin{equation}
    \partial(\overline{\partial}g g^{-1})=0 \,, \qquad \overline{\partial}(g^{-1}\partial g)=0 \,,
\end{equation}
imply the conservation of the holomorphic and anti-holomorphic currents defined below,
\begin{equation}
    J(z)=k g^{-1}\partial g \,, \qquad \overline{J}(\overline{z})=k \partial g g^{-1} \,.
\end{equation}
The highest weight components when expanded using the sl(2) generators and using the Gauss decomposition give
\begin{equation} \label{WZWcurrents}
    J^+=4k e^{-\Phi}\partial X \,, \qquad \overline{J}^-=4k e^{-\Phi}\overline{\partial}Y \,.
\end{equation}
The Drinfeld-Sokolov constraints that give Liouville impose that these highest weight components are constants set by the level $k$,
\begin{equation}
    J^+=k \,, \qquad \overline{J}^-=k \,.
\end{equation}
Using the forms of the currents in (\ref{WZWcurrents}), we get
\begin{equation}
    \partial X=\frac{1}{4}e^{\Phi} \,, \qquad \overline{\partial} Y=\frac{1}{4}e^{\Phi} \,.
\end{equation}
Upon substituting these constraints into the action after adding appropriate boundary terms to ensure $I[\Phi,X,Y]$ has a good variational principle gives the Liouville action,
\begin{equation}
   I[\Phi]=\frac{c}{6}S_L= \frac{k}{4\pi}\int_{\Gamma} d^2z\left( \partial \Phi \bar{\partial}\Phi +e^{\Phi}\right) \,.
\end{equation}
The level $k$ is related to the Liouville coupling by $k=\frac{1}{b^2}=\frac{c}{6}$. On shell, when the Liouville equation is satisfied, we can express the auxiliary fields in terms of the derivative of the Liouville field as
\begin{equation}
    \partial\overline{\partial}\Phi=\frac{e^\Phi}{2}\implies X=\frac{1}{2}\overline{\partial}\Phi\,, \quad Y=\frac{1}{2}\partial \Phi \,.
\end{equation}

Now, we interpret the Liouville defect in terms of a domain wall separating two copies of the WZW model.
Concretely, we need to express the jump condition (\ref{jumpcond}) in terms of the WZW field. To this end, observe that
\begin{equation}
    \text{Tr}(\Omega g)=-2i e^{-\frac{1}{2}\Phi}X+ 2i e^{-\frac{1}{2}\Phi}Y = e^{-\frac{1}{2}\Phi}\partial_y \Phi \,\qquad{\rm when}\ \Omega=\begin{bmatrix}
        0 & i\\
        -i & 0
    \end{bmatrix}\,,
\end{equation}
where we have written $z=x+iy$ and $y$ is the normal coordinate to the domain wall. The above expression suggests that we can interpret the matrix $\Omega$ as a projector onto the normal direction of the domain wall. Note that $\Omega$ behaves as a parity reversal matrix since it satisfies,
\begin{equation}
    \Omega^2=1 \,,\qquad \Omega^{-1} L_{\pm}\Omega=-L_{\mp} \,,\qquad \Omega^{-1}L_0\Omega=-L_0 \,.
\end{equation}
Observe that conjugation by $\Omega$ leaves the sl(2) algebra invariant. We see that the Liouville jump condition can be expressed in terms of the WZW fields on either side of the domain wall denoted by $g_\pm$ as 
\begin{equation}
    \text{Tr}(\Omega g_+)-\text{Tr}(\Omega g_-)=2\mu \,.
\end{equation}
Notice that the above jump condition explicitly breaks the $SL(2,\mathbb{R})\times SL(2,\mathbb{R})$ symmetry of non-chiral WZW model to a diagonal SL(2,$\mathbb{R}$) subgroup twisted by $\Omega$. This means that the original global symmetry encoded by the invariance of the action under $g\to h^{-1}g\overline{h}$ is broken to invariance under $g\to h^{-1}g\tau(h)$ where $\tau(h)=\Omega^{-1}h\Omega$.

Now, we derive the above condition by adding a term to the WZW action which is local on the wall. A natural choice is to introduce a Lagrange multiplier that enforces the above condition,
\begin{equation}
    I=I_{\text{WZW}}[g_+]+I_{\text{WZW}}[g_-]+\int dx \Lambda(x)\left(\text{Tr}(\Omega g_+)-\text{Tr}(\Omega g_-)-2\mu\right) \,.
\end{equation}
Here $x$ is a coordinate along the wall and $\Lambda(x)$ is a Lagrange multiplier enforcing the jump condition. Varying wrt $g_{\pm}$ gives the following equations,
\begin{equation}
    \partial_y g_{\pm}\pm\Lambda \Omega g_{\pm}=0 \,.
\end{equation}
Using the Gauss decomposition of $g_{\pm}$ and imposing the DS constraints, these equations reduce to the following ODE that relates $\Lambda$ to the Liouville field,
\begin{equation}
    \partial_y^2\Phi =\frac{1}{2}(\partial_y \Phi)^2+2\Lambda^2 \,.
\end{equation}
This is a Riccati-type ODE $v'=\frac{1}{2}v^2+f$ with $v=\partial_y\Phi$ and $f=2\Lambda^2$. This is also closely related to the Schwarzian equation via $\partial_y \Phi =\frac{w''}{w'}$ giving $\{w,y\}=2\Lambda^2$.

\bibliographystyle{ourbst}
\bibliography{ref.bib}

\end{document}